\documentclass[a4paper,titlepage,12pt]{book}

\usepackage{cite}                      
\usepackage{amsmath,amssymb,latexsym}  
\usepackage[all]{xy}                   
\usepackage[hyperref]{collect}         

\newif\ifPDF\ifx\pdfoutput\undefined\PDFfalse
            \else\ifnum\pdfoutput > 0\PDFtrue
                 \else\PDFfalse
                 \fi
            \fi

\makeatletter
\def\NAT@parse{\typeout{This is a fake Natbib command to fool Hyperref.}}
\makeatother

\usepackage{fancyhdr}                  

\DeclareMathOperator{\tr}{tr}
\DeclareMathOperator{\str}{Str}
\DeclareMathOperator{\coker}{coker}
\DeclareMathOperator{\Imag}{Im}
\DeclareMathOperator{\Real}{Re}
\DeclareMathOperator{\Ext}{Ext}

\newcommand{\dd}{\textrm{d}}
\newcommand{\ee}{\textrm{e}}
\newcommand{\ii}{\textrm{i}}

\newcommand{\id}{\mathbf{1}}
\newcommand{\vol}{\textrm{vol}}
\newcommand{\spinrep}[1]{\boldsymbol{#1}}
\newcommand{\rep}[1]{\boldsymbol{#1}}

\newcommand{\rel}[1]{\underline{#1}}
\newcommand{\ins}[1]{\mathbf{i}_{#1}}
\newcommand{\norm}[1]{\lVert #1 \rVert}
\newcommand{\abs}[1]{\lvert #1 \rvert}
\newcommand{\ket}[1]{\lvert #1 \rangle}
\newcommand{\vev}[1]{\langle #1 \rangle}

\newcommand{\tbi}[3]{#1^{#2}_{\hphantom{#2}#3}}
\newcommand{\bti}[3]{#1_{#2}^{\hphantom{#2}#3}}

\newcommand{\trans}[1]{{#1}^\textbf{T}}

\newcommand{\tbundle}[1]{\textrm{T}#1}
\newcommand{\tbundlep}[2]{\textrm{T}_{#1}{#2}}
\newcommand{\nbundle}[1]{\textrm{N}#1}
\newcommand{\sheaf}[1]{\mathcal{#1}}
\newcommand{\inv}[1]{{#1}^\mathbf{-1}}
\newcommand{\Riem}[4]{R_{#1\hphantom{#2}#3#4}^{\hphantom{#1}#2}}
\newcommand{\Ricci}[2]{R_{#1#2}}
\newcommand{\Rscalar}{R}
\newcommand{\leftup}[2]{\vphantom{#2}^{#1}{#2}}     
\newcommand{\leftupn}[2]{\vphantom{#2}^{#1}\!\!{#2}}
\newcommand{\diag}[1]{\text{diag}\left({#1}\right)}
\newcommand{\ac}[2]{\{{#1},{#2}\}}
\newcommand{\com}[2]{[{#1},{#2}]}
\newcommand{\super}[1]{\boldsymbol{#1}}
\newcommand{\pair}[1]{\vec{#1}}
\newcommand{\dist}[2]{\textrm{dist}(#1,#2)}
\newcommand{\hc}{\text{h.c.}}
\newcommand{\hodge}[1]{\makebox[3.2ex]{$#1$}}

\newcommand{\Gflux}{\tilde G}           
\newcommand{\bg}[1]{#1_\text{bg}}       
\newcommand{\soft}[1]{\leftup{s}{#1}}   

\newcommand{\WV}{\mathcal{W}}           
\newcommand{\dbem}{\mathrm{X}}          
\newcommand{\dbps}{\mathcal{X}}         
\newcommand{\dbpn}{\mathcal{X}}         
\newcommand{\dbt}{\phi}                 
\newcommand{\dbs}{\zeta}                
\newcommand{\fcharge}{Q_{\tilde f}}     

\newcommand{\dbbf}{\mathcal{B}}         

\newcommand{\covdb}{\text{D}}           
\newcommand{\covdbz}{\mathcal{D}}       
\newcommand{\spinder}{\mathcal{D}}      

\newcommand{\singspin}{\xi}             

\newcommand{\cs}{\mathcal{J}}           

\begin{document}


\begin{titlepage}
\begin{center}
\hfill hep-th/0507042\\
\hfill ZMP-HH/05-13\\
\vskip 1cm
{\Large\bf The effective action of D-branes \\[1.5ex]
           in Calabi-Yau orientifold 
           compactifications\footnote{Based on the author's Ph.D.~thesis,
           defended on June 21,2005.}}

\vskip 1cm

{\large\bf Hans Jockers\footnote{From September 1, 2005:
                                 CERN, CH-1211 Genève 23, Switzerland. \\
                                 E-mail: {\tt hans.jockers@cern.ch}}} \\ 

\vskip 0.5cm

{II. Institut f¨ur Theoretische Physik\\
 Universit¨at Hamburg\\
 Luruper Chaussee 149\\
 D-22761 Hamburg, Germany}\\

\vskip 0.5cm

{\tt hans.jockers@desy.de}

\vskip 1cm

{\bf\Large Abstract}
\end{center}
\bigskip
\noindent
In this review article${}^1$ we study type~IIB superstring compactifications in the presence of space-time filling D-branes while preserving $\mathcal{N}=1$ supersymmetry in the effective four-dimensional theory. This amount of unbroken supersymmetry and the requirement to fulfill the consistency conditions imposed by the space-time filling D-branes lead to Calabi-Yau orientifold compactifications. For a generic Calabi-Yau orientifold theory with space-time filling D3- or D7-branes we derive the low-energy spectrum. In a second step we compute the effective $\mathcal{N}=1$ supergravity action which describes in the low-energy regime the massless open and closed string modes of the underlying type~IIB Calabi-Yau orientifold string theory. These $\mathcal{N}=1$ supergravity theories are analyzed and in particular spontaneous supersymmetry breaking induced by non-trivial background fluxes is studied. For D3-brane scenarios we compute soft-supersymmetry breaking terms resulting from bulk background fluxes whereas for D7-brane systems we investigate the structure of D- and F-terms originating from worldvolume D7-brane background fluxes. Finally we relate the geometric structure of D7-brane Calabi-Yau orientifold compactifications to $\mathcal{N}=1$ special geometry.
\end{titlepage}


\pagenumbering{roman}


\thispagestyle{plain}
\vspace*{\fill}
\noindent
{\bf\Large Acknowledgments}\par
\bigskip
\noindent
This review article is based on my Ph.D.~thesis. First of all I am deeply grateful to my supervisor Jan Louis not only for teaching me countless many things in physics but also for his constant support, guidance and encouragement during the last three years. I am also thankful to Mariana Gra\~na and Thomas Grimm for many fruitful discussions and for the enjoyable collaboration during my Ph.D.~period. I want to say thank you to Iman Benmachiche, David Cerde\~no, Paul Heslop, Olaf Hohm, Anke Knauf, Boris K\"ors, Paolo Merlatti, Andrei Micu, Henning Samtleben, Michele Salvadore, Sakura Sch\"afer-Nameki, Michelle Trapletti, Silvia Vaula, Martin Weidner and Mattias Wohlfarth for many explanations and for creating a nice atmosphere here in Hamburg. I am also thankful to Jacques Distler, Michael Haack, Wolfgang Lerche, Peter Mayr, Stefan Pokorski, Uwe Semmelmann, Stephan Stieberger and Angel Uranga for interesting and stimulating discussions. 

This work is supported by the DFG -- The German Science Foundation, the DAAD -- the German Academic Exchange Service, and the European RTN Program MRTN-CT-2004-503369.
\bigskip
\newpage
\thispagestyle{plain}
\cleardoublepage

\tableofcontents
\cleardoublepage


\pagenumbering{arabic}


\chapter{Introduction} \label{ch:intro}



\section{Towards string theory compactifications} \label{sec:intro_str}


Quantum field theories have emerged as the appropriate framework to describe high energy particle physics. The strong, weak and electromagnetic interaction of quarks, leptons and gauge bosons, which constitute all present-day known elementary particles, are realized in the $SU(3)\times SU(2)\times U(1)$ standard model of particle physics. The standard model is in fantastic agreement with experimental observations made at today's generation of accelerators and describes adequately interactions of quarks and leptons mediated by the gauge bosons up to energy scales of approximately one hundred GeV. However, despite of its success there are also many questions which cannot be addressed in the context of the standard model. On the one hand there are experimental data, for instance small neutrino masses or the presence of dark matter in the universe, which is not captured within the standard model. On the other hand there are also theoretical aspects of the standard model, which are not answered satisfactorily, such as the origin of the observed particle spectrum or the presence of many parameters which are not predicted by theory.

The forth known interaction in nature is gravity which is described by Einstein's theory of general relativity. It is a description of gravity by means of geometry that is to say the Einstein equations relate the energy momentum tensor of matter with the Ricci curvature tensor of the space-time geometry. Due to its geometric origin general relativity is a classical theory in the sense that it does not capture any quantum effects in gravity. As of today general relativity is also experimentally well-tested and serves as the basis of our present understanding of cosmology. However, also general relativity faces difficulties. Often solutions to the Einstein equation are plagued with singularities, which cannot be treated within general relativity itself. Moreover, as already mentioned general relativity is conceptually rather different from the approach taken by quantum field theories. But eventually for highly curved space-times and at length scales where the quantum nature becomes relevant quantum effects are also expected to enter the description of gravity. This clash of classical versus quantum mechanical physics can already be observed at the level of the Einstein equation because it equates the Einstein tensor with the energy momentum tensor. In general relativity the former tensor is a purely geometrical and hence a classical quantity whereas the energy momentum tensor of matter arises from quantum field theory and as such is a quantum mechanical object.

Thus eventually the three interactions of the standard model should be unified with the gravitational interaction of general relativity, and then in the resulting theory both the standard model and classical general relativity have to emerge as appropriate limits. This can partially be achieved in the context of quantum field theories on curved space-times, where the mediator of gravity is a spin~2 particle called graviton. Unfortunately these quantum gravity field theories are usually non-renormalizable. This indicates that they do not provide for a valid description beyond the Planck scale which is the energy scale characteristic for gravitational interactions. However, quantum gravity field theories can still serve as an effective low energy description of some more fundamental theory which regulates the theory in the ultraviolet regime.

A (perturbatively) consistent regulator of quantum gravity field theories is given by string theory \cite{Green:1981yb,Gross:1985rr,D'Hoker:2001nj}. The basic idea of string theory is to replace the worldline of a particle by the worldsheet of a spatial one-dimensional quantized string. At energy scales corresponding to the string length this procedure smears out the ultraviolate divergencies encountered in the point-like description of quantum field theory, whereas at low energy scales the spatial extension of the string becomes invisible and then the effective theory reduces to the particle behavior of quantum field theory. Moreover the low energy spectrum of string theories always contains a graviton and thus describes (at least perturbatively) gravity consistently. 

The stable tachyon-free string theories are supersymmetric and are called superstring theories, which predict in the weak coupling limit a ten-dimensional space-time target space. The class of ten-dimensional consistent superstring theories is comprised of five different types. There are the oriented closed type~IIA and type~IIB superstring theories with $\mathcal{N}=2$ supersymmetry. The closed oriented heterotic superstring with $\mathcal{N}=1$ supersymmetry, which contains matter fields transforming either under $E_8\times E_8$ or $SO(32)$. Finally the type~I superstring with $\mathcal{N}=1$ supersymmetry, which is unoriented and consists of closed and open strings. In type~I the open string states transform in the adjoint representation of $SO(32)$. All these superstring theories do not have any free parameters and are related by a web of string dualities \cite{Sen:1998kr,Polchinski:1998rq,Polchinski:1998rr}.

From a theoretical and mathematical point of view the unique structure of superstring theories in ten-dimensions is very appealing. Furthermore the framework of superstrings allows for many attractive ideas such as supersymmetry, grand unification or Kaluza-Klein reduction of extra dimensions. On the other hand in order to make contact with the success of the standard model in particle phenomenology it is necessary to specify a string theory which allows for an effective four-dimensional description, which resembles the standard model or some generalization thereof. In the weak coupling regime of string theory this can be attempted by compactifying the ten-dimensional superstring theory on a six-dimensional compact manifold. In the ten-dimensional field theory limit this amounts to a Kaluza-Klein reduction to four space-time dimensions.

Unfortunately the guiding principles for constructing string theories which are effectively four-dimensional are not at all as stringent as the construction of the five superstring theories in ten space-time dimensions. This is due to the fact that there are many possible choices for the compact internal space and at present there is no theoretical argument which singles out one internal space in favor of the others. This huge amount of possibilities to compactify string theory to four dimensions has even led to statistical examinations of the set of all (metastable) four-dimensional vacua called the landscape of string theory \cite{Douglas:2003um,Ashok:2003gk,Denef:2004dm}. However, the derivation and investigation of particular four-dimensional string compactifications teaches us how to overcome generic difficulties in obtaining phenomenologically viable theories. Hopefully the tools developed along these lines also improve our at present limited understanding of this string theory landscape. 


\section{D-branes in orientifold compactifications} \label{sec:intro_db}


In order to obtain phenomenological interesting four-dimensional theories the string compactification should fulfill certain criteria. First of all the compactification space should be of the form $M_4\times K_6$ where $M_4$ is a maximally symmetric four-dimensional space-time manifold and $K_6$ is some compact six-dimensional internal space. Furthermore as non-supersymmetric models generically suffer from instabilities, some supersymmetry should remain unbroken at the Planck or compactification scale. The current paradigm of particle phenomenology prefers an $\mathcal{N}=1$ supersymmetric matter sector spontaneously broken at low energies. Finally the four-dimensional low energy effective spectrum should have a fermionic matter spectrum which comes in representations of realistic gauge groups.

Traditionally the contact of string theory with four-dimensional particle physics focused on the compactification of the heterotic superstring as first proposed in ref.~\cite{Gross:1984dd,Candelas:1985en}. This is due to the fact that the gauge group $E_8\times E_8$ of the heterotic string is phenomenological interesting as it has $SU(3)\times E_6$ as a subgroup and $E_6$ is a natural grand unified group in four dimensions \cite{Gursey:1975ki,Gursey:1976dn,Ramond:1976jg}.\footnote{In compactifications on manifolds with $SU(3)$ holonomy the $SU(3)$ factor is needed to embed the spin connection in the gauge connection.} 

However, after the discovery of D-branes as non-perturbative BPS objects in string theory \cite{Polchinski:1995mt} it has been realized that D-branes also serve as new ingredients in constructing four-dimensional phenomenological interesting models. Since D-branes are part of the non-perturbative spectrum of type~II superstrings, compactifications of these theories in the presence of D-branes have emerged as a possible alternative to the earlier heterotic superstring models. This is due to the fact that D$p$-branes are extended objects with gauge theories localized on their $(p+1)$-dimensional worldvolume, that is to say a stack of $N$ D-branes which fill the space-time manifold $M_4$ gives rise to non-Abelian gauge groups such as $U(N)$ in four dimensions.\footnote{In unoriented string theories also the gauge groups $SO(N)$ and $USp(N)$ can appear.} Utilizing this property it is possible to construct out of several stacks of D-branes models with chiral fermions in representations similar to the standard model group $SU(3)\times SU(2)\times U(1)$ \cite{Kiritsis:2003mc,Uranga:2003pz,Lust:2004ks,Blumenhagen:2004vz,Blumenhagen:2005mu}. 

In order to engineer four-dimensional effective theories which leave $\mathcal{N}=1$ supersymmetry unbroken it is necessary for the compactification space $K_6$ and the D-branes to fulfill certain conditions. Type II superstring theories compactified on a Calabi-Yau manifold result in $\mathcal{N}=2$ supersymmetric four-dimensional theories \cite{Greene:1996cy}. Introducing space-time filling BPS D-branes reduces supersymmetry further to $\mathcal{N}=1$. However, as soon as D-branes are added consistency requires also the presence of appropriate orientifold planes, which are in contrast to D-branes non-dynamical extended objects in string theory \cite{Sagnotti:1987tw,Dai:1989ua,Leigh:1989jq,Pradisi:1988xd,Bianchi:1990yu,Bianchi:1990tb,Horava:1989vt,Gimon:1996rq}. These orientifold planes constitute unoriented strings, which are capable to cancel tadpoles originating from space-time filling D-branes \cite{Gimon:1996rq,Giddings:2001yu}. Therefore consistent $\mathcal{N}=1$ setups with space-time filling D-branes can be arranged by compactifying on six-dimensional Calabi-Yau orientifolds \cite{Acharya:2002ag,Brunner:2003zm,Brunner:2004zd,Grimm:2004uq,Grimm:2004ua}, which are compact six-dimensional Calabi-Yau manifolds with orientifold planes.

The complications arising from space-time filling D-branes in compact Calabi-Yau spaces are often circumvented by focusing on the local geometry in the vicinity of space-time filling D-branes \cite{Douglas:1996sw,Douglas:1997de,Berenstein:2002fi,Uranga:2002pg,Lerche:2002ck,Lerche:2002yw}. This approach is sometimes referred to `non-compact compactifications'. In these scenarios gravity is decoupled and it is therefore not possible to deduce a low energy effective description of (super-)gravity. Instead one focuses on the derivation of holomorphic $\mathcal{N}=1$ supersymmetric terms such as the superpotential \cite{Lerche:2002ck,Lerche:2002yw,Dijkgraaf:2002vw,Dijkgraaf:2002dh,Cachazo:2002ry}. Space-time filling D-branes in the context of compact internal spaces are mainly discussed in the literature in terms of toroidal orbifold/orientifold compactifications where the spectrum and other features are extracted \cite{Blumenhagen:1999ev,Cvetic:2000st,Blumenhagen:2002gw,Kachru:2002he,Tripathy:2002qw,Cascales:2003zp,Cascales:2003pt}, or more recently where the low energy effective supergravity action is computed \cite{Frey:2002hf,D'Auria:2003jk,Angelantonj:2003rq,Berg:2003ri,Kors:2003wf,Lust:2004cx,Lust:2004fi,Lust:2004dn}. In this work we focus on space-time filling D-brane configurations in the context of more general $\mathcal{N}=1$ superstring compactifications, namely on D3/D7-branes in generic type~IIB Calabi-Yau orientifolds. First we derive the four-dimensional massless D-brane spectrum and then we turn to the computation of the low energy effective supergravity action depending also on the bulk moduli fields \cite{Grana:2003ek,Jockers:2004yj}. For a generic Calabi-Yau orientifold theory this has not been spelled out in detail previously. 

However, eventually for phenomenological reasons $\mathcal{N}=1$ supersymmetry needs to be broken spontaneously, which can be achieved by turning on background fluxes in the orientifold bulk \cite{Bachas:1995ik,Polchinski:1995sm,Michelson:1996pn,Gukov:1999ya,Dasgupta:1999ss,Taylor:1999ii,Mayr:2000hh,Curio:2000sc,Becker:2001pm,Giddings:2001yu,Haack:2001jz,Becker:2002nn,Giryavets:2003vd}. These bulk background fluxes are vacuum expectation values for anti-symmetric tensor fields arising from the closed string sector or in other words arising from the bulk theory. In the presence of these supersymmetry breaking fluxes the couplings of the bulk fields to the D-brane matter fields communicate the breaking of supersymmetry to the matter sector and soft supersymmetry breaking terms are generated \cite{Grana:2002tu,Grana:2002nq,Kors:2003wf,Camara:2003ku,Grana:2003ek,Lawrence:2004zk,Lust:2004fi,Lust:2004dn,Camara:2004jj,Font:2004cx,Lust:2005bd}. Here we also thoroughly discuss the flux-induced soft-terms for the D3-brane matter fields for generic Calabi-Yau orientifold compactification \cite{Grana:2003ek}.

Typically superstring compactifications do not only yield matter fields but also feature numerous neutral fields which parametrize a continuous family of vacua. These fields are called moduli and their target space or the set of vacua is called the moduli space. In order to obtain a realistic four-dimensional vacuum these moduli fields must be stabilized because within today's bounds the observed particle spectrum does not contain any massless neutral scalar fields. Bulk background fluxes do not only break supersymmetry but also provide for a mechanism to stabilize these moduli fields \cite{Polchinski:1995sm,Taylor:1999ii,Mayr:2000hh,Curio:2000sc,Becker:2001pm,Giddings:2001yu,Kachru:2002he,Blumenhagen:2002mf,Blumenhagen:2003vr,Cascales:2003wn,Cascales:2003pt}. However, generically not all flat directions of the moduli are lifted by background fluxes. The remaining moduli can be fixed, for example, by non-perturbative contributions such as gaugino condensation on a stack of hidden sector D7-branes or by Euclidean D3-brane instantons \cite{Kachru:2003aw,Kachru:2003sx,Denef:2004dm,Gorlich:2004qm,Choi:2004sx}.

Interesting string compactifications should not only meet the requirements of particle physics but in addition should also address the demands of cosmology such as the appearance of a deSitter ground state. More recently space-time filling D-branes have also been introduced in string cosmology \cite{Linde:2004kg,Balasubramanian:2004wx,Burgess:2004jk}. In particular D3-branes and anti-D3-branes in type IIB Calabi-Yau compactifications can lead to (metastable) deSitter vacua \cite{Kachru:2003aw,Kachru:2003sx}. However, the simultaneous inclusion of D-branes and anti-D-branes breaks supersymmetry explicitly. Alternatively it is possible to replace anti-D3-branes by D7-branes with internal background fluxes. These fluxes are non-trivial vacuum expectation values for the field strength of the gauge theory localized on the worldvolume of the D7-brane. Turning on these fluxes can break supersymmetry spontaneously and has been anticipated to provide for another mechanism to generate the positive energy needed for (metastable) deSitter vacua \cite{Burgess:2003ic}. With the low energy effective supergravity action of D7-brane Calabi-Yau orientifold compactifications at hand we check this suggestion explicitely and find that under certain circumstances D7-brane fluxes do indeed provide for a positive energy contribution \cite{Jockers:2005zy}.


\section{Outline of this review article} \label{sec:intro_struct}


The main focus of this review is to derive the low energy effective $\mathcal{N}=1$ supergravity action for Calabi-Yau orientifold compactifications with space-time filling D-branes. We concentrate on type~IIB Calabi-Yau orientifolds and discuss space-time filling D3- and D7-branes \cite{Grana:2003ek,Jockers:2004yj,Jockers:2005zy}. Instead of specifying a particular Calabi-Yau orientifold with a specific D-brane configuration, the computations are performed for generic setups. The resulting four-dimensional field theories are analyzed and discussed.

In chapter~\ref{ch:Dbranes} we introduce D-branes in a ten-dimensional Minkowski background from an open string perspective. This guides us towards a low energy effective description of the D-brane spectrum. In particular we discuss the bosonic effective action of a single D-brane and its non-Abelian generalization describing a stack of D-branes \cite{Fradkin:1985qd,Leigh:1989jq,Myers:1999ps}, and finally we introduce the supersymmetric effective D-brane action \cite{Cederwall:1996pv,Cederwall:1996ri,Bergshoeff:1996tu}.

Then in chapter~\ref{ch:DbranesinCY} we turn to space-time filling D-branes in the context of Calabi-Yau compactifications. Consistency conditions take us to the discussion of orientifold planes and in particular we analyze the requirements for unbroken $\mathcal{N}=1$ supersymmetry in Calabi-Yau orientifold compactifications \cite{Marino:1999af,Jockers:2005zy}. Then finally the massless four-dimensional D-brane spectrum for D3- and D7-branes in Calabi-Yau orientifold compactifications is derived.

In chapter~\ref{ch:D3D7spec} the four-dimensional low energy effective $\mathcal{N}=1$ supergravity actions for D3- and D7-brane Calabi-Yau orientifold compactifications are computed. This is achieved by performing a Kaluza-Klein reduction of the ten-dimensional type~IIB supergravity bulk theory and by a normal coordinate expansion of the D3- and D7-brane action \cite{Grana:2003ek,Grimm:2004uq,Jockers:2004yj}. The resulting four-dimensional supergravity theories are discussed in terms of the K\"ahler potential, the gauge kinetic coupling functions and the superpotentials.

Then in chapter~\ref{ch:fluxes} we turn to the analysis of non-trivial background fluxes and their role for spontaneous supersymmetry breaking. First of all in Calabi-Yau orientifold compactifications with a stack of D3-branes the effects of bulk background fluxes are discussed. By performing a soft-term analysis we examine how supersymmetry breaking is communicated to the charged `matter sector' arising form the stack of D3-branes following ref.~\cite{Grana:2003ek}. In the second part we switch to the discussion of D7-branes with non-trivial worldvolume fluxes in the context of Calabi-Yau orientifolds. We find that there are different kind of D7-brane fluxes which generate D- and F-terms \cite{Lust:2005bd,Jockers:2005zy,Berglund:2005dm}, and we briefly comment on their relevance for string cosmology \cite{Kachru:2003aw,Kachru:2003sx,Burgess:2003ic,Jockers:2005zy,Berglund:2005dm}.

The geometric structure arising from D7-branes in Calabi-Yau orientifolds is related to the variation of Hodge structure of relative forms in chapter~\ref{ch:geom} along the lines of refs.~\cite{Mayr:2001xk,Lerche:2001cw,Lerche:2002yw}. We proceed in two steps, namely first we introduce the necessary mathematical machinery, which is then applied to the D7-brane supergravity theory of the previous chapters \cite{Jockers:2004yj}.

In chapter~\ref{ch:conc} we present our conclusions and suggest some directions for future investigations. Finally some further background material and some technical details are relegated to several appendices.


\chapter{D-branes in ten dimensions} \label{ch:Dbranes}


The existence of D$p$-branes in string theory has been an essential constituent in studying string dualities and in gaining insight into non-perturbative aspects of string theories. On the other hand D$p$-branes have also provided for new ingredients in constructing phenomenological appealing string models as they give rise to non-Abelian gauge groups in type~II superstring theories. 

Also in this work D-branes play a central role. The aim of this chapter is to introduce D-branes in the context of type~II superstring theory in ten-dimensions. In order to set the stage the massless ten-dimensional type~II spectrum of closed superstring excitations is reviewed in section~\ref{sec:typeIIAB}. Then in section~\ref{sec:openstr} we turn to the open superstring sector, which allows us to introduce D-branes embedded in ten-dimensional Minkowski space. It is further argued that D-branes can be interpreted as dynamical objects itself. Therefore we describe of D-branes in terms of their effective action in section~\ref{sec:Dpaction}. We discuss the action of a single D-brane and of a stack of D-branes, and finally the supersymmetric D-brane action.


\section{Type~II superstring spectrum in D=10} \label{sec:typeIIAB}


The spectra of type~IIA and type~IIB superstring theories in ten space-time dimensions are supersymmetric and consist of a finite number of massless states and an infinite tower of massive states which all originate from closed superstring excitations. In the low energy regime, that is to say below the string scale which is the mass scale for the higher string excitations, the massive tower of string states is not relevant. Therefore for these energies we can concentrate on the massless string modes, which for type~II superstring theories are encoded in a single $\mathcal{N}=2$ supermultiplet of type~IIA or type~IIB supergravity respectively. 

The fermionic fields of the $\mathcal{N}=2$ supermultiplet of type~IIA are given by two Majorana-Weyl spinors and two Majorana-Weyl gravitinos both of opposite chirality, whereas their bosonic superpartners are the graviton, one scalar field~$\phi_{10}$ called dilaton, one vector boson~$C^{(1)}$, one anti-symmetric two-tensor~$B$, and one anti-symmetric three-tensor~$C^{(3)}$. This spectrum can also be deduced by dimensional reduction of the maximal supersymmetric $\mathcal{N}=1$ supermultiplet of eleven-dimensional supergravity.

In type~IIB on the other hand the $\mathcal{N}=2$ supermultiplet consists of two Majo\-ra\-na-Weyl gravitinos and two Majorana-Weyl spinors all of the same chirality in the fermionic sector, whereas their bosonic superpartners are comprised of the graviton, two scalar fields called dilaton~$\phi_{10}$ and axion~$C^{(0)}$, two anti-symmetric two-tensor~$B$ and $C^{(2)}$, and the self-dual four-form tensor~$C^{(4)}$. Note that this spectrum is chiral and hence cannot be obtained by dimensional reduction of the eleven-dimensional $\mathcal{N}=1$ supermultiplet.

In quantizing the superstring worldsheet of type~II superstring theories in ten dimensions using the Ramond-Neveu-Schwarz formalism the fermionic fields arise from the NS-R and the R-NS sector and the bosonic fields from the NS-NS and the RR sector. Here we concentrate on the bosonic field content because the fermions can always be reconstructed by supersymmetry. For both type~IIA and type~IIB superstrings the RR~sector gives rise to the ten-dimensional metric~$g_{10}$, the anti-symmetric two-tensor~$B$ and the dilaton $\phi_{10}$. The RR~sector differs for type~IIA and type~IIB. In the type~IIA case the RR~sector contributes all odd dimensional anti-symmetric tensors, i.e. the form fields $C^{(1)}$, $C^{(3)}$, $C^{(5)}$, and $C^{(7)}$, whereas in type~IIB the RR~sector provides for all even dimensional anti-symmetric tensors, namely the form fields $C^{(0)}$, $C^{(2)}$, $C^{(4)}$, $C^{(6)}$, and $C^{(8)}$. Note that in the RR~sector we have obtained more anti-symmetric tensors than previously described in the ten-dimensional $\mathcal{N}=2$ multiplets. The reason is that in order to obtain the right number of physical degrees of freedom we need to take into account the duality relations among the RR anti-symmetric tensors, which we further discuss in section~\ref{sec:IIBsugra}. Here we just state that in type~IIA the dual pairs are $C^{(1)}\sim C^{(7)}$, $C^{(3)}\sim C^{(5)}$, whereas in type~IIB the pairs $C^{(0)}\sim C^{(8)}$, $C^{(2)}\sim C^{(6)}$ are dual and the tensor~$C^{(4)}$ is self-dual. 


\section{Open superstrings and D-branes} \label{sec:openstr}


One of the first encounters of D-branes in string theory is obtained from an open string perspective. Open strings are spatially one-dimensional intervals which in time sweep out a two-dimensional surface in the target space-time manifold. These surfaces are called worldsheets. Since we are interested in supersymmetric space-time theories the starting point are open superstrings, which exhibit supersymmetry on these two-dimensional worldsheets. Then this worldsheet supersymmetry is indirectly linked to space-time supersymmetry \cite{Gliozzi:1976qd}.

In order to illustrate the relation of open superstrings to D-branes we start with a ten-dimensional Minkowski space-time background. This Minkowski space-time background is parametrized by the flat space-time coordinates~$X^M$, $M=0,\ldots,9$, and the dynamics of open superstrings is governed by a worldsheet action recorded in \eqref{eq:WSaction}. The main difference of closed superstrings compared to open superstrings is the fact that one has to specify boundary conditions for their endpoints. The effects of these boundary conditions are carefully examined in appendix~\ref{app:OpenWS}, where one finds that one can either have Dirichlet or Neumann boundary conditions \eqref{eq:WSBdrycon}. For concreteness let us assume that the open superstrings under consideration obey Neumann boundary conditions along the coordinate directions~$X^a$ with $a=0,\ldots,p$ and Dirichlet boundary conditions along the remaining ten-dimensional space-time coordinates. This means that their endpoints are attached to $X^n=x^n$ with $n=p+1,\ldots,9$ and as a consequence the open superstrings are confined to the $(p+1)$-dimensional plane $\mathcal{W}=\{X^M|X^n=x^n\}$. This plane is called D$p$-brane and the geometric locus~$\mathcal{W}$ is called worldvolume on which open superstring endpoints can freely propagate.\footnote{Note that $p$ denotes only the number of spatial dimensions of the D$p$-brane, whereas the space-time dimensions of the worldvolume is $(p+1)$-dimensional.} The quantization of the open superstring worldsheet action~\eqref{eq:WSaction} is sketched in some detail in appendix~\ref{app:OpenWS}. The massless open superstring modes for open strings attached to a single D$p$-brane turn out to be (c.f.~eq.~\eqref{eq:masslessstates})
\begin{align} \label{eq:Dbranestates}
   A^a(k)\equiv\hat\psi^a_{-1/2}\ket{k}_\text{NS} \ , && 
   \dbps^n(k)\equiv\hat\psi^n_{-1/2}\ket{k}_\text{NS} \ , &&
   \Theta(k)\cong\ket{k,\theta}_\text{R} \ ,
\end{align} 
with contributions from both the Neveu-Schwarz~(NS) and the Ramond~(R) sector of the open superstring. Due to eq.~\eqref{eq:momentummodes} the momentum~$k$ of these open-string modes has only non-zero components~$k^a$ in the worldvolume directions, and hence this confirms on the quantum level that open strings do only propagate in the worldvolume of the D$p$-brane. In this work most of the time the higher string excitation are not considered as we mainly work in the supergravity limit~$\alpha'\rightarrow 0$, in which these states become heavy and hence are negligible in the low energy regime of supergravity.

The first Neveu-Schwarz state~$A^a$ in \eqref{eq:Dbranestates} corresponds to a massless vector on the worldvolume of the D$p$-brane. It gives rise to the $U(1)$~gauge theory localized on the D$p$-brane.\footnote{In the BRST quantization formulation of the open superstring, the local $U(1)$ gauge freedom of the vector boson corresponds to the choice of the BRST cohomology representative.} The other Neveu-Schwarz states~$\dbps^n$ carry space-time indices in the normal direction of the D$p$-brane. These fields are often denoted as the D$p$-brane `matter fields'. 

Finally let us turn to the massless states~$\Theta$ in the open superstring Ramond sector. As demonstrated in appendix~\ref{app:OpenWS} the modes~$\Theta$ transform in the Majorana-Weyl representation $\spinrep{16'}$ of the ten-dimensional space-time Lorentz group~$SO(9,1)$ and due to the restriction of the momentum~$k$ to the worldvolume directions they give rise to worldvolume spinors, which are obtained by decomposing $\Theta$ into representations of 
\begin{equation} \label{eq:decompDp}
   SO(9,1)\rightarrow SO(p,1)\times SO(9-p) \ .
\end{equation}
Here $SO(p,1)$ is the worldvolume Lorentz group and $SO(9-p)$ is the symmetry group for the normal directions of the D$p$-brane. For D$p$-branes with even-dimensional worldvolume, i.e. for $p=2l+1$, which are the relevant cases in this work, the spinor representation~$\spinrep{16'}$ decompose under \eqref{eq:decompDp} as
\begin{equation} \label{eq:Dpfermionicspec}
   \spinrep{16'}\rightarrow (\spinrep{2^l},\spinrep{{2'}^{3-l}})
      \oplus (\spinrep{{2'}^l},\spinrep{2^{3-l}}) \ .
\end{equation}
Thus the fermionic field~$\Theta$ becomes a worldvolume spinors $\chi^\alpha$, where the spinor index~$\alpha$ transforms under the symmetry group~$SO(9-p)$. 

The bosonic and fermionic worldvolume fields~$A^a$, $\dbps^n$ and $\chi^\alpha$ furnish precisely the spectrum of a single $(p+1)$-dimensional vector multiplet associated to a $(p+1)$-dimensional supersymmetric theory with $16$ supercharges. Indeed the worldvolume theory of a D$p$-brane in flat Minkowski space gives rise to a super $U(1)$ Yang-Mills theory with $16$ supercharges \cite{Polchinski:1998rq,Polchinski:1998rr}, in which $SO(9-p)$ becomes the R-symmetry group acting on the spinors~$\chi^\alpha$ and the scalars~$\dbps^n$. 

However, before the relation of space-time and worldvolume supersymmetry is further investigated, the open superstring spectrum \eqref{eq:Dbranestates} is generalized to the case of several D$p$-branes. For $N$ D$p$-branes which are all spatially separated we obtain for each D$p$-brane a copy of the massless states \eqref{eq:Dbranestates}. In addition there are open superstrings stretching from one D$p$-brane to another D$p$-brane. However, due to the mass formula \eqref{eq:mass} all the resulting modes of these open superstring are massive as long as the D$p$-branes are separated by the order of length scale~$\sqrt{\alpha'}$ and thus these fields are not considered in the supergravity limit $\alpha'\rightarrow 0$. Then the gauge symmetry for these $N$ D$p$-branes is $U(1)^N$. If, instead, these $N$ D$p$-branes coincide, i.e. if one considers a stack of $N$ D$p$-branes, there do arise additional massless modes from open strings stretching among different D$p$-branes in the stack. To keep track of these different open string modes, the endpoints of open superstrings are marked by labels $i=1,\ldots,N$ called Chan-Paton indices. They specify the D-branes to which the open-superstrings are attached. Then including the Chan-Paton indices the massless spectrum \eqref{eq:Dbranestates} becomes
\begin{align} \label{eq:DbranestatesN}
   A^a_{ij}(k)\equiv\hat\psi^a_{-1/2}\ket{k,ij}_\text{NS} \ , && 
   \dbpn^n_{ij}(k)\equiv\hat\psi^n_{-1/2}\ket{k,ij}_\text{NS} \ , &&
   \Theta_{ij}(k)\cong\ket{k,\theta,ij}_\text{R} \ .
\end{align} 
A closer look at open superstring scattering amplitudes reveals that these non-dynamical Chan-Paton labels respect a global $U(N)$ symmetry on the worldsheet \cite{Polchinski:1998rq,Polchinski:1998rr}. Under this symmetry the massless states~\eqref{eq:DbranestatesN} transform in the adjoint representation. This global $U(N)$ worldsheet symmetry becomes a local $U(N)$ worldvolume symmetry \cite{Sagnotti:1987tw,Dai:1989ua}, such that the $U(1)^N$ gauge symmetry for $N$ spatially separated D$p$-branes is enhanced to $U(N)$ for $N$ coinciding D$p$-branes \cite{Witten:1995im}. 

As we have seen the D$p$-brane `matter fields'~$\dbps^n(k)$ in the massless spectrum carry space-time indices in the normal direction of the worldvolume. These modes parametrize the shape of the D$p$-brane or in other words these states correspond to marginal operators \cite{Katz:2002gh}, which deform the worldvolume~$\mathcal{W}$ of the D$p$-brane in the normal direction. This observation implies that the D$p$-brane is not a rigid object in string theory but instead should be viewed as a dynamical extended object itself \cite{Dai:1989ua}. In the absence of D-branes the ten-dimensional space-time manifold is governed by the type~II closed string sector, and the ten-dimensional space-time theory has $\mathcal{N}=2$ supersymmetry which corresponds to $32$ supercharges. Now we should view D-branes as additional non-perturbative states in type~II superstring theory \cite{Polchinski:1995mt}. In the presence of a D$p$-brane some of the symmetries of the bulk theory are spontaneously broken. On the one hand the D$p$-brane breaks translational invariance for the directions normal to its worldvolume, and the open string states~$\dbps^n$ can be viewed as the goldstone bosons of these spontaneously broken translation symmetries. On the other hand it is also expected that D-branes break supersymmetry. Then analogously to the goldstone bosons we can view the fermions~$\chi^\alpha$ in the open superstring spectrum as the goldstinos for spontaneously broken supersymmetry. However, these fermionic modes can only account for $16$ broken supercharges, which is an indication that a D$p$-brane in the ten-dimensional Minkowski space breaks not all but only half of the amount of the bulk supersymmetry. It indeed turns out that the discussed D$p$-branes saturate a BPS bound and hence correspond to non-perturbative BPS states of the type~II superstring theory \cite{Polchinski:1995mt}. 

Since we have just anticipated that D$p$-branes are BPS states, it should be possible to identify the corresponding BPS charges. As discussed in section~\ref{sec:typeIIAB} the perturbative type~II superstring spectrum contains in the RR~sector of type~IIA superstring theory all odd-degree anti-symmetric tensors~$C^{(2q+1)}$, and in the RR~sector of type~IIB superstring theory all even-degree anti-symmetric tensors~$C^{(2q)}$. Note that a D$p$-brane as a $(p+1)$-dimensional extended object naturally couples to such a $(p+1)$-form tensor~$C^{(p+1)}$ via
\begin{equation} \label{eq:RRcharge}
   \mu_p=\int_{\mathcal{W}_{p+1}} C^{(p+1)} \ .
\end{equation}
This worldvolume integral coupling is indeed identified as the BPS-charge~$\mu_p$ of the D$p$-brane \cite{Polchinski:1995mt}. Note that this fact also allows us to specify the type of D$p$-branes which appear in closed string theories. Due to the presence of odd-degree RR-forms in type~IIA superstring theory only odd-dimensional D$p$-branes can appear, namely D$0$-, D$2$-, D$4$-, D$6$- and D$8$-branes, whereas in type~IIB superstring theory one has D$1$-, D$3$-, D$5$- and D$7$- and D$9$-branes, which couple to the even-degree RR-forms. In this work we focus on type~IIB superstring theory and hence on the even-dimensional D$p$-branes.

To conclude this section let us emphasis again the important observation of ref.~\cite{Polchinski:1995mt} that D$p$-branes should be viewed as dynamical objects itself. As a consequence from a space-time perspective these extended objects should also be describable by a low energy effective action, which we discuss in the next section. From the open string point of view this low energy effective action for D$p$-branes reproduces the string scattering amplitudes of open string states \cite{Leigh:1989jq}. 


\section{Dp-brane action} \label{sec:Dpaction}


As dynamical objects D-branes have a low-energy effective description which can be captured by a low energy effective action. The action of D-branes in type~II superstring theories consists of the Dirac-Born-Infeld and the Chern-Simons action. In section~\ref{sec:abelaction} we review the bosonic Abelian Dirac-Born-Infeld action and Abelian Chern-Simons action. These two actions describe the dynamics of the bosonic part of a single D-brane in a type~II string theory background. In section~\ref{sec:nonabelaction} the non-Abelian extension of the Dirac-Born-Infeld and Chern-Simons action is introduced following ref.~\cite{Myers:1999ps}. The non-Abelian D-brane action describes not just a single D-brane but a stack of D-branes. Finally in section~\ref{sec:superaction} we discuss the super Dirac-Born-Infeld and super Chern-Simons action of a super-D-brane \cite{Cederwall:1996pv,Cederwall:1996ri}. This amounts to introducing an action for both the bosonic and the fermionic D-brane degrees of freedom.


\subsection{Abelian Dp-brane action} \label{sec:abelaction}


The bosonic part of the low-energy effective action of a single D$p$-brane in a type~II superstring background consists of two parts. The Dirac-Born-Infeld action captures the couplings of the D-brane fields to the closed string NS-NS fields, whereas the Chern-Simons action describes the couplings to the closed string RR-form fields of type~II string theories. 

First we turn to the Dirac-Born-Infeld action of a single D$p$-brane, which reads in the string frame \cite{Fradkin:1985qd,Leigh:1989jq}
\begin{equation} \label{eq:DBIab}
   S_\text{DBI}^\text{sf}
      =-T_p\int_\WV\dd^{p+1}\xi\:\ee^{-\varphi^*\phi_{10}}
       \sqrt{-\det\left(\varphi^*(g_{10}+B)_{ab}-\ell F_{ab}\right)}\ ,
\end{equation}
where $\ell=2\pi\alpha'$ and the constant $T_p$ is the D$p$-brane tension. The Dirac-Born-Infeld action is integrated over the $(p+1)$-dimensional worldvolume~$\WV$, which is embedded in the ten-dimensional space-time manifold $M$ via the embedding map 
\begin{equation} \label{eq:emb}
   \varphi:\WV\hookrightarrow M,\: \xi^a \mapsto\dbem^M(\xi) \ , 
       \qquad a=0,\ldots,p \ , \qquad M=0,\ldots,9 \ , 
\end{equation}
where $\dbem^M$ are the coordinates of the ten-dimensional ambient bulk space whereas $\xi^a$ parametrize the worldvolume of the D$p$-brane.  In the Dirac-Born-Infeld action \eqref{eq:DBIab} the massless bulk NS-NS fields of type~II string theory, i.e. the dilaton $\phi_{10}$, the metric $g_{10}$ and the anti-symmetric two-tensor $B$, are pulled back with $\varphi^*$ to the worldvolume of the D$p$-brane, i.e.
\begin{align} \label{eq:PBgB}
   \left(\varphi^*g_{10}\right)_{ab}={g_{10}}_{MN}(\dbem) 
       \frac{\partial\dbem^M}{\partial\xi^a}\frac{\partial\dbem^N}{\partial\xi^b} \ , &&
   \left(\varphi^*B\right)_{ab}=B_{MN}(\dbem) 
       \frac{\partial\dbem^M}{\partial\xi^a}\frac{\partial\dbem^N}{\partial\xi^b} \ . 
\end{align}
As explained in section~\ref{sec:openstr} the D$p$-brane is a dynamical extended object, which means geometrically that the embedding map $\varphi$ is not rigid, but changes dynamically. In the Dirac-Born-Infeld action \eqref{eq:DBIab} this dynamics is encoded implicitly in the pull-back of the NS-NS fields to the worldvolume~$\WV$ of the D$p$-brane. Therefore the coordinates $\dbem^M(\xi)$ in \eqref{eq:emb} should be viewed as dynamical target-space fields of the action \eqref{eq:DBIab} which parametrize the geometric shape of the worldvolume~$\WV$ and which correspond to geometric deformations of the embeddings map $\varphi$. The independent degrees of freedom of these deformations are extracted systematically by performing a normal coordinate expansion. This procedure is studied thoroughly in section~\ref{sec:normal}. The massless modes of the deformations correspond to bosonic worldvolume scalar fields~$\dbps^n(k)$ in eq.~\eqref{eq:Dbranestates}.

In addition to the couplings to NS-NS bulk fields the action \eqref{eq:DBIab} contains a $U(1)$ field strength $F$, which describes the $U(1)$ gauge theory of the worldvolume gauge boson~$A^a(k)$ defined in \eqref{eq:Dbranestates}. The non-linear action \eqref{eq:DBIab} captures the couplings of the field strength to all orders in $\alpha'F$ \cite{Leigh:1989jq}. To leading order in $F$, the gauge theory reduces to a $U(1)$ Yang-Mills theory on the worldvolume $\WV$ of the brane.

As noted in section \ref{sec:openstr} D$p$-branes carry RR charges \cite{Polchinski:1995mt}, and they couple as extended objects to appropriate RR forms of the bulk, namely the $(p+1)$-dimensional worldvolume couples naturally to the bulk RR form~$C^{(p+1)}$. These couplings are captured in the Chern-Simons action of the D$p$-brane which in the case of a single D-brane reads\footnote{The exponential function is a formal power series in the worldvolume two-form~$\ell F-\varphi^*B$, and the integrand of the Chern-Simons action is a formal sum of forms of various degree. The integral taken over the worldvolume~$\WV$, however, is only non-vanishing for $(p+1)$-forms.}
\begin{equation} \label{eq:CSAb}
   \mathcal{S}_\text{CS} = \mu_p \int_\WV \sum_q \varphi^*C^{(q)}\:
     \ee^{\ell F-\varphi^* B} \ ,
\end{equation}
where the coupling constant $\mu_p$ is the RR charge of the brane. Moreover, generically D-branes contain lower dimensional D-brane charges, and hence interact also with lower degree RR-forms \cite{Douglas:1995bn}. All these couplings to the bulk are implemented in the Chern-Simons action in a way compatible with T-duality. In type~IIA string theory there are only odd degree RR-forms $C^{(q)}$ and hence the sum in \eqref{eq:CSAb} runs only over odd forms, whereas in type~IIB string theory with all even degree RR-forms in the massless RR spectrum the sum in \eqref{eq:CSAb} is taken over even forms.

For BPS D$p$-branes the energy density, i.e. the D$p$-brane tension $T_p$, is entirely determined by its RR charge $\mu_p$. Two static and parallel BPS D-branes of the same charge do not feel a net force. Therefore the attraction due to the exchange of closed string NS-NS modes must be canceled by the repulsion resulting from closed string RR modes. This requirement relates the D-brane tension~$T_p$ to its RR-charge \cite{Polchinski:1998rq,Polchinski:1998rr}
\begin{equation} \label{eq:DBcharge}
   \mu_p^2=T_p^2=\frac{\pi}{\kappa^2}\left(4\pi\alpha\right)^{3-p} \ .
\end{equation}
 
 
\subsection{Non-Abelian Dp-brane action} \label{sec:nonabelaction}


In order to describe a stack of D$p$-branes the Abelian action \eqref{eq:DBIab} and \eqref{eq:CSAb} must be generalized. For coinciding D-branes the $U(1)$ gauge theory on the worldvolume of a single D-brane is enhanced to a non-Abelian gauge theory localized on the worldvolume of the stack of D-branes \cite{Witten:1995im}. As proposed in ref.~\cite{Douglas:1997ch,Douglas:1997sm} the target space coordinates $\dbem^M(\xi)$ of \eqref{eq:DBIab} should be promoted to matrix valued coordinates $\dbem^M_{ij}(\xi)$ and the derivatives $\partial_a\dbem^\mu$ should be replaced by appropriate covariant derivatives $\covdb_a\dbem^\mu_{ij}$ \cite{Hull:1997jc}. These modifications can be confirmed by studying open string scattering amplitudes \cite{Garousi:1998fg,Garousi:2000ea} and therefore must be implemented into the non-Abelian generalization of the Dirac-Born-Infeld \eqref{eq:DBIab} and Chern-Simons action \eqref{eq:CSAb}.

In type IIB compactifications with orientifolds, the possible gauge groups turn out to be $U(N)$, $SO(N)$ and $USp(N)$ and the matrix valued target-space coordinates $\dbem^M_{ij}$ are in the adjoint representation of the respective gauge group. For D-branes which coincide with orientifold planes the gauge groups are either $SO(N)$ or $USp(N)$ depending on the type of orientifold plane \cite{Brunner:2003zm}. D-branes that are not on top of an orientifold plane have $U(N)$ as a gauge group. In this work most of the time we limit our analysis to the latter case. 
 
The non-Abelian extension of the Dirac-Born-Infeld action \eqref{eq:DBIab}, which reproduces open string scattering amplitudes and which is also in agreement with T-duality, was derived in ref.~\cite{Myers:1999ps} and reads in the string frame
\begin{align} \label{eq:DBInonab}
   S_\text{DBI}^\text{sf}= 
     -T_p\int_\WV d^{p+1}\xi\: \str\: \ee^{-\varphi^*\phi} 
     \sqrt{-\det\left[ (\varphi^*P)_{ab}-\ell F_{ab}\right]\: \det Q^M_N } \ ,
\end{align}
where
\begin{align} \label{eq:PQterm} 
   Q^M_N  = \delta^M_N+\frac{\ii}{\ell}\com{\dbpn^M}{\dbpn^L} E_{LN} \ , &&
   P_{MN} = E_{MN} + E_{MP} \left(\inv{Q}-\id\right)^P_N \ ,
\end{align}
with $E_{MN}={g_{10}}_{MN}+B_{MN}$. This expression of the non-Abelian Dirac-Born-Infeld action needs some explanation. First of all the terms \eqref{eq:PQterm} are expressed in `static gauge'. This means that the coordinates~$\dbem^M$ are chosen (locally) in such a way that the center of mass of the stack of branes is located at $\dbem^M=0$. Then the `matrix valued fluctuations' $\dbpn^M$ parametrize the deformations of the stack of branes around this static center of mass configuration at $\dbem^M=0$. Now the non-Abelian Dirac-Born-Infeld action \eqref{eq:DBInonab} should be viewed as an action which can be expanded into the non-Abelian D$p$-brane fluctuations $\dbpn^M$ which transform in the adjoint representation of the gauge theory localized on the stack of D-branes and which parametrize the deformations of the worldvolume of the stack in the normal direction. These degrees of freedom correspond to the modes~$\dbpn^n_{ij}(k)$ in \eqref{eq:DbranestatesN}. $F_{ab}$ is the field strength for this gauge theory and hence also matrix valued, namely $F_{ab}$ transforms also in the adjoint representation. This is the field strength of the gauge boson~$A^a_{ij}(k)$ in \eqref{eq:DbranestatesN}.

Finally due to the non-Abelian nature of the deformation parameters $\dbpn^M$ and the field strength $F_{ab}$ it is necessary to specify an ordering prescription in \eqref{eq:DBInonab} for these matrix valued fields. This is achieved by using the symmetrized trace `$\str$' with respect to the matrix valued terms $\covdb_a\dbpn^M$, $\com{\dbpn^M}{\dbpn^N}$ and $F_{ab}$ \cite{Tseytlin:1997cs}, which means that these matrix valued terms are first symmetrized and then in a second step the trace of the resulting expression is taken.

In the same spirit the Abelian Chern-Simons action \eqref{eq:CSAb} needs to be modified and becomes
\cite{Myers:1999ps}
\begin{equation} \label{eq:CSnonab}
   \mathcal{S}_\text{CS}=\mu_p \int_\WV \str \:
     \varphi^*\left(\ee^{\frac{\ii}{\ell}\ins{\dbpn}\ins{\dbpn}}\sum_q C^{(q)} \ee^{-B}\right)
     \ee^{\ell F} \ .
\end{equation}
Here $\ins{\dbpn}$ denotes the interior multiplication of a form with $\dbpn$ which yields, for instance, for the $q$-form $C^{(q)}$ in local coordinates
\begin{equation} \label{eq:intmul}
   \ins{\dbpn}C^{(q)}=\frac{1}{q!}\sum_{k=1}^q (-1)^{k+1}\dbpn^M 
      C^{(q)}_{N_1\ldots N_{k-1}\:M\:N_{k+1}\ldots N_q}
      \dd x^{N_1}\wedge\ldots\widehat{\dd x^{N_k}} \ldots\wedge \dd x^{N_q}  \ ,
\end{equation}
where the differential with the hat $\widehat{\quad}$ is omitted. Also the non-Abelian generalization of the Chern-Simons action \eqref{eq:CSnonab} should be viewed as an expression in `static gauge' which can be expanded into the matrix valued fields $\dbpn^M$ and $F_{ab}$ and for which also the symmetrized trace gives the appropriate ordering prescription. In the Abelian limit $\ins{\dbpn}\ins{\dbpn}$ yields always zero since the interior multiplication has odd degree. However, this is not the case for non-Abelian~$\dbpn$. Therefore in reducing $U(N)$ to $U(1)$, i.e. in the transition from a stack of $N$ D-branes to a single D-brane, the non-Abelian enhancement disappears and the Chern-Simons action~\eqref{eq:CSnonab} simplifies to the Abelian expression~\eqref{eq:CSAb}.

Already in the Abelian case one expects additional corrections in $\alpha'$ involving derivative terms beyond second order. However, the Abelian Dirac-Born-Infeld action is expected to capture all $\alpha'$ corrections in $F_{ab}$ for `slowly-varying' $F_{ab}$ (i.e. all derivative independent terms in $F_{ab}$). In the non-Abelian case the distinction between the field strength and its covariant derivative is ambiguous since $\com{\covdb_a}{\covdb_b}F_{cd}=\com{F_{ab}}{F_{cd}}$. The symmetrized trace proposal of ref.~\cite{Tseytlin:1997cs} treats the matrices $F_{ab}$ (as well as $\covdb_a\dbpn^M$ and $\com{\dbpn^M}{\dbpn^N}$) as if they were commuting, leaving out all commutators among these. This proposal is shown to be reliable only up to fourth order in $F_{ab}$ \cite{Hashimoto:1997gm}, but this is sufficient for our purposes.


\subsection{Super Dp-brane action} \label{sec:superaction}


In this section we introduce the effective action of a single D$p$-brane, which in addition to the bosonic modes also captures the fermionic degrees of freedom~$\Theta(k)$ of the open superstring spectrum~\eqref{eq:Dbranestates}. The resulting effective action is called the super-D$p$-brane action. A generalization to the non-Abelian case as for the bosonic part in section~\ref{sec:nonabelaction} is not known for the supersymmetrized version.

Here we concentrate on D$p$-branes in type~IIB superstring theory as in this work we focus on the type~IIB side. In superspace the corresponding type~IIB supergravity theory is formulated on the supermanifold $M^{9,1|2}$ with ten `bosonic' dimensions and two `fermionic' dimensions. Locally this supermanifold is described by the superspace coordinates $Z^{\check M}=(x^M,\pair{\theta})$ with the ten bosonic coordinates $x^M$ and the pair of fermionic coordinates $\pair{\theta}=\left(\theta^1,\theta^2\right)$. As type~IIB string theory is chiral with $\mathcal{N}=2$ supersymmetry in ten dimensions the pair of fermionic coordinates $\pair{\theta}$ consists of two Majorana-Weyl spinors of $SO(9,1)$ with the same chirality.\footnote{For type~IIA string theories the two fermionic coordinates have opposite chirality.} Hence $\theta^1$ and $\theta^2$ are related by the $SO(2)$ R-symmetry of type~IIB supergravity \cite{Howe:1983sr}.

Now in this formulation the super-D$p$-brane appears as the embedding of the $(p+1)$-dimensional worldvolume $\mathcal{W}$ in the supermanifold $M^{9,1|2}$. The embedding $\super{\varphi}:\mathcal{W}\hookrightarrow M^{9,1|2}$ is now described by the supermap $\super{\varphi}$ which maps a point in the worldvolume~$\mathcal{W}$ to a superpoint in the target space supermanifold $M^{9,1|2}$.

The super Dirac-Born-Infeld action for a single super-D$p$-brane becomes in the string frame \cite{Cederwall:1996pv,Cederwall:1996ri,Bergshoeff:1996tu}
\begin{equation} \label{eq:SDBI}
   \mathcal{S}^\text{sf}_\text{DBI}=-T_p \int_\mathcal{W}\dd^{p+1}\xi\ee^{-\super{\varphi^*\phi}}
       \sqrt{-\det\left(\super{\varphi^*}\left(\super{g_{10}}+\super{B}\right)_{ab}-\ell F_{ab} \right)} \ ,
\end{equation}
whereas the super Chern-Simons action reads
\begin{equation} \label{eq:SCS}
   \mathcal{S}_\text{CS}=\mu_p \int_\mathcal{W}\sum_q\super{\varphi^*}\left(\super{C^{(q)}}\right)
       \ee^{\ell F-\super{\varphi^*B}} \ .
\end{equation}
Both actions resemble their bosonic analogs~\eqref{eq:DBIab} and \eqref{eq:CSAb} but the bulk fields $\super{g_{10}}$, $\super{B}$, $\super{\phi}$ and $\super{C^{(q)}}$ have been promoted to bulk superfields with their lowest components being the corresponding bosonic fields. These superfields are then pulled back with the supermap~$\super{\varphi}$.\footnote{Note that the 
pulled-back quantities $\super{\varphi}^*(\cdot)$ contain no odd components because $\super{\varphi}$ is a map from an ordinary manifold into a supermanifold. As a consequence the integrals in \eqref{eq:SDBI} and \eqref{eq:SCS} are integrals only over bosonic coordinates.} The gauge field strength $F_{ab}$ on the D-brane, however, remains a bosonic object.

Since the super-D$p$-brane is a BPS state in the ten-dimensional $\mathcal{N}=2$ supersymmetric type~IIB bulk theory, the amount of preserved supersymmetry in the presence of a super-D$p$-brane must be $\mathcal{N}=1$. This in particular implies that the bosonic on-shell degrees of freedom arising from the D$p$-brane must be equal to the fermionic on-shell D-brane degrees of freedom. As discussed in section~\ref{sec:abelaction} the bosonic modes arise from the D$p$-brane fluctuations, which give rise to $10-(p+1)$ bosonic degrees of freedom due to the dimensionality of the normal space of the worldvolume~$\mathcal{W}$. In addition there are $p-1$ bosonic on-shell degrees of freedoms arising from the $U(1)$ gauge boson~$A^a$ localized on the worldvolume of the D$p$-brane. Thus there is a total number of $8$ bosonic degrees of freedom. In the effective action formulation of the super-D$p$-brane the fermionic degrees of freedom appear as fluctuations of the embedded worldvolume along the odd coordinates~$\pair{\theta}$, which corresponds to $32$ real parameters. The linear Dirac equation reduces the degrees of freedom by $1/2$ and thus we obtain $16$ real degrees of freedom. However, in order to obtain a supersymmetric spectrum the fermionic modes need to be further reduced. A closer examination of the super-D$p$-brane action reveals a local fermionic gauge symmetry called $\kappa$-symmetry which removes half of these remaining fermionic degrees of freedom \cite{Bergshoeff:1996tu,Bergshoeff:1997kr,Bergshoeff:2000ik}, and hence the bosonic and fermionic on-shell degrees of freedom are indeed both equal to eight. They correspond to the open-superstring fermionic modes~$\Theta(k)$ in \eqref{eq:Dbranestates}.


\chapter{D-branes in Calabi-Yau compactifications} \label{ch:DbranesinCY}


In the previous chapter D-branes are introduced as non-perturbative objects in flat ten-dimensional Minkowski space. In this chapter we want to analyze D-branes in more general backgrounds. We require the low energy effective theory to be four-dimensional and to preserve $\mathcal{N}=1$ supersymmetry. In several steps this eventually leads us to Calabi-Yau orientifold compactifications with space-time filling D-branes.

In section~\ref{sec:KKreduction} we first review the idea of Kaluza-Klein compactifications or more specifically in section~\ref{sec:CYcompact} of Calabi-Yau compactifications. Then in section~\ref{sec:BPSCY} for D-branes in Calabi-Yau spaces we discuss supersymmetry and the consistency requirements imposed by tadpole cancellation conditions. This naturally leads us to O-planes and orientifold Calabi-Yau compactifications. Then in section~\ref{sec:D3D7spec} we compute the D-brane spectrum in Calabi-Yau orientifold compactifications, which in this class of compactifications constitutes for the matter sector in the effective four-dimensional theory.


\section{Kaluza-Klein reduction} \label{sec:KKreduction}


First we briefly review the idea of Kaluza-Klein reduction. The details of this procedure are rather involved and are described thoroughly in ref.~\cite{Duff:1986hr}. Here we focus on the features relevant for the analysis in this work.

The starting point is a gravity theory on a product manifold
\begin{equation} \label{eq:topansatz}
   M^{9,1}=M^{3,1}\times K_6 \ , 
\end{equation}
where $M^{3,1}$ is called the space-time manifold and $K_6$ is the internal compact manifold. For concreteness and in order to be suitable for this work we choose the space-time manifold~$M^{3,1}$ to be four-dimensional whereas the internal manifold~$K_6$ should be six-dimensional. The ground state background metric for the ten-dimensional product manifold~$M^{9,1}$ also obeys the product structure 
\begin{equation} \label{eq:metansatz}
   \dd s_{10}^2 = {g_{10}}_{MN}(X)\:\dd X^M\dd X^N
       =h_{\mu\nu}(x)\dd x^\mu\dd x^\nu+g_{mn}(y)\dd y^m\dd y^n \ ,
\end{equation}
with the four-dimensional Lorentzian metric~$h_{\mu\nu}$ and the six-dimensional Euclidean metric~$g_{mn}$. 

The philosophy of the Kaluza-Klein reduction is now to expand the fields on the manifold~$M^{9,1}$ in appropriate Fourier modes of the internal compact manifold~$K_6$. Let us illustrate this procedure explicitly for a massless scalar field~$\Phi(x,y)$ for which the equation of motion reads
\begin{equation} \label{eq:laplacescalar}
   \square_{10}\Phi(x,y)=(\square_4+\Delta_6)\Phi(x,y) = 0 \ ,
\end{equation}
where $\square_{10}$ is the ten-dimensional d'Alembert operator, which for the metric ansatz \eqref{eq:metansatz} splits into the four-dimensional d'Alembert operator~$\square_4$ and into the six-dimensional Laplace operator~$\Delta_6$. Then the Hodge theorem guarantees that the scalar field $\Phi(x,y)$ can be uniquely expanded into eigenfunctions of $\Delta_6$ of the internal manifold as
\begin{equation} \label{eq:fouriermodes}
   \Phi(x,y)=\sum_{k=0}^{\infty} \phi_k(x)\cdot f_k(y) \ ,
\end{equation}
where $f_k(y)$ are the eigenfunctions of $\Delta_6$ with eigenvalue~$m^2_k$. Note that the constant function~$f_0(y)\equiv 1$ is always an eigenfunction of $\Delta_6$ with eigenvalue zero. The Hodge theorem further implies that the eigenvalues on compact manifolds are always discrete and non-negative, and therefore we assume the order $0=m^2_0< m^2_1 \le m^2_2 \ldots$ for the eigenvalues. Inserting the ansatz~\eqref{eq:fouriermodes} into \eqref{eq:laplacescalar} yields for the space-time modes $\phi_k(x)$ the equations of motions
\begin{equation}
   \square_4\:\phi_k(x)+m^2_k\:\phi_k(x) = 0 \ . 
\end{equation}
Thus the different Fourier modes are governed by Klein-Gordon equations in which the eigenvalues~$m^2_k$ appear as masses, and in particular $\phi_0(x)$ becomes a massless scalar field.

The idea is now to perform this Kaluza-Klein expansion for all fields in the ten-dimensional theory, that is to say for all bosonic fields, including the $p$-form fields and the metric perturbations~$\delta {g_{10}}_{MN}$ of the background ansatz~\eqref{eq:metansatz}, as well as for the fermionic fields in the theory. Let us further assume that the ten-dimensional gravity theory is formulated in terms of an action functional~$\mathcal{S}_{10}[\Phi^{(\alpha)}(x,y)]$ where $\Phi^{(\alpha)}(x,y)$ symbolically denotes all the fermionic and bosonic fields in the theory. Then by inserting the Kaluza-Klein expansion for all these fields~$\Phi^{(\alpha)}(x,y)$ and performing the six-dimensional integral in the action~$\mathcal{S}_{10}$ we arrive at an four-dimensional action functional~$\mathcal{S}_{4}[\phi^{(\alpha)}_k(x)]$ where now $\phi^{(\alpha)}_k(x)$ refer again symbolically to all the Kaluza-Klein modes in the expansion. Thus we have arrived at a four-dimensional formulation of the original ten-dimensional gravity theory, at the cost of introducing an infinite Kaluza-Klein tower of massive Kaluza-Klein modes. 

Let us analyze the scaling behavior of these Kaluza-Klein masses. For simplicity we return to the simple example of the scalar field~$\Phi(x,y)$ introduced in the previous paragraph. The internal Laplace operator~$\Delta_6$ scales with $R^{-2}$ in terms of the `radius'~$R$ of the internal manifold~$K_6$, i.e.~$\vol{(K_6)}\sim R^6$. Thus this implies for the Kaluza-Klein masses~$m^2_k$ 
\begin{equation}
   m^2_k \sim R^{-2} \ \ \text{for} \ k>0 \ .
\end{equation} 
Therefore in the limit of small internal manifolds~$K_6$ the higher Fourier modes, also denoted higher Kaluza-Klein modes, of the scalar field become very massive. This scaling behavior is not just typical for the scalar field Kaluza-Klein masses but is a generic feature of the masses of more general fields. Therefore it is tempting to truncate the theory in the limit of small internal manifolds~$K_6$ to the zero-modes in the Kaluza-Klein expansion, that is to say we take the action~$\mathcal{S}_{4}$ but neglect all the massive Kaluza-Klein modes in this theory, and work instead with the four-dimensional effective action~$\mathcal{S}^{\text{eff}}_{4}$ of the massless Kaluza-Klein modes. Physically, however, it is not quite correct to simply set the massive modes to zero \cite{Duff:1986hr}. Instead all the massive modes should be integrated out, which generically modifies the interaction terms of the massless fields. In principal these modifications are calculable. In practice, however, in most cases it is inconceivable to compute these non-linear interactions. In particular for Calabi-Yau compactifications, which we discuss in the next section, the explicit expression for the internal Ricci flat metric is not known, and thus it seems impossible to perform such a computation. However, in the context of Calabi-Yau compactifications it has been argued in refs.~\cite{Pope:1987ad,Duff:1989cr} that due to the structure imposed by supersymmetry the truncation is consistent in the low energy effective description of $M^{3,1}$. In addition the consistency is also confirmed by string computations \cite{Gross:1986mw,Dixon:1989fj}. Therefore in this work we take the practical approach and perform the Kaluza-Klein reduction of all fields in the ten-dimensional theory and truncate the spectrum to the massless Kaluza-Klein modes. Then we derive the effective four-dimensional theory on $M^{3,1}$ by inserting the Kaluza-Klein zero-modes into the effective ten-dimensional action and integrate out the internal space~$K_6$.


\section{Calabi-Yau compactifications} \label{sec:CYcompact}


In the previous section we have discussed general features of Kaluza-Klein reductions. Here, we want to consider a Kaluza-Klein ansatz~\eqref{eq:topansatz} for type~IIB supergravity which leaves some amount of supersymmetry unbroken. The amount of supersymmetry preserved by a given background is determined by analyzing the supersymmetry variations of the fermions. The amount of unbroken supersymmetry in the vacuum is given by the number of independent supersymmetry parameters for which the fermionic supersymmetry variations vanish. For type~IIB supergravity compactifications without bulk background fluxes the non-trivial fermionic variation arises form the ten-dimensional gravitinos \cite{Polchinski:1998rq,Polchinski:1998rr}
\begin{equation} \label{eq:susyvarCY}
   \delta_{\pair{\epsilon}}\pair{\Psi}_M = \nabla_M\pair{\epsilon} \ .
\end{equation}
Here $\nabla$ is the spinor covariant derivative with respect to the background metric~\eqref{eq:metansatz} and $\pair{\Psi}_M=(\Psi^1_M,\Psi^2_M)$ is the ten-dimensional Majorana-Weyl gravitino pair of the type~IIB spectrum introduced in section~\ref{sec:typeIIAB}, which transforms under the $SO(2)$ R-symmetry of type IIB supergravity \cite{Howe:1983sr}. The two Majorana-Weyl gravitinos of type~IIB string theory have the same chirality. In other words 
\begin{equation} \label{eq:Psichiral}
   \Gamma \pair{\Psi}_M = - \pair{\Psi}_M \ ,
\end{equation}
in terms of the ten-dimensional chirality gamma matrix~$\Gamma$ defined in eq.~\eqref{eq:G11}. The supersymmetry parameters~$\pair{\epsilon}$ in \eqref{eq:susyvarCY} are also Majorana-Weyl spinors both transforming in the spinor representation~$\spinrep{16'}$, i.e. $\Gamma \pair{\epsilon}=-\pair{\epsilon}$. 

In order to find the number of unbroken supercharges we need to evaluate the condition for vanishing gravitino variation. According to eq.~\eqref{eq:susyvarCY} this amounts to determining the number of covariantly constant spinors in the Kaluza-Klein background. For the Kaluza-Klein ansatz~\eqref{eq:topansatz} the structure group $SO(9,1)$ is reduced to $SO(3,1)\times SO(6)$ and the spinor representation~$\spinrep{16'}$ decomposes into
\begin{equation}
   \spinrep{16'}\rightarrow (\spinrep{2},\spinrep{\bar 4})\oplus
                            (\spinrep{\bar 2},\spinrep{4}) \ .
\end{equation}
Thus on a generic internal manifold~$K_6$ there are no spinor singlets in the internal space, which potentially give rise to covariantly constant supersymmetry parameters in the effective four-dimensional Minkowski space. In order to obtain spinor singlets~$\spinrep{1}$ the internal manifold must have $SU(3)$ structure since the $\spinrep{4}$ of $SO(6)$ decomposes for the subgroup~$SU(3)$ into $\spinrep{3}\oplus\spinrep{1}$. However, for a vacuum with unbroken supersymmetry we need not only a globally defined spinor singlet but moreover a globally covariantly constant spinor \cite{Gross:1984dd,Candelas:1985en}. This is fulfilled if the internal manifold has not only $SU(3)$ structure but even $SU(3)$ holonomy, which guarantees that the globally defined spinor singlet is covariantly constant \cite{Joyce:2000}.

The requirement for the six-dimensional internal manifold to have $SU(3)$ holonomy is very strong, namely this is the case if and only if it is K\"ahler and Ricci-flat \cite{Yau:1977}. Such manifolds are called Calabi-Yau threefolds and therefore we choose as our compactification ansatz
\begin{equation} \label{eq:CYansatz}
   M^{9,1}=\mathbb{R}^{3,1}\times Y \ ,
\end{equation}
with the internal six-dimensional Calabi-Yau manifold~$Y$. The corresponding metric ansatz takes the form\footnote{The hat~$\hat{}$ indicates that the quantity is in the string frame.}
\begin{equation} \label{eq:metCalabiYau}
   \dd s^2_{10}=\hat\eta_{\mu\nu}\dd x^\mu\dd x^\nu+2\hat g_{i\bar\jmath} \dd y^i \dd\bar y^{\bar\jmath} \ ,
\end{equation}
with the four-dimensional Minkowski metric~$\hat\eta$ and the six-dimensional internal Ricci-flat metric $\hat g_{i\bar\jmath}$ of $Y$. As Calabi-Yau manifolds are K\"ahler the metric $\hat g_{i\bar j}$ is also related to the closed K\"ahler form~$\hat J$ of $Y$ by
\begin{equation} \label{eq:Kform}
   \hat J_{i\bar\jmath}=\ii\hat g_{i\bar\jmath} \ .
\end{equation}

As discussed in the previous paragraph this choice for the Kaluza-Klein ansatz guarantees the existence of covariantly constant spinors. In order to obtain the least amount of preserved supersymmetry we require that the Calabi-Yau threefolds has exactly~$SU(3)$ holonomy and not a subgroup thereof.\footnote{In the following Calabi-Yau threefolds always refer to six-dimensional manifolds which have exactly $SU(3)$ holonomy.} In this case there is a single covariantly constant spinor~$\check\singspin$ and its conjugate~$\check\singspin^\dagger$ \cite{Joyce:2000}. As a consequence the variation~\eqref{eq:susyvarCY} vanishes for the supersymmetry parameters
\begin{equation} \label{eq:susyparaCY}
   \pair{\epsilon}=\pair{\bar\eta}\otimes\check\singspin(y)+
      \pair{\eta}\otimes\check\singspin^\dagger(y) \ ,
\end{equation}   
where $\pair{\eta}=(\eta^1,\eta^2)$ are the four-dimensional supersymmetry parameters obeying $\hat\gamma\pair{\eta}=+\pair{\eta}$ and $\hat\gamma\pair{\bar\eta}=-\pair{\bar\eta}$ in terms of the four-dimensional chirality gamma matrix $\hat\gamma$ defined in eq.~\eqref{eq:G5}. 

Since one finds in Calabi-Yau compactifications two complex four-dimensional Weyl supersymmetry parameters~$\pair{\eta}$, Calabi-Yau compactifications preserve eight supercharges or in other words $\mathcal{N}=2$ supersymmetry in four dimensions. That is to say the complex supersymmetry parameters~$\eta_1,\eta_2$ fulfill $\delta_{\eta^1}(\text{fermions})=0$ and $\delta_{\eta^2}(\text{fermions})=0$. 

Before we conclude this section let us summarize some properties of Calabi-Yau threefolds which are relevant in this work. A Calabi-Yau threefold has exactly one holomorphic $(3,0)$-form~$\Omega$ which is nowhere vanishing and is covariantly constant with respect to the Ricci-flat K\"ahler metric. For Calabi-Yau threefolds the first Betti number vanishes which implies that also the Hodge numbers $h^{1,0}$ and $h^{0,1}$ vanish.\footnote{The Hodge number $h^{p,q}$ is the dimension of the Dolbeault cohomology group $H_{\bar\partial}^{(p,q)}(Y)$ defined on complex manifolds with respect to the Dolbeault operator~$\bar\partial$.} Moreover the Hodge numbers in Calabi-Yau threefolds obey
\begin{align} 
   h^{p,0}=h^{3-p,0} \ , && h^{p,q}=h^{3-p,3-q} \ .
\end{align}
This implies that $h^{1,1}$ and $h^{2,1}$ determine all Hodge numbers of a Calabi-Yau threefold, which can be conveniently summarized in the Hodge diamond of Calabi-Yau threefolds 
\begin{equation} \label{eq:HodgeDiamond}
\begin{aligned}
   &\hodge{ }\hodge{ }\hodge{ }\hodge{h^{0,0}}\hodge{ }\hodge{ }\hodge{ } \\
   &\hodge{ }\hodge{ }\hodge{h^{1,0}}\hodge{ }\hodge{h^{0,1}}\hodge{ }\hodge{ } \\
   &\hodge{ }\hodge{h^{2,0}}\hodge{ }\hodge{h^{1,1}}\hodge{ }\hodge{h^{0,2}}\hodge{ } \\
   &\hodge{h^{3,0}}\hodge{ }\hodge{h^{2,1}}\hodge{ }\hodge{h^{1,2}}\hodge{ }\hodge{h^{0,3}} \\
   &\hodge{ }\hodge{h^{3,1}}\hodge{ }\hodge{h^{2,2}}\hodge{ }\hodge{h^{1,3}}\hodge{ } \\
   &\hodge{ }\hodge{ }\hodge{h^{3,2}}\hodge{ }\hodge{h^{2,3}}\hodge{ }\hodge{ } \\
   &\hodge{ }\hodge{ }\hodge{ }\hodge{h^{3,3}}\hodge{ }\hodge{ }\hodge{ } 
\end{aligned}\qquad\xrightarrow[\ \text{threefold}\ ]{\ \text{Calabi-Yau}\ }\qquad
\begin{aligned}
   &\hodge{ }\hodge{ }\hodge{ }\hodge{1}\hodge{ }\hodge{ }\hodge{ } \\
   &\hodge{ }\hodge{ }\hodge{0}\hodge{ }\hodge{0}\hodge{ }\hodge{ } \\
   &\hodge{ }\hodge{0}\hodge{ }\hodge{h^{1,1}}\hodge{ }\hodge{0}\hodge{ } \\
   &\hodge{1}\hodge{ }\hodge{h^{2,1}}\hodge{ }\hodge{h^{2,1}}\hodge{ }\hodge{1} \\
   &\hodge{ }\hodge{0}\hodge{ }\hodge{h^{1,1}}\hodge{ }\hodge{0}\hodge{ } \\
   &\hodge{ }\hodge{ }\hodge{0}\hodge{ }\hodge{0}\hodge{ }\hodge{ } \\
   &\hodge{ }\hodge{ }\hodge{ }\hodge{1}\hodge{ }\hodge{ }\hodge{ } 
\end{aligned} \ .
\end{equation}
From a physics point of view the Hodge numbers are important as they are in one-to-one correspondence with the number of linear independent harmonic forms, which then are needed to determine the massless modes in the Kaluza-Klein expansion of the anti-symmetric tensor fields (c.f.~section~\ref{sec:KKreduction}).


\section{BPS-Dp-branes in Calabi-Yau orientifolds} \label{sec:BPSCY}


So far D$p$-branes have been discussed as extended objects embedded in ten-dimen\-sional Minkowski space. In this ten-dimensional context D$p$-branes saturate a BPS-bound and hence reduce the amount of supersymmetry by one-half. Ultimately, however, we are interested in space-time filling D-branes which preserve some amount of supersymmetry in the context of Calabi-Yau compactifications. Therefore in section~\ref{sec:calcond} we examine the conditions for preserved supersymmetries in the presence of D-branes in Calabi-Yau manifolds and find that as in the flat case BPS-D-branes reduce supersymmetry by a factor one-half while fulfilling certain calibration conditions. In section~\ref{sec:Oplanes} the consistency conditions for space-time filling D-branes in Calabi-Yau manifolds are stated. Consistency requires the introduction of an additional ingredient, namely orientifold planes, which then further modify the D-brane calibration conditions.

\subsection{$\kappa$-Symmetry, BPS-branes and calibration conditions}
\label{sec:calcond}

In section~\ref{sec:CYcompact} we have argued that if the supersymmetry variations of all fermions vanish for a supersymmetry parameter then the chosen background preserves some amount of supersymmetry. If a super-D$p$-brane is included into a Calabi-Yau compactifications, the new fermionic degrees of freedom $\pair{\Theta}=(\Theta^1,\Theta^2)$ of the D-brane also vary with the supersymmetry transformations~$\pair{\eta}$ of eq.~\eqref{eq:susyparaCY}, but in general obey neither $\delta_{\epsilon^1}\pair{\Theta}=0$ nor $\delta_{\epsilon^2}\pair{\Theta}=0$. However, as discussed in section~\ref{sec:superaction} the super-D$p$-brane has an extra local fermionic gauge symmetry called $\kappa$-symmetry \cite{Hughes:1986fa,Aganagic:1996pe,Cederwall:1996ri,Bergshoeff:1996tu}. Hence supersymmetry is unbroken if it is possible to compensate the supersymmetry variation by a $\kappa$-symmetry gauge transformation \cite{Becker:1995kb,Bergshoeff:1997kr}
\begin{equation} \label{eq:BPSbrane} 
   \delta\pair{\Theta}=\delta_\epsilon\pair{\Theta}+\delta_\kappa\pair{\Theta}=0 \ ,
\end{equation} 
where $\epsilon$ is some linear combination of $\epsilon^1$ and $\epsilon^2$. If this condition can be fulfilled for some parameter $\epsilon$ then the super-D$p$-brane breaks only half of the supercharges and saturates a BPS bound. The details of this condition~\eqref{eq:BPSbrane} are elaborated in appendix~\ref{app:BPS}. The result of this analysis yields that for space-time filling\footnote{This work focuses on space-time filling D-branes in order to maintain four-dimensional Poincar\'e invariance.} D3- and D7-branes the condition \eqref{eq:BPSbrane} can be fulfilled for the linear combination of supersymmetry parameters given by the projection~$\mathcal{P}_\text{D3/D7}$
\begin{align} \label{eq:prD3D7}
   \eta=\mathcal{P}_\text{D3/D7}\pair{\eta} \ , &&
   \mathcal{P}_\text{D3/D7}\equiv\tfrac{1}{2}(\id+\check\sigma^2) \ , 
\end{align}
in terms of the Pauli matrices $\check\sigma^1$, $\check\sigma^2$, $\check\sigma^3$ which act on the four-dimensional supersymmetry parameter pair~$\pair{\eta}$ introduced in eq.~\eqref{eq:susyparaCY}. On the other hand for space-time filling D3- and D9-branes the preserved linear combination is given by the projection~$\mathcal{P}_\text{D5/D9}$
\begin{align} \label{eq:prD5D9}
   \eta=\mathcal{P}_\text{D5/D9}\pair{\eta} \ , &&
   \mathcal{P}_\text{D5/D9}\equiv\tfrac{1}{2}(\id+\check\sigma^1) \ , 
\end{align}
Note that the projectors~$\mathcal{P}_\text{D3/D7}$ and $\mathcal{P}_\text{D5/D9}$ are not compatible, e.g. they do not commute. Physically this means that it is not possible to preserve some supersymmetry in a Calabi-Yau background which contains simultaneously D3-/D7-branes and D5-/D9-branes. On the other hand for backgrounds with either D3-/D7-branes or D5-/D9-branes there is a linear combinations of $\pair{\eta}$ namely $\eta$ given by either \eqref{eq:prD3D7} or \eqref{eq:prD5D9} which fulfills the supersymmetry conditions~\eqref{eq:BPSbrane}.

As further analyzed in appendix~\ref{app:BPS} in addition to the projections~\eqref{eq:prD3D7} or \eqref{eq:prD5D9} the BPS D-branes must also be calibrated with respect to the K\"ahler form~\eqref{eq:Kform} of the ambient Calabi-Yau manifolds\footnote{For ease of notation we use the letter $\hat J$ for both the K\"ahler form $\hat J$ of the Calabi-Yau manifold~$Y$ and for the pull-back K\"ahler form $\iota^*\hat J$ with respect to the embedding map~$\iota$ of the internal worldvolume cycle. Similarly the metric $\hat g$ stands for $\hat g$ as well as $\iota^*\hat g$.}
\begin{equation} \label{eq:calisec}
   \dd^{p-3}\xi\sqrt{\det{\hat g}}=\frac{1}{\left(\tfrac{p-3}{2}\right)!} \hat J^\frac{p-3}{2} \ .
\end{equation}
This is a non-trivial condition on the D5- and D7-branes and which requires the internal cycles of the D5- and D7- to be holomorphically embedded into the ambient Calabi-Yau manifold~$Y$. 

So far the BPS conditions are derived without including the couplings to the closed string anti-symmetric NS-NS two-tensor~$B$ and without turning on internal D-brane fluxes~$f$. D-brane fluxes are topologically non-trivial backgrounds for the field strength~$F$ of the worldvolume gauge theory. Since we want to maintain four-dimensional Poincar\'e invariance we only turn on background fluxes~$f$ for the internal part of the D-brane worldvolume. Here we include the couplings to the NS-NS $B$-field and the possibility of non-trivial internal background fluxes~$f$ in the calibration condition, but postpone a detailed discussion of D-brane fluxes to section~\ref{sec:D7fluxes}. 

In this work we mainly concentrate on the D3/D7-brane case and state explicitly the calibration condition for BPS-D7-branes wrapped on the internal four-cycle~$S$. For space-time filling D7-branes the calibration condition including background fluxes~$f$ on the internal four-cycle~$S$ is worked out in detail in ref.~\cite{Marino:1999af}
\begin{align} \label{eq:cali2}
   \dd^4\xi \sqrt{\det \left(\hat g+\mathcal{F} \right)}
    =\frac{1}{2}e^{-\ii\theta}\left(\hat J+\ii\mathcal{F}\right)
     \wedge\left(\hat J+\ii\mathcal{F}\right) \ , &&
   \mathcal{F}=B-\ell f \ .
\end{align}
Here the real constant $\theta$ parametrizes the unbroken supersymmetry variation as a linear combination of $\epsilon^1$ and $\epsilon^2$ defined in eq.~\eqref{eq:susyparaCY}. $B$ is the anti-symmetric NS-NS two-tensor pulled back to the internal cycle~$S$, and $f$ are the internal D7-brane fluxes associated to the $U(1)$ field strength~$F$. The condition~\eqref{eq:cali2} must be realized on the whole cycle~$S$ in order for the D7-brane to be a BPS state.

Note that the left hand side of \eqref{eq:cali2} is real, and hence for vanishing $\mathcal{F}$, i.e. for vanishing $B$ and trivial background fluxes $f$, we find $\theta=0$ and therefore recover the calibration condition \eqref{eq:calisec}. For this particular case the unbroken supersymmetries are given by \eqref{eq:prD3D7}. Therefore the parameter~$\theta$ measures the flux-induced deviation from the condition \eqref{eq:prD3D7}. 

\subsection{Tadpoles and O-planes} \label{sec:Oplanes}

In the presence of space-time filling D$p$-branes their arise to two kinds of tadpoles, namely RR~tadpoles and NS-NS~tadpoles. The RR~tadpoles are due to their RR~charge~\eqref{eq:RRcharge} or more generally due to the couplings to the bulk RR~fields as captured in the Chern-Simons action of the D$p$-brane~\eqref{eq:CSAb}. These RR~tadpoles for D3- and D7-branes are readily deduces from the Chern-Simons action~\eqref{eq:CSAb} \cite{Blumenhagen:2002wn}. For space-time filling D7-branes wrapped on the internal Calabi-Yau four-cycles $S_i^{(7)}$ with internal two-form fluxes~$f_i^{(7)}$ and for space-time filling D3-branes located at $s^{(3)}_k$ in the internal Calabi-Yau space the tadpoles read\footnote{Bulk background fluxes do also contribute to the RR~tadpoles.}
\begin{equation} \label{eq:RRtadpoles}
\begin{split}
   &\sum_i \mu_7 \int_{\mathbb{R}^{3,1}\times S_{i}^{(7)}} C^{(8)} \ , \qquad\qquad
    \sum_i \mu_7\ell \int_{\mathbb{R}^{3,1}\times S_{i}^{(7)}} C^{(6)}\wedge f_i^{(7)} \ , \\
   &\sum_i \mu_7\ell^2 \int_{\mathbb{R}^{3,1}\times S_{i}^{(7)}} C^{(4)}\wedge
   f_i^{(7)}\wedge f_i^{(7)}+
   \sum_k \mu_3 \int_{\mathbb{R}^{3,1}\times \{s_k^{(3)}\}} C^{(4)} \ .
\end{split}
\end{equation}
These are RR~eight-form, RR~six-form and RR~four-form tadpoles resulting from D3- and D7-branes. Note that the internal D7-brane fluxes~$f_i^{(7)}$ contribute to the D5-brane and D3-brane RR~charge. Therefore D7-brane fluxes should be seen as D3-brane charges smeared out over the worldvolume of the D7-branes \cite{Douglas:1995bn}. 

Consistency requires these RR~tadpoles~\eqref{eq:RRtadpoles} to disappear \cite{Blumenhagen:2002wn}. Physically this can be understood as follows. Since the considered D-branes are space-time filling extended objects, their RR-charges are sources in the compact internal Calabi-Yau space. However, due to the compactness of the internal space such a configuration is only consistent if the RR~sources are compensated by corresponding RR~sinks. This is a generalization of the Gauss' law in electrodynamics applied to charges on compact manifolds.

In addition to RR~tadpoles these space-time filling D-branes also generate NS-NS~tadpoles. While the appearance of the RR~tadpoles render the theory inconsistent, the divergencies of NS-NS~tadpoles give rise to potentials for NS-NS~fields \cite{Dudas:2000ff,Blumenhagen:2001te} and can be absorbed in the background fields via the Fischler-Susskind mechanism \cite{Fischler:1986ci,Fischler:1986tb}. The NS-NS tadpoles arise form the couplings of D-branes to the NS-NS~graviton and the NS-NS~dilaton due to their energy density. If the NS-NS tadpoles do not vanish it is an indication that the chosen background does not correspond to a stable ground state. In this case the Fischler-Susskind mechanism is a process which drives the theory towards a stable vacuum configuration.

Thus in order to cancel both the RR~tadpoles and the NS-NS~tadpoles we need sources with negative RR~charges and negative tension, i.e. `negative NS-NS charges'. Fortunately string theory provides for extended objects with these properties, which are called orientifold planes or O-planes \cite{Sagnotti:1987tw,Horava:1989vt,Dai:1989ua,Gimon:1996rq}. Analogously to the nomenclature of D$p$-branes an O$p$-plane denotes a $(p+1)$-dimensional orientifold plane.

These O-planes arise in orientifold superstring theories through gauging a discrete $\mathbb{Z}_2$-symmetry which contains the worldsheet parity transformation~$\Omega_p$ \cite{Gimon:1996rq}. This parity transformation exchanges right- and left-movers of the closed superstring worldsheets. Due to the same chiralities in the right- and left-moving sectors of type~IIB superstring theory the parity transformation~$\Omega_p$ is a symmetry of type~IIB superstring theory itself. Gauging this symmetry leads to the unoriented type~I superstring theory \cite{Polchinski:1998rq,Polchinski:1998rr}, which can be viewed as an orientifold superstring theory with O9-planes.

However, for type~IIB string theory compactified on Calabi-Yau manifolds generically there is a greater variety of discrete $\mathbb{Z}_2$ parity symmetries which can be gauged.\footnote{In general the orientifold group can be more complicated \cite{Gimon:1996rq}.} For the $\mathbb{Z}_2$-generator we take as an ansatz $\mathcal{O}=\Omega_p h$, where $h$ in general also contains a non-trivial $\mathbb{Z}_2$ action acting geometrically on the internal Calabi-Yau space~$Y$. This geometric part in the following is specified by the diffeomorphic map~$\sigma$ on $Y$. In order for this map to generate a $\mathbb{Z}_2$-action the diffeomorphism~$\sigma$ must clearly be an involution, i.e. $\sigma^2=\id$. In addition in order for the orientifold action to preserve some space-time supersymmetry, the involution must be isometric and holomorphic \cite{Brunner:2003zm}. This in particular implies for the K\"ahler form of the Calabi-Yau manifold~$Y$ to obey 
\begin{equation}
   \hat J=\sigma^* \hat J \ .
\end{equation}
As a consequence of the holomorphicity of $\sigma$ the pullback~$\sigma^*$ always maps $(p,q)$-forms of $Y$ to $(p,q)$-forms. This is also true on the level of cohomology as the Dolbeault operator~$\bar\partial$ commutes with the pullback of $\sigma^*$. This implies
\begin{equation} \label{eq:CohomSplit}
   H^{(p,q)}_{\bar\partial}(Y) =
   H^{(p,q)}_{\bar\partial,+}(Y)\oplus H^{(p,q)}_{\bar\partial,-}(Y)\ .
\end{equation}
In particular the unique holomorphic $(3,0)$-form~$\Omega$ must be an eigenform of $\sigma^*$ and due to the involutive property of $\sigma^*$ with possible eigenvalues $\pm 1$. To infer the relevance of the sign of the eigenvalue it is necessary to go through some rather technical considerations which are relegated to appendix~\ref{app:orient}. Here, instead, we present the result of this analysis.

The precise structure of the orientifold projection~$\mathcal{O}$ is determined by the eigenvalue of the holomorphic three-form with respect to $\sigma^*$ according to \eqref{eq:orientproj}. Furthermore, in order to preserve some supersymmetry in the four-dimensional effective theory it is necessary that the orientifold projection is in accord with the supersymmetry preserved by the space-time filling D-branes. This implies that the orientifold projection must keep the four-dimensional gravitinos in the spectrum which correspond to the supersymmetries preserved by the D-branes. By comparing the projections \eqref{eq:prD3D7} and \eqref{eq:prD3D7} with the gravitino projection~\eqref{eq:4Dgravproj} one deduces that for D3/D7-brane systems the $\mathbb{Z}_2$ orientifold generator should read
\begin{align} \label{eq:projO3O7}
   \mathcal{O}=(-1)^{F_L}\Omega_p\sigma^* \ , && \Omega=-\sigma^*\Omega \ ,
\end{align}
whereas for D5/D9-systems one obtains
\begin{align}
   \mathcal{O}=\Omega_p\sigma^* \ , && \Omega=\sigma^*\Omega \ .
\end{align}
Here $F_L$ denotes the space-time fermion number for the left-movers.

The O-planes in these theories arise as the fix-point locus of the geometric involution~$\sigma$. In the vicinity of such a fixed point~$p_f$ in the internal Calabi-Yau space and by choosing appropriate coordinates the action of $\sigma^*$ on $\Omega=\dd z^i\wedge\dd z^j\wedge\dd z^k|_{p_f}$ can be written as
\begin{equation}
   \sigma^* (\dd z^i\wedge\dd z^j\wedge\dd z^k)|_{p_f}=\pm\dd z^i\wedge\dd z^j\wedge\dd z^k|_{p_f} \ .
\end{equation}
This means that at the fixed point~$p_f$ either an even or an odd number of differentials~$\dd z^i$ have to change the sign. As a consequence for an even number, that is for $\sigma^*\Omega=\Omega$, the internal fixed point locus is two- or six-dimensional, whereas for an odd number corresponding to $\sigma^*\Omega=-\Omega$, the fixed point locus is zero- or four-dimensional. Therefore the orientifold projection in accord with D5/D9-branes generates O5/O9-planes whereas the orientifold projection for D3/D7-branes allows for O3/O7-planes. In this work we mainly examine D3/D7-brane systems and therefore from hereon we concentrate on the O3/O7-plane case, which means that we use the orientifold projection \eqref{eq:projO3O7}. 

Originally the appearance of tadpoles in the presence of space-time filling D3/D7-branes led us to Calabi-Yau orientifold compactifications. With the introduction of O3/O7-planes there are additional contributions to the RR~tadpoles \eqref{eq:RRtadpoles} and the NS-NS tadpoles due to the charge and negative tension of the orientifold planes~\cite{Giddings:2001yu}. The contributions of the O3/O7-planes to the RR~tadpole condition~\eqref{eq:RRtadpoles} is obtained form the analog of the Chern-Simons action for orientifold planes \cite{Stefanski:1998yx,Scrucca:1999uz}, which encodes their share of RR~charges also in terms of topological expressions. For O7-planes wrapped on $O_j^{(7)}$ and O3-planes located at $o_l^{(3)}$ we find altogether two tadpole cancellation conditions\footnote{The indices $i$, $j$ , $k$ and $l$ account for several D7-branes, O7-planes, D3-branes and O3-planes respectively. $S_i^{(7)}$ and $O_j^{(7)}$ are four-cycles whereas $s_k^{(3)}$ and $o_l^{(3)}$ are points in the Calabi-Yau manifold.} \cite{Blumenhagen:2002wn}
\begin{align} \label{eq:RRtad}
   0&=\sum_i \mu_7 \int_{\mathbb{R}^{3,1}\times S_{i}^{(7)}} C^{(8)}+
   \sum_j \nu^{j}_7 \int_{\mathbb{R}^{3,1}\times O_j^{(7)}} C^{(8)} \ , \nonumber \\
   0&=\sum_i \mu_7\ell^2 \int_{\mathbb{R}^{3,1}\times S_{i}^{(7)}} C^{(4)}\wedge
   f_i^{(7)}\wedge f_i^{(7)} \\ 
   &\qquad\qquad\qquad+\sum_k \mu_3 \int_{\mathbb{R}^{3,1}\times \{s_k^{(3)}\}} C^{(4)} 
   +\sum_l \nu^{l}_3 \int_{\mathbb{R}^{3,1}\times \{o_l^{(3)}\}} C^{(4)} \ . \nonumber
\end{align}
Here $\nu^{j}_7$ and $\nu^{l}_3$ are the RR~charges of the O-planes. Note that there are no six-form tadpoles anymore because the transition to orientifold Calabi-Yau compactifications also requires for each D-brane to include its image-D-brane with respect to the involution~$\sigma$.\footnote{This is the orientifold picture in the covering space.} This requirement automatically cancels all RR~six-form tadpoles in eq.~\eqref{eq:RRtadpoles} \cite{Blumenhagen:2001te}. 

In the following as we are not considering a specific orientifold compactification we cannot explicitly check the conditions \eqref{eq:RRtad}. Instead we assume that we have appropriately chosen a Calabi-Yau manifold~$Y$ with involution $\sigma$ and with D3/D7-branes such that the RR~tadpole conditions \eqref{eq:RRtad} are fulfilled.

In ref.~\cite{Blumenhagen:2001te} it is argued that all NS-NS~tadpoles arise as derivatives of a D-term scalar potential with respect to the corresponding NS-NS~fields. In the supersymmetric case the NS-NS~tadpoles vanish as they are related to the RR~tadpole conditions via supersymmetry. This corresponds to the vanishing of the D-term \cite{Blumenhagen:2002wn,Lust:2004fi} and thus the potential arising form the Dirac-Born-Infeld action \eqref{eq:DBIab} of the D-branes has to be canceled by the negative tension of the orientifold planes. If the NS-NS~tadpoles do not vanish a D-term is induced leading generically to an unstable background. We discuss the appearance of various D-terms as we go along.

Before concluding this section we come back to the calibration condition~\eqref{eq:cali2} for space-time filling D7-branes. In section~\ref{sec:calcond} we argued that the real parameter~$\theta$ in \eqref{eq:cali2} parametrizes the linear combination of supersymmetry parameters. In the absence of O-planes the constant $\theta$ can freely be adjusted such that half of the amount of supersymmetry is preserved. However, as we have seen in the presence of orientifold planes only a particular linear combination of supercharges is preserved and which corresponds to $\theta=0$. Therefore the calibration condition \eqref{eq:cali2} for orientifold theories becomes \cite{Jockers:2004yj}
\begin{align} \label{eq:cali3}
   \dd^4\xi\sqrt{\det\left(\hat g+\mathcal{F}\right)}
     =\tfrac{1}{2} \hat J\wedge \hat J -\tfrac{1}{2}\mathcal{F}\wedge\mathcal{F} \ , &&
   \mathcal{F}=B-\ell f \ .
\end{align}
Note, however, that this calibration condition in the context of orientifolds is only valid for
\begin{equation} \label{eq:Dterm0}
   \hat J\wedge\mathcal{F}=0 \ ,
\end{equation}
which is the condition for the D7-brane to be calibrated with $\theta=0$ or in other words to be calibrated with \eqref{eq:cali3}. If \eqref{eq:Dterm0} is not fulfilled than supersymmetry is spontaneously broken by the D7-brane configuration. This is related to the $\omega$-stability condition $\hat J\wedge\mathcal{F}=\text{const.}\ \hat J\wedge \hat J$ of refs.~\cite{Brunner:1999jq,Harvey:1996gc} which is imposed by supersymmetry. In orientifolds $\mathcal{F}$ is odd and $\hat J$ is even with respect to the involution~$\sigma$ whence $\hat J\wedge\mathcal{F}=0$ becomes the $\omega$-stability condition. Moreover it is argued in ref.~\cite{Brunner:1999jq}, that $\omega$-stability gives rise to a D-term constraint in the low energy effective action, i.e. if supersymmetry is broken $\omega$-stability is not fulfilled and the non-vanishing D-term breaks supersymmetry spontaneously. Conversely, a $\omega$-stable configuration corresponds to a vanishing D-term in field theory. Thus in the low energy effective theory we expect a D-term of the form 
\begin{equation} \label{eq:Dterm1}
   \text{D}\sim \int_{S} \hat J\wedge\mathcal{F} \ ,
\end{equation}
where $S$ is the four-cycle in the internal space~$Y$, which is wrapped by the D7-brane. The corresponding D-term scalar potential $V_\text{D}$ has the form
\begin{equation} \label{eq:D1}
   V_\text{D} \sim \left(\int_S \hat J\wedge\mathcal{F}\right)^2 \ .
\end{equation}

In the forthcoming derivation of the low energy effective action we use the calibration condition~\eqref{eq:cali3}, but we are aware that we do not capture terms proportional to $\hat J\wedge\mathcal{F}$, in particular we do not obtain directly the D-term scalar potential~\eqref{eq:D1}, but instead have to rely on supersymmetry to rigorously compute this potential.


\section{D-brane spectrum in Calabi-Yau orientifolds} \label{sec:D3D7spec}


In this section we compute the spectrum arising from D-branes in orientifold theories, namely we examine the spectra arising from D3-brane and D7-brane systems. The low energy spectrum of a stack of D3-branes does not dependent on the precise structure of the internal space, which is due to the fact that space-time filling D3-branes only constitute a point in the internal compactification space. On the other hand, however, the spectrum of a D7-brane depends on the topology of the four-cycle in the internal space, which is wrapped by the internal part of the D7-brane.

\subsection{D3-brane spectrum in Calabi-Yau orientifolds}

Before entering the discussion of the spectrum arising from a stack of $N$ D3-branes let us clarify the geometric picture which one should have in mind. The internal part of the stack of D3-branes comprises just a point~$p$ in the Calabi-Yau orientifold~$Y$. However, in addition to the point~$p$ there is also an image stack of $N$ D3-branes at the orientifold image $\tilde p=\sigma(p)$ of $p$. If the stack of D3-branes resides on an orientifold fixed point, i.e. $\tilde p=p$, then the gauge group on the worldvolume of the D3-brane is either $SO(N)$ or $USp(N)$.\footnote{The precise gauge group depends on the type of the orientifold plane \cite{Witten:1997bs,Brunner:2003zm}.} Otherwise the gauge group of the stack of $N$ D3-branes is $U(N)$. In the following we concentrate on the latter case and further assume that the D3-branes are separated far enough from their image D3-branes, i.e. $\dist{p}{\tilde p}\gg \sqrt{\alpha'}$, such that no additional light modes arise from open strings stretching from the D3-branes to the image D3-branes. Note that the above description captures the geometry in the covering space of the orientifold theory. On the orientifold space $Y/\mathcal{O}$ the D3-branes coincide with their image D3-branes and combine to a single stack of D3-branes. 

For simplicity we first focus on the four-dimensional massless spectrum of a single space-time filling D3-brane located at $p$. First of all the four-dimensional bosonic spectrum consists of the vector boson $A_\mu(x)$ of the $U(1)$ gauge theory localized on the worldvolume of the D3-brane. Second there arise D3-brane matter fields which describe the geometric fluctuations~$\dbt(x)$ of the locus of the D3-brane in the internal Calabi-Yau space~$Y$. These fluctuations~$\dbt(x)$ are space-time dependent vectors in the tangent space $\tbundlep{p}{Y}$ of the internal space.\footnote{Alternatively, we can also view the D3-brane fluctuations as section of $H^0(p,\nbundle{p})$. This interpretations is useful as it naturally generalizes to space-time filling D7-brane (c.f. section~\ref{sec:D7branespectrum}).} Since $\dim_\mathbb{R} \tbundlep{p}{Y}=6$ the fluctuations~$\dbt(x)$ gives rise to six real scalar fields~$\dbt^n(x)$ in four dimensions, i.e.
\begin{equation} \label{eq:D3Phireal}
   \dbt(x)=\dbt^n(x)\:\partial_n|_p \ , \qquad n=1,\ldots,6 \ .
\end{equation}
Here $\partial_n|_p$ denotes a basis of $\tbundlep{p}{Y}$.

According to section~\ref{sec:Oplanes} D3-branes in Calabi-Yau orientifolds with O3/O7-planes preserve $\mathcal{N}=1$ supersymmetry and thus the bosonic fields appear in $\mathcal{N}=1$ supermultiplets. The $U(1)$ gauge boson $A_\mu(x)$ is the bosonic part of a $U(1)$ vector multiplet whereas the six real scalar fields $\dbt^n(x)$ combine to three complex scalar fields $\dbt^i(x)$ with respect to the complex structure inherited from the ambient Calabi-Yau space~$Y$.\footnote{The precise complex structure for the fields~$\dbt^n$ will be discussed in section~\ref{sec:D3chiral}.} Then in terms of complex fields \eqref{eq:D3Phireal} becomes
\begin{equation}
   \dbt(x)=\dbt^i(x)\:\partial_i|_p+\bar\dbt^{\bar\jmath}(x)\:\partial_{\bar\jmath}|_p \ ,
           \qquad i,\bar\jmath=1,2,3 \ .
\end{equation}
As discussed in section~\ref{sec:CYcompact} the internal Calabi-Yau manifold~$Y$ has $SU(3)$-structure, the frame bundle of $Y$ admits a $SU(3)$ principal subbundle. Then one can view the complex fields~$\dbt^i(x)$ as defined with respect to a basis $\partial_i|_p$ of this subbundle at $p$. The complex fields~$\tilde\dbt^i(x)$ with respect to a different basis $\tilde\partial_i|_p$ are related to $\dbt^i(x)$ via a $SU(3)$ transformation. Therefore the fields $\dbt^i(x)$ transform in the fundamental representation $\rep{3}$ (and analogously $\bar\dbt^{\bar\jmath}$ in the anti-fundamental representation $\rep{\bar 3}$) with respect to the $SU(3)$ structure group on the Calabi-Yau space~$Y$.

The next task is to determine the fermionic fields in the $\mathcal{N}=1$ supermultiplets. They arise from fermionic fluctuations of the embedding of the super-D3-brane action \eqref{eq:SDBI} in the type~IIB supergravity supermanifold $M^{9,1|2}$. Therefore for the ansatz \eqref{eq:CYansatz} the fermionic spectrum is obtained by decomposing the odd coordinates~$\pair{\theta}$ transforming as $\spinrep{16'}$ of $SO(9,1)$ into representations of the subgroups
\begin{equation} \label{eq:decompCY}
   SO(9,1)\rightarrow SO(3,1)\times SO(6)
          \rightarrow SO(3,1)\times SU(3) \ ,
\end{equation}
where the first decomposition of the ten-dimensional Lorentz group is due to the compactification ansatz while the $SU(3)$ structure group of complex threefolds brings about the second decomposition. Correspondingly the spinor representation $\spinrep{16'}$ splits into 
\begin{equation} \label{eq:CYdecomp16}
   \spinrep{16'}\rightarrow (\spinrep{2},\spinrep{\bar 4})
                            \oplus (\spinrep{\bar 2},\spinrep{4})
                \rightarrow (\spinrep{2},\spinrep{\bar 3})
                            \oplus (\spinrep{2},\spinrep{\bar 1}) 
                            \oplus (\spinrep{\bar 2},\spinrep{\bar 3})
                            \oplus (\spinrep{\bar 2},\spinrep{1}) \ .
\end{equation}
$\spinrep{2}$, $\spinrep{\bar 2}$ are the two Weyl spinors of $SO(3,1)$, $\spinrep{4}$, $\spinrep{\bar 4}$ are the two Weyl spinors of $SO(6)$, $\spinrep{3}$, $\spinrep{\bar 3}$ are the fundamentals and anti-fundamentals of $SU(3)$, and finally $\spinrep{1}$, $\spinrep{\bar 1}$ are $SU(3)$ singlets.

Note, however, as discussed in section~\ref{sec:superaction} the super-D3-brane action possesses the local fermionic $\kappa$-symmetry, which reduces the fermionic degrees of freedom by one-half. As a consequence the physical fermionic degrees of freedoms arise from the decomposition of a single odd coordinate transforming in the $\spinrep{16'}$ of $SO(9,1)$ in agreement with the fermionic open string spectrum \eqref{eq:Dpfermionicspec}. Hence, from \eqref{eq:CYdecomp16} one readily reads off the massless fermionic four-dimensional spectrum arising from the D3-branes. On the one hand there is a Weyl fermion~$\lambda(x)$, which arises from the $SU(3)$ singlet and is identified as the superpartner of $A_\mu(x)$ with the gaugino in the $\mathcal{N}=1$ vector multiplet. On the other hand there are the three Weyl fermions~$\psi^i(x)$ which appear in the fundamental $\rep{3}$ of $SU(3)$. They are the superpartners of the bosonic fields~$\dbt^i$ and form three $\mathcal{N}=1$ chiral multiplets. Hence the explicit decomposition of fermionic D3-brane fluctuations~$\Theta(\xi)$ reads
\begin{equation} \label{eq:D3fermions}
   \Theta(\xi)=N_\lambda\:\lambda(x)\otimes\check\singspin^\dagger_p
                +\bar N_{\bar\lambda}\:\bar\lambda(x)\otimes\check\singspin_p 
               +N_{\chi^i}\:\chi^i(x)\otimes\gamma_i\singspin_p
                +\bar N_{\bar\chi^{\bar\jmath}}\:\bar\chi^{\bar\jmath}(x)\otimes
                \gamma_{\bar\jmath}\singspin^\dagger_p \ ,
\end{equation}
where $\check\singspin_p$ and $\check\singspin^\dagger_p$ are the covariantly constant spinor singlets \eqref{eq:checksingspin} of the Calabi-Yau manifold~$Y$ restricted to the D3-brane locus~$p$. $N_\lambda$ and $N_{\chi^i}$ are normalization constants to be determined in section~\ref{sec:D3fermionic}.

This completes the analysis of a single space-time filling D3-brane. For a stack of $N$ D3-branes the $U(1)$ gauge theory is enhanced to $U(N)$ \cite{Witten:1995im}, and then the gauge boson $A_\mu(x)$ and the gaugino $\lambda(x)$ transform in the adjoint representation of $U(N)$. Moreover, the bosonic and fermionic fluctuations become `matrix valued fields' and also transform in the adjoint representation of $U(N)$ \cite{Douglas:1997ch,Douglas:1997sm}. This enhancement of the D3-brane fields is discussed in some detail in section~\ref{sec:openstr} from an open string perspective.

\begin{table}
\begin{center}
\begin{tabular}{|c|c|c|c|}
   \hline
      \bf bosonic fields &  \bf fermionic fields &  \bf multiplet  & \bf multiplicity
      \rule[-1.5ex]{0pt}{4.5ex} \\
   \hline
   \hline
      $A_\mu$ & $\lambda, \bar\lambda$ & vector & $1$ \rule[-1.5ex]{0pt}{4.5ex} \\
   \hline
      $\dbt^i, \bar\dbt^{\bar\jmath}$ & $\psi^i, \bar\psi^{\bar\jmath}$ & chiral & 
      $3$ \rule[-1.5ex]{0pt}{4.5ex} \\ 
   \hline
\end{tabular} 
\caption{Massless D3-brane spectrum in $\mathcal{N}=1$ multiplets} \label{tab:D3spec} 
\end{center}
\end{table}
In Table~\ref{tab:D3spec} the massless four-dimensional spectrum is summarized in terms of the resulting four-dimensional $\mathcal{N}=1$ multiplets, namely the vector multiplet $(A_\mu,\lambda)$ and the three chiral multiplets $(\dbt^i,\psi^i)$, which all transform in the adjoint representation of $U(N)$. Thus the D3-brane spectrum is non-chiral and can be assembled in a single $U(N)$ $\mathcal{N}=4$ gauge multiplet.\footnote{$\mathcal{N}=4$ gauge multiplets in four dimensions consist of a vector, four Weyl fermions and six real scalars, which all transform in the adjoint representation of the gauge group. Under the $\mathcal{N}=4$ R-symmetry group $SU(4)_R$ the vector is a singlet, the Weyl fermions transform in the fundamental~$\rep{4}$, and the scalars transform in the rank 2 anti-symmetric~$\rep{[6]}$.} This is not very surprising as type~IIB orientifold compactifications on a six torus with space-time filling D3-branes preserve $\mathcal{N}=4$ in four dimensions. In this case both the bulk spectrum and the D3-brane spectrum assemble themselves into $\mathcal{N}=4$ supergravity multiplets. The spectrum of a stack of D3-branes depends only on the tangent space in the vicinity of $p$. Locally, however, the tangent spaces of all six-dimensional manifolds are isomorphic, and hence the spectrum arising from a stack of space-time filling D3-branes in type IIB orientifolds is always $\mathcal{N}=4$ supersymmetric.\footnote{This does not hold for D3-branes at singularities\cite{Douglas:1996sw,Douglas:1997de}.} However, as we will see in the next chapter this does not imply that the Yang-Mills theories on the worldvolume of the D3-brane is $\mathcal{N}=4$ since also the metric of the manifold in the vicinity of $p$ enters in the effective gauge theory.

\subsection{D7-brane spectrum in Calabi-Yau orientifolds} \label{sec:D7branespectrum}

The geometry of space-time filling D7-brane in Calabi-Yau orientifold compactifications is somewhat richer compared to the geometry of space-time filling D3-branes. This is due to the fact that the D7-brane wraps a four-cycle, whereas a D3-brane simply constitutes a point in the internal space. The internal four-cycle in the Calabi-Yau manifold~$Y$, which is wrapped by the worldvolume of the D7-brane, is denoted by $S^{(1)}$. Since we are working with an orientifold theory we must in addition to the D-brane on $S^{(1)}$ also include its image under the orientifold involution~$\sigma$, i.e. we have an image D7-brane on the Calabi-Yau manifold~$Y$ wrapped on the four-cycle $S^{(2)}=\sigma(S^{(1)})$. Hence it is convenient to introduce the four-cycle $S^\Lambda$ which is the union of the cycles $S^{(1)}$ and $S^{(2)}$ in the Calabi-Yau manifold~$Y$. $S^\Lambda$ obeys
\begin{equation}
  \sigma(S^\Lambda)=S^\Lambda \ .
\end{equation}
The Poincar\'e dual two-form $\omega_\Lambda$ of $S^{\Lambda}$ is an element of $H^2_+(Y)$ in terms the parity graded cohomology groups defined in eq.~\eqref{eq:CohomSplit}. By referring to the D7-brane we mean in the following the object which wraps the internal cycle $S^{\Lambda}$ and thus describes both the D7-brane and its image of the Calabi-Yau orientifold. For later convenience we further define $S^P$ as the union of the cycle $S^{(1)}$ and its orientation reversed image $-S^{(2)}$. This cycle obeys 
\begin{equation}
   \sigma(S^P)=-S^P \ ,
\end{equation}
and has a Poincar\'e dual two-form $\omega_P$ in $H^2_-(Y)$. In Table~\ref{tab:cycles} all these different D7-brane four-cycles are listed with their associated Poincar\'e dual two-forms.
\begin{table}
\begin{center}
\begin{tabular}{|c|c|c|c|}
   \hline
      \bf description  &  \bf cycle  &  \bf relation  &  \bf Poincar\'e dual
      \rule[-1.5ex]{0pt}{4.5ex} \\
   \hline
   \hline
      D7-brane (cov. space)  &  $S^{(1)}\in H_4(Y,\mathbb{Z})$  
      &  $\tfrac{1}{2}(S^\Lambda + S^P)$ 
      &  $\omega_{(1)}\in H_{\bar\partial}^{(1,1)}(Y)$ 
      \rule[-1.5ex]{0pt}{4.5ex} \\
   \hline
      D7-image-brane (cov. space)  &  $S^{(2)}\in H_4(Y,\mathbb{Z})$  
      &  $\tfrac{1}{2}(S^\Lambda - S^P)$ 
      &  $\omega_{(2)}\in H_{\bar\partial}^{(1,1)}(Y)$
      \rule[-1.5ex]{0pt}{4.5ex} \\
   \hline
      D7-brane (orientifold $Y$)  &  $S^\Lambda\in H_4(Y,\mathbb{Z})$  
      &  $S^{(1)}+S^{(2)}$ 
      &  $\omega_\Lambda\in H_{\bar\partial,+}^{(1,1)}(Y)$ 
      \rule[-1.5ex]{0pt}{4.5ex} \\
   \hline
      D7-pair (opp. orientation)  &  $S^P\in H_4(Y,\mathbb{Z})$  
      &  $S^{(1)}-S^{(2)}$ 
      &  $\omega_P\in H_{\bar\partial,-}^{(1,1)}(Y)$
      \rule[-1.5ex]{0pt}{4.5ex} \\      
   \hline
\end{tabular} 
\caption{D7-brane cycles} \label{tab:cycles} 
\end{center}
\end{table}
Note that this is the geometry in the covering space of the orientifold theory. As in the case of D3-branes the D7-brane and its image D7-brane coincide and represent a single object in the orientifold covering space~$Y/\mathcal{O}$.

In the above analysis we have implicitly assumed that the D7-brane does not coincide with any orientifold O7-plane, because this would imply that $S^{(1)}$ and $S^{(2)}$ represent the same cycle and that the gauge group of the worldvolume theory is $SO(N)$ or $USp(N)$ \cite{Polchinski:1998rq,Polchinski:1998rr}. Additionally we require that the involution $\sigma$ does not have any fixed points in $S^{(1)}$ since this would give rise to extra massless states in the twisted open string sector \cite{Berkooz:1996km}.

The bosonic spectrum of the D7-brane is comprised of two parts. The first part corresponds to fluctuations of the embedding of the internal four-cycle $S^\Lambda$ in the Calabi-Yau orientifold~$Y$, and the second part describes Wilson lines of the $U(1)$ gauge field on the four-cycle $S^\Lambda$. Both types of degrees of freedom give rise to bosonic components of chiral multiplets in the effective four dimensional low energy theory. The former complex bosons are members of the `matter' multiplets and we denote them by $\dbs(x)$. The latter Wilson line moduli are denoted by $a(x)$.

The fluctuations of the D7-brane arise as deformations of the worldvolume in the normal directions with respect to the ambient space. Moreover, since the D7-brane is space-time filling all normal directions lie in the internal Calabi-Yau space~$Y$. Therefore the `matter' multiplets arise from an appropriate expansion of the sections $\dbs(x,y)$ of the normal bundle $\nbundle{S^\Lambda}$. On the other hand the Wilson line moduli fields arise from non-trivial background configurations of the gauge boson $A(x,y)$ and therefore are related to non-contractible loops in the internal cycle~$S^\Lambda$. The massless modes of the fields~$\dbs(x,y)$ and $A(x,y)$ are in one-to-one correspondence with global sections of the cohomology groups\footnote{In general, D-branes are described by sheaves that are supported on the worldvolume of the brane. Then the spectrum of marginal open string modes of strings stretching from a D-brane specified by the sheaf $\sheaf{E}$ to a D-brane specified by the sheaf $\sheaf{F}$ is given by the $\Ext$-group $\Ext^1(\sheaf{E},\sheaf{F})$ \cite{Katz:2002gh}. However, if the tangent bundle of the Calabi-Yau restricted to the wrapped four-cycle splits holomorphically, which we always assume in the following, the $\Ext$-group reduces to the sheaf cohomology description, and in the case of open strings with both ends on a single brane, we have the above spectrum \eqref{eq:D7spec0}.} \cite{Witten:1992fb,Katz:2002gh,Lerche:2002yw,Jockers:2004yj}
\begin{align} \label{eq:D7spec0}
   \dbs\in H^0(S^\Lambda,\nbundle{S^\Lambda}) \ , &&
   a \in H^{(0,1)}_{\bar\partial}(S^\Lambda) \ ,
\end{align}
where $H^0(S^\Lambda,\nbundle{S^\Lambda})$ is the space of global sections of the holomorphic normal bundle of $S^\Lambda$. 

Note, however, that in \eqref{eq:D7spec0} the truncation due to the orientifold projection is not yet taken into account. In order to truncate the spectrum consistent with the orientifold projection $\mathcal{O}$ the action of $\mathcal{O}$ on the open string states has to be examined. The result of this technical analysis yields the truncation of the cohomology groups~\eqref{eq:D7spec0} to \cite{Pradisi:1988xd,Gimon:1996rq,Douglas:1996sw,Jockers:2004yj}
\begin{align} \label{eq:D7spec}
   \dbs\in H^0_+(S^\Lambda,\nbundle{S^\Lambda}) \ , &&
   a \in H^{(0,1)}_{\bar\partial,-}(S^\Lambda) \ .
\end{align}
More generally for a stack of $N$ D7-branes wrapped on $S^\Lambda$ the gauge theory on the worldvolume of these branes is enhanced to $U(N)$ \cite{Witten:1995im}. As a consequence the massless fields $\dbs$ and $a$ transform in the adjoint representation of $U(N)$, i.e. $\dbs$ is a $U(N)$ Lie algebra valued section of the normal bundle and $a$ is a $U(N)$ Lie algebra valued one-form \cite{Douglas:1997zw,Douglas:1997ch,Douglas:1997sm}. 

In the following we consider just a single D7-brane wrapped on $S^\Lambda$ with the spectrum given by \eqref{eq:D7spec}. In this case the expansion of $\dbs(x,y)$ into massless four-dimensional modes yields
\begin{equation} \label{eq:D7fluct}
   \dbs(x,y)=\dbs^A(x)\:s_A(y)+\bar\dbs^{\bar A}(x)\:\bar s_{\bar A}(y) \ ,
   \quad A=1,\ldots,\dim H^0_+(S^\Lambda,\nbundle{S^\Lambda}) \ , 
\end{equation}
where $s_A$ is a basis of $H^0_+(S^\Lambda,\nbundle{S^\Lambda})$. As discussed in refs.~\cite{Griffiths:1978,Lerche:2002yw}, we can map sections $\dbs$ of $H^0_+(S^\Lambda,\nbundle{S^\Lambda})$ isomorphically to $H_{\bar \partial,-}^{(2,0)}(S^\Lambda)$ via the Poincar\'e residue map. In practice this map is simply given by contracting $\dbs$ with the holomorphic three-form $\Omega$ of the Calabi-Yau manifold~$Y$ \cite{Lerche:2002yw}
\begin{equation} \label{eq:PR}
   \Omega : \ H^0_+(S^\Lambda,\nbundle{S^\Lambda}) \rightarrow
      H_{\bar\partial,-}^{(2,0)}(S^\Lambda),\ \dbs \mapsto \ins{\dbs} \Omega \ .
\end{equation}
Hence the Poincar\'e residue map also allows us to rewrite the basis elements $s_A$ and $\bar s_{\bar A}$ into basis elements $\tilde s_A$ and $\tilde s_{\bar B}$ of $H_{\bar\partial,-}^{(2,0)}(S^\Lambda)$ and $H_{\bar\partial,-}^{(0,2)}(S^\Lambda)$, and then the expansion \eqref{eq:D7fluct} can be mapped to an expansion in two-forms of $S^\Lambda$. 

Analogously for the $U(1)$ gauge boson $A(x,y)$ the expansion becomes 
\begin{equation} \label{eq:D7wilson}
   A(x,y)=A_\mu(x)\dd x^\mu\:P_-(y)+a_I(x)\:A^I(y)+\bar a_{\bar J}(x)\:\bar A^{\bar J}(y) \ , 
\end{equation}
with $A^I$ as a basis of $H^{(0,1)}_{\bar\partial,-}(S^\Lambda)$. $P_-$ is the harmonic zero form of $S^\Lambda$ given by
\begin{align} \label{eq:P}
   P_-(y)=\begin{cases} 1 & y\in S^{(1)} \\ -1 & y\in S^{(2)} \end{cases} \ , &&
   P_-\in H_-^0(S^\Lambda) \ .
\end{align}
Note that in addition to the Wilson line moduli~$a_I(x)$ the expansion yields the four-dimensional gauge boson~$A_\mu(x)$.

In Table~\ref{tab:spec} the massless open string spectrum resulting from the D7-brane is summarized. The table shows the cohomology groups, which describe the spectrum in geometric terms, and lists the basis elements thereof. Note that for the massless `matter fields' $\dbs$ there are the two alternative descriptions related by \eqref{eq:PR} as described above.
\begin{table}
\begin{center}
\begin{tabular}{|c|c|c|c|}
   \hline
      \bf multiplet  &  \bf bosonic fields  
         &  \bf geometric space &  \bf basis \rule[-1.5ex]{0pt}{4.5ex} \\
   \hline
   \hline
  matter  &  $\dbs^A$, $A=1,\ldots, \dim H^0_+(S^\Lambda,\nbundle{S^\Lambda})$ 
         &  $H^0_+(S^\Lambda,\nbundle{S^\Lambda})$  &  $s_A$ 
         \rule[-1.5ex]{0pt}{4.5ex} \\
   &  &  $H_{\bar\partial,-}^{(2,0)}(S^\Lambda)$  &  $\tilde s_A$ 
         \rule[-1.5ex]{0pt}{4.5ex} \\
   \hline
  vector  & $A_\mu$ & $H^0_-(S^\Lambda)$ & $P_-$ \rule[-1.5ex]{0pt}{4.5ex} \\
   \hline
  Wilson lines &  $a_I$, $I=1,\ldots, \dim H_{\bar\partial,-}^{(0,1)}(S^\Lambda)$
         &  $H_{\bar\partial,-}^{(0,1)}(S^\Lambda)$   &  $A^I$ 
         \rule[-1.5ex]{0pt}{4.5ex} \\
   \hline
\end{tabular} 
\caption{Massless D7-brane spectrum} \label{tab:spec} 
\end{center}
\end{table}

After the discussion of the bosonic spectrum we now turn to the fermionic fields arising form the D7-brane. As discussed in section~\ref{sec:superaction} the fermionic spectrum can again be understood from the embedding of the super-D7-brane into the type~IIB supergravity superspace~$M^{9,1|2}$ parametrized by the superspace coordinates~$Z^{\check M}=(x^M,\pair{\theta})$. Then the fermionic spectrum arises from fluctuations of the embedding supermap~$\super{\phi}$ along the odd superspace coordinates $\pair{\theta}$, which for the fermionic fluctuations takes the simple form \cite{Grisaru:1997ub,Marolf:2003vf}
\begin{equation}
   \super{\varphi}:\mathcal{W}\hookrightarrow M^{9,1|2},\qquad
      \xi\mapsto\left(\varphi(\xi),\pair{\Theta}(\xi)\right) \ .
\end{equation}
Hence the pullback with respect to this supermap simply amount to replacing the superspace coordinates~$\pair\theta$ by $\pair{\Theta}(\xi)$ \cite{Millar:2000ib,Grana:2002tu,Grana:2002nq}. Due to the dependence on $\xi$ the fermionic fields~$\pair\Theta(\xi)$ are localized on the worldvolume~$\mathcal{W}$ of the D7-brane and contain all the fermionic degrees of freedom of the super-D7-brane.  Note that $\pair\Theta(\xi)$ has the same transformation behavior as the odd supercoordinates~$\pair\theta$, that is to say they are Majorana-Weyl spinors transforming in the Weyl representation $\spinrep{16'}$ of $SO(9,1)$.  

The next task is to study how $\spinrep{16'}$ decomposes on the worldvolume of the D7-brane into representations of appropriate subgroups of $SO(9,1)$. For the fields localized on the D7-brane the structure group $SO(6)$ of the tangent bundle of the internal six-dimensional space splits into $SO(4)\times SO(2)$. Here $SO(4)$ is the structure group of the tangent bundle of the four-dimensional internal D7-brane cycle $S^\Lambda$, whereas $SO(2)$ is the structure group of the two-dimensional normal bundle of $S^\Lambda$. Moreover for the D7-branes under consideration we always assume that the pullback tangent bundle $\iota^*\tbundle{Y}$ splits holomorphically into the direct sum $\tbundle{S^\Lambda}\oplus\nbundle{S^\Lambda}$, i.e. the structure group of the tangent and normal bundle reduces to $U(2)\times U(1)$. Hence we have the following chain of subgroups 
\begin{multline} \label{eq:gr}
   SO(3,1)\times SO(6)\xrightarrow{\ \iota^*\tbundle{Y}\rightarrow\tbundle{S^\Lambda}\oplus\nbundle{S^\Lambda}\ } 
      SO(3,1)\times SO(4) \times SO(2) \\
      \xrightarrow{\ \text{holomorphicity}\ } SO(3,1)\times\left(SU(2)\times U(1)\right)\times U(1) \ ,
\end{multline}
which tells us that the D7-brane worldvolume fields are appropriate representations of $SO(3,1)\times SU(2)\times U(1)\times U(1)$. Then the ten-dimensional spinor representation decomposes under $SO(3,1)\times SO(4)\times SO(2)$ according to
\begin{equation} \label{eq:D10_SO4}
   \spinrep{16'}\rightarrow \left(\spinrep{2},\spinrep{2},\spinrep{\bar 1}\right)\oplus
   \left(\spinrep{2},\spinrep{2'},\spinrep{1}\right)\oplus
   \left(\spinrep{\bar 2},\spinrep{2},\spinrep{1}\right)\oplus
   \left(\spinrep{\bar 2},\spinrep{2'},\spinrep{\bar 1}\right) \ ,
\end{equation}
with the two Weyl spinors $\spinrep{2}$ and $\spinrep{2'}$ of $SO(4)$ and the $SO(2)$ complex conjugated singlets $\spinrep{1}, \spinrep{\bar 1}$.

As reviewed in appendix~\ref{app:Spinors} in the context of complex manifolds the $SO(4)$ gamma matrices~$\gamma^n$ corresponding to the tangent bundle~$\tbundle{S^\Lambda}$ can be combined to complex gamma matrices~$\gamma^i, \gamma^{\bar\jmath}$ and then they are interpreted as raising and lowering operators which are used to construct the `ground states'~$\singspin$ and the `conjugate ground state'~$\singspin^\dagger$ 
\begin{align} \label{eq:singspin}
   \gamma^i\singspin=0 \ , && \singspin^\dagger\gamma^{\bar\imath}=0 \ .
\end{align}
These `ground states' are $SU(2)$ spinor singlets of the same chirality\footnote{The singlets $\singspin$ and $\singspin^\dagger$ have the same chirality for structure groups $U(2k)$ and different chiralities for structure groups $U(2k+1)$ because the `conjugate ground states'~$\singspin^\dagger$ is also obtained by acting with all raising operators on the `ground state' $\singspin$. Thus for the group $SU(2k)$ there is an even number of raising operators which yields for $\singspin$ and $\singspin^\dagger$ the same chirality, whereas for $SU(2k+1)$ the odd number of raising operators results in different chiralities for $\singspin$ and $\singspin^\dagger$.}
\begin{align} \label{eq:G5Ch}
   \gamma\singspin=+\singspin \ , && \singspin^\dagger\gamma=+\singspin^\dagger \ ,
\end{align}
where $\gamma$ is the chirality gamma matrix of $SO(4)$.

In order to study the spinors on the D7-brane cycle~$S^\Lambda$, we need to analyze sections of the $U(2)\cong SU(2)\times U(1)$ tangent bundle. Locally there are two spinor singlets~$\singspin\otimes\mathcal{L^*}$ and $\singspin^\dagger\otimes\mathcal{L}$, where $\mathcal{L}$ and $\mathcal{L^*}$ denote sections of the line bundles associated to the $U(1)$ part in $U(2)$, or in other words to the $U(1)$ part of the spin connection. In general these line bundles are non-trivial as $S^\Lambda$ need not be a Calabi-Yau manifold with trivial first Chern class. However, the spinors of $S^\Lambda$ relevant for our analysis transform under the induced spin connection of the ambient space~$Y$ which is a connection with respect to both the tangent and the normal bundle of $S^\Lambda$. Therefore due to the triviality of the $U(1)$ part of the spin connection in the ambient space~$Y$ the overall $U(1)$ `charge' of the induced spinors must also be trivial and as a consequence the line bundles associated to the structure group $U(1)$ of the normal bundle~$\nbundle{S^\Lambda}$ must be dual to the line bundles of the $U(1)$ part in the structure group of the tangent bundle~$\tbundle{S^\Lambda}$. Hence there are two induced spinors which are global sections of $SU(2)\times U(1)\times U(1)$, i.e.
\begin{align} 
   (\singspin\otimes\mathcal{L}^*)\otimes\mathcal{L} \cong \singspin\ , &&
   (\singspin^\dagger\otimes\mathcal{L})\otimes\mathcal{L}^* \cong \singspin^\dagger\ ,
\end{align}
which behave like `neutral' spinors with respect to the $U(1)$ part of the induced spin connection.\footnote{Mathematically this neutrality is a consequence of the Whitney formula for the first Chern class. Since $\iota^*\tbundle{Y}\cong\tbundle{S^\Lambda}\oplus\nbundle{S^\Lambda}$ holomorphically, the Whitney formula applied to the Calabi-Yau manifold~$Y$ yields $0=\iota^*c_1(\tbundle{Y})=c_1(\tbundle{S^\Lambda})+c_1(\nbundle{S^\Lambda})$. Thus the line bundle associated to the tangent bundle must be dual to the line bundle associated to the normal bundle.} Note that the chiralities of the sections~$\mathcal{L}$ and $\mathcal{L^*}$ with respect of the $SO(2)$ structure group of the normal bundle~$\nbundle{S^\Lambda}$ are 
\begin{align} \label{eq:G3Ch}
   \tilde\gamma \mathcal{L} = +\mathcal{L} \ , && \tilde\gamma \mathcal{L}^* = -\mathcal{L}^* \ ,
\end{align}
in terms of the chirality gamma matrix~$\tilde\gamma$ of $SO(2)$ as introduced in appendix~\ref{app:Spinors}.

The next task is to determine the massless Kaluza-Klein modes resulting from the Majorana-Weyl spinors $\pair\Theta(\xi)$ compactified on the worldvolume $\mathbb{R}^{3,1}\times S^\Lambda$. However, only one linear combination of $\pair\Theta(\xi)$ is invariant with respect to the orientifold projection~$\mathcal{O}$. The fermions~$\pair\Theta(\xi)$ are the fluctuations of the odd superspace coordinates~$\pair\theta$, which in turn correspond to the infinitesimal supersymmetry parameters for supersymmetry variations. Since the gravitinos are the gauge fields for local supersymmetry the supersymmetry parameters $\pair\theta$ and their fluctuations $\pair\Theta$ transform exactly like the gravitinos under $\mathcal{O}$. Thus the projector $\tfrac{1}{2}\left(\id+\mathcal{O}\right)$ acting on $\pair\Theta$ becomes $\tfrac{1}{2}\left(\id+\check\sigma^2\right)$ as in \eqref{eq:prD3D7}, and we define
\begin{equation} \label{eq:inv3}
   \Theta(\xi)=\tfrac{1}{2}(\id+\mathcal{O})\:\pair\Theta(\xi) \ .
\end{equation} 
Then the projected ten-dimensional Majorana-Weyl spinor~$\Theta(\xi)$ transforming as $\spinrep{16'}$ of $SO(9,1)$ needs to be decomposed into representations of the subgroups in eq.~\eqref{eq:gr} according to \eqref{eq:D10_SO4}. 

Since $\singspin$ and $\singspin^\dagger$ are constant sections on $S^\Lambda$ (or zero-forms) we can identify $\gamma^{\bar\imath}\singspin$ and $\singspin^\dagger\gamma^i$ with $(0,1)$-forms and $(1,0)$-forms and $\gamma^{\bar\imath}\gamma^{\bar\jmath}\singspin$ and $\singspin^\dagger\gamma^i\gamma^j$ with $(0,2)$-forms and $(2,0)$-forms. Furthermore, we need to expand $\Theta(\xi)$ into fermionic modes which are invariant under the orientifold projection $\mathcal{O}$. By supersymmetry we already know that the invariant fermionic modes are only identified with negative parity forms in order to match the negative parity of their bosonic superpartners (c.f.~\ref{tab:spec}). Finally, we only keep massless fermionic excitations, which are zero-modes of the internal Dirac operator. As explained in the previous paragraph the relevant Dirac operator is induced from the ambient Calabi-Yau space~$Y$ for which the $U(1)$ part of the spin connection is trivial, and in this case the square of the Dirac operator can be identified with the Laplace operator. This implies that the massless fermionic excitations are in one-to-one correspondence with the global odd harmonic $(p,q)$-forms. Using \eqref{eq:G5Ch} and \eqref{eq:G3Ch} this leads to the Kaluza-Klein expansion
\begin{equation} \label{eq:decomptheta}
\begin{split}
   \Theta(\xi)\ =\  &\
       N_\lambda\:\lambda(x)\otimes P_-\singspin^\dagger
          +\bar N_{\bar\lambda}\:\bar\lambda(x)\otimes
           P_-\singspin\\
& +N_{\chi_I}\:\chi_I(x)\otimes A^I_{\bar\imath}\gamma^{\bar\imath}\singspin
          +\bar N_{\bar\chi_{\bar I}}\:\bar\chi_{\bar I}(x)
           \otimes\bar A^{\bar I}_i\singspin^\dagger\gamma^i   \\
     &+N_{\chi^A}\:\chi^A(x)\otimes\tfrac{1}{2} \tilde s_{A\: ij} 
          \singspin^\dagger\gamma^j\gamma^i
          +\bar N_{\bar\chi^{\bar A}}\:\bar\chi^{\bar A}(x)\otimes
           \tfrac{1}{2}\tilde s_{\bar A\:\bar\imath\bar\jmath} 
          \gamma^{\bar\imath}\gamma^{\bar\jmath}\singspin\ ,
\end{split}
\end{equation}
where $\lambda(x), \chi_I(x), \chi^A(x)$ are four-dimensional Weyl spinors. $P_-$ is the harmonic zero form of $S^\Lambda$ defined in \eqref{eq:P}. $A^I$ is a basis of odd (0,1)-forms on $S^\Lambda$ while $\tilde s_A$ is a bases of odd (2,0)-forms, both of which we already introduced in Table~\ref{tab:spec}. For the moment we also included a set of normalization constants $N_\lambda$, $N_{\chi^A}$ and $N_{\chi_I}$ which are determined in the next chapter.  Note that the Majorana property of the spinor $\Theta(\xi)$ implies that the decomposition \eqref{eq:D10_SO4} must fulfill a reality condition. This is reflected in the expansion \eqref{eq:decomptheta} in that for each term the complex conjugate term also appears.

Thus altogether we conclude that the four-dimensional massless fermionic modes invariant under $\mathcal{O}$ are identified with the negative harmonic forms of the cycle~$S^\Lambda$, which justifies the expansion of \eqref{eq:decomptheta} into the forms $P_-$, $\tilde s_A$ and $A^I$. As a consequence the massless fermionic modes in \eqref{eq:decomptheta} pair up with the bosonic fields in Table~\ref{tab:spec} to form four-dimensional $\mathcal{N}=1$ multiplets. Note that the D7-brane spectrum is not $\mathcal{N}=4$ supersymmetric anymore as in the D3-brane case, since it depends on the topological data of the cycle~$S^\Lambda$.  
\begin{table}
\begin{center}
\begin{tabular}{|c|c|c|c|}
   \hline
      \bf bos. fields  &  \bf ferm. fields  &  \bf multiplet & \bf multiplicity
      \rule[-1.5ex]{0pt}{4.5ex} \\
   \hline
   \hline
      $A_\mu$  &  $\lambda$, $\bar\lambda$  &  vector  &  $1$
      \rule[-1.5ex]{0pt}{4.5ex} \\
   \hline
      $\dbs^A$, $\bar\dbs^{\bar A}$  &  $\chi^A$, $\bar\chi^{\bar A}$  &  
      chiral  &  $\dim H_{\bar\partial,-}^{(2,0)}(S^\Lambda)$ 
      \rule[-1.5ex]{0pt}{4.5ex} \\
   \hline
      $a_I$, $\bar a_{\bar I}$  &  $\chi_I$, $\bar\chi_{\bar I}$  &  
      chiral  &  $\dim H_{\bar\partial,-}^{(0,1)}(S^\Lambda)$ 
      \rule[-1.5ex]{0pt}{4.5ex} \\
   \hline
\end{tabular} 
\caption{D7-brane spectrum in four dimensions and $\mathcal{N}=1$ multiplets} \label{tab:D7spec} 
\end{center}
\end{table}



\chapter{Effective supergravity action} \label{ch:D3D7spec}


The aim of this chapter is to compute the low energy effective action for Calabi-Yau orientifold with D3- or D7-branes. In section~\ref{sec:IIBsugra} the ten-dimensional type~IIB supergravity action is reviewed as it serves as the starting point for the discussed compactifications, and finally in section~\ref{sec:4DSUGRA} the generic form of a $\mathcal{N}=1$ supergravity action is given because this is the class of low energy effective theories obtained from compactifications of orientifold theories with D-branes. In section~\ref{sec:Bulkaction} the Kaluza-Klein reduction of the bulk theory is carried out. In order to include D-branes a normal coordinate expansion as described in section~\ref{sec:normal} as to be employed. This technique allows us to enhance the bulk effective action by the D3-brane fields in section~\ref{sec:D3braneSUGRA} and by the D7-brane fields in section~\ref{sec:D7action}. The resulting Lagrangians are $\mathcal{N}=1$ supersymmetric and therefore they are treated in terms of the specifying data of a $\mathcal{N}=1$ (gauged) supergravity in four dimensions. 


\section{Type IIB supergravity in D=10} \label{sec:IIBsugra}


The spectrum of type~IIB supergravity, which consists of the massless modes of type~IIB superstring theory, has been introduced in section~\ref{sec:typeIIAB}. The aim of this section is to describe the low energy dynamics of these fields in terms of the type~IIB low energy effective supergravity action in ten dimensions. 

The massless modes of the RR sector of type~IIB superstring theory are the even dimensional anti-symmetric tensors, i.e. the form fields $C^{(0)}$, $C^{(2)}$, $C^{(4)}$, $C^{(6)}$ and $C^{(8)}$. The field strengths of these form fields, as they appear in the low energy effective action, are given by \cite{Bergshoeff:2001pv}
\begin{equation} \label{eq:fs}
   G^{(p)}=\begin{cases} 
              \dd C^{(0)} & p=1 \\ 
              \dd C^{(p-1)}-\dd B\wedge C^{(p-2)} & \text{else} \ .
           \end{cases} 
\end{equation}
Note, however, in order to obtain the right number of bosonic degrees of freedom as encoded in the ten-dimensional $\mathcal{N}=2$ multiplet we need to impose duality conditions on the RR~field strength\footnote{In ten dimensions the massless spectrum falls into representations of the little group~$SO(8)$. The massless RR~fields of type~IIB string theory are given by the tensor product of the two Weyl spinors of $\spinrep{8}\otimes\spinrep{8}$ which decomposes into the irreducible anti-symmetric tensors $\rep{[0]}\oplus\rep{[2]}\oplus\rep{[4]_+}$ or equivalently $\rep{[4]_-}\oplus\rep{[6]}\oplus\rep{[8]}$, where $\rep{[4]_\pm}$ are the self-dual and anti-self-dual anti-symmetric four-tensor of $SO(8)$.} 
\begin{align} \label{eq:dual}
   G^{(1)}=*_{10}G^{(9)} \ , && G^{(3)}=(-1)*_{10}G^{(7)} \ , && G^{(5)}=*_{10}G^{(5)} \ .
\end{align}
These duality conditions tell us that for counting physical degrees of freedom 
one should only keep the self-dual part of the RR~four-form $C^{(4)}$, and one
can eliminate the RR~eight-form $C^{(8)}$ in favor of its dual axion $C^{(0)}$ and the RR~six-form $C^{(6)}$ in favor of its dual RR~two-form $C^{(2)}$. Then one obtains together with the massless NS-NS~fields the bosonic spectrum of the ten-dimensional $\mathcal{N}=2$ multiplets stated in the conventional form.

However, we do not yet impose the duality conditions \eqref{eq:dual} and proceed with the type~IIB supergravity action in ten dimensions, which contains all RR~forms of IIB~supergravity $C^{(0)}$, $C^{(2)}$, the self-dual RR~four-form $C^{(4)}$, and the dual RR~forms $C^{(6)}$ and $C^{(8)}$. This action is called the democratic formulation of type~IIB supergravity in ten dimensions as state in ref.~\cite{Bergshoeff:2001pv}. The bosonic part of the action reads
\begin{equation} \label{eq:IIBdemo}
\begin{split}
   \mathcal{S}_\text{IIB}^\text{sf}
    =& \frac{1}{2\kappa^2}\int\dd^{10}x\sqrt{-g_{10}}\:\ee^{-2\phi_{10}} \Rscalar 
     - \frac{1}{4\kappa^2}\int\ee^{-2\phi_{10}}
       \left(8\: \dd\phi\wedge *_{10}\dd\phi-H\wedge *_{10}H\right) \\
     +& \frac{1}{8\kappa^2}\int\sum_{p=1,3,5,7,9} G^{(p)}\wedge *_{10}G^{(p)} \ ,
\end{split}
\end{equation}
where $H$ is the field strength $H = \dd B$. Note that $\kappa$ is not the gravitational coupling constant but instead $\kappa_{10}=\kappa\:\ee^{\phi_0}=\kappa\: g_\text{s}$, which depends on the string coupling constant, or in other word on the vacuum expectation value of the dilaton $\phi_0$.\footnote{The coupling $\kappa$ is proportional ${\alpha'}^2$ and the proportionality constant depends on the conventions.}

In this democratic formulation the equations of motion resulting from this action must be combined with the duality constraints \eqref{eq:dual}. Note that already for the conventional used ten-dimensional IIB~supergravity the same phenomenon appears, namely for the four-form $C^{(4)}$ the self-duality condition on its five-form field strength must be imposed by hand \cite{Marcus:1982yu,Schwarz:1983qr,Dall'Agata:1998va}. 


\section{N=1 supergravity action in four dimensions} 
\label{sec:4DSUGRA}


Eventually we are interested in the effective four-dimensional description of the Kaluza-Klein reduced orientifold theory with space-time filling D$p$-branes. As presented in section~\ref{sec:calcond} and section~\ref{sec:Oplanes} we always choose a configuration which preserves some space-time supersymmetry in the resulting four-dimensional effective theory. In this work we focus on theories with $\mathcal{N}=1$ supergravity in four dimensions. Therefore we recall here the generic form of a four-dimensional $\mathcal{N}=1$ supergravity action.

A $\mathcal{N}=1$ supergravity theory with chiral multiplets~$\boldsymbol{M}^M$ and vector multiplets~$\boldsymbol{V}^\Gamma$ is completely specified in terms of the K\"ahler potential~$K$, the holomorphic superpotential~$W$ and the holomorphic gauge kinetic coupling functions~$f_{\Gamma\Delta}$ \cite{Cremmer:1982en,Wess:1992cp}. The on-shell degrees of freedom of one chiral multiplet~$\boldsymbol{M}^M=(M^M,\chi^M)$ arise from one complex scalar~$M^M$ and one complex Weyl fermion~$\chi^M$, whereas the vector multiplet~$\boldsymbol{V}^\Gamma=(V^\Gamma,\lambda^\Gamma)$ consists on-shell of one vector boson~$V^\Gamma$ and one complex Weyl fermion~$\lambda^\Gamma$. In addition to these multiplets there is always the gravity multiplet, which contains the graviton and a complex Weyl gravitino~$\psi_\mu$ as fermionic superpartner. Note, however, that the defining date namely the K\"ahler potential~$K$, the superpotential~$W$ and the gauge kinetic coupling functions~$f_{\Gamma\Delta}$ can already be unambiguously read off from the bosonic part of this supergravity action
\begin{equation} \label{eq:4Dbos}
\begin{split}
   \mathcal{S}_\text{Bosons}=&-\frac{1}{2\kappa_4^2}\int\dd^4x\:\sqrt{-\eta}\left( R 
      + 2\:K_{M\bar N}\nabla_\mu M^M\nabla^\mu \bar M^{\bar N} + V_\text{D} + V_\text{F} \right) \\
     &-\frac{1}{4\kappa_4^2}\int\dd^4x\:\sqrt{-\eta}\:(\Real f)_{\Gamma\Delta}
        F_{\mu\nu}^\Gamma F^{\mu\nu\:\Delta} 
      + \frac{1}{2\kappa_4^2}\int(\Imag f)_{\Gamma\Delta} F^\Gamma\wedge F^\Delta \ .
\end{split}
\end{equation}
Here $F^\Gamma$ refers to the field strength of the vector boson~$V^\Gamma$ and $K_{M\bar N}=\partial_M\partial_{\bar N}K$ is the K\"ahler metric of the K\"ahler potential~$K$. The F-term scalar potential $V_\text{F}$ and the D-term scalar potential $V_\text{D}$ is given by
\begin{align} \label{eq:spot}
   V_\text{F}=e^K\left(K^{M\bar N}\mathcal{D}_M W\mathcal{D}_{\bar N}\bar W -3|W|^2\right) \ , &&
   V_\text{D}=\frac{1}{2}\left(\Real{f}\right)^{\Gamma\Delta} \text{D}_\Gamma\text{D}_\Delta \ ,
\end{align}
where $V_\text{F}$ is expressed in terms of the K\"ahler covariant derivatives $\mathcal{D}_M W=\partial_M W + \left(\partial_MK\right) W$ of the superpotential. 

The D-term potential~$V_\text{D}$ involves the inverse matrix~$\left(\Real f\right)^{\Gamma\Delta}$ of the real part of the coupling matrix~$f_{\Gamma\Delta}$. The D-term itself is computed from the equation \cite{Wess:1992cp}
\begin{equation} \label{eq:Killing}
   \partial_N\partial_{\bar M}K\:\bar X_\Gamma^{\bar M} = \ii\partial_N \text{D}_\Gamma \ .
\end{equation}
Here $X_\Gamma=X^M_\Gamma \partial_M$ is the holomorphic Killing vector field of the corresponding gauged isometry of the target space K\"ahler manifold.

Although the bosonic part of the $\mathcal{N}=1$ supergravity action determines by supersymmetry the whole $\mathcal{N}=1$ supergravity Lagrangian, some data are encoded in the fermionic part of the supergravity action in a more direct way. For instance the superpotential~$W$ and the D-terms enter the bosonic scalar potential \eqref{eq:spot} quadratically whereas they appear linearly in the fermionic terms.

In the conventions of ref.~\cite{Wess:1992cp} the kinetic terms of the fermions in the supergravity action read
\begin{align} \label{eq:4Dferm}
    \mathcal{S}_\text{Fermions}=& - \frac{1}{\kappa_4^2}\int\dd^4x\:\sqrt{-\eta} 
     \left[\vphantom{\frac{1}{2}}
     \left(-\epsilon^{\mu\nu\rho\tau}\bar\psi_\mu\bar\sigma_\nu\nabla_\rho\psi_\tau +
     \ii K_{M\bar N}\bar\chi^{\bar N}\bar\sigma^\mu\nabla_\mu\chi^M \right)\right. \\
    &+\left.\frac{\ii}{2} 
     (\Real f)_{\Gamma\Delta}\left(\lambda^\Gamma\sigma^\mu\nabla_\mu\bar\lambda^\Delta+ 
     \bar\lambda^\Gamma\bar\sigma^\mu\nabla_\mu\lambda^\Delta\right)
     -\frac{1}{2}(\Imag f)_{\Gamma\Delta}
     \nabla_\mu\left(\lambda^\Gamma\sigma^\mu\bar\lambda^\Delta\right)\right] \ , \nonumber
\end{align}
with appropriate covariant derivatives $\nabla_\mu$. Out of the fermionic couplings we only record here those which are relevant for this work, namely these are the couplings of the gravitinos $\psi_\mu$ to gauginos $\lambda^\Delta$ and to the fermionic matter fields $\chi^M$, and the mass terms of the matter fields~$\chi^M$ and of the gauginos~$\lambda^\Delta$ 
\begin{align} \label{eq:4Dcoupl}
   \mathcal{S}_\text{Couplings}=&-\frac{1}{2\kappa_4^2}\int\dd^4x\:\sqrt{-\eta} 
     \left[\text{D}_\Gamma\,\psi_\mu\sigma^\mu\bar\lambda^\Gamma 
          -\text{D}_\Gamma\,\bar\psi_\mu\bar\sigma^\mu\lambda^\Gamma\vphantom{\frac{a}{2}}
          \vphantom{\frac{1}{4}}\right.\nonumber \\
     &+\ee^{K/2}\left(\sqrt{2}\ii\,\mathcal{D}_MW\chi^M\sigma^\mu\bar\psi_\mu
          +2\,\bar W\,\psi_\mu\sigma^{\mu\nu}\psi_\nu+\hc\right) \\
     &\left.+\ee^{K/2}\left(\frac{1}{2}\mathcal{D}_M\mathcal{D}_NW\chi^M\chi^N
          -\frac{1}{4}K^{M\bar N}\mathcal{D}_{\bar N}\bar W\partial_M f_{\Gamma\Delta}+\hc\right)
      +\ldots\right] \ , \nonumber
\end{align}
where $\ldots$ denotes all the omitted fermionic coupling terms.


\section{Orientifold bulk for O3/O7-planes} \label{sec:Bulkaction}


In this section the closed string sector, i.e. the bulk theory, of Calabi-Yau orientifold compactifications of the type~IIB superstring is examined in order to set the stage for introducing space-time filling D-branes in the forthcoming sections. As argued in section~\ref{sec:Oplanes} in order to preserve $\mathcal{N}=1$ supersymmetry in the presence of D3/D7-branes the four-dimensional bulk theory must be an orientifold compactification with O3/O7 planes.

The first task is to perform the Kaluza-Klein reduction of the ten-dimensional theory on a Calabi-Yau orientifold space. As we work in the supergravity limit, that is to say in the limit where the string coupling constant~$\alpha'$ becomes small, the starting point for the Kaluza-Klein reduction is the ten-dimensional type~IIB supergravity of section~\ref{sec:IIBsugra}. In a second step the four-dimensional effective theory for the massless Kaluza-Klein modes are determined by inserting the massless Kaluza-Klein modes into the ten-dimensional supergravity action and by integrating out the internal space.

\subsection{Massless bulk spectrum of O3/O7 orientifolds}

In order to perform a Kaluza-Klein reduction of the ten-dimensional spectrum the background ansatz \eqref{eq:topansatz} with its metric \eqref{eq:metansatz} for the ten-dimensional type~IIB supergravity theory needs to be specified, i.e.
\begin{align} \label{eq:prodmet}
   \mathbb{R}^{3,1}\times Y/\mathcal{O} \ , &&
   \dd s_{10}^2 = \hat\eta_{\mu\nu}\:\dd x^\mu\dd x^\nu+
                  2\:\hat g_{i\bar\jmath}(y)\:\dd y^i \dd \bar y^{\bar\jmath} \ .
\end{align}
Here $\mathbb{R}^{3,1}$ is the four-dimensional Minkowski space while $Y/\mathcal{O}$ denotes the internal Calabi-Yau orientifold. That is to say we consider the Calabi-Yau manifold~$Y$ moded out by the orientifold projection~$\mathcal{O}$ introduced in eq.~\eqref{eq:projO3O7} \cite{Acharya:2002ag,Brunner:2003zm}. Recall that $\mathcal{O}$ contains a geometric action~$\sigma^*$, where $\sigma$ is an isometric holomorphic involution on the Calabi-Yau manifold~$Y$, which has the O3/O7-planes as its fixed point locus. In the ansatz for the background metric \eqref{eq:prodmet}, $\hat\eta_{\mu\nu}$ denotes the flat metric of the four-dimensional Minkowski space and $\hat g_{i\bar\jmath}(y)$ is the metric of the internal Calabi-Yau manifold~$Y$. This ansatz for the metric, however, is a little subtle. One really has to make a warped ansatz for the metric so as to capture the back-reaction of these localized sources to geometry \cite{Giddings:2001yu,DeWolfe:2002nn,deAlwis:2003sn,Buchel:2003js,deAlwis:2004qh}. This approach is beyond the scope of this work. Instead, as mentioned in the previous section and as further discussed in appendix~\ref{app:warp}, the analysis is performed in a regime, where the internal space is large enough such that this back-reaction can be treated as a negligible perturbation to our product ansatz \eqref{eq:prodmet}. 

As we have seen in section~\ref{sec:KKreduction} the Kaluza-Klein zero modes are governed by the harmonic forms of the compactification space~$Y$.\footnote{In the following for ease of notation the Calabi-Yau orientifold and the Calabi-Yau manifold are both denoted by~$Y$.} These harmonic forms of the Calabi-Yau manifold~$Y$ are in one-to-one corresponds to elements of the cohomology groups $H^{(p,q)}_{\bar\partial}(Y)$, which due to the holomorphicity and the involutive property of $\sigma$ split into even and odd eigenspaces~$H^{(p,q)}_{\bar\partial,+}(Y)$ and $H^{(p,q)}_{\bar\partial,-}(Y)$ according to eq.~\eqref{eq:CohomSplit}. In order to carry out the compactification we need to choose basis elements for the various harmonic forms, which are listed in Table~\ref{tab:coh}.
\begin{table}
\begin{center}
\begin{tabular}{|c|c|c||c|c|c|}
   \hline
      \bf space  &  \bf basis  &  \bf dimension  &
      \bf space  &  \bf basis  &  \bf dimension  
      \rule[-1.5ex]{0pt}{4.5ex} \\
   \hline
   \hline
      $H^{(1,1)}_{\bar\partial,+}(Y)$  &  $\omega_\alpha$
      &  $\alpha=1,\ldots,h^{1,1}_+$ 
      &  $H^{(1,1)}_{\bar\partial,-}(Y)$  &  $\omega_a$
      &  $a=1,\ldots,h^{1,1}_-$
      \rule[-1.5ex]{0pt}{4.5ex} \\
   \hline
      $H^{(2,2)}_{\bar\partial,+}(Y)$  &  $\tilde\omega^\alpha$
      &  $\alpha=1,\ldots,h^{2,2}_+$ 
      &  $H^{(2,2)}_{\bar\partial,-}(Y)$  &  $\tilde\omega^a$
      &  $a=1,\ldots,h^{1,1}_-$
      \rule[-1.5ex]{0pt}{4.5ex} \\
   \hline
      $H^{3}_+(Y)$  &  $\alpha_{\hat\alpha},\beta^{\hat\alpha}$ 
      &  $\hat\alpha=1,\ldots,h^{2,1}_+$
      &  $H^{3}_-(Y)$  &  $\alpha_{\hat a},\beta^{\hat a}$ 
      &  $\hat a=0,\ldots,h^{2,1}_-$
      \rule[-1.5ex]{0pt}{4.5ex} \\
    \hline
      $H^{(2,1)}_{\bar\partial,+}(Y)$  &  $\chi_{\tilde\alpha}$  
      &  $\tilde\alpha=1,\ldots,h^{2,1}_+$
      &  $H^{(2,1)}_{\bar\partial,-}(Y)$  &  $\chi_{\tilde a}$
      &  $\tilde a=1,\ldots,h^{2,1}_-$
      \rule[-1.5ex]{0pt}{4.5ex} \\
    \hline  
      $H^{(1,2)}_{\bar\partial,+}(Y)$  &  $\bar\chi_{\tilde\alpha}$  
      &  $\tilde\alpha=1,\ldots,h^{2,1}_+$
      &  $H^{(1,2)}_{\bar\partial,-}(Y)$  &  $\bar\chi_{\tilde a}$
      &  $\tilde a=1,\ldots,h^{2,1}_-$
      \rule[-1.5ex]{0pt}{4.5ex} \\
    \hline  
      $H^{(3,0)}_{\bar\partial,+}(Y)$  &  --
      &  $0$
      &  $H^{(3,0)}_{\bar\partial,-}(Y)$  &  $\Omega$  
      &  $1$
      \rule[-1.5ex]{0pt}{4.5ex} \\
    \hline  
      $H^{(0,3)}_{\bar\partial,+}(Y)$  &  --
      &  $0$
      &  $H^{(0,3)}_{\bar\partial,-}(Y)$  &  $\bar\Omega$  
      &  $1$
      \rule[-1.5ex]{0pt}{4.5ex} \\
    \hline  
\end{tabular} 
\caption{Cohomology basis of the Calabi-Yau orientifold~$Y$} \label{tab:coh} 
\end{center}
\end{table}

Now with all ingredients for the Kaluza-Klein reduction at hand the next task is to expand type~IIB supergravity modes into harmonics invariant under the orientifold projection $\mathcal{O}=(-1)^{F_L}\Omega_p\sigma^*$ of eq.~\eqref{eq:projO3O7}. Among the NS-NS~fields the anti-symmetric two-tensor~$B$ has odd parity with respect to the operator $(-1)^{F_L}\Omega_p$, whereas the dilaton~$\phi_{10}$ and the metric~$g_{10}$ are even \cite{Grimm:2004uq}. Hence also the K\"ahler form $J$ which is proportional to the internal metric $g_{i\bar\jmath}$ is even.\footnote{$J$ and $g_{i\bar\jmath}$ are the K\"ahler form and the Calabi-Yau metric in the ten-dimensional Einstein frame.} Therefore the expansion of the NS-NS~fields invariant under the involutive orientifold symmetry \eqref{eq:projO3O7} yields
\begin{align} \label{eq:NS}
   J&=v^\alpha(x)\:\omega_\alpha \ , & B&=b^a(x)\:\omega_a \ , & \phi_{10}&=\phi_0+\phi(x) \ .
\end{align}
Note that we have included a vacuum expectation value~$\phi_0$ for the dilaton~$\phi_{10}$, which determines the string coupling constant $g_\text{s}$ by\footnote{In the following we set $g_\text{s}=1$. In the ten-dimensional supergravity action~\eqref{eq:IIBdemo} the dimensionless constant $g_\text{s}$ can be absorbed into the coupling constant~$\kappa$ if $G^{(p)}$ is rescaled accordingly.}
\begin{equation}
   g_\text{s}=\ee^{\phi_0} \ .
\end{equation}
Analogously to the NS-NS~fields the RR~fields are also expanded into appropriate harmonic forms. For Calabi-Yau orientifolds with O3/O7-planes the RR~zero-form $C^{(0)}$, four-form $C^{(4)}$ and eight-form $C^{(8)}$ are even with respect to the operator $(-1)^{F_L}\Omega_p$ and the two-form $C^{(2)}$ and six-form $C^{(6)}$ are odd \cite{Grimm:2004uq,Jockers:2004yj}. As a consequence the RR~fields enjoy the expansion
\begin{equation} \label{eq:C}
\begin{split}
   &\begin{aligned}
        C^{(0)}&=l(x) \ , &\qquad\qquad
        C^{(8)}&=\tilde l^{(2)}(x)\wedge\frac{\Omega\wedge\bar\Omega }
                 {\int_Y\Omega\wedge\bar\Omega} \ , \\
        C^{(2)}&=c^a(x)\:\omega_a \ , &\qquad\qquad
        C^{(6)}&=\tilde c^{(2)}_a(x)\wedge\tilde\omega^a
         \vphantom{\frac{\Omega}{\int_Y\Omega}} \ , \\
     \end{aligned} \\ 
   & \begin{aligned}
   C^{(4)}&=D_{(2)}^\alpha(x)\wedge\omega_\alpha
            +V^{\hat\alpha}(x)\wedge \alpha_{\hat\alpha}
            +U_{\hat\alpha}(x)\wedge \beta^{\hat\alpha}
            +\rho_\alpha(x)\wedge\tilde\omega^\alpha \ . 
     \end{aligned}
\end{split}
\end{equation}
In the effective four-dimensional theory $v^\alpha(x)$, $b^a(x)$, $\phi(x)$, $\rho_\alpha(x)$, $c^a(x)$ and $l(x)$ are scalar fields, $V^{\hat\alpha}(x)$ and $U_{\hat\alpha}(x)$ are vector fields, and $\tilde l^{(2)}(x)$, $\tilde c_a^{(2)}(x)$ and $D^\alpha_{(2)}(x)$ are two-form tensor fields. 

In addition to the above fields we have complex scalars, which arise from the complex structure deformations of the internal Calabi-Yau orientifold space. For the case of O3/O7 orientifold compactifications the complex structure deformations are in one-to-one correspondence with the elements of $H_{\bar\partial,-}^{(2,1)}(Y)$ \cite{Brunner:2003zm,Candelas:1990pi}, and we denote the corresponding four-dimensional scalar fields by $z^{\tilde a}$. To lowest order a complex structure deformation changes the internal Calabi-Yau metric \eqref{eq:prodmet} by
\begin{align} \label{eq:CSdef}
   \delta g_{\bar\imath\bar\jmath}(z^{\tilde a})= - \frac{\ii}{\norm{\Omega}^2}
     \:z^{\tilde a}\:\chi_{\tilde a\:\bar\imath jk}\:
     \bar\Omega^{jkl} g_{l\bar\jmath} \ , &&
   \delta g_{ij}(\bar z^{\tilde a}) = \frac{\ii}{\norm{\Omega}^2}
     \:\bar z^{\tilde a}\:\bar\chi_{\tilde a\:i\bar\jmath\bar k}\: 
     \Omega^{\bar\jmath\bar k\bar l} g_{j\bar l} \ ,
\end{align}
where $\norm{\Omega}^2=\tfrac{1}{3!} \Omega_{ijk}\bar\Omega^{ijk}$.

The resulting four-dimensional $\mathcal{N}=1$ supergravity spectrum of the O3/O7 orientifold model is summarized in Table~\ref{tab:sp}. Note that in this table only the physical degrees of freedom are listed, i.e.\ the duality conditions \eqref{eq:dual} are taken into account. In four dimensions this duality relates a massless scalar to a massless two-form or a gauge boson to its magnetic dual. In terms of the fields given in eqs.~\eqref{eq:C} the duality \eqref{eq:dual} corresponds to the dual pairs $\tilde l^{(2)}(x)\sim l(x)$, $\tilde c^{(2)}_a(x)\sim c^a(x)$, $D^\alpha_{(2)}(x)\sim \rho_\alpha(x)$, $V^{\hat\alpha}(x)\sim U_{\hat\alpha}(x)$.
\begin{table}
\begin{center}
\begin{tabular}{|c|c|c||c|c|c|}
   \hline
      \bf multiplet  &  \bf multi. &  \bf bos. fields &
      \bf multiplet  &  \bf multi. &  \bf bos. fields \rule[-1.5ex]{0pt}{4.5ex} \\
   \hline
   \hline
      gravity  &  $1$  &  $g_{\mu\nu}$  &
      chiral &  $h_-^{1,1}$  &  $(b^a,c^a)$ \rule[-1.5ex]{0pt}{4.5ex} \\
   \hline
      vector  &  $h_+^{2,1}$  &  $V^{\hat\alpha}_\mu$  & 
      chiral  &  $h_+^{1,1}$  &  $(\rho_\alpha,v^\alpha)$ \rule[-1.5ex]{0pt}{4.5ex} \\
   \hline
      chiral &  $1$  &  $(l,\phi)$  &
      chiral &  $h_-^{2,1}$  &  $z^{\tilde a}$ \rule[-1.5ex]{0pt}{4.5ex} \\
   \hline
\end{tabular} 
\caption{$\mathcal{N}=1$ multiplets} \label{tab:sp} 
\end{center}
\end{table}

\subsection{Democratic low energy effective action} \label{sec:bulkaction}

In this section we want to derive the effective four-dimensional action for the bosonic bulk spectrum of Calabi-Yau orientifolds with O3/O7 planes. Instead of computing the effective action for only the physical degrees of freedom \cite{Grimm:2004uq}, we choose to compute the four-dimensional effective action in the democratic formulation \cite{Jockers:2004yj}. This means we seek an action in four dimensions which describes both the fields $l(x)$, $c^a(x)$, $\rho_\alpha(x)$, $V^{\hat\alpha}(x)$ and their dual partners $\tilde l^{(2)}(x)$, $\tilde c_a^{(2)}(x)$, $D^\alpha_{(2)}(x)$ simultaneously. Then the equations of motion of this democratic action need to be supplemented by the duality conditions of these fields. The advantage of this formulation is that it facilitates to couple D-branes to this democratic bulk theory because D-branes generically couple to all RR~forms.

In order to compute the effective action for the massless four-dimensional fields arising from the Calabi-Yau orientifold compactification on $Y$ we insert the massless Kaluza-Klein modes~\eqref{eq:NS} and \eqref{eq:C} into the ten-dimensional democratic supergravity action \eqref{eq:IIBdemo} and then in a second step by performing the internal six-dimensional part of the ten-dimensional integral in \eqref{eq:IIBdemo}. This procedure yields an effective four-dimensional action, which, however, is not given in the Einstein frame. In order to cast the four-dimensional action in the Einstein frame one needs to perform a Weyl rescaling of the metric \eqref{eq:prodmet}
\begin{align} \label{eq:Weyl}
   \hat\eta\ =\ \frac{6}{\mathcal{K}}\, \ee^{\phi/2}\,  \eta \ ,
   && \hat g\ =\ \ee^{\phi/2}\, g \ .
\end{align}
Then the effective four-dimensional action Weyl-rescaled to the Einstein frame is found to be
\begin{align} \label{eq:4Dbulk}
   \mathcal{S}_\text{Bulk}^\text{E} 
   =&\frac{1}{2\kappa_4^2}\int \left[-R\:*_4 1
        +2\mathcal{G}_{\tilde a\tilde b}\dd z^{\tilde a}\wedge*_4\dd\bar z^{\tilde b}
        +2G_{\alpha\beta}\dd v^\alpha \wedge *_4 \dd v^\beta \right. \nonumber \\
   &+\frac{1}{2}\dd(\ln \mathcal{K})\wedge *_4 \dd(\ln \mathcal{K})
        +\frac{1}{2}\dd\phi\wedge *_4 \dd\phi  
        +2\ee^{-\phi}G_{ab}\dd b^a\wedge\dd b^b \nonumber \\
   &+\frac{1}{4}\ee^{2\phi} \dd l\wedge *_4\dd l 
        +\ee^\phi G_{ab}(\dd c^a-l\dd b^a) \wedge *_4(\dd c^b-l\dd b^b) \nonumber \\
   &+\frac{9}{4\mathcal{K}^2}G^{\alpha\beta}
        \left(\dd\rho_\alpha-\mathcal{K}_{\alpha bc} \dd b^b\wedge c^c\right)
        \wedge *_4\left(\dd \rho_\beta-\mathcal{K}_{\beta de}\dd b^d\wedge c^e\right) \\
   &+\frac{1}{4}\ee^{-2\phi} 
        \left(\dd\tilde l^{(2)}+\dd b^a\wedge\tilde c^{(2)}_a\right)
        \wedge *_4\left(\dd\tilde l^{(2)}+\dd b^b\wedge\tilde c^{(2)}_b\right) \nonumber \\
   &+\frac{1}{16}\: \ee^{-\phi} G^{ab}
        \left(\dd\tilde c^{(2)}_a-\mathcal{K}_{ac\gamma}
              \dd b^c\wedge D^\gamma_{(2)}\right)
        \wedge *_4\left(\dd\tilde c^{(2)}_b-\mathcal{K}_{bd\delta}
              \dd b^d\wedge D^\delta_{(2)}\right) \nonumber \\
   &+\frac{  \mathcal{K}^2}{36} G_{\alpha\beta} \dd D_{(2)}^\alpha\wedge 
        *_4 \dd D_{(2)}^\beta \nonumber \\
   &+\left.\frac{1}{4} B_{\hat\alpha\hat\beta} 
        \dd V^{\hat\alpha}\wedge *_4\dd V^{\hat\beta}-
        \frac{1}{4} C^{\hat\alpha\hat\beta} 
        \dd U_{\hat\alpha}\wedge *_4\dd U_{\hat\beta} 
        -\frac{1}{2}\bti{A}{\hat\beta}{\hat\alpha} 
        \dd U_{\hat\alpha}\wedge *_4\dd V^{\hat\beta} \right] \ . \nonumber
\end{align}
$\kappa_4$ is now the four-dimensional gravitational coupling constant related to the ten-dimensional coupling constant $\kappa_{10}$ by $\kappa_4=\kappa_{10}\:{\vol(Y)}^{-1/2}$. The triple intersection numbers $\mathcal{K}_{\alpha\beta\gamma}$ and $\mathcal{K}_{ab\gamma}$, the metrics $G_{ab}$, $G_{\alpha\beta}$ and $\mathcal{G}_{\tilde a\tilde b}$, the matrices $\bti{A}{\hat\beta}{\hat\alpha}$, $B_{\hat\alpha\hat\beta}$ and $C^{\hat\alpha\hat\beta}$, and $\mathcal{K}$ arise from certain integrals over the internal Calabi-Yau space~$Y$. Their definitions are spelled out in appendix~\ref{app:CYorient}.  

In the four-dimensional democratic action \eqref{eq:4Dbulk} we could now impose the (dimensional reduced) duality condition \eqref{eq:dual} by adding Lagrangian multiplier terms and then integrate out the redundant degrees of freedom as demonstrated in appendix~\ref{app:dual}. Depending on whether one eliminates the space-time two-forms or their dual scalars, one obtains the effective orientifold action of ref.~\cite{Grimm:2004uq} either in terms of chiral multiplets or in terms of linear multiplets \cite{Binetruy:2000zx,Grimm:2004uq,Louis:2004xi}. However, as there arise additional couplings of the D-brane fields to the RR~forms, we postpone this dualization procedure until we have added the D-brane effective action to the orientifold bulk theory.

Before this section is concluded let us pause to consider the mass scales relevant so far \cite{Dine:1985kv,Kaplunovsky:1985yy,DeWolfe:2002nn}. A priori in string theory the only parameter is $\alpha'$ with dimension $[\text{length}]^2$. It gives rise to the string scale $M_\text{s}\sim {\alpha'}^{-1/2}$, which is the mass scale for the massive modes in the tower of string excitations. Therefore $M_\text{s}$ is a natural cut-off scale for the low energy effective description of the massless string modes. On the other hand in order to treat string theory perturbatively such that the low energy effective action \eqref{eq:IIBdemo} yields a reliable description, string theory must be weakly coupled. This in turn implies for the string coupling constant $g_\text{s}$ to satisfy $g_\text{s}\ll 1$, which due to $\phi_0=\ln g_\text{s}$ is a condition on the vacuum expectation value~$\phi_0$ of the ten-dimensional dilaton~$\phi_{10}$. In this weakly coupled regime one finds with $\kappa_{10}\sim g_\text{s} {\alpha'}^2$ for the ten-dimensional Planck mass $M_\text{p}^{(10)}={\kappa_{10}}^{-1/4}$ the relation
\begin{equation}
   M^{(10)}_\text{p}\gg M_\text{s} \ .
\end{equation}
This confirms that $M_\text{s}$ is indeed a good cut-off scale for the ten-dimensional effective supergravity action \eqref{eq:IIBdemo}.

The compactification on the Calabi-Yau orientifold~$Y$ introduces yet another scale, namely the Kaluza-Klein scale~$M_\text{KK}$, which is the mass scale for the tower of the massive Kaluza-Klein modes. In terms of the `radius'~$R$ of the internal space, i.e. $\vol(Y)\sim R^6$, the Kaluza-Klein masses are of the order $M_\text{KK}\sim R^{-1}$. In order to be able to treat the internal manifold semi-classically and in order for string winding modes to be irrelevant it is necessary to work in the large radius limit. This implies that the string length~$\sqrt{\alpha'}$ must be small compared to the size of the internal compactification space, that is to say $\sqrt{\alpha'}\ll R$. As $\kappa_4\sim\kappa_{10}R^{-3}$ the four-dimensional Planck scale~$M^{(4)}_\text{p}$ is given by
\begin{equation}
   M^{(4)}_\text{p} \sim \frac{R^3}{g_s\: {\alpha'}^2} \ .
\end{equation}
Altogether one finds for a regime, where the low energy effective description \eqref{eq:4Dbulk} is applicable, that the discussed mass parameters obey 
\begin{equation} \label{eq:scales1}
   M_\text{KK}\ll M_\text{s}\ll M^{(4)}_\text{p} \ .    
\end{equation}


\section{Normal coordinate expansion} \label{sec:normal}


In section~\ref{sec:KKreduction} we have described the Kaluza-Klein reduction of the ten-dimensional field theory to an effective four-dimensional field theory. In this work, however, in addition to the bulk action we also want to discuss the effective action of D-branes. In principal localized sources such as D-branes and O-planes cannot simple be added to the bulk theory, as they cause a back-reaction to the bulk geometry and hence also alter the Kaluza-Klein reduction alluded in section~\ref{sec:KKreduction}. However, in this work we consider the localized sources in the probe limit. That is we add the D-branes and O-planes and neglect the back-reaction to geometry. This approach is valid as long as the localized sources are added in a controlled fashion that is to say in a way such that the consistency conditions of section~\ref{sec:Oplanes} are fulfilled. Secondly, the internal Calabi-Yau space needs to be large enough such that back-reaction to geometry only effects the vicinity of the localized sources and not the whole internal bulk space.

In the orientifold Calabi-Yau limit under consideration the O-planes arise as the fixed point locus of the involution~$\sigma$. Even though these O-planes contribute tension and hence also trigger a back-reaction to geometry, they are not dynamical objects by themselves and hence do not give rise to additional four-dimensional effective fields \cite{Brunner:2003zm}. On the other hand as argued in section~\ref{sec:openstr} D-branes are dynamical objects which (again in the probe limit) are described by the effective action introduced in section~\ref{sec:Dpaction}.

In the process of performing a Kaluza-Klein reduction as performed in the previous section there appear in addition to the Kaluza-Klein modes of the bulk theory also the Fourier modes of the D-brane worldvolume fields. The D-brane worldvolume gauge boson Kaluza-Klein-reduced along the lines of the bulk fields yields as the massless modes the four-dimensional vector boson and possibly Wilson line moduli fields arising in the case of non-trivial fundamental groups for the internal D-brane cycle. The other D-brane fields parametrize deformations of the embedding of the D-brane into the bulk theory. That is to say we start with an ansatz for the D-brane configuration given by the `embedding map'~$\varphi:\mathcal{W}\hookrightarrow M^{9,1}$ of the worldvolume~$\mathcal{W}$ into the bulk theory as discussed in section~\ref{sec:Dpaction}. In the bulk theory this corresponds to choosing the ground state background metric~\eqref{eq:metansatz}. But in order to capture all the fields arising from the space-time filling D-branes one also has to allow for perturbations to the `background map'~$\varphi$, namely
\begin{equation}
   \varphi \rightarrow \varphi+\delta\varphi \ .
\end{equation}
The normal coordinate expansion then amounts to expressing a general fluctuation~$\delta\varphi$ in terms of a vector field in the normal direction of the worldvolume of the D-brane. This procedure is further discussed in appendix~\ref{app:normal}. The normal coordinate expansion yields the kinetic terms for the fields parametrizing the fluctuations around the `background embedding'~$\varphi$. Now for obtaining an effective four-dimensional description of the massless modes one has to perform a Kaluza-Klein reduction of these kinetic terms to four dimensions.

In the above analysis we have only discussed the kinetic terms arising from the D-branes. In principal there are also additional potential terms arising from obstructions to deformations in the normal direction due to the neglected back-reaction to geometry and due to higher order obstruction to the deformations \cite{Brunner:1999jq,Kachru:2000ih}. As we go along some of these potential terms are derived by means of supersymmetry.


\section{Calabi-Yau orientifolds with D3-branes} \label{sec:D3braneSUGRA}


Now we turn to the discussion of the low energy effective action of the massless D3-brane spectrum. The first task is to derive the four-dimensional effective action for the bosonic fields which are obtained by a normal coordinate expansion and a subsequent Kaluza-Klein reduction of the Dirac-Born-Infeld action~\eqref{eq:DBIab} and the Chern-Simons action~\eqref{eq:CSAb} or, as adequate for a stack of D3-branes, from the reduction of the non-Abelian generalized actions \eqref{eq:DBInonab} and \eqref{eq:CSnonab}. We identify the bosonic components of the D3-brane `matter fields'. Finally the effective D3-brane action is combined with the bulk and is cast into the standard~$\mathcal{N}=1$ supergravity form introduced in section~\ref{sec:4DSUGRA}.

\subsection{Bosonic D3-brane action} \label{sec:D3bosaction}

The bosonic D3-brane action consists of the Dirac-Born-Infeld action and the Chern-Simons action. First the terms resulting from the Dirac-Born-Infeld action are discussed. For simplicity we start with the Abelian action \eqref{eq:DBIab} and then include the modifications which are necessary to implement the non-Abelian nature of a stack of D3-branes.

In order to derive the kinetic terms resulting from the action \eqref{eq:DBIab} we first need to evaluate the pullback of the metric \eqref{eq:prodmet} and of the anti-symmetric two-tensor~$B$ to the worldvolume~$\WV$ of the D3-brane in a way that captures the dynamics of the D3-brane. This is achieved by the normal coordinate expansion of section~\ref{sec:normal} with the technical details collected in appendix~\ref{app:normal}. Here we apply the pullback formulae \eqref{eq:PB} to the metric and \eqref{eq:PB2} to the two-form field~$B$ which yield up to second order in derivatives\footnote{By slight abuse of notation the worldvolume indices~$a$, $b$ of D3-brane are identified with the indices $\mu$, $\nu$ of the space-time manifold.}
\begin{equation} \label{eq:gBexp}
\begin{split}
   \varphi^*g_{10}&=
      \hat\eta_{\mu\nu}\dd x^\mu\dd x^\nu+
      \hat g_{mn} \covdb_\mu\dbt^m\covdb_\nu\dbt^n \dd x^\mu\dd x^\nu+\ldots \ , \\
   \varphi^*B&= 
      B_{mn} \covdb_\mu\dbt^m\covdb_\nu\dbt^n\dd x^\mu\dd x^\nu+\ldots \ .
\end{split}
\end{equation}
Note that in this expansion many terms of the formulae \eqref{eq:PB} and \eqref{eq:PB2} do not appear due to the product ansatz \eqref{eq:prodmet} and due to the form of the expansion of the $B$ field according to \eqref{eq:NS}. 

The next task is to expand the square root of the determinant in the Dirac-Born-Infeld action \eqref{eq:DBIab} by using the Taylor series
\begin{equation} \label{eq:dettaylor}
   \frac{\sqrt{\det\left(\mathfrak{A}+t\mathfrak{B}\right)}}{\sqrt{\det\mathfrak{A}}} 
    = 1+\frac{t}{2}\tr \mathfrak{A}^{-1}\mathfrak{B}
      +\frac{t^2}{8}\left[\left(\tr\mathfrak{A}^{-1}\mathfrak{B}\right)^2
      -2\:\tr \left(\mathfrak{A}^{-1}\mathfrak{B}\right)^2\right]+\cdots \ .
\end{equation}  
Here according to \eqref{eq:gBexp} and \eqref{eq:DBIab} one sets 
\begin{align}
   \mathfrak{A}=-\eta_{\mu\nu} \ , && 
   \mathfrak{B}=-\hat g_{mn} \covdb_\mu\dbt^m\covdb_\nu\dbt^n-
                B_{mn} \covdb_\mu\dbt^m\covdb_\nu\dbt^n+\ell F_{\mu\nu} \ ,
\end{align}
and by just keeping terms up to second order in derivatives one obtains the kinetic terms for the D3-brane fields. However, before we perform this step let us comment on the modifications necessary for a stack of D3-branes.

Instead of applying a normal coordinate expansion to the Abelian Dirac-Born-Infeld action \eqref{eq:DBIab} the same techniques are adopted to the non-Abelian Dirac-Born-Infeld action \eqref{eq:DBInonab}. The non-Abelian nature of the $\dbt$ and $F$ are taken into account by utilizing a non-Abelian Taylor expansion for the background fields \cite{Douglas:1997ch,Garousi:1998fg,Myers:1999ps}, which yields for the determinant of~$Q^M_N$ in the non-Abelian Dirac-Born-Infeld action \cite{Grana:2002nq,Grana:2003ek}
\begin{equation} \label{eq:Qexpand}
   \sqrt{\det Q^M_N} = 
      1+\frac{\ii}{2\ell}\com{\dbt^m}{\dbt^n}\dbt^p\partial_p B_{nm}
       +\frac{1}{4\ell^2} \hat g_{mn}\hat g_{pq}\com{\dbt^p}{\dbt^m}\com{\dbt^n}{\dbt^q}
       +\ldots\  ,
\end{equation}
where $\ldots$ denotes terms which vanish after taking the symmetrized trace in the action \eqref{eq:DBInonab}.\footnote{The non-Abelian enhancement of $E_{MN}={g_{10}}_{MN}+B_{MN}$ to $P_{MN}$ as defined in eq.~\eqref{eq:PQterm} does not give a contribution at the order considered in this expansion.} Assembling all terms we arrive after Weyl-rescaling with \eqref{eq:Weyl} at the action in the four-dimensional Einstein frame \cite{Grana:2002nq}
\begin{multline}  \label{eq:DBID3}
   S_{\mathrm{DBI}}^{\mathrm{E}}
     =-\mu_3   \int \dd^4\xi\sqrt{-\det\eta}\:\tr\left(\frac{36}{\mathcal{K}^2}\:\id
       -\frac{\ell^2}{4}\:\ee^{-\phi} F_{\mu\nu} F^{\mu\nu}\right. \\ 
       +\left.\frac{3}{\mathcal{K}}\, g_{mn} \covdb_\mu\dbt^m \covdb^\mu\dbt^n 
       +\frac{9\:\ee^\phi}{\mathcal{K}^2\ell^2 } g_{mn}g_{pq}
        \:\com{\dbt^m}{\dbt^p}\com{\dbt^q}{\dbt^n} \right) \ ,
\end{multline}
where for the D3-brane coupling constant the BPS relation \eqref{eq:DBcharge} has been inserted. The first term, not containing any derivatives, is the NS-NS tadpole of the stack of D3-branes. As we will see in the next two paragraphs it is compensated in the supersymmetric case due to the presence of O3-planes. Then there is the standard kinetic term for the D3-brane fields $\dbt$ and the kinetic term for the $U(N)$ field strength $F$. Finally there appear quartic couplings of $\dbt$, which vanish in the Abelian limit of a single D3-brane. 

Our next task is to expand the non-Abelian Chern-Simons action \eqref{eq:CSnonab} for which we need the pull-back formula \eqref{eq:PB2} and as before the non-Abelian nature of the D3-brane fields with the symmetrized trace over appropriate non-commuting quantities must be taken into account.\footnote{c.f.~section~\ref{sec:nonabelaction}.} Altogether this yields the expanded Chern-Simons action in the four-dimensional Einstein frame \cite{Grana:2003ek} 
\begin{equation} \label{eq:CSD3}
\begin{split}
   S_\text{CS}^\text{E}
    =&\mu_3\int  \dd^4\xi \sqrt{-\det\eta}\:
       \tr\left(\frac{36 }{\mathcal{K}^2} \id\right) \\
    +&\frac{\mu_3}{4}\int\:\tr\left(\phi^m\covdb_\mu\dbt^n\right)
       \omega_{\alpha\:mn}\:\dd x^\mu\wedge\dd D_{(2)}^\alpha 
       +\frac{\mu_3\ell^2}{2}\int l\:\tr\left(F\wedge F\right) \ .
\end{split}
\end{equation}
The first term in \eqref{eq:CSD3} requires some further explanation. Recall that the presence of localized sources such as D3-branes actually requires a modification of the metric ansatz \eqref{eq:prodmet} by including a non-trivial warp factor \eqref{eq:metwarp}. This warped ansatz is valid if the tadpole cancellation conditions of section~\ref{sec:Oplanes} are fulfilled \cite{Giddings:2001yu}. However, as argued in appendix~\ref{app:warp} in the large radius limit the warp factor approaches~$1$. But the analysis of the warped compactification tells us that for a consistent setup the appearance of the warp factor requires the simultaneous introduction of a flux parameter $\alpha$ for the self-dual five-form field strength~$G^{(5)}$ as stated in eq.~\eqref{eq:G5bg}. This parameter enter also in the effective D3-brane action via the RR couplings in the Chern-Simons action \eqref{eq:CSnonab}. In the large radius regime $\alpha$ becomes constant according to \eqref{eq:warpcon} and generates the first term in \eqref{eq:CSD3} as a remnant of the warped ansatz \eqref{eq:metwarp}.

Now we can combine the Dirac-Born Infeld action \eqref{eq:DBID3} and the Chern-Simons action \eqref{eq:CSD3} to arrive at
\begin{equation}
\begin{split} \label{eq:D3action}
   S_\text{bos}^\text{E}
    =&-\frac{\mu_3\ell^2 }{4} \int \dd^4\xi \sqrt{-\det\eta}\:  
      \ee^{-\phi} \tr F^{\mu\nu}F_{\mu\nu}
      +\frac{\mu_3\ell^2}{2}\int l\:\tr\left(F\wedge F\right) \\
    &+\mu_3  \int \dd^4\xi\sqrt{-\det\eta}\:
      \tr\left(\frac{3}{\mathcal{K}}\, g_{mn} \covdb_\mu\dbt^m \covdb^\mu\dbt^n \right) \\
    &+\frac{\mu_3}{4}\int\tr\left(\dbt^m\covdb_\mu\dbt^n\right)\omega_{\alpha\:mn}\:
      \dd x^\mu\wedge\dd D_{(2)}^\alpha \\
    &-\mu_3\int \dd^4\xi\sqrt{-\det\eta}\: 
      \tr\left(\frac{9 \:\ee^\phi}{\mathcal{K}^2\ell}\,
      g_{mn} g_{pq} \com{\dbt^m}{\dbt^p}\com{\dbt^q}{\dbt^n} \right) \ .
\end{split}   
\end{equation}
Note that the tadpole terms in \eqref{eq:DBID3} and \eqref{eq:CSD3} exactly cancel each other as a consequence of expanding around a consistent background. The computed action \eqref{eq:D3action} for a stack of $N$ D3-branes contains the standard kinetic term and $\Theta$-angle term for $U(N)$ field strength~$F$. The next terms is the kinetic term for the six real scalar matter fields~$\dbt^n$ transforming in the adjoint representation of $U(N)$. These matter fields couple to the bulk RR two-form~$D^\alpha_{(2)}$ and appear in the non-Abelian quartic couplings.

\subsection{Bosonic D3-brane action in chiral coordinates} \label{sec:D3chiral}

Since the $\dbt^m$ are the scalar components of the $\mathcal{N}=1$ chiral multiplets given in Table~\ref{tab:D3spec} they have to combine to complex variables. Then we also need to rewrite the action \eqref{eq:D3action} in terms of these complex fields. Or in other words we have to find the complex structure compatible with $\mathcal{N}=1$ supersymmetry. From the action \eqref{eq:D3action} we see that the $\sigma$-model metric of the $\dbt^m$ coincides with the Calabi-Yau metric $g_{mn}$. Thus a natural guess is to choose the complex structure $\cs$ of $Y$ also as the complex structure for the $\sigma$-model metric of the low energy effective action. For fixed complex structure we just rewrite all equations in terms of complex indices, i.e. we choose a basis in which the complex structure $\cs$ is block diagonal
\begin{equation}
   \cs=\begin{pmatrix} +\ii\id & \\ & -\ii\id \end{pmatrix}  .
\end{equation}
Including the complex structure deformations to lowest order we have to perturb $\cs$ according to \cite{Green:1987mn}
\begin{equation}
   \tilde{\cs}(z)= \cs+\delta\cs(z) = 
      \begin{pmatrix} 
          +\ii\id & z^{\tilde a}\wp_{\tilde a} \\ 
          \bar z^{\tilde a}\bar\wp_{\tilde a} & -\ii\id  
      \end{pmatrix} 
   \quad\text{with}\quad
   {\wp_{\tilde a}}_{\bar\jmath}^i
      = \frac{1}{\norm{\Omega}^2}\:\bar\Omega^{ikl} \chi_{\tilde a\: kl\bar\jmath} \ ,
\end{equation}
where $\wp_{\tilde a}$ is an element of $H^1_+(Y,\tbundle{Y})$ related to the basis elements $\chi_{\tilde a}$ of $H^{(2,1)}_{\bar\partial,-}(Y)$ defined in Table~\ref{tab:coh}. 

As we perturb the complex structure $\cs$ to $\tilde\cs$ the eigenvectors of $\cs$ are also modified. To first order the perturbed eigenvectors read
\begin{align}
   \begin{pmatrix} \dbt \\ 0 \end{pmatrix} \rightarrow 
      \begin{pmatrix} 
         \dbt \\ 
         -\frac{\ii}{2}\bar z^{\tilde a}\bar\wp_{\tilde a}\dbt 
      \end{pmatrix} \ , &&
   \begin{pmatrix} 0 \\ \bar\phi \end{pmatrix} \rightarrow
      \begin{pmatrix} 
         \frac{\ii}{2}z^{\tilde a}\wp_{\tilde a}\bar\dbt \\ 
         \bar\dbt 
      \end{pmatrix} \ ,
\end{align}
with $\dbt$ a vector of $\text{T}^{(1,0)}Y$ and $\bar\dbt$ a vector of $\text{T}^{(0,1)}Y$ with respect to the fixed complex structure $\cs$. Furthermore $\wp_{\tilde a}$ maps a tangent vector of type $(0,1)$ to a tangent vector of type $(1,0)$ and $\bar\wp_{\tilde a}$ vice versa. Hence the complex structure deformations act (up to first order) on the total vector $\dbt^n\partial_n=\dbt^i\partial_i+\bar\dbt^{\bar\jmath}\partial_{\bar\jmath}$ in component notation as 
\begin{align} \label{eq:dbtper}
   \dbt^i \rightarrow 
     \dbt^i+\frac{\ii}{2} z^{\tilde a}
     {\wp_{\tilde a}}_{\bar\jmath}^i \:\bar\dbt^{\bar\jmath} \ , &&
   \bar\dbt^{\bar\jmath} \rightarrow 
     \bar\dbt^{\bar\jmath}-\frac{\ii}{2}{\bar z}^{\tilde a}\bar\wp_{\tilde a}
     \vphantom{\wp_{\tilde a}}_l^{\bar\jmath} \:\dbt^l  \ .
\end{align}

Now we are ready to rewrite the action \eqref{eq:D3action} in terms of the complex fields $\dbt^i$ and simultaneously include the complex structure deformations $z^{\tilde a}$ up to linear order. For the kinetic term of the `matter fields'~$\dbt$ the target-space metric~$g_{mn}$ and the fields~$\dbt^n$ need to be expressed in terms of complex coordinates. Then the complex structure deformations~$z^{\tilde a}$ are included to linear order by perturbing the metric $g_{i\bar\jmath}$ with eq.~\eqref{eq:CSdef} and by substituting the complex `matter fields'~$\dbt^i$ by eq.~\eqref{eq:dbtper}. Altogether these steps amount to \cite{Grana:2003ek}
\begin{equation} \label{eq:D3kinetic}
   \frac{1}{2}g_{mn}\covdb_\mu\dbt^m\covdb^\mu\dbt^n \rightarrow
     g_{i\bar\jmath}\covdbz_\mu\dbt^i\covdbz^\mu\bar\dbt^{\bar\jmath}
     =-\ii v^\alpha(x)\omega_{\alpha\:i\bar\jmath}\covdbz_\mu\dbt^i\covdbz^\mu\bar\dbt^{\bar\jmath} \ ,
\end{equation}
where in the last step we also used $g_{i\bar\jmath}=-\ii v^\alpha(x)\omega_{\alpha\:i\bar\jmath}$. The covariant derivatives are defined as
\begin{align} \label{eq:VectcovderivTM}
   \mathcal{D}_\mu\dbt^i = \covdb_\mu\phi^i+\frac{i}{2}\partial_\mu z^{\hat a}
       (\chi_{\hat a})_{\bar l}^i\,\bar\phi^{\bar l}\ , &&
   \mathcal{D}_\mu\bar\dbt^{\bar\jmath}=\covdb_\mu\bar\phi^{\bar\jmath}
      -\frac{i}{2}\partial_\mu{\bar z}^{\hat a}(\bar\chi_{\hat a})_l^{\bar\jmath}\,\dbt^l \ ,
\end{align}
and hence introduce additional (derivative) couplings to the complex structure deformations~$z^{\tilde a}$. Note that the definition of the covariant derivative $\covdbz$ contains both a connection of the gauge group $U(N)$ and the newly added connection to include complex structure fluctuations~$z^{\tilde a}$. 

Similarly, the derivative couplings of $\dbt$ to the bulk RR two-forms $D^\alpha_{(2)}$ are modified accordingly
\begin{equation}
   \tr\left(\dbt^m\covdb_\mu\dbt^n\right)\omega_{\alpha\:mn}\:\dd x^\mu\wedge\dd D_{(2)}^\alpha 
     \rightarrow\tr\left(\dbt^i\covdbz_\mu\bar\dbt^{\bar\jmath}-\bar\dbt^{\bar\jmath}\covdbz_\mu\dbt^i\right)
                \omega_{\alpha\:i\bar\jmath}\:\dd x^\mu\wedge\dd D_{(2)}^\alpha \ ,
\end{equation}     
whence there also appear couplings between complex structure deformations~$z^{\tilde a}$, the `matter fields'~$\dbt$ and the RR two-forms $D^\alpha_{(2)}$.

The final chore is to express the non-Abelian quartic terms in the action \eqref{eq:D3action} in terms of complex chiral variables
\begin{equation} \label{eq:phi4}
   g_{mn}g_{pq}\com{\dbt^m}{\dbt^p}\com{\dbt^q}{\dbt^n} 
      \rightarrow 2 g_{i\bar\jmath} g_{k\bar l} 
                     \com{\dbt^i}{\dbt^k}\com{\bar\dbt^{\bar l}}{\bar\dbt^{\bar\jmath}}
                + 2 g_{i\bar\jmath} g_{k\bar l} 
                     \com{\dbt^k}{\bar\dbt^{\bar\jmath}}\com{\dbt^i}{\bar\dbt^{\bar l}} \ .
\end{equation}
Note that we have not included any dependence on the complex structure in this expression, simply because we have performed the Kaluza-Klein reduction only to fourth order in $\dbt^n$, which are considered to be small. Due to the linear approximation of the complex structure deformations in eq.~\eqref{eq:dbtper}, $z^{\tilde a}$ is also taken to be small and hence the inclusion of the complex structure in the quartic terms would not be consistent with the order of the Kaluza-Klein reduction.

Finally, substituting eqs.~\eqref{eq:D3kinetic} and \eqref{eq:phi4} into the D3-brane action \eqref{eq:D3action} yields in terms of the complex fields~$\dbt^i$ and the complex structure deformations~$z^{\tilde a}$ 
\begin{align} \label{eq:D3actioncomp}
   S_\text{bos}^\text{E}
    =&-\frac{\mu_3\ell^2 }{4} \int \dd^4\xi \sqrt{-\det\eta}\:  
      \ee^{-\phi} \tr F^{\mu\nu}F_{\mu\nu}
      +\frac{\mu_3\ell^2}{2}\int l\:\tr\left(F\wedge F\right) \nonumber \\
    +&\mu_3  \int \dd^4\xi\sqrt{-\det\eta}\:
      \tr\left(\frac{6\ii}{\mathcal{K}}\, v^\alpha \omega_{\alpha\:i\bar\jmath} 
      \covdbz_\mu\dbt^i \covdbz^\mu\bar\dbt^{\bar\jmath} \right) \\
    +&\frac{\mu_3}{4}\int\tr \left(\dbt^i\covdbz_\mu\bar\dbt^{\bar\jmath}
      -\bar\dbt^{\bar\jmath}\covdbz_\mu\dbt^i\right)
      \omega_{\alpha\:i\bar\jmath}\dd x^\mu\wedge \dd D_{(2)}^\alpha \nonumber \\
    -&\mu_3\int \dd^4\xi\sqrt{-\det\eta}\:\tr\left[
      \frac{18 \:\ee^\phi}{\mathcal{K}^2\ell^2}
      \left(g_{i\bar\jmath}g_{k\bar l}\com{\dbt^i}{\dbt^k}
      \com{\bar\dbt^{\bar l}}{\bar\dbt^{\bar\jmath}}
      +g_{i\bar\jmath}g_{k\bar l}\com{\dbt^k}{\bar\dbt^{\bar\jmath}}
      \com{\dbt^i}{\bar\dbt^{\bar l}}\right)\right] \ . \nonumber 
\end{align}   

Now this action has to be added to the democratic bulk action \eqref{eq:IIBdemo} and in order to obtain a conventional effective action the duality conditions \eqref{eq:dual} must be imposed to eliminate redundant degrees of freedoms. This is achieved by applying the techniques described in appendix~\ref{app:dual} to the two-forms $\tilde l^{(2)}$, $\tilde c^{(2)}_a$ and $D_{(2)}^\alpha$ succinctly. The resulting low energy effective action is reads \cite{Grana:2003ek}
\begin{align} \label{eq:D3SUGRAaction} 
   \mathcal{S}^\text{E}_\text{D3}
      &=\frac{1}{2\kappa_4^2}\int \left[-\,*_4 R
          +2\mathcal{G}_{\tilde a\tilde b}\dd z^{\tilde a} \wedge *_4 \dd\bar z^{\tilde b}
          +2G_{\alpha\beta}\dd v^\alpha \wedge *_4 \dd v^\beta \right. \nonumber \\
      &+\frac{1}{2}\dd(\ln \mathcal{K})\wedge *_4 \dd(\ln \mathcal{K})
          +\frac{1}{2}\dd\phi\wedge *_4 \dd\phi 
          +\frac{\ee^{2\phi}}{2} \dd l \wedge *_4 \dd l \nonumber \\
      &+\frac{12\ii}{\mathcal{K}}\kappa_4^2\mu_3\,v^\alpha\omega_{\alpha\,i\bar\jmath}
          \tr \mathcal{D}\dbt^i\wedge *_4 \mathcal{D}\bar\dbt^{\bar\jmath} \nonumber \\
      &+2\ee^\phi G_{ab}\dd b^a\wedge *_4 \dd b^b 
          +2\ee^\phi G_{ab} \left(\dd c^a-l\dd b^a \right)\wedge 
          *_4 \left(\dd c^b-l\dd b^b \right) \nonumber \\
      &+\frac{9}{2\mathcal{K}^2}G^{\alpha\beta}
          \left(\dd\rho_\alpha-\mathcal{K}_{\alpha bc} c^b\dd b^c
          +\mu_3\kappa_4^2\,\omega_{\alpha\:i\bar\jmath} 
          \tr\left(\bar\dbt^{\bar\jmath}\mathcal{D}\dbt^i
          -\dbt^i\mathcal{D}\bar\dbt^{\bar\jmath}\right)\right)\wedge \nonumber \\
      &\qquad\qquad *_4 \left(\dd\rho_\beta-\mathcal{K}_{\beta ab} c^a\dd b^b
          +\mu_3\kappa_4^2\,\omega_{\alpha\:i\bar\jmath} 
          \tr\left(\bar\dbt^{\bar\jmath}\mathcal{D}\dbt^i
          -\dbt^i\mathcal{D}\bar\dbt^{\bar\jmath}\right)\right) \nonumber \\
      &+\kappa_4^2\mu_3\ell^2\,\ee^{-\phi}\tr F\wedge *_4 F
       +\kappa_4^2\mu_3\ell^2\,l \tr F\wedge F \nonumber \\
      &+\frac{1}{2}(\Imag \mathcal{M})_{\hat\alpha\hat\beta} 
          \dd V^{\hat\alpha}\wedge *_4 \dd V^{\hat\beta}
          +\frac{1}{2}(\Real \mathcal{M})_{\hat\alpha\hat\beta}
          \dd V^{\hat\alpha}\wedge\dd V^{\hat\beta} \nonumber \\
      &\left.-*_4\frac{36\kappa_4^2\mu_3\ee^\phi}{\mathcal{K}^2\ell^2}
          \tr\left(g_{i\bar\jmath}g_{k\bar l}\com{\dbt^i}{\dbt^k}
          \com{\bar\dbt^{\bar l}}{\bar\dbt^{\bar\jmath}}
          +g_{i\bar\jmath}g_{k\bar l}\com{\dbt^k}{\bar\dbt^{\bar\jmath}}
          \com{\dbt^i}{\bar\dbt^{\bar l}} \right) \right] \ .
\end{align}
Note that the D3-brane action \eqref{eq:D3actioncomp} contains a coupling to the space-time two-forms~$D_{(2)}^\alpha$. Eliminating these two-forms amounts to a shift of the derivative of the dual scalars $\rho_\alpha$ according to
\begin{equation} \label{eq:Denhance}
   \partial_\mu\rho_\alpha \rightarrow
   \partial_\mu\rho_\alpha+\mu_3\kappa_4^2\,\omega_{\alpha\:i\bar\jmath}
   \tr\left(\bar\dbt^{\bar\jmath}\covdbz_\mu\dbt^i-
            \dbt^i\covdbz_\mu\bar\dbt^{\bar\jmath}\right) \ .
\end{equation}
This modification to the derivatives~$\partial_\mu\rho_\alpha$ was also introduced in ref.~\cite{Frey:2002hf}, where it was argued to emerge from a modified five-form Bianchi identity due to the source terms of the charged D3-branes. In our analysis the enhancement of these derivatives arise naturally from RR~couplings of the Chern-Simons action~\eqref{eq:CSnonab} to the space-time two-forms $D_{(2)}^\alpha$. Furthermore, here the modification also includes couplings to the complex structure deformations. 

\subsection{N=1 supergravity description} \label{sec:D3SUGRA}

In the previous section the derivation of the effective action for the D3-branes in Calabi-Yau orientifolds is sketched. The resulting four-dimensional bosonic low energy effective supergravity action is recorded in eq.~\eqref{eq:D3SUGRAaction}. From the general consideration about supersymmetry in chapter~\ref{ch:Dbranes} we already know that the derived action exhibits $\mathcal{N}=1$ supersymmetry, and therefore it must be possible to cast the action into the standard $\mathcal{N}=1$ supergravity form reviewed in section~\ref{sec:4DSUGRA}.

In order to specify the K\"ahler potential in the standard form, we must first identify the correct K\"ahler variables, which are the lowest components in the $\mathcal{N}=1$ chiral multiplets. Then in terms of these variables the metric of the scalar fields in \eqref{eq:D3SUGRAaction} becomes manifest K\"ahler. Geometrically this step corresponds to identifying the correct complex structure of the K\"ahler manifold, which is the target space of the scalar fields. We know already that the metric~$G_{\tilde a\tilde b}$ of the complex structure moduli fields~$z^{\tilde a}$ defined in \eqref{eq:CSt} is K\"ahler \cite{Candelas:1990pi}, and thus the complex scalar fields~$z^{\tilde a}$ are good K\"ahler coordinates. As analyzed in section~\ref{sec:D3chiral} the fields~$\dbt^i$ also serve as K\"ahler coordinates. For the remaining fields it is not so obvious how they combine to K\"ahler variables. However, guided by refs.~\cite{Haack:1999zv,Becker:2002nn,Grimm:2004uq} altogether the set of chiral fields turn out to be $\tau$, $G^a$, $T_\alpha$, $z^{\tilde a}$ and $\dbt^i$, where $\tau$, $G^a$ and $T_\alpha$ are defined as 
\begin{equation} \label{eq:D3KCoord}
\begin{split}
   \tau&=l+\ii\ee^{-\phi} \ , \qquad\qquad  G^a = c^a-\tau b^a  \ , \\
   T_\alpha&=\frac{3\ii}{2}\left(\rho_\alpha-\tfrac{1}{2}\mathcal{K}_{\alpha bc}c^bb^c
            +\mu_3\kappa_4^2\,\omega_{\alpha\:i\bar\jmath}\:
            \tr \dbt^i \left(\bar\dbt^{\bar\jmath}-\tfrac{\ii}{2}\bar z^{\tilde a}
            {\bar\wp}_{\tilde a}\vphantom{\wp_{\tilde a}}^{\bar\jmath}_l\dbt^l\right)\right) \\
   &\qquad\qquad\qquad\qquad
            +\frac{3}{4}\mathcal{K}_\alpha-\frac{3\ii}{4(\tau-\bar\tau)}
            \mathcal{K}_{\alpha bc}G^b(G-\bar G)^c \ .
\end{split}
\end{equation}

In terms of these K\"ahler coordinates the K\"ahler potential for the supergravity action \eqref{eq:D3SUGRAaction} is found to be
\begin{equation} \label{eq:KD3SUGRA}
   K(\tau,G,T,z,\dbt)=K_\text{CS}(z)-\ln\left[-\ii(\tau-\bar\tau)\right]
   -2 \ln\left[\tfrac{1}{6}\mathcal{K}(\tau,G,T,\dbt,z)\right] \ .
\end{equation}
Here $\mathcal{K}\equiv\mathcal{K}_{\alpha\beta\gamma}v^\alpha v^\beta v^\gamma$ is proportional to the volume of the internal Calabi-Yau space and $K_\text{CS}(z)$ is the K\"ahler potential for the complex structure deformations \cite{Candelas:1990pi}
\begin{equation} \label{eq:KCS}
   K_\text{CS}(z,\bar z) = -\ln \left(-\ii\int_Y\Omega(z) \wedge \bar\Omega(\bar z)\right) \ .
\end{equation}
The K\"ahler potential~\eqref{eq:KD3SUGRA} reproduces all the kinetic terms of \eqref{eq:D3SUGRAaction}. However, it is given as an implicit expression since $\mathcal{K}$ is explicitly only known in terms of $v^\alpha$ which are no K\"ahler coordinates. Instead they are determined in terms of $\tau$, $G^a$, $T_\alpha$, $z^{\tilde a}$ and $\dbt^i$ by solving \eqref{eq:D3KCoord} for $v^\alpha(\tau,G^a,T_\alpha,\dbt^i)$. Unfortunately, this solution cannot be given explicitly in general.

Note that the definition of the K\"ahler variables strongly resembles those specified in ref.~\cite{Grimm:2004uq} for the pure bulk theory. The new feature due to the presence of the stack of D3-branes is the modification of the definition of the K\"ahler variables~$T_\alpha$ by the D3-brane `matter fields'~$\dbt^i$. This adjustment is due to the fact that the D3-branes couple to the RR four-form~$C^{(4)}$, which in the expansion \eqref{eq:C} gives rise to the four-dimensional scalar fields~$\rho_\alpha$ and their dual two-form fields~$D_{(2)}^\alpha$. Hence, after eliminating the two-forms~$D_{(2)}^\alpha$, the scalar fields~$\rho_\alpha$ are shifted according to \eqref{eq:Denhance}, which is reflected in the definition of the chiral variables~$T_\alpha$. 

In order to gain further insight into the implicit definition of the K\"ahler potential \eqref{eq:KD3SUGRA} let us consider a simple model with a single (radial) K\"ahler modulus $v^1(x)$. Then one has $\mathcal{K}=\mathcal{K}_{111}(v^1)^3$ and eq.~\eqref{eq:D3KCoord} can be solved explicitly, namely
\begin{multline}
  2 \ln \mathcal{K}= 
     3 \ln\frac{2}{3}\left[T_1 + \bar T_1
     +\frac{3\ii}{4(\tau-\bar\tau)}\mathcal{K}_{1ab}(G^a-\bar G^a)(G^b-\bar G^b)\right.\\
     +3\ii\mu_3\omega_{1\:i\bar\jmath}\tr (\dbt^i\bar\dbt^{\bar\jmath})
     \left.+\frac{3}{4}\mu_3\left(\omega_{1\:i\bar\jmath}\bar z^{\tilde a}
     \bar\wp_{\tilde a}\vphantom{\wp_{\tilde a}}_l^{\bar\jmath}
     \tr (\dbt^i\dbt^l)+\hc \right)\right] \ .
\end{multline}
Note that in this case for frozen complex structure moduli fields~$z^{\tilde a}$ the K\"ahler potential is of the sequestered form, which means that $\mathcal{K}$ splits into a sum $\mathcal{K}(\tau,G,T,\dbt)=\mathcal{K}_\text{hidden}(\tau,G,T)+\mathcal{K}_\text{obs}(\dbt)$ \cite{Randall:1998uk}, where $\mathcal{K}_\text{hidden}$ depends only on the hidden sector fields whereas $\mathcal{K}_{obs}$ on the observable fields. Moreover, for vanishing couplings $G^a$ a similar structure of the K\"ahler potential has been suggested in ref.~\cite{DeWolfe:2002nn,Kachru:2003sx}.

The kinetic terms for the $\mathcal{N}=1$ vector multiplets are governed by the holomorphic gauge kinetic coupling functions. From the action~\eqref{eq:D3SUGRAaction} we can read off the gauge kinetic functions for the bulk vector fields~$V^{\hat\alpha}$ arising form the expansion of the bulk RR four-form~\eqref{eq:C}. Their gauge kinetic coupling matrix~$f_{\hat\alpha\hat\beta}$ is given by
\begin{equation} \label{eq:bulkf}
   f_{\hat\alpha\hat\beta}=\left.-\frac{\ii}{2}\mathcal{\bar M}_{\hat\alpha\hat\beta}
                           \right|_{z^{\tilde\alpha}=\bar z^{\tilde\beta}=0} \ .
\end{equation}
Here $\mathcal{M}$ denotes the $\mathcal{N}=2$ gauge kinetic matrix defined in \eqref{eq:GaugeMatrix}. Due to the fact that the subset~$z^{\tilde\alpha}$ of bulk complex structure deformations of Calabi-Yau manifolds is projected out by the orientifold involution~$\sigma$, it is necessary to evaluate in the $\mathcal{N}=1$ orientifold context the matrix~$\mathcal{M}$ at $z^{\tilde\alpha}=\bar z^{\tilde\alpha}=0$. The coupling matrix does not depend on the D3-brane `matter fields'~$\dbt^i$ and appears already in this form in the Calabi-Yau orientifold compactifications. In ref.~\cite{Grimm:2004uq} it is demonstrated that \eqref{eq:bulkf} is indeed a holomorphic matrix depending on the complex K\"ahler coordinates~$z^{\tilde a}$. 

The holomorphic coupling constant for the four-dimensional $U(N)$ gauge boson arising from the $U(N)$ gauge theory of the stack of D3-branes can be read off from the action~\eqref{eq:D3action}
\begin{equation} \label{eq:D3f}
   f^\text{D3}=-\ii \kappa_4^2\mu_3 \ell^2 \tau \ .
\end{equation} 
Note that due to \eqref{eq:DBcharge} the proportionality constant $\mu_3\ell^2$ is dimensionless. Furthermore recall that for a stack of $N$ D3-branes the `matter fields'~$\dbt^i$ transform in the adjoint representation of $U(N)$ and hence gives rise to a D-term potential. Using \eqref{eq:Killing} one determines
\begin{equation}
   \text{D}=-\frac{6\ii}{\mathcal{K}}\, v^\alpha \omega_{\alpha\:i\bar\jmath}
            \com{\dbt^i}{\bar\dbt^{\bar\jmath}} \ .
\end{equation}
According to eq.~\eqref{eq:spot} this D-term gives rise to a non-Abelian scalar potential, which, after using the Jacobi identity, can be identified with some of the quartic couplings in the action~\eqref{eq:D3action}. The remaining quartic terms are traced back to F-terms, which appear from the non-Abelian superpotential
\begin{equation} \label{eq:D3suppot}
   W(\dbt)=\frac{\mu_3}{\ell}\, \Omega_{ijk} \tr \dbt^i\dbt^j\dbt^k \ .
\end{equation}
This superpotential can be checked explicitly by computing the scalar potential via \eqref{eq:spot}. Instead of going through this computation we explicitly rederive this superpotential in section~\ref{sec:D3softterms} and in section~\ref{sec:FCS} by two independent methods. 


\section{Calabi-Yau orientifolds with D7-branes} \label{sec:D7action}


The derivation of the bosonic D7-brane action proceeds in the same spirit as the computation of the bosonic D3-brane action discussed in section~\ref{sec:D3braneSUGRA}. The worldvolume of a space-time filling D3-branes appeared as a point in the compactification space~$Y$, whereas the space-time filling D7-brane wraps a non-trivial four-cycle~$S^\Lambda$ in the internal Calabi-Yau space~$Y$, i.e. the worldvolume of the D7-brane is given by $\mathcal{W}=\mathbb{R}^{3,1}\times S^\Lambda$. For simplicity in section~\ref{sec:D7actionandbulk} we concentrate on the computation of the effective action of single D7-brane and as a consequence need not to expand the complicated non-Abelian D-brane actions described in section~\ref{sec:nonabelaction}. In section~\ref{sec:D7SUGRA} we specify for the computed action the defining data of the underlying $\mathcal{N}=1$ supergravity theory, and finally apply our results to simple examples in section~\ref{sec:D7examples}.

\subsection{Bosonic D7-brane and bulk action} \label{sec:D7actionandbulk}

The first task is the reduction of the bosonic part of the Abelian Dirac-Born-Infeld action \eqref{eq:DBIab} for the space-time filling D7-brane. Analogously to the D3-branes in order to obtain the effective four-dimensional fields describing the fluctuations of the internal cycle $S^\Lambda$ in the compactified six dimensions, we perform a normal coordinate expansion of the pullback metric $\varphi^*g_{10}$ and the pullback two-form $\varphi^*B$ as described in appendix~\ref{app:normal}. Applying \eqref{eq:PB} and \eqref{eq:PB2} to the metric~\eqref{eq:prodmet} and the two-from~$B$ respectively we obtain up to second order in the fluctuations $\dbs$ in the string frame
\begin{equation}\label{eq:nc}
\begin{split}
   \varphi^*g_{10}&=\hat\eta_{\mu\nu}\:\dd x^\mu\dd x^\nu
      +2\hat g_{i\bar\jmath}\:\dd y^i\dd\bar y^{\bar\jmath}
      +2\hat g_{i\bar\jmath}\:\partial_\mu\dbs^i\partial_\nu\bar\dbs^{\bar\jmath}
      \:\dd x^\mu\dd x^\nu+\ldots \ , \\
   \varphi^*B&=b^a\:\iota^*\omega_a
      +b_{i\bar\jmath}\:\partial_\mu\dbs^i\partial_\nu\bar\dbs^{\bar\jmath}\:\dd x^\mu\dd x^\nu
      +\ldots \ . 
\end{split}
\end{equation}
Note that $\dbs$ and $\bar\dbs$ are vector fields in the normal bundle~$\nbundle{S^\Lambda}$ and give rise to the massless D7-brane `matter fields'. Now we insert eq.~\ref{eq:nc} into the Dirac-Born-Infeld action~\eqref{eq:DBIab} and expand the spare root of the determinant with the Taylor series~\eqref{eq:dettaylor}. This yields the effective four dimensional action of the Dirac-Born-Infeld Lagrangian for the massless bosonic Kaluza-Klein modes in Table~\ref{tab:spec}. The next task is to insert the BPS calibration condition \eqref{eq:cali3} and to rescale the resulting action to the four-dimensional Einstein frame with \eqref{eq:Weyl}. Then one obtains 
\begin{equation} \label{eq:DBID7}
\begin{split}
   \mathcal{S}^\text{E}_\text{DBI}
      =&\mu_7  \ell^2\int\left[\frac{1}{4}\left(\mathcal{K}_\Lambda
        -\ee^{-\phi}\mathcal{K}_{\Lambda ab} b^a b^b\right) F\wedge *_4 F 
        +\frac{12}{\mathcal{K}}\ii\mathcal{C}^{I\bar J}_\alpha v^\alpha
        \dd a_I\wedge *_4\dd\bar a_{\bar J}\right] \\
      +&\mu_7  \int\left[\ii\mathcal{L}_{A\bar B}\left(e^\phi-G_{ab} b^a b^b\right)
        \dd\dbs^A\wedge *_4\dd\bar\dbs^{\bar B}+\frac{18}{\mathcal{K}^2} 
        \left(e^\phi\mathcal{K}_\Lambda-\mathcal{K}_{\Lambda ab}b^a b^b \right) *_4 1
        \right] \ ,
\end{split}
\end{equation}
with
\begin{equation} \label{eq:LCint}
\begin{split}
   \mathcal{L}_{A\bar B}&=\frac{\int_{S^\Lambda}\tilde s_A\wedge\tilde s_{\bar B}}
                               {\int_Y\Omega\wedge\bar\Omega}  \ , \\
   \mathcal{C}_\alpha^{I\bar J}&=\int_{S^\Lambda}\iota^*\omega_\alpha\wedge 
     A^I\wedge\bar A^{\bar J} \ .
\end{split}
\end{equation}
The details of the derivation of \eqref{eq:DBID7} are presented in ref.~\cite{Jockers:2004yj}.
The first term in \eqref{eq:DBID7} is the kinetic term of the field strength~$F$ of the $U(1)$ gauge boson arising from the gauge theory of the space-time filling part of the worldvolume~$\mathcal{W}$ of the D7-brane. The next two terms are the kinetic terms for the Wilson line moduli of the D7-brane and the matter fields (see \eqref{eq:D7spec}). Finally the last term is a potential term, which arises from the NS-NS tadpole of the D7-brane discussed in section~\ref{sec:Oplanes}. Note that it is proportional to the inverse square of the gauge coupling and thus can be identified as a D-term potential. In supersymmetric configurations this tadpole term is canceled by the negative tension contributed from the orientifold planes.

The next step is to describe the couplings of the D7-brane fields to the bulk RR~fields as captured in the Chern-Simons action \eqref{eq:CSAb} for the D7-brane. The normal coordinate expansion \eqref{eq:PB2} applied to the anti-symmetric pullback tensors of \eqref{eq:CSAb} the four-dimensional effective Chern-Simons action
\begin{align} \label{eq:CSD7}
   \mathcal{S}_\text{CS}&=
      \mu_7\int\left(\tfrac{1}{4}\dd\tilde l^{(2)}
        -\dd\left(\tilde c^{(2)}_a\:b^a\right)+\tfrac{1}{2}\mathcal{K}_{\alpha bc}
        \dd\left(D^\alpha_{(2)}\: b^b b^c\right)\right)
        \wedge\mathcal{L}_{A\bar B}
        \left(\dd\dbs^A\bar\dbs^{\bar B}-\dd\bar\dbs^{\bar B}\dbs^A\right) \nonumber \\ 
      &-\mu_7\ell^2\int\left[\tfrac{1}{2}\mathcal{C}_\alpha^{I\bar J}\dd D_{(2)}^\alpha\wedge
        \left(\dd a_I\bar a_{\bar J}-\dd\bar a_{\bar J}a_I\right) 
        +\dd\left(\tilde c^{(2)}_P-\mathcal{K}_{\alpha bP}
        D^\alpha_{(2)} b^b\right)\wedge A\right]  \nonumber \\
      &+\mu_7\ell^2\int\frac{1}{2}
        \left(\rho_\Lambda-\mathcal{K}_{\Lambda ab}c^a b^b
        +\tfrac{1}{2}\mathcal{K}_{\Lambda ab}b^a b^bl\right)F\wedge F  \ . 
\end{align}
Again the details of this computation are assembled in refs.~\cite{Jockers:2004yj,Jockers:2005zy}. However, there is one aspect which should be pointed out. In the derivation of the Chern-Simons action there seem to appear couplings between the field strength~$F$ and the field strength of the bulk vector fields $V^{\hat\alpha}$ and $U_{\hat\alpha}$, which appear in the reduction as integrals of the general form
\begin{equation}
   \int_{S^\Lambda} \iota^*\eta\wedge\theta \ ,
\end{equation}
where $\eta$ is a generic closed Calabi-Yau three-form and $\theta$ a generic closed one-form of $S^\Lambda$. However, all these integrals vanish \cite{Jockers:2005zy}, because there are no harmonic one-forms in the ambient Calabi-Yau manifold~$Y$.

The four-dimensional Chern-Simons action \eqref{eq:CSD7} contains the topological Yang-Mills term $F\wedge F$, with a field dependent $\Theta$-angle, which due to supersymmetry must eventually be given as the imaginary part of the holomorphic gauge coupling function. All the other terms in \eqref{eq:CSD7} involve the space-time two-forms $\tilde l^{(2)}$, $\tilde c^{(2)}$ or $D_{(2)}$, resulting from the expansion of the ten-dimensional RR~fields \eqref{eq:C}. The third term in \eqref{eq:CSD7} is known as a Green-Schwarz term in that, after integrating by parts, the $U(1)$ field strength $F$ couples linearly to the space-time two-forms $\tilde c^{(2)}$ and $D_{(2)}$. 

The final chore is now to add the D7-brane action \eqref{eq:DBID7} and \eqref{eq:CSD7} to the bulk orientifold action \eqref{eq:4Dbulk}. This yields the four-dimensional effective action of the orientifold bulk theory combined with a D7-brane. However, the obtained action is still in the democratic formulation introduced in section~\ref{sec:bulkaction}. Therefore the equations of motion resulting from this action must be supplemented by the four-dimensional analog of the duality conditions~\eqref{eq:dual}. In order to obtain a conventional action, that is an action which does not contain the fields and their dual fields simultaneously, the redundant degrees of freedom must be removed. This is again achieved by applying the dualization procedure outlined in appendix~\eqref{app:dual} succinctly to the dual pairs $(\tilde l^{(2)}, l)$, $(\tilde c^{(2)}_a,c^a)$, $(D^\alpha_{(2)},\rho_\alpha)$ and $(U_{\hat\alpha},V^{\hat\alpha})$. In order to obtain a four-dimensional effective action in terms of chiral multiplets we choose to eliminate the space-time two-forms $\tilde l^{(2)}$, $\tilde c^{(2)}_a$, $D^\alpha_{(2)}$ and the vectors $U_{\hat\alpha}$ in favor of their dual scalars and vectors $V^{\hat\alpha}$. The resulting four-dimensional effective bosonic action reads \cite{Jockers:2004yj}
\begin{align} 
   \mathcal{S}^\text{E}_\text{D7} \label{eq:BulkD7action}
     =&\frac{1}{2\kappa_4^2}\int \left[-R\:*_41
          +2\mathcal{G}_{\tilde a\tilde b}\dd z^{\tilde a} \wedge *_4 \dd\bar z^{\tilde b}
          +2G_{\alpha\beta}\dd v^\alpha \wedge *_4 \dd v^\beta \right. \nonumber \\
      &+\frac{1}{2}\dd(\ln \mathcal{K})\wedge *_4 \dd(\ln \mathcal{K})
          +\frac{1}{2}\dd\phi\wedge *_4 \dd\phi 
          +2\ee^\phi G_{ab}\dd b^a\wedge *_4 \dd b^b \nonumber \\
      &+2\ii\kappa_4^2\mu_7\mathcal{L}_{A\bar B}\left(\ee^\phi+4 G_{ab} b^a b^b\right)
          \dd\dbs^A\wedge *_4\dd\bar\dbs^{\bar B} 
          +\frac{24}{\mathcal{K}}\kappa_4^2\mu_7\ell^2\ii\mathcal{C}^{I\bar J}_\alpha v^\alpha
          \dd a_I\wedge *_4\dd\bar a_{\bar J} \nonumber \\
      &+\frac{\ee^{2\phi}}{2}
          \left(\dd l+\kappa_4^2\mu_7\mathcal{L}_{A\bar B}
          \left(\dd\dbs^A\bar\dbs^{\bar B}-\dd\bar\dbs^{\bar B}\dbs^A\right)\right) \wedge \nonumber \\
      &\qquad\qquad*_4\left(\dd l+\kappa_4^2\mu_7\mathcal{L}_{A\bar B}
          \left(\dd\dbs^A\bar\dbs^{\bar B}-\dd\bar\dbs^{\bar B}\dbs^A\right)\right) \nonumber \\
      &+2\ee^\phi G_{ab}
          \left(\nabla c^a-l\dd b^a-\kappa_4^2\mu_7 b^a \mathcal{L}_{A\bar B}
          \left(\dd\dbs^A\bar\dbs^{\bar B}-\dd\bar\dbs^{\bar B}\dbs^A\right)\right)\wedge \nonumber \\
      &\qquad\qquad *_4 \left(\nabla c^b-l\dd b^b-\kappa_4^2\mu_7 b^b 
          \mathcal{L}_{A\bar B} 
          \left(\dd\dbs^A\bar\dbs^{\bar B}-\dd\bar\dbs^{\bar B}\dbs^A\right)\right) \nonumber \\
      &+\frac{9}{2\mathcal{K}^2}G^{\alpha\beta}
          \left(\nabla\rho_\alpha-\mathcal{K}_{\alpha bc}
          c^b\dd b^c-\tfrac{1}{2}\kappa_4^2\mu_7
          \mathcal{K}_{\alpha bc} b^b b^c \mathcal{L}_{A\bar B}
          \left(\dd\dbs^A\bar\dbs^{\bar B}-\dd\bar\dbs^{\bar B}\dbs^A\right)\right. \nonumber \\
      &\qquad\qquad\qquad\left. +2\kappa_4^2\mu_7\ell^2\mathcal{C}_\alpha^{I\bar J}
          \left(a_I\dd \bar a_{\bar J}-\bar a_{\bar J}\dd a_I\right)\right)\wedge \nonumber \\
      &\qquad\qquad *_4 \left(\nabla\rho_\beta-\mathcal{K}_{\beta ab} c^a\dd b^b
          -\tfrac{1}{2}\kappa_4^2\mu_7 \mathcal{K}_{\beta bc} b^b b^c 
          \mathcal{L}_{A\bar B} \left(\dd\dbs^A\bar\dbs^{\bar B}
          -\dd\bar\dbs^{\bar B}\dbs^A\right) \right. \nonumber \\
      &\qquad\qquad\qquad \left. +2\kappa_4^2\mu_7\ell^2\mathcal{C}_\beta^{I\bar J}
          \left(a_I\dd\bar a_{\bar J}-\bar a_{\bar J}\dd a_I\right)\right) \nonumber \\
      &+\kappa_4^2\mu_7\ell^2 \left(\tfrac{1}{2}\mathcal{K}_\Lambda
          -\tfrac{1}{2}\ee^{-\phi}\mathcal{K}_{\Lambda ab} b^a b^b \right) F\wedge *_4F \nonumber \\
      &+\kappa_4^2\mu_7\ell^2\left(\rho_\Lambda-\mathcal{K}_{\Lambda ab}c^a \dbbf^b
          +\tfrac{1}{2}\mathcal{K}_{\Lambda ab}b^a b^bl \right) F\wedge F \nonumber \\
      &\left.+\frac{1}{2}(\Imag \mathcal{M})_{\hat\alpha\hat\beta} 
          \dd V^{\hat\alpha}\wedge *_4 \dd V^{\hat\beta}
          +\frac{1}{2}(\Real \mathcal{M})_{\hat\alpha\hat\beta}
          \dd V^{\hat\alpha}\wedge\dd V^{\hat\beta} \right] \ . 
\end{align} 

As observed before the Chern-Simons action \eqref{eq:CSD7} contains also a Green-Schwarz term involving the space-time two-forms $\tilde c^{(2)}_P$ and $D^\alpha_{(2)}$. After removing these two-forms in favor of their dual scalars these Green-Schwarz terms are responsible for gauging an isometry of the K\"ahler target space manifold. Note that the bulk theory has a set of global shift symmetries
\begin{align} \label{eq:shift}
   &c^a\rightarrow c^a+\theta^a \ , 
   &\rho_\alpha\rightarrow\rho_\alpha+\mathcal{K}_{\alpha bc}\dbbf^b\theta^c \ .
\end{align}
In the presence of a D7-brane wrapped on the cycle~$S^\Lambda$ one of these symmetries is gauged due to the dualization of the Green-Schwarz term \cite{Cremades:2002te}. Therefore the action \eqref{eq:BulkD7action} contains covariant derivatives for the charged fields $c^P$ and $\rho_\alpha$, i.e.
\begin{align} \label{eq:cd1}
   &\nabla c^a=\partial_\mu c^a\:\dd x^\mu-4\kappa_4^2\mu_7\ell\delta^a_P A \ ,
   &\nabla \rho_\alpha=\partial_\mu\rho_\alpha\:\dd x^\mu-4\kappa_4^2\mu_7
     \ell\mathcal{K}_{\alpha bP} b^b A \ .
\end{align} 
In the next section we further examine the relevance of these gauge covariant derivatives in the context of gauged isometries in $\mathcal{N}=1$ supergravity.

\subsection{N=1 supergravity description} \label{sec:D7SUGRA}

Just as in the case of the D3-brane Calabi-Yau orientifold compactifications we now also want to describe the effective four-dimensional theory \eqref{eq:BulkD7action} of the D7-brane Calabi-Yau orientifold compactification in the $\mathcal{N}=1$ supergravity language. 

Similarly as in section~\ref{sec:D3SUGRA} the first task is to identify the complex structure of the target space K\"ahler manifold by specifying the set of K\"ahler variables \cite{Grimm:2004uq,Jockers:2004yj}. The result of this analysis shows that $S$, $G^a$, $T_\alpha$, $z^{\tilde a}$, $\dbs^A$ and $a_I$ are the appropriate K\"ahler variables, where $S$, $G^a$, $T_\alpha$ are defined as
\begin{align}
   S&=\tau-\kappa_4^2\mu_7\mathcal{L}_{A\bar B}\dbs^A\bar\dbs^{\bar B}  \ , \qquad\qquad
   G^a=c^a-\tau b^a \ , \label{eq:G} \\ 
   T_\alpha&=\frac{3i}{2}\left(\rho_\alpha
      -\tfrac{1}{2}\mathcal{K}_{\alpha bc}c^b b^c\right) +\frac{3}{4}\mathcal{K}_\alpha
      +\frac{3i}{4(\tau-\bar\tau)} \mathcal{K}_{\alpha bc}G^b(G^c-\bar G^c)
      +3i\kappa_4^2\mu_7\ell^2 \mathcal{C}^{I\bar J}_\alpha a_I \bar a_{\bar J} \ ,\nonumber
\end{align}
with $\tau=l+\ii\ee^{-\phi}$. Note, however, that $\tau$ is not a K\"ahler variable anymore, but instead is modified by the D7-brane `matter fields'~$\dbs^A$ to form the `new dilaton'~$S$, which is a proper K\"ahler variable. The adjustment of the dilaton arises due to the fact that the D7-brane worldvolume~$\mathcal{W}$ couples via the ten-dimensional RR-form~$C^{(8)}$, or equivalently to the space-time two-form~$\tilde l^{(2)}$ dual to the axion~$l$ in the four-dimensional theory.

In terms of these K\"ahler coordinates the K\"ahler potential for the low energy effective supergravity action is found to be \cite{Jockers:2004yj}
\begin{multline} \label{eq:K1}
   K(S, G, T, z, \dbs, a)=K_\text{CS}(z) 
   -\ln\left[-i\left(S-\bar S\right)-2i\kappa_4^2\mu_7\mathcal{L}_{A\bar B}
   \dbs^A\bar\dbs^{\bar B}\right] \\
     -2 \ln\left[\tfrac{1}{6}\mathcal{K}(S,G,T,\dbs,a)\right] \ ,
\end{multline}
where $K_\text{CS}(z)$ is the K\"ahler potential of the complex structure moduli~$z^{\tilde a}$ defined in eq.~\eqref{eq:KCS}. As in the case of the D3-brane Calabi-Yau orientifold compactifications it is in general not possible to express the K\"ahler potential explicitly in terms of the K\"ahler variables. Instead $\mathcal{K}\equiv\mathcal{K}_{\alpha\beta\gamma} v^\alpha v^\beta v^\gamma$ is implicitly given in terms of the Kaluza-Klein variables $v^\alpha(S,G,T,\dbs,a)$ as a function of $S$, $G^a$, $T_\alpha$, $\dbs^A$ and $a_I$ \cite{Haack:1999zv,Becker:2002nn,Grana:2003ek,Grimm:2004uq,Jockers:2004yj}. However, in section~\ref{sec:D7examples} we give two examples for a specific orientifold compactification where the K\"ahler potential can be stated explicitly in terms of the chiral variables.

The final chore in describing the effective theory~\eqref{eq:BulkD7action} in terms of the $\mathcal{N}=1$ supergravity data is to specify the gauge sector. The gauge kinetic coupling functions of the bulk vector fields are not effected by the presence of the D7-brane, and their holomorphic coupling constant is still given in terms of the gauge kinetic coupling matrix \eqref{eq:bulkf}. Therefore we turn now to the discussion of the gauge kinetic coupling function of the $U(1)$ D7-brane gauge boson. If the D7-brane has no Wilson line moduli $a_I$, we readily extract from the action \eqref{eq:BulkD7action} using \eqref{eq:G} the coupling function 
\begin{equation} \label{eq:fbrane}
   f^\text{D7}=\frac{2\kappa_4^2\mu_7\ell^2}{3}\ T_\Lambda \ ,
\end{equation}
which is clearly holomorphic in the chiral fields. As expected the gauge coupling of the D7-brane is not the dilaton but the modulus controlling the size of the wrapped four-cycle \cite{Kakushadze:1998wp,Ibanez:1998rf}. If, however, the internal brane cycle $S^\Lambda$ has one-forms, which give rise to Wilson line moduli~$a_I$, the reduction of the D7-brane action \eqref{eq:DBIab} does not reproduce the Wilson line term in the gauge kinetic coupling function \eqref{eq:fbrane} which appears in the definition of $T_\Lambda$ \eqref{eq:G}. This mismatch has already been observed in \cite{Hsu:2003cy}. The reason for this seeming discrepancy is due to the fact, that the Dirac-Born-Infeld action is only an effective description comprising the open string tree level amplitudes \cite{Berg:2004ek}. Using a CFT approach, the open string one loop amplitudes for toroidal orientifolds with branes are computed in ref.~\cite{Berg:2004ek}, and the analysis shows that the missing quadratic Wilson line terms in the gauge kinetic coupling function of the D7-brane vector fields do indeed appear at the open string one loop level. In our case, we also expect that the coupling function \eqref{eq:fbrane} is corrected in the presence of Wilson line moduli, and that the missing terms are also generated at the one loop level of open string amplitudes.

We have seen that in the action~\eqref{eq:BulkD7action} there appear also charged scalar fields with respect to the $U(1)$ gauge theory localized on the D7-brane. In terms the corresponding charged K\"ahler variable is $G^P$. Recall that this complex scalar arises from the Kaluza-Klein reduction along the $(1,1)$-form $\omega_P$ which is dual to the four-cycle~$S^P$ in Table~\ref{tab:cycles}. By assembling the derivatives \eqref{eq:cd1} the gauge covariant derivative of chiral variable~$G^P$ becomes
\begin{equation} \label{eq:cd2}
   \nabla_\mu G^P=\partial_\mu G^P - 4\kappa_4^2\mu_7\ell A_\mu \ .
\end{equation}
Hence the holomorphic Killing vector of the associated gauged isometry is easily determined to be
\begin{equation} \label{eq:Killing1}
   X=4\kappa_4^2\mu_7\ell\partial_{G^P} \ .
\end{equation}
Gauged isometries in $\mathcal{N}=1$ supergravity theories give rise to D-terms according to \eqref{eq:Killing}. Here the gauged shift symmetry generates the D-term
\begin{equation} \label{eq:D}
   \text{D}=\frac{12\kappa_4^2\mu_7\ell}{\mathcal{K}}\ {\mathcal{K}_{Pa}b^a} 
           =\frac{12\kappa_4^2\mu_7\ell}{\mathcal{K}} \int_{S_P} J\wedge B \ ,
\end{equation} 
which can either be expressed in terms of $\mathcal{K}_{Pa}$ or in terms of the integral representation over the four-cycle~$S^P$. Finally this D-term induces the D-term scalar potential~$V_\text{D}$ according to eq.~\eqref{eq:spot}
\begin{equation}
   V_\text{D}=\frac{108\kappa_4^2\mu_7}{\mathcal{K}^2\Real{T_\Lambda}}
     \left(\mathcal{K}_{Pa}b^a\right)^2 \ .
\end{equation}
Note that $V_\text{D}$ is minimized for $b^a=0$ where the D-term and $V_\text{D}$ itself vanish and hence where one obtains a supersymmetric ground state.

Note that we have recovered the D-term, which we have derived in eq.~\eqref{eq:Dterm1} by analyzing the D7-brane calibration conditions. The supergravity analysis, however, has also specified the field dependent proportionality constant in the expression~\eqref{eq:Dterm1}.

\subsection{Instructive examples} \label{sec:D7examples}

In order to shed some more light on the structure of the implicitly defined K\"ahler potential~\eqref{eq:K1} we consider as a simple example a Calabi-Yau orientifold with $h_+^{1,1}=1$, i.e. we have a single harmonic two-form $\omega_\Lambda$ with positive parity under the involution~$\sigma$. By Poincar\'e duality we associate to $\omega_\Lambda$ the four-cycle $S^\Lambda$. Let us further assume that this cycle $S^\Lambda$ is suitable to wrap a D7-brane. Then with a D7-brane wrapped on $S^\Lambda$ we obtain a model with K\"ahler variables $S$, $G^a$, $\dbs^A$, $a_I$ and a single $T_\Lambda$. This implies that $\mathcal{K}=\mathcal{K}_{\Lambda\Lambda\Lambda}(v^\Lambda)^3$, and thus \eqref{eq:G} can be solved for $v^\Lambda$, i.e. 
\begin{equation}
   2\ln\mathcal{K}=3\ln \left[T_\Lambda+\bar T_\Lambda
     -\frac{3\ii\mathcal{K}_{\Lambda ac} (G^a-\bar G^a)(G^c-\bar G^c)}
      {4(S-\bar S-2\kappa_4^2\mu_7\mathcal{L}_{A\bar B}\dbs^A\bar\dbs^{\bar B})}
     -6\ii\kappa_4^2\mu_7\ell^2\mathcal{C}^{I\bar J}_\Lambda a_I\bar a_{\bar J}\right] 
     +\text{const} \ .
\end{equation}
This expression is further simplified if there are no chiral multiplets $G^a$ and $a_I$, namely 
\begin{equation}
   K(S,T_\Lambda,z,\dbs)=K_\text{CS}(z)-3\ln\left[T_\Lambda+\bar T_\Lambda\right]
   -\ln\left[-\ii\left(S-\bar S\right)+2\ii\kappa_4^2\mu_7\mathcal{L}_{A\bar B}
   \dbs^A\bar\dbs^{\bar B}\right] \ .
\end{equation}
The K\"ahler metric resulting from this K\"ahler potential is block diagonal in the modulus $T_\Lambda$ and the brane fluctuations $\dbs$. This particular feature of the K\"ahler potential was already anticipated for D7-brane models in ref.~\cite{Hsu:2003cy}, although we stress that it does not hold in the general case \eqref{eq:K1}.

As a second limit of \eqref{eq:K1} we consider the case $h_-^{1,1}=0$ and $h_+^{1,1}=3$ with a suitable four-cycle $S^\Lambda$ wrapped by a D7-brane. Then the K\"ahler variables of this example are $S$, $T_\alpha$, $z^{\tilde a}$, $\dbs^A$ and $a_I$ with $\alpha=\Lambda,1,2$. Moreover we suppose in analogy to the six dimensional torus that $C^{I\bar J}_\Lambda=0$ and that $\mathcal{K}_{\Lambda 12}$ is up to permutations the only non-vanishing triple intersection number, i.e. $\mathcal{K}=6\:\mathcal{K}_{\Lambda 12}v^\Lambda v^1 v^2$. Then as in the previous example we can specify $v^\alpha(S,T_\alpha,\dbs,a)$ explicitly and the K\"ahler potential \eqref{eq:K1} becomes
\begin{multline} \label{eq:Ktor}
   K(S,T,z,\dbs,a)=K_\text{CS}(z)-\ln\left[-\ii\left(S-\bar S\right)
      +2\ii\kappa_4^2\mu_7\mathcal{L}_{A\bar B}\dbs^A\bar\dbs^{\bar B}\right]
      +\ln\left[T_\Lambda+\bar T_\Lambda\right] \\
      +\ln\left[T_1+\bar T_1-6\ii\kappa_4^2\mu_7\ell^2
       \mathcal{C}_1^{I\bar J}a^I\bar a^{\bar J}\right]
      +\ln\left[T_2+\bar T_2-6\ii\kappa_4^2\mu_7\ell^2
       \mathcal{C}_2^{I\bar J}a^I\bar a^{\bar J}\right] \ .
\end{multline}
We can expand this K\"ahler potential up to second order in the D7-brane fields $\dbs$ and $a$, and obtain
\begin{multline} \label{eq:Kexp}
   K(S,T_\Lambda,z,\dbs,a)=K_\text{CS}(z)-\ln\left[-\ii\left(S-\bar S\right)\right]
    +\sum_{\alpha=\Lambda,1,2}\ln\left(T_\alpha+\bar T_\alpha\right) \\
    +\frac{\kappa_4^2\mu_7\mathcal{L}_{A\bar B}}{S-\bar S}\ \dbs^A\bar\dbs^{\bar B}
    -\frac{3\ii\kappa_4^2\mu_7\ell^2\mathcal{C}^{I\bar J}_1}{T_1+\bar T_1}\
    a_I\bar a_{\bar J} 
    -\frac{3\ii\kappa_4^2\mu_7\ell^2\mathcal{C}^{I\bar J}_2}{T_2+\bar T_2}\
    a_I\bar a_{\bar J} \ .
\end{multline}
This expansion agrees with the result of ref.~\cite{Lust:2004fi}, where the K\"ahler potential of a certain toroidal orientifold was derived to second order in the brane fields $\dbs^A$ and the Wilson line moduli $a_i$ by computing string scattering amplitudes. 


\section{Discussion} \label{sec:DisGenericK}


The computed K\"ahler potentials is the main result in this chapter. We observe that the structure of the K\"ahler potential has in all cases the generic form
\begin{equation} \label{eq:genericK}
   K=-\ln\left[-\ii\int_Y\Omega\wedge\bar\Omega\right]-\ln\left[-\ii(\tau-\bar\tau)\right]
     -2\ln\vol(Y) \ ,
\end{equation}
independently of the presence of D3- and/or D7-branes. However, the non-trivial task for a specific setup is to determine the K\"ahler coordinates and their precise relation to the holomorphic three-form~$\Omega$, to the dilation~$\tau$ and to the volume~$\vol(Y)$. 

The appearance of this generic K\"ahler potential can be justified by the following argument. Let us assume that the theory also has a superpotential~$W$. The kind of superpotential we have in mind is linear in $\Omega$ and is expressed as an integral over the internal space. These are the kind of superpotentials we encounter in the next chapter. By supersymmetry we already know from eq.~\eqref{eq:4Dcoupl} that the gravitino mass in the presence of a superpotential has the form $m_{3/2}\sim\ee^{K/2}W$. On the other hand such a four-dimensional mass term can only arise from the kinetic term of the ten-dimensional gravitino. The reduction of the ten-dimensional gravitino term, however, generates schematically the four-dimensional gravitino mass term
\begin{equation} \label{eq:gravmass}
   m_{3/2}\sim \frac{\ee^{\phi/2}}{\vol(Y)\sqrt{-\ii\int_Y \Omega\wedge\bar\Omega}}\,W \ .
\end{equation}
In the ten-dimensional string frame the kinetic term of the gravitino comes with a factor of $\ee^{-2\phi}$ since it is a closed string tree-level amplitude. Rescaling to the ten-dimensional Einstein frame yields a factor of $\ee^{5/2\phi}$, and hence altogether we obtain in the Einstein frame the dilaton factor $\ee^{\phi/2}$. Compactifying the ten-dimensional theory to four dimensions we integrate out the internal space and obtain the volume factor $\vol(Y)$. However, we need to Weyl rescale again to the four-dimensional Einstein frame which according to \eqref{eq:Weyl} gives rise to an extra factor $\vol(Y)^{-2}$. Once again Weyl rescaling explains the overall volume factor of $\vol(Y)^{-1}$. The dependence on $\Omega$ is not so obvious. However, in the analysis we have implicitly assumed that $W$ does not contribute to all these rescalings. But this is not quite true since $W$ depends linearly on $\Omega$, and hence $W/\sqrt{-\ii\int\Omega\wedge\bar\Omega}$ is really the quantity which does not participate in the rescaling. Therefore we also need to include the factor $1/\sqrt{-\ii\int\Omega\wedge\bar\Omega}$ in the gravitino mass term. This motivates the appearance of all the factors in eq.~\eqref{eq:gravmass} and hence due to $m_{3/2}\sim\ee^{K/2}W$ also the generic form of the K\"ahler potential~\eqref{eq:genericK}.

Note that we have discussed the K\"ahler potentials in the lowest order in $\alpha'$ and have not included any quantum corrections. Thus including these effects one definitely expects further adjustments to the K\"ahler potential in eq.~\eqref{eq:genericK}.


\chapter{Background fluxes and supersymmetry breaking} \label{ch:fluxes}


Up to now we have analyzed D-branes in Calabi-Yau orientifolds which yield a $\mathcal{N}=1$ supergravity description in the effective four-dimensional low-energy regime. The computed effective actions describe the low energy dynamics of the massless bulk moduli fields and the massless D-brane `matter fields'. These models introduced in the previous chapter serve as the basic ingredients to engineer more general configurations, which ultimately lead to standard model like gauge groups.

However, for phenomenological viable setups eventually mechanisms for supersymmetry breaking at low energy scales must be employed and in addition the neutral bulk moduli fields need to be stabilized. The introduction of background fluxes provides generically for both features \cite{Bachas:1995ik,Polchinski:1995sm,Michelson:1996pn,Gukov:1999ya,Dasgupta:1999ss,Taylor:1999ii,Mayr:2000hh,Curio:2000sc,Becker:2001pm,Giddings:2001yu,Haack:2001jz,Kachru:2002he,Becker:2002nn,Blumenhagen:2002mf,Blumenhagen:2003vr,Giryavets:2003vd,Cascales:2003wn,Cascales:2003pt}. Hence in this chapter we add non-trivial fluxes to the previous analysis and discuss their relevance for supersymmetry breaking.

First of all in section~\ref{sec:D3fluxes} we turn on bulk background fluxes and analyze their implications for D3-branes in Calabi-Yau orientifolds. In particular we compute flux-induced soft-terms which in the limit were gravity is decoupled are phenomenological signatures for supersymmetry breaking. Complementary to the bulk background fluxes in section~\ref{sec:D7fluxes} we turn to the discussion of D7-brane fluxes in D7-brane orientifold models. These fluxes are also capable to break supersymmetry and we derive the corresponding D- and F-terms. Finally in section~\ref{sec:FCS} we come back to the general structure of D-brane superpotentials and their relation to the holomorphic Chern-Simons theory of ref.~\cite{Witten:1992fb}.


\section{D3-branes and bulk background fluxes} \label{sec:D3fluxes}


In this section we focus on spontaneous supersymmetry breaking due to non-trivial bulk background fluxes in Calabi-Yau orientifolds with a stack of D3-branes. These fluxes generate non-vanishing F-term expectation values which break the computed $\mathcal{N}=1$ supergravity action of section~\ref{sec:D3braneSUGRA} spontaneously. Then in the limit where gravity decouples, that is to say in the limit $M_\text{p}^{(4)}\rightarrow \infty$ with the gravitino mass~$m_{3/2}$ fixed, the supergravity reduces to a globally supersymmetric effective field theory for the D3-brane `matter fields', which is then broken by flux-induced soft-terms \cite{Ibanez:1992hc,Kaplunovsky:1993rd,Brignole:1993dj}. Decoupling gravity is interesting from a phenomenological point of view as in this limit one makes contact with standard model like scenarios, in which the soft-terms reveal the spontaneous supersymmetry breaking of the underlying supergravity theory. As a matter of fact in the minimal supersymmetric standard model the appearance of soft-terms is crucial in order to allow for the embedding of the standard model in this supersymmetric extension \cite{Louis:1998rx}.

\subsection{Bulk background fluxes} \label{sec:bulkfluxes}

In order to set the stage for the discussion of supersymmetry breaking in orientifold theories with space-time filling D3-branes we first introduce three-form background fluxes in the bulk theory. Along the lines of ref.~\cite{Giddings:2001yu} we define the combined three-form flux $\Gflux$ as
\begin{equation} \label{eq:Gflux}
   \Gflux= \bg{F^{(3)}}-\tau \bg{H} \ ,
\end{equation}
with $\tau=l + \ii\ee^{-\phi}$. $\bg{F^{(3)}}$ is a non-trivial internal background flux for the field strength $F^{(3)}=\dd C^{(2)}$, whereas $\bg{H}$ is a non-trivial internal background flux for the field strength $H=\dd B$. In order to preserve four-dimensional Poincar\'e invariance only background fluxes in the internal space are turned on. The equations of motions and the Bianchi identities imply that $\bg{F}^{(3)}$ and $\bg{H}$ should be harmonic three-forms of the Calabi-Yau manifold~$Y$. Moreover, since $C^{(2)}$ and $B$ have odd parity with respect to the orientifold involution, the background flux~$\Gflux$ corresponds to an elements of $H^3_-(Y)$.

\begin{table}
\begin{center}
\begin{tabular}{|c|c||c|c|}
   \hline
      \bf background flux &  \bf eigenvalue &  \bf $(p,q)$-form & \bf eigenvalue
      \rule[-1.5ex]{0pt}{4.5ex} \\
   \hline
   \hline
      $\Gflux^+$ & $*_6\Gflux^+=+\ii\Gflux$ & 
         $\Omega$ & $*_6\Omega=-\ii\Omega$ \rule[-1.5ex]{0pt}{4.5ex} \\
   \hline
      $\Gflux^-$ & $*_6\Gflux^-=-\ii\Gflux$ & 
         $\chi_{\tilde a}$ & $*_6\chi_{\tilde a}=+\ii\chi_{\tilde a}$ \rule[-1.5ex]{0pt}{4.5ex} \\
   \hline
      $\bar\Gflux^+$ & $*_6\bar\Gflux^+=-\ii\bar\Gflux^+$ & 
          $\bar\chi_{\tilde a}$ & $*_6\bar\chi_{\tilde a}=-\ii\bar\chi_{\tilde a}$ \rule[-1.5ex]{0pt}{4.5ex} \\
   \hline
      $\bar\Gflux^-$ & $*_6\bar\Gflux^-=+\ii\bar\Gflux^-$ 
          & $\bar\Omega$ & $*_6\bar\Omega=+\ii\bar\Omega$ \rule[-1.5ex]{0pt}{4.5ex} \\
   \hline
\end{tabular} 
\caption{Imaginary and anti-imaginary self-dual fluxes and forms} \label{tab:SDforms} 
\end{center}
\end{table}
As the internal Hodge-star operator~$*_6$ of the internal six-dimensional space~$Y$ acting on three-forms squares to $-1$, the background fluxes~$\Gflux$ decompose into an imaginary self-dual part~$\Gflux^-$ and an anti-imaginary self-dual part~$\Gflux^+$ according to 
\begin{align} \label{eq:Gfluxsplit}
   \Gflux=\Gflux^+ + \Gflux^- \ , && *_6 \Gflux^\pm=\mp\ii\Gflux^\pm \ .
\end{align}
In Calabi-Yau threefolds the harmonic $(p,3-p)$-forms are also eigenforms of the Hodge-star operator~$*_6$ and are specified in Table~\ref{tab:SDforms}.\footnote{The eigenvalue can readily be computed by acting with the definition of the Hodge-star operator~$*_6$ on the local expression of a $(p,3-p)$-form and by taking into account that the harmonic forms in a Calabi-Yau threefold are always primitive.} Hence the background fluxes~$\Gflux^+$ can be expanded into appropriate three-forms \cite{Giddings:2001yu,DeWolfe:2002nn}
\begin{align} \label{eq:fluxexpand}
   \Gflux^+=-\frac{1}{\int_Y \Omega\wedge\bar\Omega}\left(\Omega\,\mathcal{I}
     +\bar\chi_{\tilde a}\mathcal{G}^{\tilde a\tilde b}\mathcal{I}_{\tilde b}\right) \ , &&
   \bar\Gflux^+=\frac{1}{\int_Y \Omega\wedge\bar\Omega}\left(\bar\Omega\,\mathcal{\bar I}
     +\chi_{\tilde a}\mathcal{G}^{\tilde a\tilde b}\mathcal{\bar I}_{\tilde b}\right) \ ,
\end{align}
with the inverse metric~$\mathcal{G}^{\tilde a\tilde b}$ of \eqref{eq:CSt} and with
\begin{align} \label{eq:FluxInt}
   \mathcal{I}(\bar z,\tau)=\int_Y \bar\Omega(\bar z)\wedge\Gflux(\tau) \ , &&
   \mathcal{I}_{\tilde a}(z,\bar z,\tau)=\int_Y \chi_{\tilde a}(z,\bar z)\wedge\Gflux(\tau) \ .
\end{align}

In ten-dimensional type~IIB supergravity the flux~$\Gflux$ enters in the effective action written in the Einstein frame as \cite{Giddings:2001yu}
\begin{equation} \label{eq:fluxaction}
   \mathcal{S}^\text{E}_\text{flux}
     = \frac{1}{4\kappa^2_{10}}\int \ee^\phi\:\Gflux\wedge *_{10}\bar\Gflux \ .
\end{equation}
The decomposition~\eqref{eq:Gfluxsplit} allows us to rewrite \eqref{eq:fluxaction} in terms of $\Gflux^+$. Then the ten-dimensional background flux term~\eqref{eq:fluxaction} corresponds in the four-dimensional effective theory rescaled to the four-dimensional Einstein frame to the flux-induced scalar potential~$V$ given by\footnote{Actually there arises an additional topological term $\sim\int \bg{H}\wedge\bg{F^{(3)}}$. Consistency requires this term to be canceled by Wess-Zumino like couplings of localized sources as discussed in section~\ref{sec:Oplanes}.}
\begin{equation} \label{eq:fluxspot}
   V = \frac{18\ee^\phi}{\mathcal{K}^2} \int_Y \Gflux^+\wedge *_6 \bar\Gflux^+ \ .
\end{equation}
This expression can be further rewritten by expanding the flux~$\Gflux^+$ into appropriate anti-imaginary self-dual three-forms of the Calabi-Yau according to the expansion~\eqref{eq:fluxexpand}. Then in terms of the integral representations~\eqref{eq:FluxInt} of the flux~$\Gflux$ the four-dimensional scalar potential becomes \cite{Taylor:1999ii,Giddings:2001yu,Becker:2002nn,DeWolfe:2002nn,Grimm:2004uq}
\begin{equation} \label{eq:bulkpot}
   V = \ee^K\,\left(\abs{\mathcal{I}}^2
       +\mathcal{G}^{\tilde a\tilde b}\:\mathcal{I}_{\tilde a}\mathcal{\bar I}_{\tilde b}
        \right) \ .
\end{equation}

In $\mathcal{N}=1$ string compactifications with background fluxes, such as orientifold compactifications, this scalar potential \eqref{eq:bulkpot} originates from the flux induced superpotential \cite{Taylor:1999ii,Gukov:1999ya}
\begin{equation} \label{eq:GWPot}
   \hat W(z,\tau)=\int_Y\Omega(z)\wedge\Gflux(\tau) \ .
\end{equation}
For orientifold Calabi-Yau compactification with O3/O7-planes the K\"ahler covariant derivatives of the superpotential \eqref{eq:GWPot} read \cite{Grimm:2004uq,Grana:2003ek}
\begin{equation} \label{eq:KDPot}
\begin{aligned}
   \mathcal{D}_{\tau}\hat W&=\frac{\ii}{2}\ee^{\phi}\mathcal{\bar I}
             +\ii G_{ab}b^a b^b \hat W \ , &
   \mathcal{D}_{G^a}\hat W&=2\ii G_{ab}b^b\hat W \ , \\
   \mathcal{D}_{T_{\alpha}}\hat W&=-2\frac{v^\alpha}{\mathcal{K}}\hat W \ , &
   \mathcal{D}_{z^{\tilde a}}\hat W&=\mathcal{I}_{\tilde a} \ .
\end{aligned}
\end{equation}
The conditions for unbroken supersymmetry, namely $\mathcal{D}_I\hat W=0$, imply $\mathcal{I}=\mathcal{I}_{\tilde a}=\hat W=0$. The relations $\mathcal{I}=\mathcal{I}_{\tilde a}=0$ are fulfilled for vanishing imaginary-anti-self-dual fluxes~$\Gflux$, that is for $\Gflux^+=0$, whereas $\hat W\ne 0$ tells us that supersymmetry is broken for non-vanishing imaginary-self-dual $(0,3)$-fluxes. In the context of Calabi-Yau orientifolds with D3-branes we come back to a detailed discussion of supersymmetry breaking in section~\ref{sec:D3softterms}.

\subsection{Flux-induced D3-brane couplings} \label{sec:D3fermionic}

To examine the effect of the bulk background fluxes introduced in the previous section, we first derive the flux induced couplings to the D3-brane `matter fields'. In the presence of non-trivial background fluxes~$\Gflux$ the second term in the expansion~\eqref{eq:Qexpand} couples to the background flux~$\bg{H}$ in \eqref{eq:Gflux} and yields in the non-Abelian Dirac-Born-Infeld action~\eqref{eq:DBInonab} trilinear couplings to the `matter fields'~$\dbt^n$.  Analogously there is a contribution to these trilinear couplings in the non-Abelian Chern-Simons action~\eqref{eq:CSnonab} arising from the background flux~$\bg{F^{(3)}}$ in \eqref{eq:Gflux}. Altogether the flux-induced trilinear terms in the effective D3-brane action become \cite{Grana:2003ek}
\begin{equation}
   \mathcal{S}^\text{E}_\text{tri,flux}=-\frac{18\mu_3}{\ell}
      \int\dd^4\xi\sqrt{-\det\eta} \:\frac{\ee^\phi}{\mathcal{K}^2}\:
      \left(\Gflux^+_{mnp}\:\tr\dbt^n\dbt^m\dbt^p +\hc\right) \ .
\end{equation}
In the spirit of section~\ref{sec:D3chiral} we rewrite the trilinear terms in $\dbt^n$ in terms of the chiral variables $\dbt^i$. For fixed complex structure the trilinear coupling in \eqref{eq:D3action} is first rewritten with \eqref{eq:fluxexpand} as
\begin{equation} \label{eq:Gfluxcomplex}
   \Gflux^+_{mnp}\dbt^m\dbt^n\dbt^p \rightarrow
    -\frac{1}{\int_Y\Omega\wedge\bar\Omega}\left(\mathcal{I}\Omega_{ijk}\dbt^i\dbt^j\dbt^k+
     \mathcal{I}_{\tilde a}\mathcal{G}^{\tilde a\tilde b}\bar\chi_{\tilde b\:i\bar\jmath\bar k}
     \dbt^i\bar\dbt^{\bar\jmath}\bar\dbt^{\bar k} \right) \ .
\end{equation}
However, in order to obtain an expansion up to fourth order in $\dbt^i$ and $z^{\tilde a}$ we also need to include the complex structure deformations $z^{\tilde a}$ linearly. Together with \eqref{eq:dbtper} and \eqref{eq:Kodaira} one obtains up to linear order in the complex structure deformations~$z^{\tilde a}$ \cite{Grana:2003ek}
\begin{equation} \label{eq:LinComp}
\begin{split}
   \frac{1}{3!}\Omega_{ijk}\dbt^i\dbt^j\dbt^k
     &\rightarrow \frac{1}{3!}(1+k_{\tilde a} z^{\tilde a})\Omega_{ijk}\dbt^i\dbt^j\dbt^k \ , \\
   \frac{1}{2!}\chi_{\tilde a\:ij\bar k}\:\dbt^i\dbt^j\bar\dbt^{\bar k}
     &\rightarrow\frac{1}{2!}(1+k_{\tilde b}z^{\tilde b}+\bar k_{\tilde b}\bar z^{\tilde b})
      \chi_{\tilde a\:ij\bar k}\:\dbt^i\dbt^j\bar\dbt^{\bar k} \ ,
\end{split}
\end{equation}
and with \eqref{eq:Gfluxcomplex} the trilinear couplings become linearly in the complex structure deformations~$z^{\tilde a}$
\begin{multline} \label{eq:trilinear}
   \mathcal{S}^\text{E}_\text{tri,flux}=-\frac{\ii\mu_3}{\ell}
      \int\dd^4\xi\sqrt{-\det\eta} \: \ee^K \: 
      \left(\mathcal{I}(1+k_{\tilde a} z^{\tilde a})\Omega_{ijk}\dbt^i\dbt^j\dbt^k \right. \\
      \left. +3\mathcal{I}_{\tilde a}\mathcal{G}^{\tilde a\tilde b}
      (1+k_{\tilde c} z^{\tilde c}+\bar k_{\tilde c}\bar z^{\tilde c})  
      \bar\chi_{\tilde b\:\bar\imath\bar\jmath k}
      \bar\dbt^{\bar\imath}\bar\dbt^{\bar\jmath}\dbt^k+\hc\right) \ ,
\end{multline}
where we inserted \eqref{eq:LinComp} into \eqref{eq:Gfluxcomplex}.

Recall that in the computation of the D3-brane action~\eqref{eq:D3action} the tadpole terms in the Chern-Simons and the Dirac-Born-Infeld action canceled. This is the consequence of expanding around a stable vacuum. However, if we want to include imaginary-anti-self-dual fluxes~$\Gflux^+$ into the analysis this cancellation does not occur anymore. Instead one expects small deviations, which also depend on the `matter fields'~$\dbt^i$. In particular expanding these deviations to second order in $\dbt^i$ generates flux-induced masses for some of the `matter fields'~$\dbt^i$ \cite{Camara:2003ku,Grana:2003ek}. Since these masses are interlinked to the back-reaction of geometry, which is not captured in our derivation, it is hard to derive their precise form from the Kaluza-Klein reduction. However, instead of just relying on the bosonic fields, we can also determine the masses of the fermionic superpartners and then try to gain further insight into the structure of these bosonic masses by using supergravity methods. Therefore we now come to the discussion of the fermionic effective action and address the supergravity analysis in section~\ref{sec:D3softterms}.

The kinetic terms and the interaction terms of the D3-brane fermions originate from the supersymmetric version of the Dirac-Born-Infeld and the Chern-Simons action discussed in section~\ref{sec:superaction}.  The expansion of the super Dirac-Born-Infeld action \eqref{eq:SDBI} and super Chern-Simons action \eqref{eq:SCS} is now obtained by a normal coordinate expansion of the D3-brane embedded into the ten-dimensional type~IIB superspace \cite{Grisaru:1997ub,Marolf:2003vf}. The non-Abelian version of the super Dirac-Born-Infeld and super Chern-Simons action is not known yet. However, for a stack of $N$ D3-branes one can perform the normal coordinate expansion for the Abelian case and in a second step adjust the expanded action to the non-Abelian $U(N)$ case. Then this amounts to rendering the resulting Lagrangian gauge invariant by including $U(N)$ gauge traces since the D3-brane fields transform in the adjoint representation of $U(N)$. In addition one expects the appearance of additional non-Abelian terms. Due to the cyclic property of the gauge trace these extra terms, however, must contain at least three (adjoint-valued) D3-brane fields. In this section we concentrate on the derivation of the D3-brane fermionic kinetic terms and their background flux induced masses. As all these terms are of second order in the D3-brane fields the analysis is insensitive to this ambiguity arising at third order in the D3-brane fields.

The normal coordinate expansion for space-time filling D3-branes as described above has been carried out in detail in ref.~\cite{Grana:2002tu}. Here we use the result of this computation for the bilinear fermionic D3-brane terms, which in the string frame reads
\begin{equation} \label{eq:D3fermbilinear}
   \mathcal{S}_\text{ferm}^\text{sf}=-\mu_3\int\dd^4\xi\sqrt{-\det\hat\eta}\:
      \tr\left(-\frac{\ii}{2}\ee^{-\phi}\bar\Theta\Gamma^\mu\covdb_\mu\Theta
      +\frac{1}{24}\bar\Theta\Gamma^{mnp}\Theta\:\Imag\Gflux^+_{mnp}\right) \ .
\end{equation}
Here $\Theta(\xi)$ is the fermionic D3-brane fluctuation \eqref{eq:D3fermions} and the covariant derivative $\covdb_\mu$ contains a spin connection and in the non-Abelian generalization also the gauge connection of the $U(N)$ gauge group.

In order to derive the kinetic terms for the fermionic fields~$\chi^i$ and $\lambda$ the expansion \eqref{eq:D3fermions} is inserted into the first term of eq.~\eqref{eq:D3fermbilinear}. After Weyl rescaling to the four-dimensional Einstein frame one arrives at
\begin{multline} \label{eq:D3fermkin}
   \mathcal{S}_\text{kin}^\text{E}=-\mu_3  \int\dd^4\xi\sqrt{-\det\eta}\:\tr\left(
      \ii\abs{N_{\chi^i}}^2\left(\frac{12}{\mathcal{K}}\ee^\phi\right)
      g_{i\bar\jmath}\bar\chi^{\bar\jmath}\bar\sigma^\mu\nabla_\mu\chi^i \right. \\
      \left.-\frac{\ii}{2}\abs{N_\lambda}^2\left(\frac{6}{\mathcal{K}}\ee^{\phi/2}\right)
      \left(\lambda\sigma^\mu\nabla_\mu\bar\lambda+\bar\lambda\bar\sigma^\mu\nabla_\mu\lambda\right)
      \right) \ .
\end{multline}      
Now we compare these terms with the kinetic terms of their bosonic superpartners $\dbt^i$ and $A_\mu$. Due to $\mathcal{N}=1$ supersymmetry the target-space K\"ahler metric is the same for both the fermionic and the bosonic fields in the chiral multiplets (c.f.~\eqref{eq:4Dbos} and \eqref{eq:4Dferm}). Similarly the real part of the gauge coupling constant~\eqref{eq:D3f} appears as the coefficient of the kinetic term for the field strength $F$ and for the gaugino~$\lambda$. Therefore supersymmetry allows us to determine the normalizations constants $N_{\chi^i}$ and $N_\lambda$ from \eqref{eq:D3fermkin}, namely
\begin{align} \label{eq:D3norm}
   N_{\chi^i}=\frac{\sqrt{2}}{2}\ee^{-\phi/2} \ , &&
   N_\lambda=\sqrt{\frac{\mathcal{K}}{6}}\ell\ee^{-3/4\phi} \ .
\end{align}

With the normalization constants at hand it is now possible to derive the flux induced fermionic masses. The relevant interaction term in \eqref{eq:D3fermbilinear} contains the combinations $\Gflux^+$, which we expand according to eq.~\eqref{eq:fluxexpand}. Then inserting \eqref{eq:D3fermions} and \eqref{eq:D3norm} into the second term of \eqref{eq:D3fermbilinear} we get the fermionic masses in the four-dimensional Einstein frame \cite{Grana:2003ek}
\begin{equation} \label{eq:D3fermmass}
   \mathcal{S}^\text{E}_\text{mass}
      =-\int\dd^4\xi\sqrt{-\det\eta}\:\ee^{\hat K/2}
       \left(\mu_3\ell^2\ee^{-\phi}\mathcal{I}\,\lambda\lambda
       +\frac{6\ii\mu_3}{\mathcal{K}}\mathcal{I}_{\tilde a}\mathcal{G}^{\tilde a\tilde b}
       g_{i\bar l}\,\bar\wp_{\tilde a}\vphantom{\wp_{\tilde a}}_j^{\bar l}\,
       \chi^i \chi^j+\hc \right)  \ . 
\end{equation}
These are the relevant fermionic couplings for the forthcoming soft-term supergravity analysis carried out in section~\ref{sec:D3softterms}. 

\subsection{Soft-terms in Calabi-Yau orientifolds with D3-branes} \label{sec:D3softterms}

With the flux-induced scalar potential terms and the fermionic masses at hand we can now compute the soft-terms resulting from the spontaneously broken supergravity theory by taking the decoupling limit of gravity~$M_\text{p}^{(4)}\rightarrow\infty$. The definition of soft-terms and their relation to spontaneously broken $\mathcal{N}=1$ supergravity theories is reviewed in appendix~\ref{app:softterms} together with the relevant formulae to carry out the details of the following discussion.

In the case of Calabi-Yau orientifolds with space-time filling D3-branes the neutral moduli fields are comprised of $M^I=(\tau,G^a,T_\alpha,z^{\tilde a})$ whereas the charged fields correspond to the D3-brane `matter fields'~$\dbt^i$. Then the first task in the soft-term analysis is the expansion of the supergravity K\"ahler potential with respect to the `matter fields'~$\dbt^i$. The K\"ahler potential~$\hat K(M,\bar M)$ of the bulk moduli fields is simply given by the K\"ahler potential of the orientifold bulk theory without space-time filling D3-branes, i.e. $\hat K(\tau,G,T,z)=K(\tau,G,T,z,\dbt=0)$, which amounts to setting $\dbt=0$ in the implicit definition of the K\"ahler coordinates~\eqref{eq:D3KCoord}. 

The K\"ahler potential~$K(\tau,G,T,z,\dbt)$ expanded to second order in the D3-brane `matter fields'~$\dbt^i$ yields the moduli dependent second order coefficients~$Z_{i\bar\jmath}$ and $H_{ij}$
\begin{align} \label{eq:ZH}
   Z_{i\bar\jmath}=-\frac{6\ii\mu_3}
     {\mathcal{K}}v^\alpha\:\omega_{\alpha\:i\bar\jmath} \ , &&
   H_{ij}=\frac{3\mu_3}{\mathcal{K}}v^\alpha
     \left(\omega_{\alpha\:i\bar l}\:\bar\wp_{\tilde a}\vphantom{\wp_{\tilde a}}_j^{\bar l}
     +\omega_{\alpha\:j\bar l}\:\bar\wp_{\tilde a}\vphantom{\wp_{\tilde a}}_i^{\bar l}\right)
     \bar z^{\tilde a} \ .
\end{align} 

The next task is to determine the quadratic and trilinear `supersymmetric couplings' $\mu_{ij}$ and $Y_{ijk}$ of the effective superpotential~$W^\text{eff}$ as given in \eqref{eq:Weff}. This is achieved by taking the decoupling limit~$M_\text{p}^{(4)}\rightarrow\infty$, $m_{3/2}=\text{const.}$ of the scalar potential of the supergravity action~\eqref{eq:D3SUGRAaction} and the flux-induced trilinear terms~\eqref{eq:trilinear} which yield 
\begin{equation} \label{eq:Veffect}
   V^\text{eff}(\dbt,\bar\dbt)
     =V^\text{eff}_\text{mass}(\dbt,\bar\dbt)+V^\text{eff}_\text{trilinear}(\dbt,\bar\dbt)
      +V^\text{eff}_\text{quartic}(\dbt,\bar\dbt) \ ,
\end{equation}
with
\begin{equation} \label{eq:Veff34}
\begin{aligned}
   V^\text{eff}_\text{trilinear}&=
     \frac{18\mu_3\ee^\phi}{\mathcal{K}^2\int_Y\Omega\wedge\bar\Omega}
     \tr\left(\mathcal{I}(1+k_{\tilde a}z^{\tilde a})\Omega_{ijk}\dbt^i\dbt^j\dbt^k\right. \\
     &\qquad\qquad\qquad\left.+\mathcal{I}_{\tilde a}\mathcal{G}^{\tilde a\tilde b}
     (1+k_{\tilde c}z^{\tilde c}+\bar k_{\tilde c}\bar z^{\tilde c}) 
     \bar\chi_{\tilde b\:\bar\imath\bar\jmath k}\bar\dbt^{\bar\imath}
     \bar\dbt^{\bar\jmath}\dbt^k+\hc\right) \ , \\
   V^\text{eff}_\text{quartic}&=
      \frac{18\mu_3 \:\ee^\phi}{\mathcal{K}^2\ell^2}
      \tr\left(g_{i\bar\jmath}g_{k\bar l}\com{\dbt^i}{\dbt^k}
      \com{\bar\dbt^{\bar l}}{\bar\dbt^{\bar\jmath}}
      +g_{i\bar\jmath}g_{k\bar l}\com{\dbt^k}{\bar\dbt^{\bar\jmath}}
      \com{\dbt^i}{\bar\dbt^{\bar l}}\right) \ .
\end{aligned}
\end{equation}
As noted already in section~\ref{sec:D3fermionic} there are also quadratic potential terms which, however, are difficult to compute from the Kaluza-Klein reduction due to the neglected back-reaction to geometry (c.f.~\ref{app:warp}). 

The first step is to analyze the quartic scalar potential~$V^\text{eff}_\text{quartic}$. With eqs.~\eqref{eq:Veff} and \eqref{eq:Weff} one observes that the quartic terms can only arise from the effective D-term~$\text{D}^\text{eff}$ and from the Yukawa couplings~$Y_{ijk}$ in the effective superpotential~$W^\text{eff}$, and therefore the effective D-term can be computed with~$Z_{i\bar\jmath}$ and becomes
\begin{equation} \label{eq:D3Deff}
   \text{D}^\text{eff}=-\frac{6\ii}{\mathcal{K}}
      \sqrt{\frac{\mu_3 }{\ell^2}}\:\ee^{\phi/2}\: 
      v^\alpha\omega_{\alpha\:i\bar\jmath}\:
      \com{\dbt^i}{\bar\dbt^{\bar\jmath}} \ ,
\end{equation}
where the real part of the gauge coupling function~\eqref{eq:D3f} is inserted.\footnote{Since the `matter fields'~$\dbt^i$ transform in the adjoint representation of $U(N)$ there appears the commutator in \eqref{eq:D3Deff}.} Now it is easy to see that the D-term \eqref{eq:D3Deff} gives rise to the second term in the quartic effective potential \eqref{eq:Veff34}. Therefore we can compute unambiguously the Yukawa couplings~$Y_{ijk}$ from the first term in $V^\text{eff}_\text{quartic}$. After a view steps of algebra the Yukawa couplings are found to be
\begin{equation} \label{eq:D3Yukawa}
   Y_{ijk}=\frac{3\mu_3}{\ell}\ee^{\hat K/2}\Omega_{ijk} \ .
\end{equation}

In our case the quadratic term $\mu_{ij}$ in the effective superpotential~$W^\text{eff}$ is most directly determined from the fermionic masses since we have not obtained the quadratic terms in the effective scalar potential~\eqref{eq:Veffect}. From the fermionic interaction terms \eqref{eq:D3fermmass} one reads off
\begin{equation} \label{eq:D3mu}
   \mu_{ij}=-\frac{3\ii\mu_3}{\mathcal{K}}\ee^{\hat K/2}
      \mathcal{I}_{\tilde a}\mathcal{G}^{\tilde a\tilde b}
      \left(g_{i\bar l}\,\bar\wp_{\tilde a}\vphantom{\wp_{\tilde a}}_j^{\bar l}
            +g_{j\bar l}\,\bar\wp_{\tilde a}\vphantom{\wp_{\tilde a}}_i^{\bar l}\right) \ .
\end{equation}
Note that the $\mu$-term is proportional to the supersymmetry breaking flux parameter $\mathcal{I}_{\tilde a}$ and thus vanishes in the supersymmetric limit. On the other hand $\mu_{ij}$ is also computed by \eqref{eq:muY}, which consists of two distinct pieces. The first term proportional to $\tilde \mu_{ij}$ in \eqref{eq:muY} survives in the supersymmetric limit while the second and third term are proportional to $m_{3/2}$ and $\mathrm{F}^I$ respectively, and hence vanish in the supersymmetric limit. Thus $\mu_{ij}$ must arise from these latter terms in eq.~\eqref{eq:muY}. In other words the masses~$\mu_{ij}$ are induced by a Giudice-Masiero mechanism \cite{Giudice:1988yz}. We conclude that $\tilde\mu_{ij}$ must vanish and using \eqref{eq:D3Yukawa} and \eqref{eq:muY} we deduce that the expansion of the supergravity superpotential \eqref{eq:KWexp} becomes
\begin{equation}
   W(\tau,z,\dbt)=\hat W(\tau,z)+\frac{\mu_3}{\ell}\Omega_{ijk}\tr \dbt^i\dbt^j\dbt^k \ .
\end{equation}
This is the superpotential already anticipated at the end of section~\ref{sec:D3SUGRA}. Now, with the precise form of the superpotential at hand one checks the $\mu$-term of \eqref{eq:D3mu} by employing the formula~\eqref{eq:muY} with \eqref{eq:ZH}. The other non-trivial check involves the trilinear scalar potential couplings~$V^\text{eff}_\text{trilinear}$. Note that due to the presence of both $\mu_{ij}$ and $Y_{ijk}$ the effective superpotential~$W^\text{eff}$ generates a mixed trilinear term, which turns out to be the second term of $V^\text{eff}_\text{trilinear}$ in \eqref{eq:Veff34} proportional to $\mathcal{I}_{\tilde a}$. These computations have been carried out in detail in ref.~\cite{Grana:2003ek}.

The final chore is to specify the soft-terms arising the bulk background fluxes \eqref{eq:FluxInt}. The A-term~$\soft{a}_{ijk}$ is directly read off from the first term in the trilinear scalar potential~$V^\text{eff}_\text{trilinear}$, whereas the gaugino mass~$\soft{m}_\text{g}$ is determined from the fermionic interaction terms \eqref{eq:D3fermmass}. On the other hand using the techniques reviewed in appendix~\ref{app:softterms} all soft-term can be derived by a supergravity analysis. This computation is straight forward but tedious and is also carried out in detail in ref.~\cite{Grana:2003ek}. In summary the soft-terms as determined by the supergravity computation and by comparing with the effective action are found to be
\begin{align} \label{eq:D3softterms}
   \soft{m}_{i\bar\jmath}=V_0Z_{i\bar\jmath} \ , &&
   \soft{a}_{ijk}=\ee^{\hat K/2}\mathcal{I}Y_{ijk} \ , &&
   \soft{b}_{ij}=V_0H_{ij} \ , &&
   \soft{m}_\text{g}=-\ee^{\hat K/2}\mathcal{I} \ .
\end{align}
Note that the soft-terms~$\soft{m}_{ij}$ and $\soft{b}_{ij}$ have not been obtained from the reduction of the D3-brane action \eqref{eq:D3actioncomp} since the masses for the `matter scalars'~$\dbt$ have not been computed by the reduction of the D3-brane action due to the undetermined quadratic terms in the effective scalar potential. Instead the soft-term supergravity analysis sketched above determines these masses indirectly and as a consequence the flux-induced masses for the `matter scalars'~$\dbt$ should be
\begin{align}
   m^2_{\text{scalar}\:ij}=\soft{b}_{ij} \ , &&
   m^2_{\text{scalar}\:i\bar\jmath}
      =Z^{l\bar k}\mu_{il}\bar\mu_{\bar\jmath\bar k}+\soft{m}_{i\bar\jmath} \ .
\end{align}

\subsection{Discussion}

In section~\ref{sec:bulkfluxes} we have seen that the background fluxes naturally split according to \eqref{eq:Gfluxsplit} into an imaginary self-dual piece~$\Gflux^-$ and into an imaginary anti-self-dual piece~$\Gflux^+$. 

In the derivation of the flux-induced terms in section~\ref{sec:D3fermionic} one notes that the scalar potential and the computed fermionic terms only depend on the imaginary anti-self-dual fluxes~$\Gflux^+$ represented by the non-vanishing integrals~$\mathcal{I}$ and $\mathcal{I}_{\tilde a}$. Hence it is natural to expect that the impact of imaginary self-dual fluxes~$\Gflux^-$ is not so severe. Indeed all the soft-terms and the $\mu$-term computed in section~\ref{sec:D3softterms} vanish for $\mathcal{I}=\mathcal{I}_{\tilde a}=0$. 

According to Table~\ref{tab:SDforms} the imaginary self-dual fluxes are either $(2,1)$- or $(0,3)$-fluxes. For $(2,1)$-fluxes the $\mathcal{N}=1$ effective supergravity theory is unbroken and hence as a consequence in the limit $M^{(4)}_\text{p}\rightarrow\infty$, $m_{3/2}=\text{const}$ the resulting global supersymmetric theory is unbroken as well and thus exhibits no soft-terms. For the $(0,3)$-fluxes the flux-induced superpotential~$\hat W$ is non-vanishing and therefore supersymmetry is broken but is not communicated to the observable sector. This property has been denoted by `no-scale supersymmetry breaking' in refs.~\cite{Cremmer:1983bf,Ellis:1983sf}. However, in this case higher order $\alpha'$ and loop corrections potentially induce soft-terms, which then spoil the `no-scale' behavior.

Let us now discuss the imaginary anti-self-dual fluxes~$\Gflux^+$. According to Table~\ref{tab:SDforms} these constitute the $(1,2)$- and $(3,0)$-fluxes and result in non-vanishing flux integrals~$\mathcal{I}$ and $\mathcal{I}_{\tilde a}$. These flux integrals also break the low energy effective $\mathcal{N}=1$ supergravity action spontaneously and hence there appear soft-terms in the corresponding global supersymmetric theory. The A-terms~$\soft{a}_{ijk}$ are proportional to the Yukawa couplings~$Y_{ijk}$. The $\soft{b}_{ij}$ terms are not proportional to the $\mu$-term, which is sometimes assumed in certain phenomenological models. However, more interesting for phenomenological applications the soft scalar masses $\soft{m}_{i\bar\jmath}$ are universal, i.e. all soft scalar masses are equal after appropriately normalizing the corresponding kinetic terms. The universality of the soft scalar masses is important in minimal supersymmetric standard model like scenarios as this property ensures that the strong experimental bounds imposed on the appearance of flavor changing neutral currents are respected \cite{Chung:2003fi}.

To conclude this discussion we complete the analysis of scales started at the end of section~\ref{sec:bulkaction}. By turning on supersymmetry breaking bulk background fluxes the supersymmetry breaking scale~$M_\text{susy}$ enters as a new mass scale. Since background fluxes are quantized in units of $2\pi\alpha'$ the three-form flux density scales as $\alpha'/R^3$ in terms of the `radius'~$R$ of the internal space \cite{Camara:2003ku}. The flux-density is also the relevant quantity for appearing in D3-brane couplings since the stack of D3-branes is localized at a point in the internal space and hence is sensitive to the flux density \cite{Camara:2003ku}. Therefore the relevant mass scale~$M_\text{susy}$ for supersymmetry breaking in the observable sector also scales like $\alpha'/R^3$. Since $\sqrt{\alpha'}\ll R$ is required for the compactification ansatz the hierarchy of scales \eqref{eq:scales1} is extended to
\begin{equation} \label{eq:scales2}
   M_\text{susy}\ll M_\text{KK}\ll M_\text{s}\ll M^{(4)}_\text{p} \ .    
\end{equation}
This result is rather remarkable. It shows that if the overall volume is stabilized in the large radius regime the Kaluza-Klein reduction is valid and in addition the setup allows for supersymmetry breaking at small scales. On the other hand it also justifies in retrospect the Kaluza-Klein ansatz for the analysis of soft-terms while neglecting all the massive Kaluza-Klein modes.

\section{D7-branes with internal D7-brane fluxes} \label{sec:D7fluxes}


In the presence of space-time filling D7-branes it is also possible to turn on background fluxes on the internal cycle of the worldvolume which are introduced in section~\ref{sec:D7fluxesintro}. These fluxes generate D- and F-terms (c.f.~section~\ref{sec:bosana} and section~\ref{sec:fermana}), which analogously to bulk background fluxes can also break supersymmetry spontaneously. But, moreover, D7-brane background fluxes have also been suggested as relevant ingredients for generating metastable deSitter vacua \cite{Burgess:2003ic}. This aspect is carried on in section~\ref{sec:disdesitter}.

\subsection{Internal D7-brane fluxes} \label{sec:D7fluxesintro}

Up to now we have only specified the homology cycle~$S^\Lambda$ which is wrapped by the D7-brane. However, this does not completely specify the geometric data of a generic D7-brane, since in addition there can arise non-trivial background fluxes for the field strength of the $U(1)$ gauge theory localized on the D7-brane worldvolume. However, in order to preserve Poincar\'e invariance of the four-dimensional effective theory we consider only background fluxes on the internal D7-brane cycle~$S^\Lambda$. These fluxes are topologically non-trivial two-form configurations for the internal $U(1)$ field strength which nevertheless satisfy the Bianchi identity and the equation of motion. Therefore the background flux $f$ is constrained to be a harmonic form on $S^\Lambda$. In section~\ref{sec:openstr} it is shown that the gauge boson is odd with respect to the orientifold involution~$\sigma$ and as a consequence the possible background fluxes~$f$ correspond to elements in $H_-^2(S^\Lambda)$. 

The D7-brane cycle $S^\Lambda$ is embedded into the ambient Calabi-Yau manifold~$Y$ via the embedding map $\iota:S^\Lambda\hookrightarrow Y$, which induces the pullback map $\iota^*$ on forms
\begin{equation}
  \iota^*:H_-^2(Y)\rightarrow H_-^2(S^\Lambda) \ .
\end{equation}
Therefore one can distinguish between two different kinds of fluxes which we denote by $\leftupn{Y}{f}$ and $\tilde f$. $\leftupn{Y}{f}$ are harmonic two-forms on $S^\Lambda$ which are inherited from the ambient Calabi-Yau space~$Y$. $\tilde f$ on the other hand correspond to harmonic forms on $S^\Lambda$, which cannot be obtained by pullback from the ambient space~$Y$. Put differently, $\leftupn{Y}{f}$ are harmonic two-forms in the image of $\iota^*$ while $\tilde f$ are harmonic two-forms in the cokernel of $\iota^*$. This amounts to the fact that the cohomology group $H_-^2(S^\Lambda)$ can be decomposed as
\begin{equation} \label{eq:FluxCoh}
   H_-^2(S^\Lambda)\cong \leftupn{Y}{H_-^2(S^\Lambda)}\oplus \tilde H_-^2(S^\Lambda) \ ,
\end{equation}
where $\leftupn{Y}{H_-^2(S^\Lambda)}=\iota^*\left(H_-^2(Y)\right)$ and $\tilde H_-^2(S^\Lambda)=\coker\left(H_-^2(Y)\xrightarrow{\iota^*} H_-^2(S^\Lambda)\right)$. 
Then the flux $f\in H_-^2(S^\Lambda)$ splits accordingly 
\begin{equation}\label{eq:fsplit}
   f=\leftupn{Y}{f}+\tilde f \ ,
\end{equation} 
with $\leftupn{Y}{f} \in \leftupn{Y}{H_-^2(S^\Lambda)}$ and $\tilde f \in \tilde H_-^2(S^\Lambda)$.

Now there is a comment in order on a technical issue concerning the distinction of the fluxes~$\leftupn{Y}{f}$ and $\tilde f$. This splitting is not unique but we choose it in such a way that the integrals 
\begin{equation} \label{eq:vanishint}
   \int_{S^\Lambda}\iota^*\omega_a \wedge \tilde f=0 
\end{equation}
vanish for all two-forms $\omega_a$ in $H^2_-(Y)$. This can alway be achieved by first choosing a basis of two-forms~$\omega_a$ for $\leftupn{Y}H_-(S^\Lambda)$ and then by choosing a basis of two-cycles~$\tilde S^{\tilde a}$ for $\ker\: (H_{2,-}(S^\Lambda)\xrightarrow{\iota_*}H_{2,-}(Y))$. Then the Poincar\'e dual basis~$\omega_{\tilde a}$ of $\tilde S^{\tilde a}$ spans the cokernel $\tilde H^2_-(S^\Lambda)$. If now the splitting is chosen in such a way that $\leftupn{Y}{f}$ can be expanded into $\omega_a$ and $\tilde f$ into $\omega_{\tilde a}$ the relation~\eqref{eq:vanishint} is fulfilled.

Before we conclude this discussion we introduce the topological flux charges, which can be built from the fluxes~$f$. The intrinsic fluxes~$\tilde f$ give rise to the topological charges
\begin{align} \label{eq:fcharges}
   Q_A = \ell \int_{S^\Lambda}\tilde s_A\wedge\tilde f \ , &&
   Q_\alpha=\ell\int_{S^\Lambda}\iota^*\omega_\alpha\wedge P_-\tilde f \ , &&
   \fcharge = \ell^2 \int_{S^\Lambda} \tilde f\wedge\tilde f \ . 
\end{align}
At first sight the charges $Q_\alpha$ always seem to vanish due to \eqref{eq:vanishint}. However, although $\tilde f$ is an element in the cokernel of $\iota^*$ obeying \eqref{eq:vanishint}, $P_-\tilde f$ need not be in the cokernel of $\iota^*$. Thus some of the charges $Q_\alpha$ can indeed become non-vanishing. For instance this situation occurs if the flux on the D7-brane and the negative value of the flux on the image-D7-brane can both be written as the pullback of the same two-form in the ambient space. In this case $\tilde f$ corresponds to a trivial and $P_-\tilde f$ to a non-trivial two-form in the ambient space~$Y$.

On the other hand the flux~$\leftupn{Y}{f}$ can be expanded into $(1,1)$-forms pulled back to the internal D7-brane cycle~$S^\Lambda$ 
\begin{equation}
   \leftupn{Y}{f}=\leftupn{Y}{f^a}\:\iota^*\omega_a \ .
\end{equation}
In the D7-brane effective action \eqref{eq:DBIab} and \eqref{eq:CSAb} the fluxes $f$ always appear in the combination $\mathcal{F}=B-\ell f$ and therefore we expect that the fluxes $\leftupn{Y}{f}$ combine to $b^a-\ell\leftupn{Y}{f^a}$ and we define
\begin{align} \label{eq:shiftB}
   \dbbf=B-\ell\leftupn{Y}{f} \ , && \dbbf^a=b^a-\ell\leftupn{Y}{f^a} \ .
\end{align}
In the following sections it is confirmed that the flux charges~\eqref{eq:fcharges} and the shifts of the bulk $b^a$-fields appear indeed in the effective description in the presence of internal D7-brane fluxes~$f$.

\subsection{Bosonic effective action with D7-brane fluxes} \label{sec:bosana}

In order to derive the additional terms induced from the D7-brane fluxes~$f$ in the four-dimensional effective action it is necessary to redo the Kaluza-Klein reduction of section~\ref{sec:D7action}. Here we do not go through this computation and state the resulting effective action in appendix~\ref{app:D7action}. In the derivation of the action~\eqref{eq:BulkD7actionf} one now uses the calibration condition~\eqref{eq:cali3} with non-trivial background fluxes~$f$ and the integral relation~\eqref{eq:vanishint}. Further details of the Kaluza-Klein reduction are given in refs.~\cite{Jockers:2004yj,Jockers:2005zy}.

In the $\mathcal{N}=1$ supergravity language the additional terms in the action~\eqref{eq:BulkD7actionf} amount to a modification of the chiral coordinates~$G^a$ and $T_\alpha$ in eq.~\eqref{eq:G}. Or in other words the complex structure of the K\"ahler target space manifold is adjusted by the D7-brane fluxes~$f$. The new definitions of the chiral fields~$G^a$ and $T_\alpha$ read 
\begin{equation} \label{eq:fluxGT}
\begin{aligned}
   G^a=&c^a-\tau\dbbf^a \ , \\
   T_\alpha=&\frac{3\ii}{2}\left(\rho_\alpha
      -\tfrac{1}{2}\mathcal{K}_{\alpha bc}c^b\dbbf^c\right) +\frac{3}{4}\mathcal{K}_\alpha 
      +\frac{3\ii}{4(\tau-\bar\tau)} \mathcal{K}_{\alpha bc}G^b(G^c-\bar G^c) \\ 
      &\qquad\qquad+3\ii\kappa_4^2\mu_7\ell^2 \mathcal{C}^{I\bar J}_\alpha a_I \bar a_{\bar J} 
                   +\frac{3\ii}{4}\:\delta_\alpha^\Lambda\:\tau\fcharge \ .
\end{aligned}
\end{equation}
The K\"ahler potential \eqref{eq:K1} is unchanged but 
the additional terms proportional to $\fcharge$ enter nevertheless once $K$ is expressed in terms of its chiral coordinates.

The covariant derivatives of the scalars $\rho_\alpha$ are also modified. This can be seen from the fact that in the Kaluza-Klein reduction of the D7-brane Chern-Simons action \eqref{eq:CSAb} additional Green-Schwarz terms are induced from the fluxes~$\tilde f$
\begin{equation}\label{eq:GSterms}
   \mu_7\int_\mathcal{W} C^{(4)}\wedge\ell\tilde f\wedge\ell P_-F
   = \mu_7\ell \,Q_\alpha \int_{\mathbb{R}^{3,1}} D^\alpha\wedge F 
   = -\mu_7\ell \,Q_\alpha \int_{\mathbb{R}^{3,1}} \dd D^\alpha\wedge A \ ,
\end{equation}
where we used the expansion \eqref{eq:C} and definition of $Q_\alpha$ in \eqref{eq:fcharges}. After eliminating the two-forms $D_\alpha$ in favor of their dual scalars~$\rho_\alpha$ by imposing the self-duality condition on the five-form field-strength of $C^{(4)}$ according to appendix~\ref{app:dual},
the Green-Schwarz terms \eqref{eq:GSterms} modify the local Peccei-Quinn symmetry discussed in \eqref{eq:shift}. The covariant derivative for $\rho_\alpha$ changes and \eqref{eq:cd1} is 
replaced by
\begin{equation}
   \nabla_\mu\rho_\alpha=\partial_\mu\rho_\alpha
    -4\kappa_4^2\mu_7\ell\mathcal{K}_{\alpha bP}\dbbf^b A_\mu
    +4\kappa_4^2\mu_7\ell Q_\alpha A_\mu \ .
\end{equation}
In terms of the chiral coordinates \eqref{eq:fluxGT} the contribution proportional to $Q_\alpha$ leave the covariant derivative of $G^P$ \eqref{eq:cd2} unchanged while the fields $T_\alpha$ become charged and a covariant derivative of the form
\begin{equation} \label{eq:cd3}
   \nabla_\mu T_\alpha=\partial_\mu T_\alpha+6\ii\kappa_4^2\mu_7\ell Q_\alpha A_\mu \ 
\end{equation}
is induced. Thus fluxes $\tilde f$ which lead to non-vanishing $Q_\alpha$ change the gauged isometry \eqref{eq:shift} in that additional fields $T_\alpha$ become charged and transform non-linearly.

As a consequence of these additional charged chiral fields the D-term is also modified. The holomorphic Killing vector field \eqref{eq:Killing1} receives an additional contribution from the $T_\alpha$ and reads
\begin{equation} \label{eq:Killing2}
   X=4\kappa_4^2\mu_7\ell \partial_{G^P}-6\ii\kappa_4^2\mu_7\ell Q_\alpha \partial_{T_\alpha} \ .
\end{equation}
This in turn adjusts the D-term via eq.~\eqref{eq:Killing} and we find
\begin{equation} \label{eq:fDterm}
   \text{D}\ =\ \frac{12\kappa_4^2\mu_7\ell}{\mathcal{K}}
     \left(\mathcal{K}_{Pa}\dbbf^a-Q_\alpha v^\alpha\right) \
=\ \frac{12\kappa_4^2\mu_7\ell}{\mathcal{K}}\int_{S_P} J\wedge \mathcal{F}\ ,
\end{equation}
where $\mathcal{F}\equiv B-\ell f = \dbbf - \ell\tilde f$. The corresponding D-term potential is given by
\begin{equation} \label{eq:Vf}
   V_\text{D}=\frac{108\kappa_4^2\mu_7}{\mathcal{K}^2\Real{T_\Lambda}}
     \left(\mathcal{K}_{Pa}\dbbf^a-Q_\alpha v^\alpha\right)^2 \ .
\end{equation}
 
\subsection{D7-brane flux induced D- and F-terms} \label{sec:fermana}

In this section we determine the D7-brane D-term and F-term couplings by computing the fermionic couplings of the gravitinos to the D7-brane fermions as stated in \eqref{eq:4Dcoupl}. As we have seen it seems difficult to determine these terms directly from the Kaluza-Klein reduction of the bosonic terms. In the bosonic scalar potential the D- and F-terms are encoded quadratically according to \eqref{eq:spot}, whereas in the fermionic action these terms appear linearly according to \eqref{eq:4Dcoupl}. This makes it easier to retrieve the D- and F-terms from the fermionic couplings \eqref{eq:4Dcoupl}. Analogously in section~\ref{sec:D3softterms} the bosonic masses were only computed indirectly via a supergravity analysis, whereas the fermionic masses are directly obtained by a Kaluza-Klein reduction of the fermionic terms.

In order to reliably derive the fermionic couplings \eqref{eq:4Dcoupl} we must fix the normalization $N_\lambda$ and $N_{\chi^A}$ in \eqref{eq:decomptheta} of the `matter fermions'~$\chi^A$ and of the gaugino~$\lambda$ and the normalization of the four-dimensional gravitino~$\psi_\mu$. The normalization of the gravitino arises from the Kaluza-Klein reduction of the ten-dimensional Rarita-Schwinger term and the canonical normalization of the four-dimensional Rarita-Schwinger term is determined by the normalization of the in section~\ref{sec:calcond} introduced internal covariantly constant spinor singlets~$\check\singspin$\footnote{Canonically in the sense that it agrees with the normalization of the Rarita-Schwinger term in ref.~\cite{Wess:1992cp}.} \cite{Jockers:2005zy}
\begin{equation} \label{eq:normsingspin}
   \check\singspin^\dagger\check\singspin = 1 \ .
\end{equation}
The normalization of the `matter fermions'~$\chi^A$ are obtained by deriving the fermionic kinetic terms from the super Dirac-Born-Infeld action~\eqref{eq:SDBI}. This is achieved by computing the pullback $\super{\varphi^*}\super{g}$ of the supermetric~$\super{g}$ with respect to the supermap $\super{\varphi}$ which yields \cite{Grana:2002tu,Grana:2002nq,Jockers:2005zy}
\begin{equation} \label{eq:smet}
\begin{split}
   \super{\varphi^*}\super{g}
     =&\hat\eta_{\mu\nu}\dd x^\mu \dd x^\nu+2\hat g_{i\bar\jmath}\dd y^i\dd y^{\bar\jmath} 
       +2\hat g_{i\bar\jmath} \partial_\mu\dbs^i\partial_\nu\bar\dbs^{\bar\jmath}
        \dd x^\mu\dd x^\nu \\
      &-\frac{\ii}{2}\bar\Theta\Gamma_\mu\spinder_\nu\Theta\dd x^\mu\dd x^\nu
       -\frac{\ii}{2}\bar\Theta\Gamma_\nu\spinder_\mu\Theta\dd x^\mu\dd x^\nu+\ldots \ .
\end{split}
\end{equation}
Note that this generalizes equation \eqref{eq:nc} by the appearance of the fermionic fluctuations $\Theta$. As discussed in detail in ref.~\cite{Jockers:2005zy} inserting the expansion \eqref{eq:smet} into the super-Dirac-Born-Infeld action \eqref{eq:SDBI} allows us to determine the normalization constants $N_\lambda$ and $N_{\chi^A}$ explicitly by comparing the coefficients of the bosonic kinetic terms in \eqref{eq:4Dbos} with the fermionic kinetic terms in \eqref{eq:4Dferm} \cite{Jockers:2005zy}, i.e.
\begin{align} \label{eq:norm}
   N_\lambda=\sqrt{\frac{\mathcal{K}}{6}}\ell\:\ee^{-\phi/4} \ , &&
   N_{\chi^A}=\frac{1}{2}\sqrt{\frac{\ii\:\mathcal{K}}
     {6\int\Omega\wedge\bar\Omega}}\:\ee^{3\phi/4} \ .
\end{align}

Now we have introduced all the ingredients to compute the couplings of the gravitino~$\psi_\mu$ with the D7-brane fermions. These couplings appear in the super-Chern-Simons action \eqref{eq:SCS} in the term involving the super RR-six-form
\begin{equation} \label{eq:SCS6}
   -\mu_7 \int_{\mathcal{W}} \super{\varphi^*}\left(\super{C^{(6)}}\right) \wedge \mathcal{F} \ ,
\end{equation}
with the worldvolume two-form $\mathcal{F}=B-\ell f = \dbbf - \ell\tilde f$. In order to evaluate this term we need to determine the super-pullback of the super-RR-six-form~$\super{C^{(6)}}$ which reads after taking into account the orientifold truncation \cite{Bergshoeff:1999bx,Jockers:2005zy}
\begin{equation}
   \super{\varphi^*}\left(\super{C^{(6)}}\right)=\varphi^*C^{(6)}
   +\frac{\ii}{5!}\ee^{-\phi}\:\bar\Theta\:\Gamma_{M_1\ldots M_5}
   \Psi_{M_6}\:\dd x^{M_1}\wedge\ldots\wedge\dd x^{M_6}+ \ldots \ .
\end{equation}
Recall that in $\Theta$ all the fermionic D7-brane fields are encoded according to eq.~\eqref{eq:decomptheta}. In particular the last equation also captures the couplings of the four-dimensional gravitinos $\psi_\mu$ to the D7-brane fermions $\Theta$, which are of the form \eqref{eq:4Dcoupl}, i.e. 
\begin{multline} 
   \frac{\ii}{2\cdot 3!} \ee^{-\phi} \bar\Theta \Gamma_{\mu_0\mu_1\mu_2 mn}\Psi_{\mu_3}\:
   \:\dd x^{\mu_0}\wedge\ldots\wedge\dd x^{\mu_3}\wedge\dd y^m\wedge \dd y^n \\ 
   =\frac{\ii}{2}\:\ee^{-\phi}\:\sqrt{-\hat\eta}\:\dd^4x\:\bar\Theta 
   \left(\hat\gamma^\mu\hat\gamma^5\otimes\gamma_{mn}\right)\Psi_\mu\:\dd y^m\wedge\dd y^n  \ .
\end{multline}
Finally inserting this expression into the super-Chern-Simons term \eqref{eq:SCS6} together with \eqref{eq:decomptheta} we arrive after Weyl rescaling with \eqref{eq:Weyl} to the four-dimensional Einstein frame and a few steps of algebra at \cite{Jockers:2005zy}
\begin{align} \label{eq:CouplCS}
   \mathcal{S}_\text{Couplings}=&-\frac{1}{2\kappa_4^2}\int\dd^4x\sqrt{-\eta} \nonumber \\
   &\cdot\left[\sqrt{2}\ii\:\ee^{K/2}\cdot \kappa_4^2\mu_7\left(
        \chi^A\sigma^\mu\bar\psi_\mu\:\int_{S^\Lambda}\tilde s_A\wedge\mathcal{F}
        +\bar\chi^{\bar A}\bar\sigma^\mu\psi_\mu 
        \int_{S^\Lambda}\tilde s_{\bar A}\wedge\mathcal{F}\right)\right. \nonumber \\
   &\quad+\left.\left(\psi_\mu\sigma^\mu\bar\lambda-\bar\psi_\mu\bar\sigma^\mu\lambda\right)
        \left(-12\kappa_4^2\mu_7\ell\frac{1}{\mathcal{K}}\int_{S^P}J\wedge\mathcal{F}\right)
        \right] \ .
\end{align}
The final chore is to compare these fermionic terms with the supergravity action~\eqref{eq:4Dcoupl} and one extracts immediately a generic expression for the D-term of the $U(1)$ D7-brane gauge theory
\begin{equation}\label{eq:Dfermionic}
   \text{D}\ =\ \frac{12\kappa_4^2\mu_7\ell}{\mathcal{K}}\int_{S^P}J\wedge\mathcal{F} \
    =\
\frac{12\kappa_4^2\mu_7\ell}{\mathcal{K}}
    \left(\mathcal{K}_{Pa}\dbbf^a-Q_\alpha v^\alpha\right) \ ,
\end{equation}
where in the last step we used again \eqref{eq:K} and \eqref{eq:fcharges}. Hence the fermionic D7-brane reduction confirms the D-term \eqref{eq:fDterm}, which is computed in the previous section by means of analyzing the bosonic part of the $\mathcal{N}=1$ supergravity action. Note that the fermionic computation has neither required the knowledge of the gauged isometries nor the structure of the K\"ahler potential. This is a consequence of the general form of the K\"ahler potential~\eqref{eq:genericK} discussed in section~\ref{sec:DisGenericK}. However, for the supergravity derivation it is crucial to know the definition of the chiral variables and the K\"ahler potential, as this information enters in the differential equation~\eqref{eq:Killing} which encodes the D-term. Therefore the fermionic computation is an alternative and more direct way to determine the D-term. For the case at hand it confirms the supergravity computation and in addition checks the structure of the K\"ahler potential \eqref{eq:genericK}. 

In a second step we match the terms in \eqref{eq:4Dcoupl} with \eqref{eq:CouplCS} and we readily determine the flux induced superpotential. The integral relation \eqref{eq:vanishint} is responsible for the superpotential to be independent of $\dbbf$ because the two-forms $\tilde s_A$ are elements of $\tilde H^2_-(S^\Lambda)$ whereas the two-form $\dbbf$ are inherited from the bulk. Therefore we arrive at the holomorphic superpotential
\begin{equation} \label{eq:W1}
   W(\dbs)=\kappa_4^2\mu_7\ell \dbs^A \int_{S^\Lambda}\tilde s_A \wedge \tilde f  
          =\kappa_4^2\mu_7\, Q_A \dbs^A \ ,
\end{equation} 
in terms of the charge $Q_A$ defined in \eqref{eq:fcharges}.

By parsing through the derivation of the D- and F-terms it is striking that both terms arise from the same Chern-Simons worldvolume term, and moreover reveal a remarkable similar structure. This feature can be traced back to the fact that the $\mathcal{N}=1$ orientifold supergravities arise as truncations of $\mathcal{N}=2$ supergravities \cite{D'Auria:2004kx}. In the case under consideration the computed D- and F-terms can be associated to the same $\mathcal{N}=2$ Killing prepotential $SU(2)$-triplet \cite{Lust:2005bd,Berglund:2005dm}, which describes scalar potential terms arising from gaugings in the hypermultiplet sector of the parent $\mathcal{N}=2$ supergravity theory \cite{Andrianopoli:1996cm}.

\subsection{Discussion} \label{sec:disdesitter}

By computing the flux-induced D- and F-terms form fermionic couplings we have circumvented the problem of extracting this information from the scalar potential, which seems difficult to derive directly from a Kaluza-Klein reduction. In particular the computation of the D-term from the fermionic terms confirms the supergravity analysis of section~\ref{sec:bosana} and therefore also constitutes for a non-trivial check on the K\"ahler potential \eqref{eq:K1}. For the F-term, on the other hand, we find an independent confirmation in the next section.

According to eq.~\eqref{eq:fsplit} the internal D7-brane fluxes naturally split into two distinct contributions. The fluxes denoted by $\leftupn{Y}{f}$ are inherited from the ambient Calabi-Yau space and adjust the chiral variables~$G^a$. However, there do not appear any new charged chiral fields and hence the D-term is simply shifted but does not receive any additional contributions. On the other hand the intrinsic fluxes~$\tilde f$ are more diverse. Potentially they both contribute to the D-term and generate F-terms. This feature, namely that a class of D7-brane fluxes generates D- and F-terms simultaneously, has also been observed in ref.~\cite{Lust:2005bd}. The modification of the D-term is also caused by the appearance of new charged chiral fields out of the set of chiral moduli~$T_\alpha$.

Note that in the case that $T_\Lambda$ becomes charged, the (classical) effective action is not gauge invariant anymore, since $T_\Lambda$ is proportional to the gauge coupling function~\eqref{eq:fbrane}. A detailed analysis as performed in ref.~\cite{Jockers:2005zy} shows that in this case there appear chiral fermions in charged $\mathcal{N}=1$ chiral multiplets~$X_i$ which are suitable to cancel this anomaly by the Green-Schwarz mechanism. These additional multiplets arise from the intersections of D7-brane with its image-D7-brane. At the same time these intersections are also necessary for the field~$T_\Lambda$ to become charged \cite{Jockers:2005zy}. 

In order to gain further insight into the effects of the flux-induced D- and F-terms, we briefly comment on the scalar potential of a Calabi-Yau orientifold~$Y$ with $h_+^{1,1}=1$ and $h_-^{1,1}=0$, that is to say we have a single harmonic two-form~$\omega_\Lambda$. Then the only K\"ahler variables in the theory are given by $S$, $\dbs^A$ and $T_\Lambda$, where we also keep the complex structure deformations fixed. Now we turn on D7-brane fluxes such that $Q_\Lambda$ and $\tilde f$ are non-zero but $\fcharge$ vanishes.\footnote{We do not have an explicit Calabi-Yau orientifold with all these required properties. However, this example is instructive because it reveals many features of the generic case.} The corresponding D-term scalar potential of eq.~\eqref{eq:Vf} reduces for this particular setup to \cite{Jockers:2005zy}
\begin{equation} \label{eq:spotD}
   V_\text{D}=\frac{6\kappa_4^2\mu_7}{T_\Lambda+\bar T_\Lambda}
      \left(\frac{9\:Q_\Lambda}{T_\Lambda+\bar T_\Lambda}-\sum_i 
      q_i \abs{X_i}^2\right)^2 \ ,
\end{equation}   
where as discussed above additional chiral matter multiplets~$X_i$ with charge~$q_i$ are present to cancel the anomaly of the gauge coupling. This form of the scalar potential precisely coincides with the potential obtained in ref.~\cite{Burgess:2003ic}. However, we should stress that it crucially depends on the existence of a  non-vanishing $Q_\Lambda$ and the absence of the fields~$\dbbf^a$ which follows from our choice $h_-^{1,1}=0$.

As already discussed in ref.~\cite{Burgess:2003ic} the  minimum of  $V_\text{D}$ depends on the properties of other couplings of $X_i$ and also on possible non-perturbative corrections.  If the vacuum expectation value of the $X_i$ is not fixed by additional F-term couplings, $V_\text{D}=0$ can be obtained by adjusting $\vev{X_i}$. If, on the other hand, F-terms impose $\vev{X_i}=0$ a vanishing D-term potential only occurs for $T_\Lambda+\bar T_\Lambda \rightarrow \infty$ resulting in a run-away behavior. However, as discussed in refs.~\cite{Kachru:2003aw,Kachru:2003sx,Gorlich:2004qm,Denef:2004dm} the K\"ahler modulus $T_\Lambda$ can possibly be stabilized by non-perturbative effects such as Euclidean-D3-brane instantons and/or gaugino condensation on a stack of D7-branes. In this case the D-term spontaneously breaks supersymmetry and can indeed provide for a mechanism to uplift an Anti-deSitter vacuum to a metastable deSitter minimum along the lines of ref.~\cite{Kachru:2003aw,Kachru:2003sx}.

Now we turn to the discussion of the F-term scalar potential, which is computed by inserting \eqref{eq:W1} into \eqref{eq:spot} \cite{Jockers:2005zy}, i.e.
\begin{equation}
   V_\text{F}=\frac{3^6}{2} \kappa_4^2\mu_7\, G^{C\bar D} Q_C Q_{\bar D} \,
     \frac{1+\kappa_4^2\mu_7 \ee^{\phi} G_{A\bar B}\:\dbs^A\bar\dbs^B}
     {(T_\Lambda+\bar T_\Lambda)^3}\, \ .
\end{equation}
Here we have defined the metric $G_{A\bar B}=\ii\mathcal{L}_{A\bar B}$ and its inverse $G^{A\bar B}$. As one can easily see the effect of the F-term scalar potential is twofold. On the one hand it also exhibits a runaway behavior but on the other hand once the K\"ahler modulus~$T_\Lambda$ is stabilized the fluxes~$\tilde f$ render some of the fluctuations~$\dbs^A$ massive and hence stabilize these D7-brane fields. Note that in this case some of the F-terms associated to $\dbs^A$ are non-vanishing, and therefore these D7-brane fluxes also break supersymmetry spontaneously. This can be traced back to the structure of the superpotential~\eqref{eq:W1}, which is linear in the D7-brane `matter fields'~$\dbs^A$. 


\section{The holomorphic Chern-Simons theory} \label{sec:FCS}


In the derivation of the effective supergravity actions we have encounter the appearance of various superpotentials. On the one hand the soft-terms computed in section~\ref{sec:D3softterms} are induced from the non-trivial bulk background fluxes and the effective action is governed by the flux-induced superpotential~\eqref{eq:GWPot}. This superpotential does not just arise in orientifold compactifications but instead it reoccurs as a generic expression for many superstring compactifications with $\mathcal{N}=1$ supersymmetry in the low energy effective action \cite{Taylor:1999ii,Giddings:2001yu,Becker:2002nn,DeWolfe:2002nn}. This is due to the fact that by general arguments the superpotential~\eqref{eq:GWPot} can be traced back to the BPS tension of D5-brane domain walls \cite{Gukov:1999ya}. This observation suggested that there is also a general argument which captures the structure of the superpotentials involving the D-brane `matter fields' fields. In the supergravity computations we find the superpotential \eqref{eq:D3suppot} for the D3-brane `matter fields'~$\dbt^i$ and the superpotential \eqref{eq:W1} for the D7-brane `matter fields'~$\dbs^A$. Both superpotentials can indeed be obtained from an appropriate reduction of the holomorphic Chern-Simons action.

In ref.~\cite{Witten:1992fb} it is shown that the topological open string disk partition function in the presence of $N$ D-branes wrapping the entire internal Calabi-Yau manifold~$Y$ is given by the holomorphic Chern-Simons action 
\begin{equation} \label{eq:HCS}
   W_Y=\int_Y\Omega\wedge\tr \left(A\wedge\bar\partial A +\frac{2}{3}A\wedge A\wedge A\right) \ , 
\end{equation}
where $A$ is the gauge field of the $U(N)$ gauge theory of the D-branes. On the other hand the open string disk partition function is the superpotential in the low energy effective action of the physical string theory \cite{Bershadsky:1993cx}.\footnote{This is only true for a constant dilaton background which we have always assumed in our analysis.} 

Thus in order to obtain the superpotential for D-branes which do not fill the entire internal Calabi-Yau space, the holomorphic Chern-Simons action~\eqref{eq:HCS} for the six-cycle~$Y$ must be dimensionally reduced to the holomorphic cycles wrapped by the lower dimensional D-branes. This is achieved by saturating the normal components of the integrand \eqref{eq:HCS} with D-brane fluctuations which as discussed in section~\ref{sec:openstr} are sections of the normal bundle of the wrapped internal cycle \cite{Kachru:2000ih,Lerche:2002yw,Lerche:2003hs,Lust:2005bd}.

For a stack of D3-branes the holomorphic internal cycle reduces to a point and consequently the normal bundle of the stack of D3-branes is identified with the tangent bundle at this point. In other words all directions of the holomorphic Chern-Simons action~\eqref{eq:HCS} for the six-cycle need to be replaced by fluctuations~$\dbt^i$ of the stack of D3-branes and the integral over the internal space disappears
\begin{equation}
   W_\text{D3}=\frac{2}{9}\Omega_{ijk}\tr\dbt^i\dbt^j\dbt^k \ .
\end{equation}
Note that this expression of the reduced holomorphic Chern-Simons action agrees with the superpotential~\eqref{eq:D3suppot} computed by means of supergravity.

For a stack of $N$ D7-branes wrapped on the cycle~$S^\Lambda$ the dimensional reduction can be performed analogously. This time, however, the resulting expression contains an integral over the internal cycle~$S^\Lambda$ and reads
\begin{equation} \label{eq:W3}
   W_{S^\Lambda}=\int_{S^\Lambda}\tilde s_A
     \tr\left(\dbs^A\:\partial_{\bar\imath} A_{\bar\jmath}
     +2\dbs^A A_{\bar\imath}A_{\bar\jmath}\right)
     \dd z^j\wedge\dd z^k\wedge\dd\bar z^{\bar\imath}\wedge\dd\bar z^{\bar\jmath} \ ,
\end{equation} 
where we used $\tfrac{1}{2}\Omega_{ijk}\dbs^i\dd z^j\wedge\dd z^k=\tilde s_A\dbs^A$. Note that for an Abelian gauge field~$A$ and Abelian `matter fields'~$\dbs^A$, that is to say for a single D7-brane wrapped on the cycle~$S^\Lambda$, the second term in \eqref{eq:W3} vanishes. Then the first term for a single D7-brane becomes  $\partial_{\bar\imath}A_{\bar\jmath}\dd\bar z^{\bar\imath}\wedge\dd\bar z^{\bar\jmath}$ and is just the $(0,2)$ part of the background flux~$f$. Therefore eq.~\eqref{eq:W3} simplifies with \eqref{eq:vanishint} to
\begin{equation} \label{eq:W5}
   W_{S^\Lambda} \sim \int_{S^\Lambda} \tilde s_A\dbs^A\wedge f \ ,
\end{equation} 
in agreement with the superpotential~\eqref{eq:W1}. 


\chapter{Geometry of the D7-brane moduli spaces} \label{ch:geom}


In section~\ref{sec:D7action} we examined the $\mathcal{N}=1$ effective action for orientifold theories with D7-branes and we have discussed the corresponding K\"ahler potential. So far in this analysis the D7-brane matter fields $\dbs$ and the complex structure deformations $z$ have been treated independently. The matter fields $\dbs$ are geometrically governed by $(2,0)$-forms of the four-cycle $S^\Lambda$. As the complex structure of the submanifold $S^\Lambda$ is induced from the complex structure of the ambient Calabi-Yau space $Y$, we expect that the moduli space of the matter fields $\dbs$ and the moduli space of the bulk complex structure deformations $z$ is not of product type \cite{Mayr:2001xk,Lerche:2001cw,Lerche:2002yw}. We should rather have in mind a common moduli space $\mathcal{M}_{\mathcal{N}=1}$, which is parametrized by both the complex structure deformations $z$ and the matter fields $\dbs$. In section~\ref{sec:var} we examine the structure of this common moduli space, which then is applied to the supergravity K\"ahler potential of section~\ref{sec:D7SUGRA} and the superpotential resulting form the D7-brane fluxes as stated in section~\ref{sec:FCS}.


\section{Variation of Hodge structure} \label{sec:var}


The complex structure deformations of the bulk theory is mathematical captured in the language of variation of Hodge structure, which describes how the definition of $(p,3-p)$-forms in $H^3(Y)$ varies over the complex structure moduli space $\mathcal{M}_\text{CS}$ \cite{Candelas:1990pi,Morrison:1992,Aspinwall:1993nu,Greene:1993vm}. The next task is to include the D7-brane fields~$\dbs$ into the concept of the variation of Hodge structure. In order to describe the bulk complex structure deformations~$z$ and the D7-brane matter fields~$\dbs$ in their common moduli space $\mathcal{M}_{\mathcal{N}=1}$ first one needs to find a formulation which captures both types of fields simultaneously. As these fields are respectively expanded into three-forms of the bulk and into two-forms of the internal D7-brane cycle~$S^\Lambda$ the relative cohomology group $H^3(Y,S^\Lambda)$ proves to be the adequate framework \cite{Mayr:2001xk,Lerche:2001cw,Lerche:2002yw}. In ref.~\cite{Brunner:2003zm} it is further argued that the deformation theory of orientifolds with holomorphic involution~$\sigma$ is unobstructed and hence the framework of variation of Hodge structure also applies for $H^3_-(Y)$. Analogously for D7-branes in orientifolds we also choose the truncated relative cohomology group $H^3_-(Y,S^\Lambda)$ as our starting point, which is isomorphic to the direct sum of cohomology groups $\tilde H^3_-(Y)$ and $\tilde H^2_-(S^\Lambda)$. The precise definition of these spaces and their relation to $H^3_-(Y,S^\Lambda)$ are spelled out in detail in appendix~\ref{app:relform}. 

The next task is to extend the concept of variation of Hodge structure to the orientifold case with D7-branes or in other words to analyze the variation of Hodge structures of $H^3_-(Y,S^\Lambda)$ \cite{Lerche:2002yw}. This means we consider (locally) the variation of relative forms over the moduli space $\mathcal{M}_{\mathcal{N}=1}$ which has the complex coordinates $(z^{\tilde a},\dbs^A)$.

Part of the definition of the Hodge structure of $H^3(Y,S^\Lambda)$ is the Hodge filtration $\{F^p\}$ \cite{Morrison:1992,Cox:1999,Lerche:2002yw}, which is a decomposition of $H^3_-(Y,S^\Lambda)$ into\footnote{In this section, we think of relative forms as a pair of a three-form on $Y$ and a two-form on $S^\Lambda$ according to \eqref{eq:RC2}.} 
\begin{equation}
   H^3(Y,S^\Lambda)=F^0\supset \ldots \supset F^3 \ ,
\end{equation}
where
\begin{equation} \label{eq:filt}
\begin{split}
   F^3&=\widetilde H^{(3,0)}_-(Y) \ , \\
   F^2&=F^3\oplus\widetilde H^{(2,1)}_{\bar\partial,-}(Y)
           \oplus\widetilde H^{(2,0)}_{\bar\partial,-}(S^\Lambda) \ , \\
   F^1&=F^2\oplus\widetilde H^{(1,2)}_{\bar\partial,-}(Y)
           \oplus\widetilde H^{(1,1)}_{\bar\partial,-}(S^\Lambda) \ , \\
   F^0&=F^1\oplus\widetilde H^{(0,3)}_{\bar\partial,-}(Y)
           \oplus\widetilde H^{(0,2)}_{\bar\partial,-}(S^\Lambda) \ .
\end{split}
\end{equation}
Note that this filtration looks almost like the Hodge filtration of $H^3_-(Y)$ for orientifold models except for the additional two-forms of $S^\Lambda$, i.e. if one considers the case of a vanishing four-cycle $S^\Lambda$ then all relative forms reduce to ordinary three-forms of $Y$ and the Hodge filtration simplifies to the orientifold case without D7-branes. 

The spaces $\widetilde H^{(3-p,p)}_{\bar\partial,-}(Y)$ for $p>3$ and $\widetilde H^{(2-q,q)}_{\bar\partial,-}(S^\Lambda)$ do not vary holomorphically with respect to $(z^{\tilde a},\dbs^A)$, instead $F^p$ are the fibers of holomorphic fiber bundles $\mathcal{F}^p$ over the moduli space $\mathcal{M}_{\mathcal{N}=1}$ \cite{Morrison:1992,Greene:1993vm,Cox:1999,Lerche:2002yw}.
\begin{table}
\begin{center}
\begin{tabular}{|c|c|c|}
   \hline
      \bf filtration  
      \rule[-1.5ex]{0pt}{4.5ex} &  \multicolumn{2}{|c|}{\bf basis} \\ \cline{2-3}
      \bf &  \bf holomorphic section  &  \bf fiber at $z^{\tilde a}=\dbs^A=0$
      \rule[-1.5ex]{0pt}{4.5ex} \\
   \hline
   \hline
      $\mathcal{F}^3$  &  $\rel{\Omega}$  &  $\Omega$
      \rule[-1.5ex]{0pt}{4.5ex} \\
   \hline
      $\mathcal{F}^2$  
        &  $\rel{\Omega}$, $\rel{\chi_{\tilde a}}$, $\rel{\tilde s_{A}}$
        &  $\Omega$, $\chi_{\tilde a}$, $\tilde s_{A}$
      \rule[-1.5ex]{0pt}{4.5ex} \\
   \hline
      $\mathcal{F}^1$
        &  $\rel{\Omega}$, $\rel{\chi_{\tilde a}}$, $\rel{\tilde s_{A}}$, 
           $\rel{\bar\chi_{\tilde a}}$, $\rel{\tilde\eta_{\tilde A}}$
        &  $\Omega$, $\chi_{\tilde a}$, $\tilde s_{A}$,
           $\bar\chi_{\tilde a}$, $\tilde\eta_{\tilde A}$
      \rule[-1.5ex]{0pt}{4.5ex} \\
   \hline
      $\mathcal{F}^0$
        &  $\rel{\Omega}$, $\rel{\chi_{\tilde a}}$, $\rel{\tilde s_{A}}$, 
           $\rel{\bar\chi_{\tilde a}}$, $\rel{\tilde\eta_{\tilde A}}$,
           $\rel{\bar\Omega}$, $\rel{\tilde s_{\bar A}}$
        &  $\Omega$, $\chi_{\tilde a}$, $\tilde s_{A}$,
           $\bar\chi_{\tilde a}$, $\tilde\eta_{\tilde A}$,
           $\bar\Omega$, $\tilde s_{\bar A}$
      \rule[-1.5ex]{0pt}{4.5ex} \\
   \hline
\end{tabular} 
\caption{D7-brane cycles} \label{tab:basis} 
\end{center}
\end{table}
For each holomorphic fiber bundle~$\mathcal{F}^p$ we choose a (local) basis of sections summarized in Table~\ref{tab:basis}. The fibers at $z^{\tilde a}=\dbs^A=0$ of these local sections coincide with the form bases of Table~\ref{tab:spec}, Table~\ref{tab:coh} and the basis $\{\eta_{\tilde A}\}$ of $\widetilde H^{(1,1)}_{\bar\partial,-}(S^\Lambda)$. Note that at a generic point in the moduli space $\mathcal{M}_{\mathcal{N}=1}$ the fibers of these sections are a mixture of various three- and two-forms due to the non-holomorphicity of the bundles $\widetilde H^{(3-p,p)}_{\bar\partial,-}(Y)$ and $\widetilde H^{(2-q,q)}_{\bar\partial,-}(S^\Lambda)$ over $\mathcal{M}_{\mathcal{N}=1}$.

As the space $H^3_-(Y,S^\Lambda)$ is purely topological the bundle $\mathcal{F}^0=H^3_-(Y,S^\Lambda)$ is locally constant over the moduli space $\mathcal{M}_{\mathcal{N}=1}$. Thus this bundle has a canonically flat connection $\nabla$ called Gauss-Manin connection and it fulfills Griffith's transversality \cite{Morrison:1992,Greene:1993vm,Cox:1999} 
\begin{equation} \label{eq:connect}
   \nabla \mathcal{F}^p \subseteq \mathcal{F}^{p-1}\otimes 
      \Omega^1\mathcal{M}_{\mathcal{N}=1} \ .
\end{equation}     
Note that the covariant derivatives $\nabla_{z^{\tilde a}}$ and $\nabla_{\dbs^A}$ acting on sections of $\mathcal{F}^p$ differ form the ordinary derivatives $\partial_{z^{\tilde a}}$ and $\partial_{\dbs^A}$ only by sections in $\mathcal{F}^p\otimes\Omega^1\mathcal{M}_{\mathcal{N}=1}$ \cite{Lerche:2002yw}. As a consequence and with eq.~\eqref{eq:connect} one reaches local sections of all $\mathcal{F}^p$ by taking derivatives of the unique section $\rel{\Omega}(z^{\tilde a},\dbs^A)$ of $\mathcal{F}^3$.
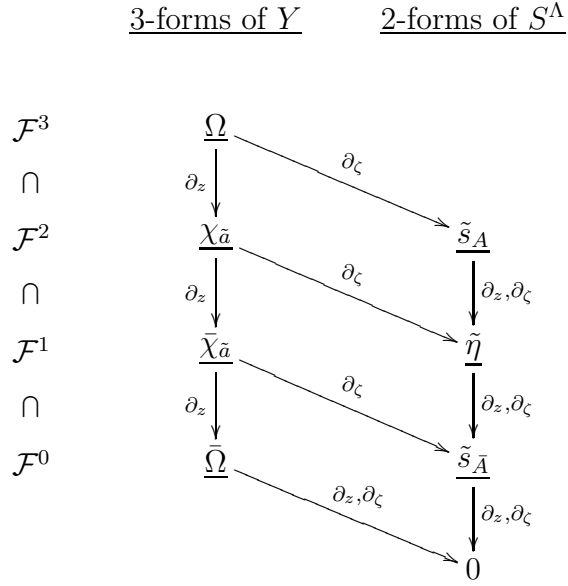
\begin{figure}
\begin{center}
\begin{displaymath}
   \xymatrix{
     &  \text{\underline{3-forms of $Y\vphantom{S^\Lambda}$}}  
        &  \text{\underline{2-forms of $S^\Lambda$}} \\
     \mathcal{F}^3 \ar@{}[d]|{\bigcap}  
        &  \rel{\Omega}\ar[d]_{\partial_z}\ar[dr]^{\partial_\dbs} \\
     \mathcal{F}^2 \ar@{}[d]|{\bigcap}  
        &  \rel{\chi_{\tilde a}}\ar[d]_{\partial_z}\ar[dr]^{\partial_\dbs}
        &  \rel{\tilde s_A}\ar[d]^{\partial_z,\partial_\dbs} \\
     \mathcal{F}^1 \ar@{}[d]|{\bigcap}  
        &  \rel{\bar\chi_{\tilde a}}\ar[d]_{\partial_z}\ar[dr]^{\partial_\dbs}
        &  \rel{\tilde\eta}\ar[d]^{\partial_z,\partial_\dbs} \\ 
     \mathcal{F}^0  &  \rel{\bar\Omega}\ar[dr]^{\partial_z,\partial_\dbs} 
        &  \rel{\tilde s_{\bar A}}\ar[d]^{\partial_z,\partial_\dbs} \\
     &&  0 } 
\end{displaymath} 
\end{center}
\caption{Variation of Hodge structure of $H^3_-(Y,S^\Lambda)$.} \label{fig:var}
\end{figure}
This procedure is schematically depicted in Figure~\ref{fig:var} and one obtains the extended
Kodaira formulae \cite{Candelas:1990pi}, i.e.
\begin{equation} \label{eq:Kod}
\begin{aligned}
   \partial_{z^{\tilde a}}\rel{\Omega}=k_{\tilde a}\rel{\Omega}+\ii \rel{\chi_{\tilde a}} \ , 
   \qquad && \qquad\partial_{\dbs^A}\rel{\Omega}=k_A\rel{\Omega}+\rel{\tilde s_A} \ .
\end{aligned}
\end{equation}

As alluded in Figure~\ref{fig:var} one generates sections of $\mathcal{F}^p$ for all $p$ by acting with the connection $\nabla$ on the unique relative form $\rel{\Omega}$. Since $H^3(Y,S^\Lambda)$ is a finite dimensional space one obtains linear relations among $\rel{\Omega}$ and its covariant derivatives \cite{Greene:1993vm,Hosono:1994av}, i.e. 
\begin{equation} \label{eq:GKZ}
   \mathcal{L}(z,\dbs,\partial_z,\partial_\dbs)\rel{\Omega}(z,\dbs) \sim 0 \ ,
\end{equation}
where $\mathcal{L}(z,\dbs,\partial_z,\partial_\dbs)$ are fourth order differential operators, and where $\sim$ means modulo exact relative forms.

Similar to the derivation of the differential equations for the bulk complex structure deformations, the system of differential equations of forms \eqref{eq:GKZ} can be transformed into a set of differential equations over relative periods \cite{Lerche:2002yw,Lerche:2003hs}. These relative periods arise as integrals of the relative three-form $\rel{\Omega}$ over a fixed homology basis of relative three-cycles. For this basis we choose $\{\rel{A^{\hat a}}, \rel{B_{\hat a}},\rel{\Gamma^{\hat A}}\}$ which is dual to the relative forms associated to the three-forms $\alpha_{\hat a}$, $\beta^{\hat b}$ and the two-forms $\gamma_{\hat A}$ where the latter forms are a basis of $\widetilde H^2_-(S^\Lambda)$. With this choice eq.~\eqref{eq:GKZ} gives rise to the system of differential equations for the relative periods
\begin{equation} \label{eq:GKZ2}
   \mathcal{L}(z,\dbs,\partial_z,\partial_\dbs) \Pi^{\hat a}(z,\dbs)= 
   \mathcal{L}(z,\dbs,\partial_z,\partial_\dbs) \Pi_{\hat a}(z,\dbs)= 
   \mathcal{L}(z,\dbs,\partial_z,\partial_\dbs)\Pi^{\hat A}(z,\dbs)=0 \ , 
\end{equation}
where 
\begin{align}
   \Pi^{\hat a}(z,\dbs)=\langle \rel{A^{\hat a}},\rel{\Omega}\rangle \ , &&
   \Pi_{\hat a}(z,\dbs)=\langle \rel{B_{\hat a}},\rel{\Omega}\rangle \ , &&
   \Pi^{\hat A}(z,\dbs)=\langle \rel{\Gamma^{\hat A}},\rel{\Omega}\rangle \ . 
\end{align}
The solution to the system of partial differential equations \eqref{eq:GKZ2} takes the form
\begin{equation} \label{eq:sol1}
   \rel{\Omega}(z,\dbs)=X^{\hat a}(z,\dbs)\rel{\alpha_{\hat a}}
      +\mathcal{F}_{\hat a}(z,\dbs)\rel{\beta^{\hat a}}
      +\mathcal{G}^{\hat A}(z,\dbs)\rel{\gamma_{\hat A}} \ ,
\end{equation}
with holomorphic functions $X^{\hat a}(z,\dbs)$, $\mathcal{F}_{\hat a}(z,\dbs)$ and $\mathcal{G}^{\hat A}(z,\dbs)$. Note that eq.~\eqref{eq:sol1} reduces for $\dbs=0$ to the known bulk part where the solution is given by the prepotential $\mathcal{F}$ of $\mathcal{N}=2$ special geometry. In general we do not expect that the solution of the system of differential equations \eqref{eq:GKZ2} can be expressed in terms of a single holomorphic function $\mathcal{F}$. This reflects the fact that the structure of $\mathcal{N}=1$ is less restrictive than $\mathcal{N}=2$ supersymmetry.  


\section{Extended K\"ahler and superpotential} \label{sec:kcs}


In this section we recall the definitions of the metrics for the bulk complex structure deformations and the D7-brane fluctuations independently. That is to say in the limit of small fields $z^{\tilde a}$ and $\dbs^A$ where the metric remains block diagonal. Then we apply the mathematical tools of the previous section in order to obtain a K\"ahler metric for the moduli space $\mathcal{M}_{\mathcal{N}=1}$ which is not block diagonal anymore but holds for higher orders in $z^{\tilde a}$ and $\dbs^A$ as well. This extension turns out to be also suitable to generalize the supergravity K\"ahler potential of section~\ref{sec:D7SUGRA}. 

In refs.~\cite{Candelas:1990pi} it is demonstrated that the metric of the bulk complex structure deformations $\delta z$ reads 
\begin{equation} \label{eq:Gz}
   \mathcal{G}_{\tilde a\tilde b}\delta z^{\tilde a}\delta\bar z^{\tilde b} 
    =\frac{3}{2\mathcal{K}}\int_Y\dd^6y \sqrt{\det g} g^{i\bar\jmath}g^{l\bar k}
      \ \delta g_{il} \delta g_{\bar\jmath\bar k} 
    =-\frac{\int_Y \chi_{\tilde a}\wedge\bar\chi_{\tilde b}}
       {\int_Y \Omega\wedge\bar\Omega}\ \delta z^{\tilde a}\delta\bar z^{\tilde b} \ .
\end{equation}
Analogously we can ask for the metric of the fluctuations $\delta\dbs$ which describe how the four-cycle $S^\Lambda$ is deformed in the normal direction of the ambient space $Y$. This metric is obtained by examining the variation of the volume element of $S^\Lambda$ with respect to $\delta\dbs$, namely one performs a normal coordinate expansion of the volume element according to \eqref{eq:PB} and with \eqref{eq:D7fluct} arrives at
\begin{equation}
   \mathcal{G}_{A\bar B}\:\delta\dbs^A\delta\dbs^{\bar B}
   =\frac{6}{\mathcal{K}}\int_{S^\Lambda}\dd^4\xi\sqrt{\det g}\:
   s^{i}_A s^{\bar\jmath}_{\bar B} g_{i\bar\jmath}\ \delta\dbs^A\delta\dbs^{\bar B} \ .
\end{equation}
which can be rewritten to \cite{Jockers:2004yj}
\begin{equation} \label{eq:Gphi}
   \mathcal{G}_{A\bar B}\:\delta\dbs^A\delta\dbs^{\bar B}
      =\ii\mathcal{L}_{A\bar B}\:\delta\dbs^A\delta\dbs^{\bar B} \ .
\end{equation}

Without any D7-brane the metric of the complex structure $\mathcal{G}_{\tilde a\tilde b}$ is K\"ahler with the K\"ahler potential \eqref{eq:CSt} \cite{Candelas:1990pi}. This expression must now be modified to take account for the generalized concept of the variation of Hodge structure of relative forms. In the limit of small complex structure deformations~$\delta z^{\tilde a}$ and small D7-brane fluctuations~$\delta\dbs^A$ the modified K\"ahler potential needs to reproduce the metrics \eqref{eq:Gz} and \eqref{eq:Gphi}. Moreover, in the limit where the D7-brane cycle~$S^\Lambda$ disappears the extended K\"ahler potential must simplify to eq.~\eqref{eq:CSt}. Guided by these observations the K\"ahler potential for the common moduli space of $z^{\tilde a}$ and $\dbs^A$ becomes 
\begin{equation} \label{eq:kcsx}
   K_{CS}(z,\bar z,\dbs,\bar\dbs)=-\ln\left[-\ii\int_{(Y,S^\Lambda)} \rel{\Omega}(z,\dbs)
      \bullet_g\rel{\bar\Omega}(\bar z,\bar\dbs)\right]  \ ,
\end{equation}
where the integral over relative three-forms $\rel{A}$ and $\rel{B}$ is defined as 
\begin{equation} \label{eq:sp}
   \int_{(Y,S^\Lambda)} \rel{A}\bullet_g\rel{B}=
      g \int_Y P^{(3)}\rel{A}\wedge P^{(3)}\rel{B}
      -\ii\int_{S^\Lambda} P^{(2)}\rel{A}\wedge P^{(2)}\rel{B} \ .
\end{equation}
Here $g$ is a coupling constant which is needed for dimensional reasons. $P^{(3)}$ and $P^{(2)}$ are projection operators that extract the three-form and the two-form part of the relative form according to eq.~\eqref{eq:RC2}. 

The next task is to check that in the limit of small bulk fields $z^{\tilde a}$ and small D7-brane matter fields $\dbs^A$ the K\"ahler potential \eqref{eq:kcsx} is contained in the supergravity K\"ahler potential \eqref{eq:K1}. In order to perform the comparison eq.~\eqref{eq:kcsx} is rewritten as
\begin{multline}
   K_{CS}(z,\bar z,\dbs,\bar\dbs)=-\ln\left[-\ii\int_Y P^{(3)}\rel{\Omega}(z,\dbs)
     \wedge P^{(3)}\rel{\bar\Omega}(\bar z,\bar\dbs)\right] \\
     -\ln\left[g-\ii
         \frac{\int_{S^\Lambda}P^{(2)}\rel{\Omega}(z,\dbs)
         \wedge P^{(2)}\rel{\bar\Omega}(\bar z,\bar\dbs)}
         {\int_Y P^{(3)}\rel{\Omega}(z,\dbs)
         \wedge P^{(3)}\rel{\bar\Omega}(\bar z,\bar\dbs)}\right] \ .
\end{multline}
Taking now the limit and using eqs.~\eqref{eq:Kod} and \eqref{eq:LCint} one finds agreement with the supergravity K\"ahler potential \eqref{eq:K1} for the coupling constant
\begin{equation} \label{eq:coupl}
   g(S)=\frac{\ii(S-\bar S)}{2\kappa_4^2\mu_7} \ .
\end{equation}

Thus on the common moduli space $\mathcal{M}_{\mathcal{N}=1}$ of the complex structure deformations~$z^{\tilde a}$ and of the D7-brane matter fields~$\dbs^A$ the supergravity K\"ahler potential \eqref{eq:K1} is modified to
\begin{multline} \label{eq:K2}
   K(S,G,T,\dbs,z,a)=\\
      -\ln\left[-\ii\int_{(Y,S^\Lambda)}\rel{\Omega}(z,\dbs)
      \:\bullet_{g(S)}\:\rel{\bar\Omega}(\bar z,\bar\dbs)\right]
      -2\ln\left[\tfrac{1}{6}\mathcal{K}(S,G,T,\dbs,z,a)\right] \ ,
\end{multline} 
with the coupling constant \eqref{eq:coupl}. Now in general this K\"ahler potential $\mathcal{K}=\mathcal{K}_{\alpha\beta\gamma}v^\alpha v^\beta v^\gamma$ also depends on the bulk complex structure deformations $z^{\tilde a}$ which enter in the process of solving for $v^\alpha(S,G^a,T_\alpha,z^{\tilde a},\dbs^A,a_I)$ because $\mathcal{L}_{A\bar B}$ has become a function of $z^{\tilde a}$. The K\"ahler potential \eqref{eq:K2} still constitutes all the scalar kinetic terms of \eqref{eq:BulkD7actionf} but in addition it generates new terms, which are of higher order in the fields $z^{\tilde a}$ and $\dbs^A$, and which are not captured by the Kaluza-Klein reduction of section~\ref{sec:D7action}.

Similarly we also need to rewrite the D7-brane flux superpotential~\eqref{eq:W5} in the language of relative forms. Let us first consider (small) bulk complex structure deformations~$z^{\tilde a}$ in the context of the reduction of the holomorphic Chern-Simons action in section~\ref{sec:FCS}. This yields with the Kodaira formula~\eqref{eq:Kodaira} the expression 
\begin{equation} \label{eq:W4}
   W_{S^\Lambda}=\frac{1}{2}\int_{S^\Lambda}\left[\left(1+z^{\tilde a}k_{\tilde a}\right)
     \Omega_{ijk}\dbs^i\:\partial_{\bar\imath} A_{\bar\jmath}
     +iz^{\tilde a}\chi_{\tilde a\:ij\bar\imath}\dbs^i\:\partial_{\bar\jmath}A_k\right]
     \dd z^j\wedge\dd z^k\wedge\dd\bar z^{\bar\imath}\wedge\dd\bar z^{\bar\jmath} \ .
\end{equation} 
This equation, however, is according to eq.~\eqref{eq:Kod} just the expansion of the relative form superpotential
\begin{equation}
   W_{S^\Lambda}(z,\dbs)=\frac{1}{2}\int_{S^\Lambda} P^{(2)}\rel{\Omega}(z,\dbs)\wedge f \ .
\end{equation}


\chapter{Conclusions and outlook}  \label{ch:conc}


In this work we discussed space-time filling D-branes in the context of Calabi-Yau compactifications. As a starting point the relation of open strings to D-branes was introduced to compute the massless spectrum of D3/D7-brane systems in Calabi-Yau compactifications. Consistency conditions, that is to say tadpole cancellation conditions, naturally guided us towards a more general compactification ansatz, namely towards Calabi-Yau orientifold compactifications. Geometrically this amounted to choosing for the internal space a Calabi-Yau threefold admitting an isometric holomorphic involution \cite{Brunner:2003zm}. Physically this ansatz truncated the effective four-dimensional bulk spectrum from $\mathcal{N}=2$ supersymmetry to $\mathcal{N}=1$ supersymmetry \cite{D'Auria:2004kx,Grimm:2004uq,Grimm:2004ua}, such that both the massless bulk and D-brane spectrum was comprised of $\mathcal{N}=1$ multiplets.

For this spectrum we then computed the four-dimensional low energy effective supergravity actions. For D-branes the Dirac-Born-Infeld and the Chern-Simons action served as a good starting point, since these actions captured the open string tree-level amplitudes and hence encoded the interactions of the $\mathcal{N}=1$ supergravity D-brane multiplets. For the bulk theory we started from the ten-dimensional type~IIB supergravity description in the democratic formulation \cite{Bergshoeff:2001pv}. This democratic approach enabled us to deduce the D-brane couplings to the bulk fields \cite{Jockers:2004yj}. Finally the resulting four-dimensional effective $\mathcal{N}=1$ supergravity actions were described in terms of their specifying $\mathcal{N}=1$ supergravity data, that is to say in terms of their K\"ahler potential, superpotential and gauge kinetic coupling functions. Although the K\"ahler potential always exhibited the same generic structure, the geometric data of the D-brane cycles and of the compactification space entered through the definition of the chiral variables \cite{Haack:1999zv,Becker:2002nn,Grana:2003ek,Grimm:2004uq,Jockers:2004yj}. This also implied that only for simple geometric compactifications the K\"ahler potential appeared in the form anticipated in refs.~\cite{Randall:1998uk,Hsu:2003cy}. Compared to the analysis of toroidal models as in refs.~\cite{Blumenhagen:1999ev,Cvetic:2000st,Blumenhagen:2002gw,Kachru:2002he,Tripathy:2002qw,Cascales:2003zp,Cascales:2003pt,Berg:2003ri,Kors:2003wf,Lust:2004cx,Lust:2004fi,Lust:2004dn}, our discussion was performed in a quite general context as we have not chosen a particular D-brane configuration for a certain compactification space but instead started with a rather generic ansatz. 

The computed effective supergravity action also served as a good starting point to address phenomenological questions such as supersymmetry breaking at low energy scales. In the context of D3-brane Calabi-Yau orientifolds we examined supersymmetry breaking induced from bulk background fluxes. In order to further specify phenomenological features of spontaneous supersymmetry breaking in the limit where gravity decouples \cite{Ibanez:1992hc,Kaplunovsky:1993rd,Brignole:1993dj} the flux-induced soft-terms were computed \cite{Camara:2003ku,Grana:2003ek}. For the examined D3-brane systems we observed that for certain bulk background fluxes supersymmetry breaking was communicated to the D3-brane `matter sector' and that the flux-induced soft masses were universal. We also noted that in the regime were our analysis was valid the supersymmetry breaking scale was generically much lower then the string scale \cite{Camara:2004jj}. Our second example of supersymmetry breaking involved D7-brane systems. Since D7-branes wrap a non-trivial four-cycle in the internal space, it was also possible to turn on non-trivial D7-brane fluxes on this internal cycle \cite{Jockers:2005zy}. We then analyzed the structure of the flux-induced scalar potential and commented on its relevance for metastable deSitter vacua as anticipated in ref.~\cite{Burgess:2003ic}. We observed that with some effort of geometric engineering certain kind of D7-brane background fluxes were indeed capable to generate a positive energy contribution~\cite{Jockers:2005zy,Berglund:2005dm}. However, in order to obtain a metastable deSitter minimum additional mechanisms such as gaugino condensation and/or Euclidean D3-brane instantons needed to be employed \cite{Kachru:2003aw,Kachru:2003sx}.

The effective four-dimensional supergravity action of Calabi-Yau orientifold with D7-branes also revealed a beautiful underlying geometric structure in terms of the variation of Hodge structure in the context of relative cohomology \cite{Mayr:2001xk,Lerche:2001cw,Lerche:2002yw,Jockers:2004yj}, which allowed us to treat the D7-brane fluctuations and the complex structure deformations on an equal footing. This behavior naturally makes contact with the variation of Hodge structure in Calabi-Yau fourfolds~\cite{Klemm:1996ts,Haack:2001jz}, and hence reflects the connection of type~IIB Calabi-Yau orientifold compactifications with F-theory compactifications on elliptically fibered Calabi-Yau fourfolds~\cite{Vafa:1996xn,Sen:1996vd,Bershadsky:1996gx,Bershadsky:1997zs}.

The presented work suggests further investigations in many different directions. For phenomenological applications the computation of the low energy effective action should be extended to more general D-brane Calabi-Yau orientifold models with several stacks of D-branes and including the low energy effective action of the matter multiplets arising from D-brane intersections \cite{Brunner:1999jq}. However, in order to reliably make phenomenological predictions relevant for particle physics and/or cosmology, it is necessary to also carefully examine and compute the corrections to the specifying data of the low energy effective $\mathcal{N}=1$ supergravity. That is to say one needs to analyze higher order $\alpha'$ corrections along the lines of refs.~\cite{Becker:2002nn,Balasubramanian:2004uy}, but also since the structure of $\mathcal{N}=1$ supergravity is not as stringent as the structure of extended supergravities, it is also necessary to consider the corrections arising from loop corrections \cite{Berg:2004ek}. Finally as discussed in refs.~\cite{Kachru:2003aw,Kachru:2003sx,Denef:2004dm,Gorlich:2004qm} the non-perturbative effects have been proven to be important ingredients in constructing viable phenomenological models, which also need to be thoroughly incorporated in the analysis of space-time filling D-branes in Calabi-Yau orientifold compactifications.

It would also be important to extended the performed analysis beyond the probe limit approximation for the localized sources, i.e. it would be interesting to include the back-reaction to geometry for the D-branes and the O-planes. For D3-branes and O3-planes the back-reaction can be taken into account by performing a Kaluza-Klein reduction with a warped compactification ansatz \cite{Giddings:2001yu,DeWolfe:2002nn}. As we have seen the corrections due to the warped ansatz become negligible in the large radius limit. The back-reaction resulting from D7-branes and O7-planes generates a dilaton gradient but at least in lowest order in $\alpha'$ does not contribute to the warp factor \cite{Giddings:2001yu}. This dilation gradient disappears if one requires local cancellations of D7-brane/O7-plane tadpoles, or in other words if the D7-branes are on top of the O7-planes. In our computations the couplings of the D7-branes to the dilaton became also apparent as the D7-brane fluctuations entered in the definition of a `new dilaton' which was identified as a proper chiral variable. The dilaton gradients are probably best addressed in the context of F-theory compactifications on elliptically fibered Calabi-Yau fourfolds, where the dilaton gradient of type IIB orientifolds is encoded in the geometry of the elliptic fibers \cite{Sen:1996vd}. Recently there has been some investigations in these directions in refs.~\cite{Lust:2004cx,Berglund:2005dm}.

Similar to the analysis carried out in this work, it would be interesting to compute the low energy effective action of D5-brane systems in Calabi-Yau orientifolds with O5-planes, or of D6-branes in type~IIA Calabi-Yau orientifolds with O6-planes. Then one can study mirror symmetry from the low energy effective point of view, which relates type~IIA orientifolds with D6-branes to type~IIB orientifolds with D3/D7-branes or D5/D9-branes \cite{Brunner:2003zm,Brunner:2004zd}. For the closed string sector type~IIA Calabi-Yau orientifolds have been related via mirror symmetry to type~IIB Calabi-Yau orientifolds in ref.~\cite{Grimm:2004ua}.

Here we have focused on D-brane compactifications in the geometric regime, or in other words on scenarios where the quantum effects were not dominant. This allowed us to compute the low energy effective action by performing a Kaluza-Klein reduction \cite{Grana:2003ek,Jockers:2004yj,Jockers:2005zy}. On the other hand the features originating from the quantum nature of string theory (such as tadpole cancellation conditions) are not so easily deduced in this regime. Therefore it is also desirable to further investigate D-brane compactifications in the quantum regime of the internal space along the lines of refs.~\cite{Brunner:2003zm,Brunner:2004zd,Blumenhagen:2003su,Blumenhagen:2004cg,Kapustin:2003rc,Herbst:2004jp,Herbst:2004zm}, and ultimately also to derive a low energy effective action for these non-geometric D-brane compactifications.



\chapter*{Appendices} 
\addcontentsline{toc}{chapter}{Appendices}


\renewcommand{\thesection}{A.\arabic{section}}
\renewcommand{\theequation}{A.\arabic{equation}}


\section{Conventions} \label{app:conv}


Throughout this work pseudo Euclidean metrics have the signature $(-++\ldots)$. As a consequence the determinant $\det g$ of the metric~$g_{MN}$ is negative in the pseudo Euclidean case. The epsilon symbol~$\epsilon$ is defined to be
\begin{align}
   \epsilon_{012\ldots}=\det g \ , && \epsilon^{012\ldots}=1 \ .
\end{align}
This allow us to define the Hodge star operator~$*_d$ acting on $p$-form~$C^{(p)}$ in a $d$-dimensional (pseudo)-Euclidean manifold
\begin{equation}
   *_d C^{(p)} = \frac{C^{(p)}_{M_1\ldots M_p}}{p!(d-p)!\sqrt{\abs{\det g}}}
       \epsilon^{M_1\ldots M_p N_1\ldots N_{d-p}}
       g_{N_1P_1}\ldots g_{N_{d-p}P_{d-p}}\dd x^{P_1}\wedge\ldots\wedge x^{P_{d-p}} \ .
\end{equation}
The square of the Hodge star operator acting on the $p$-form $C^{(p)}$ obeys
\begin{equation}
   (*_d)^2 C^{(p)} =\pm (-1)^{p(d-p)} \ ,
\end{equation}
where the top sign appears in the Euclidean case and the bottom sign in the pseudo Euclidean case. 

In the context of Kaluza-Klein reduction with background metrics of product type, it is useful to split the Hodge star operator. For concreteness let us assume we compactify a $D$-dimensional pseudo Euclidean space with Hodge star operator~$*_D$ on a $l$-dimensional manifold ($l=D-d$) with Hodge star operator~$*_l$ and obtain an effective theory with $d$ space-time dimensions with Hodge star operator~$*_d$. Then one finds for a $p$-form~$C^{(p)}$ on the $d$-dimensional space and a $q$-form~$G^{(q)}$ on the $l$-dimensional space the useful formula
\begin{equation}
   *_D\left(C^{(p)}\wedge G^{(q)}\right) = (-1)^{(d-p)q} \, *_d C^{(p)}\wedge *_l G^{(q)} \ .
\end{equation}


\section{Open superstrings} \label{app:OpenWS}


The dynamics of superstrings is governed by a two-dimensional supersymmetric $\sigma$-model with the space-time manifold as its target space. Geometrically the strings sweep out a two-dimensional surface called worldsheet which is embedded into the space-time manifold.

The worldsheet action for open superstrings in a flat ten-dimensional Minkowski space is given by \cite{Polchinski:1998rq,Polchinski:1998rr}\footnote{Here the action is given in conformal gauge. This gauge choice can (locally) always be achieved by an appropriate coordinate transformation combined with a Weyl transformation. The local residual gauge freedom corresponds to conformal transformations.}
\begin{multline} \label{eq:WSaction}
   \mathcal{S}_\text{WS}^\text{open}=-\frac{1}{4\pi\alpha'}\int\dd\tau\int_0^\pi\dd\sigma\:\eta_{MN}
   \Big((\partial_\sigma-\partial_\tau) X^M (\partial_\sigma+\partial_\tau)X^N \\
     -\ii\psi_+^M(\partial_\sigma-\partial_\tau)\psi_+^N
     -\ii\psi_-^M(\partial_\sigma+\partial_\tau)\psi_-^N \Big) \ .
\end{multline}
Here $\alpha'$ is the string coupling constant with dimension~$[\text{length}]^2$, $\sigma$ is the spatial coordinate of the open string, and $\tau$ is the worldsheet time coordinate. The worldsheet bosons $X^M$ are the space-time coordinates of the flat ten-dimensional Minkowski space with Lorentzian metric $\eta_{MN}$, whereas $\psi_\pm^M$ are the right- and left-moving worldsheet Majorana-Weyl fermions respectively. Modulo boundary terms this action is supersymmetric under the infinitesimal supersymmetry transformations
\begin{align} \label{eq:susytrans}
   \delta X^M=-\ii\epsilon_-\psi^M_+ + \ii \epsilon_+\psi^M_- \ , &&
   \delta \psi^M_\pm= (\partial_\sigma X^M\pm\partial X^M)\epsilon_\mp \ ,
\end{align}
where $\epsilon_\pm$ are two supersymmetry parameters.

Varying the worldsheet action yields the equations of motion for the worldsheet bosons~$X^M$ and fermions~$\psi_\pm^M$
\begin{align} \label{eq:WSEQM}
   (\partial_\tau^2-\partial_\sigma^2)X^M(\tau,\sigma)=0 \ , && 
   (\partial_\tau\mp\partial_\sigma)\psi_\pm^M(\tau,\sigma) = 0 \ .
\end{align}
Due to the worldsheet boundaries of open superstrings the variation of the action \eqref{eq:WSaction} yields additional boundary terms, which must vanish and therefore the equations of motion \eqref{eq:WSEQM} are supplemented by the boundary conditions
\begin{align} \label{eq:WSBdry}
   \int\dd\tau\:\left.\eta_{MN}\,
      \partial_\sigma X^M\:\delta X^N\right|_{\sigma=0}^{\sigma=\pi}=0 \ , &&
   \int\dd\tau\left.\eta_{MN}\,(\psi_+^M \delta\psi_+^N-\psi_-^N \delta\psi_-^M)
      \right|_{\sigma=0}^{\sigma=\pi}=0 \ .
\end{align}

The first task is to evaluate the boundary conditions for the bosonic worldsheet coordinate fields~$X^\mu(\tau,\sigma)$. Requiring locality the boundary condition \eqref{eq:WSBdry} on $X^\mu(\tau,\sigma)$ needs to hold independently at both endpoints, and hence there are two possibilities at each endpoint
\begin{align} \label{eq:WSBdrycon}
   \text{(N)}\quad \left.\partial_\sigma X^a(\tau,\sigma)\right|_{\sigma=0,\pi}=0 \ , &&
   \text{(D)}\quad \left.\delta X^n(\tau,\sigma)\right|_{\sigma=0,\pi}=0 \ . 
\end{align}
The first choice corresponds to Neumann boundary conditions~(N) in the directions~$X^a$ and the second choice to Dirichlet boundary conditions~(D) in the directions~$X^n$.\footnote{For a given coordinate direction in space-time we restrict our discussion to the case that both string endpoints either fulfill Neumann or Dirichlet boundary conditions. The case of mixed boundary conditions for a coordinate direction, that is Dirichlet for one endpoint and Neumann for the other endpoint, arises in intersecting brane scenarios.} The space-time coordinate system is chosen such that the open string endpoints obey Neumann boundary conditions in $(p+1)$-directions~$a=0,\ldots,p$ (including the time direction) whereas they fulfill Dirichlet boundary conditions in the remaining $(9-p)$-directions $n=p+1,\ldots,9$. For these boundary conditions the general classical solution to the equations of motion~\eqref{eq:WSEQM} reads
\begin{equation} \label{eq:XExp}
\begin{aligned}
   \text{(N)}\quad 
     X^a(\tau,\sigma)&=q^a+2\alpha' p^a\tau +\ii\sqrt{\frac{\alpha'}{2}}
     \sum_{k\ne 0} \frac{\alpha^a_k}{k}\left(\ee^{-\ii k(\tau+\sigma)}+\ee^{-\ii k(\tau-\sigma)}
     \right) \ , \\
   \text{(D)}\quad
     X^n(\tau,\sigma)&=\frac{x^n_\pi\:\sigma + x^n_0\left(1-\sigma\right)}{\pi}
     +\ii\sqrt{\frac{\alpha'}{2}}
     \sum_{k\ne 0}\frac{\alpha^n_k}{k}\left(\ee^{k\ii k(\tau+\sigma)}
     -\ee^{-\ii k(\tau-\sigma)} \right) \ .
\end{aligned}
\end{equation} 
For the $(p+1)$-dimensional subspace, which is parametrized by the `Neumann coordinates'~$X^a$, the expansion of the open string \eqref{eq:XExp} yields a position~$q^a$ and a momentum~$p^a$ in addition to an infinite set of oscillator modes~$\alpha^a_k$. On the other hand for the remaining `Dirichlet directions'~$X^n$ one finds the fixed position coordinates~$x^n_0$ and $x^n_\pi$ of the two endpoints of the open string and again an infinite number of oscillator modes~$\alpha^n_k$. This structure arising from the expansion \eqref{eq:XExp} deserves some further appreciation, namely one finds that the open string under discussion can propagate freely with momentum~$p^a$ along the Neumann directions, whereas the endpoints of the open string are confined to $x^n_0$ and $x^n_\pi$ in the Dirichlet directions. Or in other words the endpoints of the open string can move freely on the $(p+1)$-dimensional planes~$\mathcal{W}_{0,\pi}=\{X^M|X^n=x^n_{0,\pi}\}$ which are called D$p$-branes. In this new terminology the open string propagates in space-time with one endpoint attached to the D$p$-brane~$\mathcal{W}_0$ and the other endpoint attached to the D$p$-brane~$\mathcal{W}_\pi$.\footnote{For an open string attached to a single D$p$-brane one has $x^n_0=x^\pi_0$ and thus $\mathcal{W}=\mathcal{W}_0=\mathcal{W}_\pi$.} 

Let us now briefly discuss the boundary conditions imposed on the worldsheet fermions. Again requiring locality one finds two distinct possibilities\footnote{In principal one can also allow for $\psi_+^M(\tau,\pi)=-\psi_-^M(\tau,\pi)$. However, this minus sign can always be absorbed by a field redefinition of $\psi_-^M\rightarrow-\psi_-^M$ \cite{Polchinski:1998rq,Polchinski:1998rr}.} \cite{Hori:2000ck}
\begin{align} \label{eq:PsipmBdry}
   \psi_-^M(\tau,0)=\pm\psi_+^M(\tau,0) \ , && 
   \psi_-^M(\tau,\pi)=\psi_+^M(\tau,\pi) \ .
\end{align}
Due to these boundary conditions the worldsheet fermions~$\psi_+^M$ and $\psi_-^M$ can be combined to fermionic fields~$\psi^M$ defined as
\begin{equation}
   \psi^M(\tau,\sigma)=
      \begin{cases} 
         \psi_+^M(\tau,\sigma) & \text{for}\ 0\le\sigma<\pi  \\
         \psi_-^M(\tau,2\pi-\sigma) & \text{for}\ \pi\le\sigma\le 2\pi
      \end{cases} \ .
\end{equation}
Then with \eqref{eq:WSEQM} these fermions obey the equations of motion
\begin{equation} \label{eq:WSSingleEQM}
   (\partial_\tau-\partial_\sigma) \psi^M(\tau,\sigma) = 0 \ ,
\end{equation}
and the boundary conditions~\eqref{eq:PsipmBdry} become
\begin{align} \label{eq:WSSingleBdry}
   \text{(R)} \quad \psi^M(\tau,0)=\psi^M(\tau,2\pi) \ , &&
   \text{(NS)} \quad \psi^M(\tau,0)=-\psi^M(\tau,2\pi) \ . 
\end{align}
Periodic boundary conditions correspond to worldsheet fermions in the Ramond sector~(R), whereas anti-periodic boundary conditions to worldsheet fermions in the Neveu-Schwarz sector~(NS).\footnote{These two possible choices originate from the two distinct spin-structures of the circle~$S^1$.} Note that imposing \eqref{eq:WSBdrycon} and \eqref{eq:WSSingleBdry} at the boundary also respects one linear combination of the supersymmetry transformations~\eqref{eq:susytrans}, and hence half of the worldsheet supersymmetry is preserved at the boundary \cite{Hori:2000ck}. With the boundary conditions~\eqref{eq:WSSingleBdry} at hand a general solution to the equations of motion \eqref{eq:WSSingleEQM} is readily stated to be
\begin{equation} \label{eq:PsiExp}
\begin{split}
  \text{(R)}\quad&\psi^M(\tau,\sigma)=
     \sum_{r\in\mathbb{Z}} \psi^M_r\:\ee^{-\ii r(\tau+\sigma)} \ , \\
  \text{(NS)}\quad&\psi^M(\tau,\sigma)=
     \sum_{r\in\mathbb{Z}+\frac{1}{2}}\psi^M_r\:\ee^{-\ii r(\tau+\sigma)} \ , 
\end{split}
\end{equation}
with the integer oscillator modes~$\psi_r^M$ in the Ramond sector and the half-integer oscillator modes~$\psi_r^M$ in the Neveu-Schwarz sector.

So far the worldsheet of the superstring has been discussed classically. Hence the next step is to canonically quantize the worldsheet action~\eqref{eq:WSaction}. This is accomplished by introducing equal time commutators for the bosons~$X^M$ and equal time anti-commutators for the fermions~$\psi^M$
\begin{equation}
\begin{aligned}
   \com{\partial_\tau{\hat X}^M(\tau,\sigma_1)}{\hat X^N(\tau,\sigma_2)}&=
       -\ii\eta^{MN}\delta(\sigma_1-\sigma_2) \ , \\
   \ac{\hat\psi^M(\tau,\sigma_1)}{\hat\psi^N(\tau,\sigma_2)}&=
       \pi\eta^{MN}\delta(\sigma_1-\sigma_2) \ .
\end{aligned}
\end{equation}
The canonical quantization implies with \eqref{eq:XExp} and \eqref{eq:PsiExp} that the zero modes and the oscillator modes become operator valued and obey the following commutation and anti-commutation relations \cite{Polchinski:1998rq,Polchinski:1998rr}
\begin{align} \label{eq:osc}
   \com{\hat q^a}{\hat p^b}=\ii\, \eta^{ab} \ , && 
   \com{\hat\alpha_k^M}{\hat\alpha_{-l}^N}=k\,\eta^{MN}\,\delta_{kl} \ , &&
   \ac{\hat\psi^M_r}{\hat\psi^N_{-s}}=\eta^{MN}\,\delta_{rs} \ .
\end{align}
The first commutator is the usual canonical expression for the position operator~$\hat q^a$ and the momentum operator~$\hat p^a$. The oscillator operators are creation and annihilation operators acting on a Fock-space in which the vacuum is annihilated by all $\hat\alpha_k^M, k>0$ and $\hat\psi_r^M, r>0$.\footnote{Due to the Lorentz metric their appear negative norm states in the Fock space. In the Gupta-Bleuler quantization scheme one imposes physical state conditions to remove these unphysical negative norm states \cite{Goddard:1972iy,Brower:1973iz}. Also in the case of the superstring quantization unitarity is recovered by this method.} Note that in the Ramond sector there are additional zero-modes~$\psi^M_0$ which furnish (up to an unimportant factor of $2$) a Clifford algebra. Therefore these zero-modes are combined into raising and lowering operators, which then define a degenerate Clifford vacuum transforming as a Majorana-Dirac spinor~$\spinrep{32_\text{D}}$ under the space-time Lorentz group~$SO(9,1)$.

With this structure it is now possible to discuss the mass operator for the open superstring spectrum, which for the physical states takes the form
\begin{equation} \label{eq:mass}
   \hat m^2=\sum_n \left(\frac{x^n_\pi-x^n_0}{2\pi\alpha'}\right)^2 + \frac{1}{\alpha'}
       \left(\hat N_\text{B}+\hat N_\text{F}-a_0\right) \ ,
\end{equation}
with the bosonic number operator~$\hat N_\text{B}=\sum_{n=1}^{\infty}\hat\alpha_{-n}\cdot\hat\alpha_n$ and the fermionic number operator~$\hat N_\text{F}=\sum_{r>0}r\hat\psi_{-r}\cdot\hat\psi_r$. The constant $a_0$ is $1/2$ in the Neveu-Schwarz sector and $0$ in the Ramond sector. This mass formula \eqref{eq:mass} allows us to discuss the open superstring spectrum. First we turn to the spectrum of open superstrings attached to a single D$p$-brane or stretched between to coinciding D$p$-branes, or in other words we take $x^n_0=x^n_\pi$. In this case the ground state of the Ramond sector is a massless space-time Majorana-Dirac spinor, whereas the ground state of the Neveu-Schwarz sector has a negative mass squared and hence is tachyonic. The first excited state in the Neveu-Schwarz sector is obtained by acting with the raising operators~$\hat\psi^\mu_{-1/2}$, which according to eq.~\eqref{eq:mass} gives rise to massless modes transforming in the vector representation of the space-time Lorentz group. Finally the remaining infinite tower of higher excitations in both the Ramond and Neveu-Schwarz sector are massive and their mass is controlled by the string scale~$\alpha'$. Therefore in the supergravity regime, that is in the limit $\alpha'\rightarrow 0$, these higher string modes are negligible. By the same line of arguments open strings stretching between D$p$-branes, which are separated by distances of the order of $\sqrt{\alpha'}$, i.e. $\dist{x_0}{x_\pi}\gtrsim\sqrt{\alpha'}$, are also not taken into account. This is due to the first term in the mass operator \eqref{eq:mass} which generates in this case only massive modes of the order of magnitude of the higher string excitations.

In order to obtain a consistent superstring theory one needs to include worldsheet fermions in both the Ramond and the Neveu-Schwarz sector and then one has to project onto a definite space-time fermion number. This procedure is called the GSO projection \cite{Gliozzi:1976qd}. In the Ramond sector the GSO projection removes from the degenerate Fock vacuum one Weyl representation, such that there remains a space-time Majorana-Weyl spinor~$\spinrep{16'}$. For stable D-brane configurations the ground state in the open string Neveu-Schwarz sector is also projected out by the GSO projection and hence the tachyonic mode is removed.\footnote{For open strings stretching between D-branes and anti-D-branes the Neveu-Schwarz ground state is not projected out \cite{Green:1994iw,Banks:1995ch,Green:1987mn,Lifschytz:1996iq,Periwal:1996br,Witten:1998cd}. This indicates that such a setup is not a stable vacuum configuration. This instability disappears by a process called tachyon condensation \cite{Sen:1998sm}. Physically tachyon condensation corresponds to the annihilation of the D-brane anti-D-brane pair such that a stable space-time ground state is approached.} Hence after performing the GSO projection the light modes on a single D-brane are given by
\begin{align} \label{eq:masslessstates}
   \hat\psi^M_{-1/2}\ket{k}_\text{NS} \ , && \ket{k,\theta}_\text{R} \ ,
\end{align} 
where $k$ labels the momentum tangent to the D$p$-brane, or in other words labels the eigenvalue of the momentum operator~$\hat p^a$
\begin{align} \label{eq:momentummodes}
   \hat p^a\: \hat\psi^M_{-1/2}\ket{k}_\text{NS}=k^a\: \hat\psi^M_{-1/2}\ket{k}_\text{NS} \ , &&
   \hat p^a\: \ket{k,\theta}_\text{R}=k^a\: \ket{k,\theta} \ .
\end{align}


\section{Spin representations and Dirac gamma matrices} \label{app:Spinors}


Here we assemble some properties and fix the notation of Dirac gamma matrices for the dimensions relevant in this work. 

The ten-dimensional $32\times 32$ Dirac gamma matrices~$\Gamma^M$ fulfill the usual Clifford algebra
\begin{align} \label{eq:DG10}
   \ac{\Gamma^A}{\Gamma^B}=2 \eta^{AB} \ , && A,B=0,\ldots,9 \ ,
\end{align}
where $\eta^{AB}=\diag{-1,+1,\ldots,+1}$ denotes the metric tensor of ten-dimensional Minkowski space invariant under the Lorentz group $SO(9,1)$. Furthermore the ten-dimensional chirality matrix~$\Gamma$ is defined as
\begin{equation} \label{eq:G11}
   \Gamma=\Gamma^0\ldots\Gamma^9
     =-\frac{1}{10!}\:\epsilon_{A_0\ldots A_9}\Gamma^{A_0}\ldots\Gamma^{A_9} \ ,
\end{equation}
and fulfills
\begin{align}
   \Gamma^2=\id \ , && \ac{\Gamma^M}{\Gamma}=0 \ .
\end{align}
The ten-dimensional Dirac spinor $\spinrep{32_\text{D}}$ decomposes into two Weyl representations $\spinrep{16}$ and $\spinrep{16'}$ of $SO(9,1)$ with opposite chirality, i.e. $\spinrep{16}$ is in the $+1$-eigenspace with respect to the chirality matrix \eqref{eq:G11} whereas $\spinrep{16'}$ is the $-1$-eigenspace.

In the context of compactifying the ten-dimensional space-time manifold to four dimensions, the Weyl spinors of $SO(9,1)$ must be decomposed into representations of $SO(3,1)\times SO(6)$
\begin{align} \label{eq:D10_D4}
   \spinrep{16}\rightarrow (\spinrep{2},\spinrep{4})\oplus (\spinrep{\bar 2},\spinrep{\bar 4}) \ , &&
   \spinrep{16'}\rightarrow (\spinrep{2},\spinrep{\bar 4})\oplus (\spinrep{\bar 2},\spinrep{4}) \ , 
\end{align}
where $\spinrep{2}$ and $\spinrep{\bar 2}$ are the two Weyl spinors of $SO(3,1)$ and $\spinrep{4}$ and $\spinrep{\bar 4}$ are the two Weyl spinors of $SO(6)$. In both cases these representations are complex conjugate to each other.

The ten-dimensional Dirac gamma matrices~\eqref{eq:DG10} can be given in terms of tensor products of gamma matrices $\gamma^\alpha$ and $\check\gamma^a$ of the Clifford algebras associated to the groups $SO(3,1)$ and $SO(6)$ respectively, namely
\begin{align} \label{eq:G10to4}
   \Gamma^\alpha=\hat\gamma^\alpha\otimes\id \ , &&\alpha=0,\ldots,3 \ , &&&&
   \Gamma^a=\hat\gamma\otimes\check\gamma^a \ , && a=1,\ldots,6 \ ,
\end{align}
with $\ac{\hat\gamma^\alpha}{\hat\gamma^\beta}=2\eta^{\alpha\beta}$ and $\ac{\check\gamma^a}{\check\gamma^b}=2\delta^{ab}$ and where the chirality matrices are defined as
\begin{align} \label{eq:G5}
   \hat\gamma=
       \frac{\ii}{4!}\epsilon_{\alpha_0\ldots\alpha_3}
       \hat\gamma^{\alpha_0}\ldots\hat\gamma^{\alpha_3} \ , &&
   \check\gamma=
       \frac{\ii}{6!}\epsilon_{a_1\ldots a_6}
       \check\gamma^{a_1}\ldots\check\gamma^{a_6} \ .
\end{align}
Then it is easy to check that \eqref{eq:G10to4} leads to
\begin{align}
   \ac{\Gamma^\alpha}{\Gamma^\beta}=2\eta^{\alpha\beta} \ , &&
   \ac{\Gamma^a}{\Gamma^b}=2 \delta^{ab} \ , &&
   \ac{\Gamma^\alpha}{\Gamma^b}=0 \ ,
\end{align}
and the ten-dimensional chirality matrix \eqref{eq:G11} becomes
\begin{equation}\label{eq:gammadecomp}
   \Gamma=\hat\gamma^5\otimes\check\gamma \ .
\end{equation}

In the analysis of the fermions arising from the space-time filling D7-branes one considers the decomposition of the ten-dimensional Weyl representations of the Lorentz group $SO(9,1)$ into representations of the subgroup~$SO(3,1)\times SO(4)\times SO(2)$. This yields
\begin{equation} \label{eq:MWdecomp}
\begin{split}
   \spinrep{16}&\rightarrow \left(\spinrep{2},\spinrep{2},\spinrep{1}\right)\oplus
   \left(\spinrep{2},\spinrep{2'},\spinrep{\bar 1}\right)\oplus
   \left(\spinrep{\bar 2},\spinrep{2},\spinrep{\bar 1}\right)\oplus
   \left(\spinrep{\bar 2},\spinrep{2'},\spinrep{1}\right) \ , \\
   \spinrep{16'}&\rightarrow \left(\spinrep{2},\spinrep{2},\spinrep{\bar 1}\right)\oplus
   \left(\spinrep{2},\spinrep{2'},\spinrep{1}\right)\oplus
   \left(\spinrep{\bar 2},\spinrep{2},\spinrep{1}\right)\oplus
   \left(\spinrep{\bar 2},\spinrep{2'},\spinrep{\bar 1}\right) \ ,
\end{split}
\end{equation}
with the two Weyl spinors $\spinrep{2}$ and $\spinrep{2'}$ of $SO(4)$ and $\spinrep{1}$ and $\spinrep{\bar 1}$ of $SO(2)$. Note that the two Weyl spinors of $SO(2)$ are again related by complex conjugation.

Similar as before the ten-dimensional Dirac gamma matrices~\eqref{eq:DG10} can be written as a tensor product of the Dirac gamma matrices $\hat\gamma^\alpha$ of $SO(3,1)$, $\gamma^a$ of $SO(4)$ and $\tilde\gamma^{\tilde a}$ of $SO(2)$
\begin{equation} \label{eq:DG}
\begin{aligned} 
   \Gamma^\alpha&=\hat\gamma^\alpha\otimes\id\otimes\id \ ,  & \alpha&=0,\ldots,3 \ , \\
   \Gamma^a&=\hat\gamma\otimes\gamma^a\otimes\id \ , & a&=1,\ldots,4 \ , \\
   \Gamma^{\tilde a}&=\hat\gamma\otimes\id\otimes\tilde\gamma^{\tilde a} \ ,
      & \tilde a&=1,2 \ ,
\end{aligned}
\end{equation}
with $\ac{\gamma^a}{\gamma^b}=2\delta^{ab}$ and $\ac{\tilde\gamma^{\tilde a}}{\tilde\gamma^{\tilde b}}=2\delta^{\tilde a\tilde b}$. The chirality matrices $\gamma$ of $SO(4)$ and $\tilde\gamma$ of $SO(2)$ are given by
\begin{align} \label{eq:G5_3}
   \gamma = -\frac{1}{4!}\epsilon_{a_1\ldots a_4}\gamma^{a_1}\ldots\hat\gamma^{a_4} \ , &&
   \tilde\gamma = -\frac{\ii}{2}\epsilon_{\tilde a\tilde b}\tilde\gamma^{\tilde a}
      \tilde\gamma^{\tilde b} \ .
\end{align}
As before it is easy to check that the definition \eqref{eq:DG} gives rise to the desired anti-commutation relations
\begin{equation}
\begin{aligned}
   \ac{\Gamma^\alpha}{\Gamma^\beta}&=2\eta^{\alpha\beta} \ , & 
       \ac{\Gamma^a}{\Gamma^b}&=2 \delta^{ab} \ , &
   \ac{\Gamma^{\tilde a}}{\Gamma^{\tilde b}}&=2 \delta^{\tilde a\tilde b} \ , \\
   \ac{\Gamma^\alpha}{\Gamma^b}&=0 \ , & \ac{\Gamma^\alpha}{\Gamma^{\tilde b}}&=0 \ , &
   \ac{\Gamma^a}{\Gamma^{\tilde b}}&=0 \ .
\end{aligned}
\end{equation}
The ten-dimensional chirality matrix \eqref{eq:G11} is given in terms of the lower dimensional chirality matrices \eqref{eq:G5} and \eqref{eq:G5_3}, i.e. 
\begin{equation} \label{eq:G11b}
   \Gamma=\hat\gamma\otimes\gamma\otimes\tilde\gamma \ .
\end{equation}

In this work the relevant spaces are often complex manifolds. As a consequence the $SO(2n)$ structure group associated to Riemannian geometry reduces to the structure group~$U(n)$ associated to complex geometry. Therefore the Dirac gamma matrices can also be adjusted to this reduction of the structure group. More specifically this means that the vector indices of the gamma matrices can be decomposed so as to form holomorphic and anti-holomorphic Dirac gamma matrices. In particular in terms of the complex metric tensor~$g_{i\bar\jmath}$ the $SO(6)$ Dirac gamma matrices become
\begin{align} \label{eq:complgamma6}
   \ac{\check\gamma^i}{\check\gamma^{\bar\jmath}}=2 g^{i\bar\jmath} \ , &&
   \ac{\check\gamma^i}{\check\gamma^j}=0 \ , &&
   \ac{\check\gamma^{\bar\imath}}{\check\gamma^{\bar\jmath}}=0 \ , &&
   \ac{\check\gamma^i}{\check\gamma}=\ac{\check\gamma^{\bar\jmath}}{\check\gamma}=0 \ ,
\end{align} 
whereas the $SO(4)$ Dirac gamma matrices read
\begin{align} \label{eq:complgamma4}
   \ac{\gamma^i}{\gamma^{\bar\jmath}}=2 g^{i\bar\jmath} \ , &&
   \ac{\gamma^i}{\gamma^j}=0 \ , &&
   \ac{\gamma^{\bar\imath}}{\gamma^{\bar\jmath}}=0 \ , &&
   \ac{\gamma^i}{\gamma}=\ac{\gamma^{\bar\jmath}}{\gamma}=0 \ .
\end{align} 

Note that these anti-commutation relations~\eqref{eq:complgamma6} and \eqref{eq:complgamma4} correspond up to a rescaling to the fermionic harmonic oscillator algebra. This property is also used repeatedly in the main text, as it allows us to construct spinor representations of $U(n)$.


\section{BPS D-branes from $\kappa$-symmetry} \label{app:BPS}


As discussed in section~\ref{sec:CYcompact} the amount of preserved supercharges for a certain background configuration is given by the number of supersymmetry parameters for which the variation of the fermions vanishes. In the presence of a super-D$p$-branes also the supersymmetry variation of the D-bane fermionic modes~$\pair\Theta$ needs to vanish. The infinitesimal supersymmetry variation on the field~$\pair\Theta$ reads
\begin{equation} \label{eq:Branesusy}
   \delta_{\pair{\epsilon}}\pair{\Theta}=\pair{\epsilon} \ .
\end{equation} 
This variation does not vanish, but, as argued in section~\ref{sec:superaction}, the super-D$p$-brane action exhibits a local fermionic gauge symmetry called $\kappa$-symmetry. Therefore it suffices if the supersymmetry variation acting on the fermions~$\pair\Theta$ only vanishes modulo a $\kappa$-symmetry transformation as stated in eq.~\eqref{eq:BPSbrane}. A $\kappa$-symmetry transformation acting on the fields~$\pair\Theta$ can be expressed as \cite{Aganagic:1996pe,Bergshoeff:1996tu,Bergshoeff:1997kr}
\begin{equation} \label{eq:Branekappa}
   \delta_{\pair\kappa}\pair{\Theta}=(1+\leftup{p}{\Gamma})\pair{\kappa} \ ,
\end{equation}
where the fermionic $\kappa$-symmetry parameter~$\pair\kappa$ depends on the worldvolume coordinates of the D$p$-brane. The matrix~$\leftup{p}{\Gamma}$ is called the product structure matrix, it is hermitian and further obeys
\begin{align}
   \tr\leftup{p}{\Gamma}=0 \ , && \leftup{p}{\Gamma}^2=\id \ .
\end{align}

Using eqs.~\eqref{eq:Branesusy} and \eqref{eq:Branekappa} the supersymmetry variation condition \eqref{eq:BPSbrane} can be stated as
\begin{equation}
   \delta\pair\Theta=\pair{\epsilon}+(1+\leftup{p}{\Gamma})\pair\kappa \ .
\end{equation}
Finally this equation can be multiplied by $(1-\leftup{p}{\Gamma})$ and then the condition for unbroken supersymmetries becomes \cite{Becker:1995kb}
\begin{equation} \label{eq:kappasusycon}
   (1-\leftup{p}{\Gamma})\pair{\epsilon} = 0 \ .
\end{equation}

The next task is to evaluate the last equation for the D-brane geometries in Calabi-Yau threefold compactifications. In this work we are interested in space-time filling D$p$-branes in type~IIB superstring theory. This implies that we are looking for solutions of \eqref{eq:kappasusycon} for D3-, D5-, D7- and D9-branes. For these cases the product matrix~$\leftup{p}{\Gamma}$ becomes \cite{Bergshoeff:1996tu,Bergshoeff:1997kr}
\begin{equation} \label{eq:prodmat}
   \leftup{p}{\Gamma}=(\check\sigma^3)^\frac{p-3}{2}\:\check\sigma^2 \otimes \hat\gamma 
     \otimes\frac{1}{(p-3)!\sqrt{\det{\hat g}}}\epsilon^{a_1\ldots a_{p-3}}
     \check\gamma_{a_1}\ldots\check\gamma_{a_{p-3}} \ .
\end{equation}
In this expression the Pauli matrices~$\check\sigma^1$, $\check\sigma^2$, $\check\sigma^3$ act on the two entries of $\pair{\epsilon}$. $\hat\gamma$ is the four-dimensional chirality matrix acting on the four-dimensional spinors~$\pair{\eta}$ of eq.~\eqref{eq:susyparaCY}, whereas the internal gamma matrices~$\check\gamma_a$ act on the internal spinor part of~\eqref{eq:susyparaCY}. $\det{\hat g}$ is the volume measure of the internal cycle of the D$p$-brane induced from the ambient Calabi-Yau metric~$\hat g_{i\bar\jmath}$. 

In order to fulfill \eqref{eq:kappasusycon} it is necessary to find the eigenvalue~$+1$ in the product matrix~\eqref{eq:prodmat}. This is achieved for a space-time filling D$p$-brane in type~IIB string theory if 
\begin{equation} \label{eq:calcyle}
   \dd^{p-3}\xi\sqrt{\det{\hat g}}=\frac{1}{\left(\tfrac{p-3}{2}\right)!}\hat J^\frac{p-3}{2} \ ,
\end{equation}
in term of the K\"ahler form~\eqref{eq:Kform} and if 
\begin{equation} \label{eq:susyeigen}
   \pair{\eta}=\begin{cases}
                  \check\sigma^2 \pair{\eta} & \text{for}\ p=3,7 \\
                  \check\sigma^1 \pair{\eta} & \text{for}\ p=5,9 \ .
               \end{cases}
\end{equation}
The geometric condition~\eqref{eq:calcyle} states that the internal cycle of the D$p$-brane needs to be calibrated with respect the K\"ahler form.\footnote{For D3- and D9-branes this condition holds trivially.} This implies that the cycle is holomorphically embedded into the Calabi-Yau manifold. Physically this calibration condition amounts to the saturation of the BPS bound. The condition \eqref{eq:susyeigen} on the other hand specify the linear combination of supercharges, which are preserved by the BPS D$p$-brane. 

In the presence of internal background fluxes the product matrix~\eqref{eq:prodmat} is further enhanced by flux-induced contributions. However, the same analysis, although more complicated, can be performed as sketch above. This has been carried out in detail in ref.~\cite{Marino:1999af}. In the presence of bulk background fluxes the computed geometric calibration condition~\eqref{eq:calcyle} must be also modified as discussed in ref.~\cite{Cascales:2004qp}.


\section{Gravitinos in Calabi-Yau orientifold} \label{app:orient}


The number of four-dimensional massless spin-$3/2$ fields determines the amount of unbroken supersymmetry in the effective four-dimensional theory. In type~IIB supergravity the spin-$3/2$ fields arise as the massless modes in the Kaluza-Klein reduction of the ten-dimensional gravitinos~$\pair{\Psi}_M$. Here we determine their number indirectly by decomposing the ten-dimensional spinor representation $\spinrep{16'}$, which then by using geometric properties of Calabi-Yau threefolds allows us to construct the massless four-dimensional gravitinos. This approach also reviews some techniques which are necessary to compute the fermionic effective D-brane action.

In compactifying ten-dimensional superstring theory on a six-dimensional manifold the structure group $SO(9,1)$ of $M^{9,1}$ reduces to $SO(3,1)\times SO(6)$ due to the product structure \eqref{eq:prodmet}. Compactifications on a six-dimensional complex manifolds reduce the structure group $SO(6)$ further to $U(3)\cong SU(3)\times U(1)$ so that we have
\begin{equation} \label{eq:Decomp3C}
   SO(9,1)\rightarrow SO(3,1)\times SO(6)\rightarrow SO(3,1)\times SU(3)\times U(1) \ .
\end{equation}
Correspondingly the  Weyl spinor $ \spinrep{16'}$ of $SO(9,1)$ decomposes into representations of $SO(3,1)\times SO(6)$ or for complex manifolds into representations of $SO(3,1)\times SU(3)\times U(1)$ respectively
\begin{align} \label{eq:D10_SU3}
   \spinrep{16'}\rightarrow (\spinrep{2},\spinrep{\bar 4})\oplus (\spinrep{\bar 2},\spinrep{4}) 
\rightarrow (\spinrep{2},\spinrep{\bar 3_1})\oplus(\spinrep{2},\spinrep{\bar 1_{-3}})
\oplus (\spinrep{\bar 2},\spinrep{3_{-1}})\oplus (\spinrep{\bar 2},\spinrep{1_3})\ .
\end{align}
Here $\spinrep{2},\spinrep{\bar 2}$ are the two Weyl spinors of $SO(3,1)$, $\spinrep{4}, \spinrep{\bar 4}$ are the two Weyl spinors of $SO(6)$, $\spinrep{3}, \spinrep{\bar 3}$ are the fundamentals of $SU(3)$ and  $\spinrep{1}, \spinrep{\bar 1}$ are $SU(3)$ singlets (the subscript denotes their $U(1)$ charge).

For complex threefolds the Clifford algebra for the $SO(6)$ Dirac gamma matrices~$\check\gamma^m$ can be rewritten in terms of complex coordinate indices which then obey\footnote{The $\:\check{}\:$ denotes six-dimensional bulk quantities.}
\begin{align}\label{eq:gammaalgebra}
   \ac{\check\gamma^i}{\check\gamma^{\bar\jmath}}=2 g^{i\bar\jmath} \ , &&
   \ac{\check\gamma^i}{\check\gamma^j}=0 \ , &&
   \ac{\check\gamma^{\bar\imath}}{\check\gamma^{\bar\jmath}}=0 \ , &&
   \ac{\check\gamma^i}{\check\gamma}=\ac{\check\gamma^{\bar\jmath}}{\check\gamma}=0 \ ,
\end{align} 
with the six-dimensional Euclidean chirality matrix~$\check\gamma$ defined in eq.~\eqref{eq:G5}. These relations allow us to interpret the Dirac gamma-matrices with holomorphic indices as raising and lowering operator acting on some `ground state'~$\check\singspin$ and its `conjugate ground state'~$\check\singspin^\dagger$ \cite{Green:1987mn}
\begin{align} \label{eq:checksingspin}
   \check\gamma^i\check\singspin=0 \ , && 
   \check\singspin^\dagger\check\gamma^{\bar\imath}=0 \ .
\end{align}
$\check\singspin, \check\singspin^\dagger$ are the singlets $\spinrep{1}, \spinrep{\bar 1}$ of $SU(3)$ which obey the chirality property 
\begin{align}\label{eq:xichiral}
   \check\gamma\check\singspin=+\check\singspin \ , 
   && \check\singspin^\dagger\check\gamma=-\check\singspin^\dagger \ .
\end{align}
Note that the conditions \eqref{eq:checksingspin} are maintained on the whole complex manifold~$Y$ because the structure group $U(3)\cong SU(3)\times U(1)$ of complex threefolds does not mix the gamma matrices $\check\gamma^i$ with the gamma matrices $\check\gamma^{\bar\jmath}$. However, for a generic complex manifolds the $SU(3)$ singlets $\singspin$ and $\singspin^\dagger$ transform under the $U(1)$ part of \eqref{eq:Decomp3C}, which is the $U(1)$ part in the spin-connection of the internal space. Thus on complex manifolds the global sections $\singspin\otimes\mathcal{L}^*$ and $\singspin^ \dagger\otimes\mathcal{L}$ define the number of (charged) global spinors which are then tensors of the $SU(3)\times U(1)$ structure group bundle. Here $\mathcal{L}$ and $\mathcal{L}^*$ are sections of the line bundles corresponding to the $U(1)$ in \eqref{eq:Decomp3C}. For Calabi-Yau manifolds the first Chern class of the tangent bundle vanishes. This also implies that the line-bundle arising from the $U(1)$ part of the spin connection is trivial, and as a consequence one finds for a Calabi-Yau manifold two globally defined singlets $\singspin$ and $\singspin^\dagger$ which are in addition covariantly constant because of the $SU(3)$ holonomy of Calabi-Yau manifolds.

Due to the presence of the two globally defined and covariantly constant spinors $\check\singspin$ and $\check\singspin^\dagger$ on Calabi-Yau threefolds which appear as singlets in the decomposition \eqref{eq:D10_SU3}, the ten-dimensional gravitinos~$\pair{\Psi}_M$ compactified on the Calabi-Yau manifold~$Y$ give rise to a set of two massless four-dimensional Weyl gravitinos $\pair{\psi}_\mu(x)$\footnote{Since $\pair{\Psi}_M$ are Majorana-Weyl gravitinos, one can choose a Majorana basis such that $\pair{\Psi}_M^*=\pair{\Psi}_M$. This condition implies that the decomposed spinors in  \eqref{eq:D10_SU3} are complex conjugate to each other and as a consequence also the spinors in the Kaluza-Klein ansatz \eqref{eq:4Dgrav} are complex conjugates.} 
\begin{equation} \label{eq:4Dgrav}
   \pair{\Psi}_\mu
     =\pair{\bar\psi}_\mu(x)\otimes\check\singspin(y)
       +\pair{\psi}_\mu(x)\otimes\check\singspin^\dagger(y)\ .
\end{equation} 
Using \eqref{eq:Psichiral}, \eqref{eq:gammadecomp} and \eqref{eq:xichiral} the four-dimensional gravitinos have to obey $\hat\gamma^5\pair{\psi}_\mu=+\pair{\psi}_\mu$ and $\hat\gamma^5\pair{\bar\psi}_\mu=-\pair{\bar\psi}_\mu$.

The next task is to analyze the possible $\mathbb{Z}_2$ Calabi-Yau orientifold projections, and in particular the $\mathbb{Z}_2$ action on the four-dimensional gravitinos as this is important to determine the amount of space-time supersymmetry in the presence of space-time filling D-branes.  As demonstrated in section~\ref{sec:Oplanes} besides the worldsheet parity operator~$\Omega_p$ the generator of the orientifold~$\mathbb{Z}_2$ symmetry allows also for a geometric part~$\sigma$ acting on the internal Calabi-Yau threefold~$Y$. Dictated by space-time supersymmetry $\sigma$ must be an isometric holomorphic involution of $Y$. As further discussed in section~\ref{sec:Oplanes} these properties of $\sigma$ allow for two possibility for the action of $\sigma^*$ on the holomorphic $(3,0)$-form~$\Omega$ of the Calabi-Yau manifold~$Y$, namely
\begin{equation} \label{eq:OmegaEV}
   \Omega = \pm \sigma^*\Omega \ .
\end{equation}

The sign of the eigenvalue of $\Omega$ has far-reaching consequences for the resulting orientifold theory, which can be seen from the fermionic modes of the orientifold theory. In order to grasp the significance of the eigenvalue one needs to use the properties of spinors in the Calabi-Yau threefold~$Y$ as stated above. Namely in order to determine the transformation behavior of the spinors let us specify what is meant by pulling back the spinors via $\sigma^*$. Geometrically the transformation behavior~\eqref{eq:OmegaEV} corresponds at each point in the tangent space of~$Y$ to a rotation of $2\pi n$, $n\in\mathbb{Z}$ for $\Omega=+\sigma^*\Omega$ or to a rotation of $\pi+2\pi n$, $n\in\mathbb{Z}$ for $\Omega=-\sigma^*\Omega$. Since the singlets~$\check\singspin$ and $\check\singspin^\dagger$ are sections of the spin bundle of $Y$ these rotations are lifted to a phase factor $\ee^{\ii\pi n}$ and $\ee^{\ii\frac{\pi}{2}+\pi n}$ respectively. Hence a sensible definition of the pullback~$\sigma^*$ acting on the global spinor singlets~$\check\singspin$ and $\check\singspin^\dagger$ reads
\begin{equation} \label{eq:spinsigma}
   (\sigma^*\check\singspin,\sigma^*\check\singspin^\dagger) =
      \begin{cases}
         (\pm\check\singspin,\pm\check\singspin^\dagger) & \text{for}\ \sigma^*\Omega=+\Omega \\
         (\pm\ii\check\singspin,\mp\ii\check\singspin^\dagger) & \text{for}\ \sigma^*\Omega=-\Omega \ .
      \end{cases}
\end{equation}

The next task is to determine the action of $\Omega_p\sigma^*$ on the four-dimensional gravitinos~\eqref{eq:4Dgrav}. The worldsheet parity~$\Omega$ simply exchanges the two spinor $\psi_\mu^1$ and $\psi_\mu^2$ in $\pair{\psi}_\mu=(\psi_\mu^1,\psi_\mu^2)$ because it exchanges NS-R and R-NS sectors of the closed superstring which provide respectively for the two gravitinos in $\pair{\psi}_\mu$. This can conveniently be expressed in terms of the Pauli matrix~$\check\sigma^1$ and together with \eqref{eq:spinsigma} one obtains
\begin{equation}
   \Omega_p\sigma^* \pair{\psi}_\mu\otimes\check\singspin^\dagger =
      \begin{cases}
         \pair{\psi}_\mu\otimes\check\singspin^\dagger\:\check\sigma^1 & \text{for}\ \sigma^*\Omega=+\Omega \\
         \mp\ii\pair{\psi}_\mu\otimes\check\singspin^\dagger\:\check\sigma^1 
              & \text{for}\ \sigma^*\Omega=-\Omega \ .
      \end{cases}
\end{equation}
It is immediately apparent that $\Omega_p\sigma^*$ can only be an orientifold~$\mathbb{Z}_2$ symmetry for $\sigma^*\Omega=+\Omega$. In the other case it, however, can be made to a symmetry by adding the operator~$(-1)^{F_L}$ where $F_L$ is the fermion number of the left-movers. Therefore it adds an additional sign in the R-NS sector and the projection of $(-1)^{F_L}$ can be expressed in terms of $-\ii\check\sigma^2$. In summary we find that possible orientifold projections are given by the operator~$\mathcal{O}$ defined as
\begin{equation} \label{eq:orientproj}
   \mathcal{O}= 
      \begin{cases}
         \Omega_p\sigma^* & \text{for}\ \sigma^*\Omega=+\Omega \\
         (-1)^{F_L}\Omega_p\sigma^* & \text{for}\ \sigma^*\Omega=-\Omega \ .
      \end{cases}
\end{equation}
In ref.~\cite{Brunner:2003zm} it is indeed confirmed by a worldsheet analysis that \eqref{eq:orientproj} gives rise to the correct orientifold projections in the two distinct cases.

Before concluding this section we state the linear combination~$\psi_\mu$ of the four-dimensional gravitinos that is kept in the spectrum after gauging the discrete $\mathbb{Z}_2$ orientifold projection \eqref{eq:orientproj} 
\begin{equation} \label{eq:4Dgravproj}
   \psi_\mu\otimes\check\singspin^\dagger =\tfrac{1}{2}(\id+\mathcal{O})\pair{\psi}_\mu=
      \begin{cases}
         \pair{\psi}_\mu\otimes\check\singspin^\dagger\:
             \tfrac{1}{2}(\id+\check\sigma^1) & \text{for}\ \sigma^*\Omega=+\Omega \\
         \pair{\psi}_\mu\otimes\check\singspin^\dagger\:
             \tfrac{1}{2}(\id+\check\sigma^2) & \text{for}\ \sigma^*\Omega=-\Omega \ .
      \end{cases}
\end{equation}


\section{Warped Calabi-Yau compactifications} \label{app:warp}


In this appendix we briefly want to summarize the conditions for the validity of the Calabi-Yau ansatz for type~IIB Calabi-Yau compactifications in the presence of background fluxes and/or localized sources such as D-branes and orientifold planes. This analysis has been spelt out in detail in ref.~\cite{Giddings:2001yu}.

The product metric ansatz \eqref{eq:prodmet} for the compactification on the internal compact space~$Y$ can be further generalized by a warped metric ansatz\footnote{In this appendix the metrics are taken in the ten-dimensional Einstein frame.} 
\begin{equation} \label{eq:metwarp}
   \dd s_{10}^2=\ee^{2 A(y)}\eta_{\mu\nu} \dd x^\mu \dd x^\nu + \ee^{-2 A(y)} g_{mn} \dd y^m \dd y^n \ .
\end{equation}
Here $A(y)$ is the warp factor which only depends on the coordinates of the compact space~$Y$, and therefore four-dimensional Poincar\'e invariance is maintained. In addition for the self-dual five-form field strength~$G^{(5)}$ defined in \eqref{eq:fs} one can also allow non-trivial background flux of the form \cite{Giddings:2001yu}
\begin{equation} \label{eq:G5bg}
   \bg{G}^{(5)} = (1+*_{10}) \dd\alpha(y)\wedge\dd x^0\wedge\dd x^1\wedge\dd x^2\wedge\dd x^3 \ ,
\end{equation}
which again preserves four-dimensional Poincar\'e invariance and by construction is in accord with the Bianchi identity of the self-dual five-form field strength. Then the ten-dimensional Einstein field equations 
\begin{equation}
   \Ricci{M}{N} + \frac{1}{2}g_{MN} R  = \kappa_{10}^2 T_{MN} \ ,
\end{equation}
for the metric ansatz \eqref{eq:metwarp} and for the energy momentum $T_{MN}$ comprised of the bulk type~IIB supergravity energy momentum tensor $T_{MN}^\text{SUGRA}$ and the energy momentum tensor for localized sources $\leftup{l}{T}_{MN}$ yield for the warp factor $A(y)$ the equation \cite{Giddings:2001yu}
\begin{equation} \label{eq:warpeq}
   \nabla^2\ee^{4A}=\ee^{2A}\frac{\Gflux_{mnp}\Gflux^{mnp}}{12\Imag\tau}
     +\ee^{-6A}\left(\partial_m\alpha\partial^m\alpha
     +\partial_m{\ee^{4A}}\partial^m{\ee^{4A}}\right)
     +\frac{\kappa_{10}^2}{2} \ee^{2A}\left(\leftup{l}{T}_m^m-T_\mu^\mu\right) \ .
\end{equation}
This equation combined with the Bianchi identity of for the five-form field strength $G^{(5)}$ yields for a consistent setup, that is to say that RR tadpoles and NS-NS tadpoles are canceled, the following conditions (due to the fact that the internal space~$Y$ is compact)
\begin{align} \label{eq:warpcon}
   \ee^{4A}=\alpha \ , && *_6 \Gflux = \ii \Gflux \ .
\end{align}
The first condition relates the warp factor of the metric ansatz \eqref{eq:metwarp} to the background flux \eqref{eq:G5bg}, whereas the second equation states that in stable minima\footnote{Without taking into account non-perturbative string corrections and/or higher order $\alpha'$ corrections.} only imaginary self-dual fluxes can appear.

In order to determine the dependency of the warp factor $\ee^{4A}$ on the radius $R$ of the internal space one can analyze the scaling behavior of \eqref{eq:warpeq} under $g_{mn}\rightarrow \lambda^2 g_{mn}$. Then it is easy to see that the terms in \eqref{eq:warpeq} which contain derivatives acting on the warp factor~$\ee^{4A}$ scale as $\lambda^{-2}$, and due to \eqref{eq:warpcon} the same reasoning applies for the term involving derivatives acting on $\alpha$. The scaling behavior of the term resulting from localized sources is a little bit more involved. For localized sources, i.e. for the $(p+1)$-dimensional space-time filling D-branes and orientifold planes, the energy momentum tensors take the generic form \cite{Giddings:2001yu}
\begin{equation} \label{eq:Ttrace}
   \left(\leftup{l}{T}_m^m-\leftup{l}{T}_\mu^\mu\right)\sim (7-p)   T_p \delta(\Sigma) \ ,
\end{equation}
where the sign of the proportionality constant determines whether the extended object has positive or negative tension. Thus in lowest order in $\alpha'$ D7-branes and/or O7-planes do not contribute to eq.~\eqref{eq:warpeq}, whereas D3-branes and O3-planes enter in the differential equation for the warp factor. The variation of the action~$\leftup{l}{\mathcal{S}}$ determines $\leftup{l}{T}_{MN}$ via
\begin{equation}
   \leftup{l}{T}_{MN} \sim \frac{1}{\sqrt{-\det g_{10}}}
      \frac{\delta \leftup{l}{\mathcal{S}}}{\delta g_{10}^{MN}} \ ,
\end{equation} 
and hence the \eqref{eq:Ttrace} scales as $\lambda^{-6}$. Therefore the warp factor itself $\ee^{4A}$ scales to subleading order as \cite{Giddings:2001yu}
\begin{equation} \label{eq:warpexp}
   \ee^{4A}=1+O(\lambda^{-4}) \ .
\end{equation}
Note that due to eq.~\eqref{eq:warpcon} the flux parameter~$\alpha$ in \eqref{eq:G5bg} approaches $1$ in the limit where warping becomes negligible. 

From eq.~\eqref{eq:warpexp} one deduces that in the large radius regime of the internal manifold the back-reaction to geometry, which is captured by the appearance of a warp factor $\ee^{4A}$, can be treated as a small perturbation to the unwarped geometry chosen in \eqref{eq:prodmet}. To be more specific this is the case if the `radius'~$R$ of the internal space is much larger than the contributions from the $N$ localized sources which become significant at length scales $N^{1/4}\sqrt{\alpha'}$ and the contributions of $M$ units of background flux $\Gflux$ at characteristic length scales $\sqrt{M\,\alpha'}$. 


\section{Geometry of Calabi-Yau orientifolds} \label{app:CYorient}


In performing the Kaluza-Klein reduction on a six-dimensional Calabi-Yau orientifold the geometry of the internal space enters in the effective four-dimensional theory. In this appendix we have collected the relevant formulae for such a compactification. 

All harmonic forms of the Calabi-Yau manifold~$Y$ and their parity with respect to the orientifold involution~$\sigma$ are assembled in Table~\ref{tab:coh}. First we summarize the properties of various integrals taken over these forms. The bases in Table~\ref{tab:coh} are chosen in such a way that appropriate bases are dual to each other, i.e.
\begin{equation}
\begin{aligned}
   \int_Y\omega_\alpha\wedge\tilde\omega^\beta
       =\delta_\alpha^\beta \ , \qquad && \qquad
   \int_Y\omega_a\wedge\tilde\omega^b
       =\delta_a^b \ , \\
   \int_Y\alpha_{\hat\alpha}\wedge\beta^{\hat\beta}
       =\delta_{\hat\alpha}^{\hat\beta} \ , \qquad && \qquad
   \int_Y\alpha_{\hat a}\wedge\beta^{\hat b}
       =\delta_{\hat a}^{\hat b} \ ,  
\end{aligned}
\end{equation}
with all other pairings vanishing. 
\begin{align} \label{eq:triple}
   \mathcal{K}_{\alpha\beta\gamma}
     =\int_Y \omega_{\alpha}\wedge\omega_{\beta}\wedge\omega_{\gamma} \ , 
   &&\mathcal{K}_{ab\gamma}
     =\int_Y \omega_a\wedge\omega_b\wedge\omega_\gamma \ ,
\end{align}
are the non-vanishing triple intersection numbers of the Calabi-Yau manifold~$Y$. Note that these intersection numbers are topological invariants of the Calabi-Yau manifold~$Y$ which are symmetric in their indices. The intersection numbers $\mathcal{K}_{\alpha\beta c}$ vanish because the volume form $\dd\vol(Y)$ of the Calabi-Yau is even whereas $\omega_\alpha\wedge\omega_\beta\wedge\omega_c$ is odd with respect to the pullback $\sigma^*$ \cite{Grimm:2004uq,Andrianopoli:2001zh,Andrianopoli:2001gm,D'Auria:2004kx}. Additionally we define contractions of these intersection numbers with the fields $v^\alpha$ and obtain with \eqref{eq:NS} the non-vanishing combinations 
\begin{equation} \label{eq:K}
\begin{aligned}
   \mathcal{K}&=\int_Y J\wedge J\wedge J
      =\mathcal{K}_{\alpha\beta\gamma}v^\alpha v^\beta v^\gamma \ , 
   & \mathcal{K}_\alpha&=\int_Y \omega_\alpha\wedge J\wedge J
      =\mathcal{K}_{\alpha\beta\gamma}v^\beta v^\gamma \ , \\
   \mathcal{K}_{\alpha\beta}&=\int_Y \omega_\alpha\wedge\omega_\beta\wedge J
      =\mathcal{K}_{\alpha\beta\gamma}v^\gamma \ , 
   & \mathcal{K}_{ab}&=\int_Y \omega_a\wedge\omega_b\wedge J
      =\mathcal{K}_{ab\gamma}v^\gamma \ ,
\end{aligned}
\end{equation}
where $\mathcal{K}$ is proportional to the volume of the internal Calabi-Yau manifold~$Y$, i.e. $6\:\vol (Y)=\mathcal{K}$.

In the action \eqref{eq:4Dbulk} there appear also various metrics. On the space of harmonic two-forms one defines the metrics \cite{Strominger:1985ks,Candelas:1990pi}
\begin{equation} \label{eq:metK}
\begin{split}
   G_{\alpha\beta}&=\frac{3}{2\mathcal{K}}\int_Y\omega_\alpha\wedge *_6\omega_\beta 
      =-\frac{3}{2}\left(\frac{\mathcal{K}_{\alpha\beta}}{\mathcal{K}}-\frac{3}{2}
       \frac{\mathcal{K}_\alpha\mathcal{K}_\beta}{\mathcal{K}^2}\right) \ , \\
   G_{ab}&=\frac{3}{2\mathcal{K}}\int_Y\omega_a\wedge *_6\omega_b
      =-\frac{3}{2}\frac{\mathcal{K}_{ab}}{\mathcal{K}} \ , 
\end{split}
\end{equation}
which is just the usual metric for the space of K\"ahler deformations split into odd and even part with respect to the involution $\sigma$. The inverse metrics of \eqref{eq:metK} are denoted by $G^{\alpha\beta}$ and $G^{ab}$. Similarly, for the complex structure deformations $z^{\tilde a}$ of the Calabi-Yau orientifold, one defines the special K\"ahler metric \cite{Candelas:1990pi}
\begin{align} \label{eq:CSt}
   \mathcal{G}_{\tilde a\tilde b}
      =\frac{\partial^2}{\partial  z^{\tilde a}\partial \bar z^{\tilde b}} \
      K_\text{CS}(z,\bar z) \ , &&
      K_\text{CS}(z,\bar z)=-\ln \left(-\ii\int_Y\Omega \wedge \bar\Omega\right) \ ,
\end{align}
which is the metric on the complex structure moduli space of the Calabi-Yau manifold~$Y$ restricted to the complex structure deformations compatible with the holomorphic involution $\sigma$ \cite{Brunner:2003zm}. Furthermore we have the Kodaira formulae \cite{Candelas:1990pi}
\begin{align} \label{eq:Kodaira} 
   \frac{\partial\Omega}{\partial z^{\tilde a}}=k_{\tilde a}\Omega + \ii \chi_{\tilde a}\ , \qquad
   \frac{\partial\chi_{\tilde a}}{\partial z^{\tilde b}}=k_{\tilde b}\chi_{\tilde a}+
   \kappa_{\tilde a\tilde b}^{\tilde c}\bar\chi_{\tilde c}\ ,
\end{align}
where $\kappa_{\tilde a\tilde b}^{\tilde c}$ is defined in \cite{Candelas:1990pi}
but here we do not need its precise form.

Finally we introduce the coefficient matrices of the kinetic terms of the vector fields $V^{\hat\alpha}$ and $U^{\hat\alpha}$. They are given by \cite{Suzuki:1995rt,Ceresole:1995ca}
\begin{equation} \label{eq:mat3}
\begin{aligned}
   \bti{A}{\hat\beta}{\hat\alpha}
       &=-\int_Y \beta^{\hat\alpha}\wedge *_6 \alpha_{\hat\beta} \ , 
   \qquad & \qquad B_{\hat\alpha\hat\beta}
       &=\int_Y \alpha_{\hat\alpha}\wedge *_6 \alpha^{\hat\beta} \ , \\
   C^{\hat\alpha\hat\beta}
       &=-\int_Y \beta^{\hat\alpha}\wedge *_6 \beta^{\hat\beta} \ , 
   \qquad & \qquad \tbi{D}{\hat\beta}{\hat\alpha}
       &=\int_Y \alpha_{\hat\alpha}\wedge *_6 \beta^{\hat\beta} \ ,
\end{aligned}
\end{equation}
or equivalently for the three-forms $*_6\alpha_{\hat\alpha}$ and $*_6\beta^{\hat\alpha}$ we find modulo exact forms the expansion
\begin{align}
   *_6\alpha_{\hat\alpha}=\bti{A}{\hat\alpha}{\hat\beta}\alpha_{\hat\beta}
        +B_{\hat\alpha\hat\beta}\beta^{\hat\beta} \ , 
   &&*_6\beta^{\hat\alpha}=C^{\hat\alpha\hat\beta}\alpha_{\hat\beta}
        +\tbi{D}{\hat\alpha}{\hat\beta}\beta^{\hat\beta} \ .
\end{align}
It is straight forward to verify that the matrices \eqref{eq:mat3} fulfill
\begin{align}
   \trans{A}=-D \ , && \trans{B}=B \ , && \trans{C}=C \ .
\end{align}
In terms of these matrices the $\mathcal{N}=2$ gauge kinetic matrix~$\mathcal{M}_{\hat\alpha\hat\beta}$ reads \cite{Suzuki:1995rt,Ceresole:1995ca}
\begin{equation} \label{eq:GaugeMatrix}
   \mathcal{M}=A\inv{C}+\ii\inv{C} \ .
\end{equation}


\section{Normal coordinate expansion} \label{app:normal}


In the Dirac-Born-Infeld action and the Chern-Simons action of the D$p$-brane various tensors fields of the bulk theory are pulled back from the space-time manifold $M$ to the worldvolume $\mathcal{W}$ of the brane via $\varphi:\mathcal{W}\hookrightarrow M$. As the embedding map $\varphi$ is not rigid but fluctuates due to the dynamics of the brane, a normal coordinate expansion has to be performed so as to extract the couplings of these brane fluctuations to the bulk fields \cite{Friedan:1980jm,Alvarez-Gaume:1981hn,Grana:2003ek}. 

The fluctuation of the worldvolume $\mathcal{W}$ embedded in the space-time manifold $M$ can be described by considering a displacement vector field $\dbs$ in the normal bundle of the worldvolume. The worldvolume shifted in the direction~$\dbs$ is embedded via the map $\varphi_\dbs$. Note that for $\dbs=0$ the two maps $\varphi$ and $\varphi_\dbs$ coincide. For small fluctuations $\dbs$ any bulk tensor field~$T$ pulled back with the map $\varphi_\dbs$ can be expanded in terms of tensor fields pulled back with the map $\varphi$, i.e.
\begin{equation} \label{eq:normexp}
   \varphi_\dbs^*T=\varphi^*\left(e^{\nabla_{\dbs}}T\right) 
     =\varphi^*\left(T\right)+\varphi^*\left(\nabla_\dbs T\right)+
       \frac{1}{2}\varphi^*\left(\nabla_\dbs\nabla_\dbs T \right)+\ldots  \  ,
\end{equation}
where $\nabla$ is the Levi-Cevita connection of the manifold~$M$.

For local coordinates $x^a$ on the worldvolume $\mathcal{W}$ where $a=0,\ldots,\dim(\mathcal{W}-1)$, we have the associated vector fields $\partial_\mu$, and since the Levi-Cevita connection has no torsion one can show that \cite{Grana:2003ek}
\begin{align} \label{eq:normprop}
   \nabla_\dbs \partial_a = \nabla_{\partial_a} \dbs  \ , &&
   R(\dbs,\partial_a)\dbs=\nabla_\dbs\nabla_{\partial_a}\dbs
      =\nabla_\dbs\nabla_\dbs \partial_a \ ,
\end{align}
where $R(\cdot,\cdot)\cdot$ is the Riemann tensor. 

Applying \eqref{eq:normexp} to the metric tensor $g(\cdot,\cdot)$ of the manifold~$M$, we obtain the induced metric on the worldvolume of the brane subject to the fluctuations $\dbs$. With the identity \eqref{eq:normprop} the expansion up to second order in derivatives yields
\begin{multline}
   \varphi^*_\dbs\left(g(\partial_a,\partial_b)\right)
     =\varphi^*\left(g(\partial_a,\partial_b)\right)
      +\varphi^*\left(g(\nabla_{\partial_a}\dbs,\partial_b)\right)
      +\varphi^*\left(g(\partial_a,\nabla_{\partial_b}\dbs)\right) \\
      +\varphi^*\left(g(\nabla_{\partial_a}\dbs,\nabla_{\partial_b}\dbs) \right)
      +\varphi^*\left(g(R(\dbs,\partial_a)\dbs,\partial_b)\right) +\ldots \ .
\end{multline}
This index free notation translates in a slightly abusive way of notation into the component expression
\begin{equation}\label{eq:PB}
   \varphi^*_\dbs(g)_{ab}=g_{ab}+g_{a n}
     \nabla_b\dbs^n+g_{a n}\nabla_b\dbs^n \\
   +g_{nm}\nabla_a\dbs^n\nabla_b\dbs^m+g_{ac}\Riem{n}{c}{b}{m}\dbs^n\dbs^m +\ldots \ ,
\end{equation}
where now $\nabla$ is the connection of the normal bundle of the worldvolume~$\mathcal{W}$, which is induced form the Levi-Cevita connection of the ambient space~$M$. The indices $a,b,\ldots$ denote directions tangent to the worldvolume~$\mathcal{W}$ whereas the indices $n,m,\ldots$ stand for directions normal to the worldvolume~$\mathcal{W}$.

Analogously, one computes with \eqref{eq:normexp} and \eqref{eq:normprop} the pullback of a $q$-form of the manifold~$M$ to the worldvolume~$\mathcal{W}$ and obtains up to second order in derivatives
\begin{align}\label{eq:PB2}
   \varphi^*_\dbs C^{(q)}
   &=\left(\tfrac{1}{q!}C^{(q)}_{a_1\ldots a_q} 
    +\tfrac{1}{q!}\dbs^n\partial_n(C^{(q)}_{a_1\ldots a_q})
    -\tfrac{1}{(q-1)!}\nabla_{a_1}\dbs^n C^{(q)}_{n a_2\ldots a_q} \right. \nonumber \\
   &+\tfrac{1}{2q!}\dbs^n\partial_n(\dbs^m\partial_m(C^{(q)}_{a_1\ldots a_q}))
    -\tfrac{1}{(q-1)!}\nabla_{a_1}\dbs^n\cdot\dbs^m\partial_m(C^{(q)}_{n a_2\ldots a_q}) \\
   &+\tfrac{1}{2(q-2)!}\nabla_{a_1}\dbs^n\nabla_{a_2}\dbs^m C^{(q)}_{nm a_3\ldots a_q}
   \left.+\tfrac{q-2}{2q!}\Riem{n}{\tau}{a_1}{m}\dbs^n\dbs^m
    C^{(q)}_{\tau a_2\ldots a_q}\right)\dd x^{a_1}\wedge\ldots\wedge\dd x^{a_q}  \ . \nonumber
\end{align}


\section{Dualization procedure} \label{app:dual}


A convenient starting point for coupling D$p$-branes is the democratic supergravity action as introduced in ref.~\cite{Bergshoeff:2001pv}. The democratic actions describe the fields and their duals simultaneously. Since the D$p$-branes couple in the Chern-Simons action \eqref{eq:CSAb} to all RR-form fields the D-brane action is easily implemented in the context of the democratic formulation of the bulk theory. However, after the terms arising from the D-branes are added one would like to obtain an effective four-dimensional action in the conventional sense, that is to say an action where the redundant degrees of freedom are removed by incorporating their duality conditions. 

In this work this duality relation in four dimensions mainly concerns scalars dual to two-forms. Thus here we want to demonstrate how to remove systematically the space-time two-forms in favor of their dual scalars \cite{Quevedo:1996uu,Dall'Agata:2001zh,Louis:2002ny}. We start with the four-dimensional action
\begin{equation} \label{eq:SSD}
   \mathcal{S}_\text{SD}=\int\left[\frac{g}{4} \dd B^{(2)} \wedge * \dd B^{(2)}  
      +\frac{1}{4g}\dd S \wedge * \dd S \right] \ ,
\end{equation}
with the coupling constant $g$, the two-form field $B^{(2)}$ and the scalar field $S$. Moreover we impose by hand the duality condition 
\begin{equation} \label{eq:dual2}
   g\:* \dd B^{(2)} = \dd S \ .
\end{equation}
Thus $S$ is the dual scalar of the two-form $B^{(2)}$ and the action \eqref{eq:SSD} with \eqref{eq:dual2} possesses just one degree of freedom. If we introduce the field strengths $H=\dd B^{(2)}$ and $A=\dd S$, altogether we have the equations
\begin{align} \label{eq:SSDeq}
   \dd A=0 \ , && \dd H=0 \ , && \dd *A=0 \ , && \dd *H=0 \ , && g\:* H = A \ ,
\end{align}
where the first two equations are Bianchi identities, the next two equations are the equations of motion of \eqref{eq:SSD}, and the last equation is the duality condition \eqref{eq:dual2}. Now we modify the action \eqref{eq:SSD} to
\begin{equation} \label{eq:SSD2}
   \mathcal{S}_\text{SD}=\int\left[\frac{g}{4} H \wedge *  H 
      +\frac{1}{4g}\dd S \wedge * \dd S -\frac{1}{2} H \wedge \dd S - \lambda \dd H \right] \ ,
\end{equation}
where in this action $H$ is an independent three-form field and $\lambda$ is a Lagrangian multiplier. This Lagrangian also yields the equations \eqref{eq:SSDeq}, however, now only the first equation arises as a Bianchi identity. All the other equations, including the duality relation, is obtained from the equations of motion of \eqref{eq:SSD2}. In this formulation we can eliminate the three-form field~$H$ and arrive at the action for $S$
\begin{equation} \label{eq:SSD3}
   \mathcal{S}_\text{SD}=\int\frac{1}{2g}\dd S \wedge * \dd S \ ,
\end{equation}
without any redundant dual fields. The next task is to generalize this procedure in the presents of source terms $J$, which we add to \eqref{eq:SSD}
\begin{equation} \label{eq:SSD4}
   \mathcal{S}_\text{SD}=\int\left[\frac{g}{4} \dd B^{(2)} \wedge *\dd B^{(2)} 
      +\frac{1}{4g} A \wedge *A - \frac{1}{2} \dd B^{(2)} \wedge J \right] \ .
\end{equation}
Note that in order to be in accord with the duality condition $g*H=A$, the field strength $A$ must be adjusted to $A=\dd S+J$ and the new equations of this system are
\begin{align} \label{eq:SSDeqS}
   \dd A=\dd J \ , && \dd H=0 \ , && \dd *A=0 \ , && \dd *H=\dd J \ , && g\:* H = A \ .
\end{align}
As before we obtain this set of equations from the Lagrangian
\begin{equation} \label{eq:SSD5}
   \mathcal{S}_\text{SD}=\int\left[\frac{g}{4} H \wedge *  H 
      +\frac{1}{4g}(\dd S+J)\wedge *(\dd S+J) -\frac{1}{2} H \wedge(\dd S+J) 
      - \lambda \dd H \right] \ ,
\end{equation}
with the independent field $H$. Finally eliminating $H$ yields
\begin{equation} 
   \mathcal{S}_\text{SD}=\int\frac{1}{2g}(\dd S+J) \wedge *(\dd S+J) \ ,
\end{equation}
which is the reduction of the democratic action~\eqref{eq:SSD4} to the conventional action in terms of the scalar field~$S$.


\section{Soft supersymmetry breaking terms} \label{app:softterms}


In this appendix we briefly recall the definition of soft supersymmetry breaking terms. These soft-terms appear in the context of globally supersymmetric field theories. One of the attractive features of globally supersymmetric field theories are the non-renormalization theorems \cite{Affleck:1983mk}, which ensure that certain (holomorphic) quantities such as the superpotential are protected from perturbative corrections \cite{Shifman:1986zi,Shifman:1991dz}. Soft-terms are now additional terms~$\mathcal{L}_\text{soft}$ added to the Lagrangian~$\mathcal{L}_\text{susy}$ of a $\mathcal{N}=1$ globally supersymmetric field theory which break supersymmetry explicitly, but maintain the good renormalization behavior of the supersymmetric theory. The allowed terms turn out to be \cite{Girardello:1981wz}
\begin{equation} \label{eq:soft}
   \mathcal{L}_\text{soft}=
      -\soft{m}_{i\bar\jmath}^2 \phi^i\bar\phi^{\bar\jmath} 
      -\left(\soft{b}_{ij}\phi^i\phi^j + \soft{a}_{ijk} \phi^i \phi^j \phi^k  
      +\frac{1}{2}\:\soft{m}_{\text{g}\:\Lambda\Delta} \lambda^\Lambda\lambda^\Delta 
      +\hc\right) \ ,
\end{equation}      
where $\phi^i$ are the scalar fields in the $\mathcal{N}=1$ chiral multiplets and $\lambda^\Lambda$ are the gauginos in the $\mathcal{N}=1$ vector multiplets of the globally supersymmetric theory. Here $\soft{m}_{i\bar\jmath}$ and $\soft{b}_{ij}$ are scalar masses, $\soft{m}_{\text{g}\:\Lambda\Delta}$ are the gaugino masses, and $\soft{a}_{ijk}$ are scalar trilinear couplings often called A-terms.

Even though the soft terms \eqref{eq:soft} break supersymmetry explicitly in the context of global supersymmetry, the structure of these soft-terms can be understood as arising from a spontaneously broken $\mathcal{N}=1$ supergravity theory in the limit $M^{(4)}_\text{p}\rightarrow\infty$, $m_{3/2}=\text{const}$. This point of view as been thoroughly examined in refs.~\cite{Ibanez:1992hc,Kaplunovsky:1993rd,Brignole:1993dj}, and in the following we summarize the results of this analysis.

In $\mathcal{N}=1$ supergravities arising from compactification of string theory it is convenient to distinguish between the `matter fields'~$\dbt^i$ of charged chiral multiplets and the gauge neutral moduli fields~$M^I$ of neutral chiral multiplets. As long as the gauge symmetry is unbroken the vacuum expectation values of the `matter fields'~$\dbt^i$ vanish and  therefore it is convenient to expand the K\"ahler potential $K(M,\bar M,\dbt,\bar\dbt)$ and the superpotential $W(M,\dbt)$ in a power series in $\dbt^i$ 
\begin{align} \label{eq:KWexp}
   K(M,\bar M,\dbt,\bar\dbt) 
     &=\kappa_4^{-2}\hat K(M,\bar M)+Z_{i\bar\jmath}(M,\bar M)\dbt^i\bar\dbt^{\bar\jmath}
       +\left(\tfrac{1}{2}H_{ij}(M,\bar M)\dbt^i\dbt^j+\hc\right)+\ldots \nonumber \\
   W(M,\phi) 
     &=\hat W(M)+\frac{1}{2}\tilde\mu_{ij}(M)\dbt^i\dbt^j
       +\frac{1}{3}\tilde Y_{ijk}(M)\dbt^i\dbt^j\dbt^k+\ldots \ .
\end{align}
Spontaneous supersymmetry breaking occurs if a D-term or a F-term has a non-vanishing vacuum expectation value. In our applications supersymmetry is broken by non-vanishing flux-induced F-terms. Since the $\dbt^i$ vanish in the ground state they do not contribute to the F-terms and one has
\begin{equation} \label{eq:Fterm}
   \mathrm{\bar F}^{\bar I}=\kappa_4^2 \ee^{\hat K/2}\:\hat K^{\bar I J}\:\mathcal{D}_J\hat W \ ,
\end{equation}
where $\hat K^{\bar I J}$ is the inverse of the K\"ahler metric~$\hat K_{I\bar J}$. For a vanishing cosmological constant an alternative measure for supersymmetry breaking is given by the gravitino mass~$m_{3/2}$
\begin{equation} \label{eq:m32}
   m_{3/2}=\kappa_4^2\ee^{\hat K/2} \hat{W} \ .
\end{equation}

Without going through the analysis let us just state here the effective theory for the matter fields~$\dbt^i$, which one obtains by taking the decoupling limit of gravity $M_{p}^{(4)}\rightarrow\infty$, $m_{3/2}=\text{const}$. The resulting theory is a softly-broken globally supersymmetric field theory for the matter multiplets~$\dbt^i$ \cite{Girardello:1981wz,Ohta:1982wn,Soni:1983rm,Hall:1983iz,Ibanez:1992hc,Kaplunovsky:1993rd,Brignole:1993dj}. In this limit one finds for the scalar potential\footnote{Here and in the following we use the notation and the conventions of ref.~\cite{Kaplunovsky:1993rd}.}
\begin{multline} \label{eq:Veff}
   V^\text{eff}(\phi,\bar\phi)
      =\frac{1}{2}\left(\text{D}^{\text{eff}}\right)^2+Z^{i\bar\jmath}\:
      \partial_i W^\text{eff}\:\partial_{\bar\jmath}\bar W^\text{eff} \\
      +\soft{m}^2_{i\bar\jmath}\:\dbt^i\bar\dbt^{\bar\jmath} 
      +\left(\tfrac{1}{2}\:\soft{b}_{ij}\:\dbt^i\dbt^j
      +\tfrac{1}{3}\:\soft{a}_{ijk}\:\dbt^i\dbt^j\dbt^k+\hc\right) \ ,
\end{multline}
where the `supersymmetric terms' read
\begin{align} \label{eq:Weff}
   \text{D}^\text{eff}(\dbt,\bar\dbt)=-g\:\bar\dbt^{\bar\imath}Z_{\bar\imath j}\phi^j \ , && 
   W^\text{eff}(\dbt)=\frac{1}{2}\mu_{ij}\dbt^i\dbt^j+\frac{1}{3}Y_{ijk}\dbt^i\dbt^j\dbt^k \ .
\end{align}
Here $g$ is the gauge coupling function given by $g^{-2}=\Real f(M)$. The mass terms~$\mu_{ij}$ and the Yukawa couplings~$Y_{ijk}$ are defined in terms of the moduli dependent supergravity F-terms \eqref{eq:Fterm} and supergravity coefficients
\begin{align} \label{eq:muY}
   \mu_{ij}=\ee^{\hat{K}/2}\tilde{\mu}_{ij}+m_{3/2}H_{ij}
             -\mathrm{\bar F}^{\bar I}\bar{\partial}_{\bar I} H_{ij} \ , &&
   Y_{ijk}=\ee^{\hat{K}/2}\tilde{Y}_{ijk} \ .
\end{align}
On the other hand the soft supersymmetry breaking terms in eq.~\eqref{eq:Veff} turn out to be\footnote{Due to the inclusion of the possibility of a non-vanishing cosmological constant~$V_0$ the form of the soft-terms differ from the ones given in ref.~\cite{Kaplunovsky:1993rd}. Instead we use the expression for the soft-terms stated in ref.~\cite{Grana:2003ek}.}
\begin{equation} \label{eq:softterms}
\begin{split}
   \soft{m}^2_{i \bar\jmath}=&\left(\abs{m_{3/2}}^2 +V_0\right)Z_{i \bar\jmath}
      -\mathrm{F}^I\mathrm{\bar F}^{\bar J} R_{I\bar Ji\bar\jmath}  \ , \qquad\qquad 
   \soft{a}_{ijk}=\mathrm{F}^I\:\mathcal{D}_I Y_{ijk} \ , \\
   \soft{b}_{ij}=&\left(2\abs{m_{3/2}}^2 +V_0\right) H_{ij}-\bar m_{3/2} 
          \mathrm{\bar F}^{\bar J}\bar\partial_{\bar J} H_{ij}
          +m_{3/2}\mathrm{F}^I \mathcal{D}_I H_{ij} \\
         &-\mathrm{F}^I\mathrm{\bar F}^{\bar J}\mathcal{D}_I\partial_{\bar J} H_{ij}
          -\ee^{\hat K/2}\tilde\mu_{ij}\bar m_{3/2}+\ee^{\hat K/2}
          \mathrm{F}^I\mathcal{D}_I\tilde\mu_{ij} \ ,
\end{split}
\end{equation}
where $V_0=\vev{V}$ is the cosmological constant. The `curvature'~$R_{I\bar Ji\bar\jmath}$ and the K\"ahler covariant derivatives~$\mathcal{D}_I$ are given by
\begin{align} \label{eq:softcurv}
\begin{aligned}
   R_{I\bar Ji\bar\jmath} &=\partial_I\partial_{\bar J} Z_{i \bar\jmath}
      -\Gamma^{k}_{Ii}Z_{k\bar l}\Gamma^{\bar l}_{\bar J\bar\jmath} \ , \\
   \mathcal{D}_I Y_{ijk} &=\partial_I Y_{ijk}+\frac{1}{2}\hat{K}_I Y_{ijk}
      -3\Gamma ^l_{Ii} Y_{ljk} \ , \\
   \mathcal{D}_I \tilde\mu_{ij}&=\partial_I\tilde\mu_{ij}+\frac{1}{2}\hat{K}_I\tilde\mu_{ij} 
      -2\Gamma^l_{Ii}\tilde\mu_{lj} \ ,
\end{aligned} &&
   \text{with}\ \Gamma_{Ii}^l=Z^{l\bar\jmath}\partial_I Z_{\bar\jmath i} \ .
\end{align}

Finally, let us turn to the fermionic mass terms. Since $W^\text{eff}$ is the effective superpotential of the (softly broken) globally supersymmetric field theory, by supersymmetry it also determines the masses of the fermions in the charged chiral multiplets to be given by $\mu_{ij}$. The (canonically normalized) mass $\soft{m}_\text{g}$ of the gaugino in the vector multiplet is also a soft-term according to \eqref{eq:soft}, which can again be related to a spontaneously broken supergravity theory in the decoupling limit $M_\text{p}^{(4)}\rightarrow\infty$, $m_{3/2}=\text{const.}$
\begin{equation}\label{eq:mgaugino}
   \soft{m}_\text{g} = \mathrm{F}^I\partial_I\ln\left(\Real f\right) \ .
\end{equation}


\section{Effective action of D7-branes in  Calabi-Yau orientifolds} \label{app:D7action}


Here the four-dimensional effective action of a D7-brane in a Calabi-Yau orientifold with D7-brane fluxes is presented. This is the form obtained by performing a Kaluza-Klein reduction \cite{Jockers:2004yj,Jockers:2005zy}
\begin{align} \label{eq:BulkD7actionf}
   &\mathcal{S}^\text{E}_\text{D7}
     =\frac{1}{2\kappa_4^2}\int \left[-R\:*_41
          +2\mathcal{G}_{\tilde a\tilde b}\dd z^{\tilde a} \wedge *_4 \dd\bar z^{\tilde b}
          +2G_{\alpha\beta}\dd v^\alpha \wedge *_4 \dd v^\beta \right. \nonumber \\
      &+\frac{1}{2}\dd(\ln \mathcal{K})\wedge *_4 \dd(\ln \mathcal{K})
          +\frac{1}{2}\dd\phi\wedge *_4 \dd\phi 
          +2\ee^\phi G_{ab}\dd b^a\wedge *_4 \dd b^b \nonumber \\
      &+2\ii\kappa_4^2\mu_7\mathcal{L}_{A\bar B}\left(\ee^\phi+4 G_{ab} \dbbf^a \dbbf^b
          -\frac{6v^\Lambda}{\mathcal{K}}\fcharge \right)
          \dd\dbs^A\wedge *_4\dd\bar\dbs^{\bar B} \nonumber \\
      &+\frac{24}{\mathcal{K}}\kappa_4^2\mu_7\ell^2\ii\mathcal{C}^{I\bar J}_\alpha v^\alpha
          \dd a_I\wedge *_4\dd\bar a_{\bar J} \nonumber \\
      &+\frac{\ee^{2\phi}}{2}
          \left(\dd l+\kappa_4^2\mu_7\mathcal{L}_{A\bar B}
          \left(\dd\dbs^A\bar\dbs^{\bar B}-\dd\bar\dbs^{\bar B}\dbs^A\right)\right)
          \wedge
          *_4\left(\dd l+\kappa_4^2\mu_7\mathcal{L}_{A\bar B}
          \left(\dd\dbs^A\bar\dbs^{\bar B}-\dd\bar\dbs^{\bar B}\dbs^A\right)\right) \nonumber \\
      &+2\ee^\phi G_{ab}
          \left(\nabla c^a-l\dd b^a-\kappa_4^2\mu_7 \dbbf^a \mathcal{L}_{A\bar B}
          \left(\dd\dbs^A\bar\dbs^{\bar B}-\dd\bar\dbs^{\bar B}\dbs^A\right)\right)\wedge \nonumber \\
      &\qquad\qquad *_4 \left(\nabla c^b-l\dd b^b-\kappa_4^2\mu_7 \dbbf^b 
          \mathcal{L}_{A\bar B} 
          \left(\dd\dbs^A\bar\dbs^{\bar B}-\dd\bar\dbs^{\bar B}\dbs^A\right)\right) \nonumber \\
      &+\frac{9}{2\mathcal{K}^2}G^{\alpha\beta}
          \left(\nabla\rho_\alpha-\mathcal{K}_{\alpha bc}
          c^b\dd b^c-\tfrac{1}{2}\kappa_4^2\mu_7
          \left(\mathcal{K}_{\alpha bc} \dbbf^b \dbbf^c
          +\delta^\Lambda_\alpha \fcharge\right) \right. \nonumber \\
      &\qquad\qquad\qquad\left. \cdot \mathcal{L}_{A\bar B}
          \left(\dd\dbs^A\bar\dbs^{\bar B}-\dd\bar\dbs^{\bar B}\dbs^A\right)
          +2\kappa_4^2\mu_7\ell^2\mathcal{C}_\alpha^{I\bar J}
          \left(a_I\dd \bar a_{\bar J}-\bar a_{\bar J}\dd a_I\right)\right)\wedge \nonumber \\
      &\qquad\qquad *_4 \left(\nabla\rho_\beta-\mathcal{K}_{\beta ab} c^a\dd b^b
          -\tfrac{1}{2}\kappa_4^2\mu_7 \left(\mathcal{K}_{\beta bc}\dbbf^b \dbbf^c
          +\delta^\Lambda_\alpha \fcharge\right)\right. \nonumber \\
      &\qquad\qquad\qquad \left.\cdot\mathcal{L}_{A\bar B}
          \left(\dd\dbs^A\bar\dbs^{\bar B}-\dd\bar\dbs^{\bar B}\dbs^A\right)
          +2\kappa_4^2\mu_7\ell^2\mathcal{C}_\beta^{I\bar J}
          \left(a_I\dd\bar a_{\bar J}-\bar a_{\bar J}\dd a_I\right)\right) \nonumber \\
      &+\kappa_4^2\mu_7\ell^2 \left(\tfrac{1}{2}\mathcal{K}_\Lambda
          -\tfrac{1}{2}\ee^{-\phi}\mathcal{K}_{\Lambda ab} \dbbf^a \dbbf^b 
          -\tfrac{1}{2}\ee^{-\phi}\fcharge \right) F\wedge *_4F \nonumber \\
      &+\kappa_4^2\mu_7\ell^2\left(\rho_\Lambda-\mathcal{K}_{\Lambda ab}c^a \dbbf^b
          +\tfrac{1}{2}\mathcal{K}_{\Lambda ab}\dbbf^a \dbbf^bl 
          +\tfrac{1}{2}l\fcharge \right) F\wedge F \nonumber \\
      &\left.+\frac{1}{2}(\Imag \mathcal{M})_{\hat\alpha\hat\beta} 
          \dd V^{\hat\alpha}\wedge *_4 \dd V^{\hat\beta}
          +\frac{1}{2}(\Real \mathcal{M})_{\hat\alpha\hat\beta}
          \dd V^{\hat\alpha}\wedge\dd V^{\hat\beta} \right] \ , 
\end{align} 
with the gauge covariant derivatives given by
\begin{equation}
\begin{split}
   &\nabla_\mu c^a=\partial_\mu c^a-4\kappa_4^2\mu_7\ell\delta^a_P A_\mu \ , \\
   &\nabla_\mu \rho_\alpha=\partial_\mu\rho_\alpha-4\kappa_4^2\mu_7
     \ell\mathcal{K}_{\alpha bP} \dbbf^b A_\mu-4\kappa_4^2\mu_7\ell Q_\alpha A_\mu \ .
\end{split}
\end{equation}


\section{Relative cohomology and relative homology} \label{app:relform}


In this appendix we present a primer on the cohomology theory of relative forms $\rel{\Theta}\in\Omega^n(Y,S^\Lambda)$.\footnote{For an introduction on relative forms see e.g. ref.~\cite{Karoubi:1987}.} These forms are $n$-forms of the Calabi-Yau manifold~$Y$ in the kernel of $\iota^*$. Recall that the map $\iota$ embeds $S^\Lambda$ into $Y$, i.e. $\iota:S^\Lambda\hookrightarrow Y$. Hence the set of relative forms $\Omega^n(Y,S^\Lambda)$ fits into the exact sequence
\begin{equation} \label{eq:RFexact}
   0\rightarrow\Omega^n(Y,S^{\Lambda})\hookrightarrow\Omega^n(Y)
    \xrightarrow{\iota^*}\Omega^n(S^\Lambda)\rightarrow 0 \ .
\end{equation}
Then the cohomology of these relative forms with respect to the exterior differential~$\dd$ defines the relative cohomology groups $H^n(Y,S^\Lambda)$, namely
\begin{equation}
   H^n(Y,S^\Lambda)=\frac{\{\rel{\Theta}\in\Omega^n(Y,S^\Lambda)|\dd\rel{\Theta}=0\}}
     {\dd\left(\Omega^{n-1}(Y,S^\Lambda)\right)} \ .
\end{equation}
Analogously to the duality of the cohomology group $H^n(Y)$ to the homology group $H_n(Y)$, each relative cohomology group $H^n(Y,S^\Lambda)$ has a dual description in terms of a relative homology group $H_n(Y,S^\Lambda)$. The elements of $H_n(Y,S^\Lambda)$ are $n$-cycles $\rel{\Gamma}$, which are not necessarily closed anymore, but may have boundaries $\partial\rel{\Gamma}$ in $\iota(S^\Lambda)$. Furthermore the pairing of a relative $n$-cycle with a relative $n$-form is given by the integral
\begin{equation} \label{eq:rpair}
   \langle\rel{\Gamma},\rel{\Theta}\rangle\:=\:\int_{\rel{\Gamma}} \rel{\Theta} \ .
\end{equation}
Note that this bilinear product is independent of the choice of representative of the relative cohomology element $\rel{\Theta}$ and the relative homology element $\rel{\Gamma}$.

In the following we concentrate on the relative cohomology group $H^3(Y,S^\Lambda)$, which is relevant for the moduli space $\mathcal{M}_{\mathcal{N}=1}$ discussed in chapter~\ref{ch:geom}. In order to get a better handle on this space of relative three-forms one constructs from the short exact sequence \eqref{eq:RFexact} the long exact sequence
\begin{equation}
   \ldots\rightarrow H^2(Y)\xrightarrow{\iota^*} H^2(S^\Lambda)\xrightarrow{\delta}
   H^3(Y,S^\Lambda)\rightarrow H^3(Y)\xrightarrow{\iota^*}H^3(S^\Lambda)\rightarrow\ldots \ .
\end{equation} 
From this sequence one extracts
\begin{equation} \label{eq:RC}
   H^3(Y,S^\Lambda)\cong\ker\left(H^3(Y)\xrightarrow{\iota^*}H^3(S^\Lambda)\right) 
     \oplus\coker\left(H^2(Y)\xrightarrow{\iota^*}H^2(S^\Lambda)\right) \ .
\end{equation}
Thus we can think of a representative $\rel{\Theta}$ of $H^3(Y,S^\Lambda)$ as a pair of a three-form $\Theta_Y$ of $Y$ and a two-form $\theta_{S^\Lambda}$ of $S^\Lambda$, where $\Theta_Y$ is in the kernel of $\iota^*$ and $\theta_{S^\Lambda}$ is in the cokernel of $\iota^*$.

Recall that in the context of Calabi-Yau orientifolds with a holomorphic involution~$\sigma$ the complex structure deformations~$z^{\tilde a}$ and the D7-brane matter fields~$\dbs^A$ are expanded into odd forms with respect to the involution~$\sigma$. Therefore the appropriate relative cohomology space is $H^3_-(Y,S^\Lambda)$, that is to say the elements are also odd relative forms with respect to the involution~$\sigma$. In this case the relation \eqref{eq:RC} becomes
\begin{equation} \label{eq:RC2}
   H^3_-(Y,S^\Lambda)\cong \widetilde H^3_-(Y) \oplus \widetilde H^2_-(S^\Lambda) \ ,
\end{equation}
with 
\begin{equation}
\begin{split}
   \widetilde H^3_-(Y)&=\ker\left(H^3_-(Y)\xrightarrow{\iota^*}H^3_-(S^\Lambda)\right) \ , \\
   \widetilde H^2_-(S^\Lambda)&=\coker\left(H^2_-(Y)\xrightarrow{\iota^*}H^2_-(S^\Lambda)\right) \ .
\end{split}
\end{equation}


\cleardoublepage
\addcontentsline{toc}{chapter}{Bibliography}  
\bibliographystyle{utphys}
\bibliography{phd}

\end{document}